\documentclass[draft,tightenlines,nofootinbib,preprint,aps,eqsecnum,amsmath,amssymb]{revtex4}

\newcommand{\beq}{\begin{equation}}
\newcommand{\eeq}{\end{equation}}
\newcommand{\bea}{\begin{eqnarray}}
\newcommand{\eea}{\end{eqnarray}}
\newcommand{\cir}{{\buildrel \circ \over =}}
\newcommand{\sgn}{\mbox{\boldmath $\epsilon$}}

\newcommand{\on}{\stackrel{\circ}{=}}
\newcommand{\byd}{\stackrel{def}{=}}

\newcommand{\h}{h}

\begin{document}

\baselineskip 18pt

\title{Charged Particles and the Electro-Magnetic Field in Non-Inertial Frames
of Minkowski Spacetime.}

\medskip

\author{David Alba}

\affiliation{
Sezione INFN di Firenze\\Polo Scientifico, via Sansone 1\\
 50019 Sesto Fiorentino, Italy\\
 E-mail ALBA@FI.INFN.IT}

\author{Luca Lusanna}

\affiliation{ Sezione INFN di Firenze\\ Polo Scientifico\\ Via Sansone 1\\
50019 Sesto Fiorentino (FI), Italy\\ Phone: 0039-055-4572334\\
FAX: 0039-055-4572364\\ E-mail: lusanna@fi.infn.it}

\begin{abstract}

By using the 3+1 point of view and parametrized Minkowski theories
we develop the theory of {\it non-inertial} frames in Minkowski
space-time.  The transition from a non-inertial frame to another one
is a gauge transformation connecting the respective notions of
instantaneous 3-space (clock synchronization convention) and of the
3-coordinates inside them. As a particular case we get the extension
of the inertial rest-frame instant form of dynamics to the
non-inertial rest-frame one. We show that every isolated system can
be described as an external decoupled non-covariant canonical center
of mass (described by frozen Jacobi data) carrying a pole-dipole
structure: the invariant mass and an effective spin. Moreover we
identify the constraints eliminating the internal 3-center of mass
inside the instantaneous 3-spaces.

In the case of the isolated system of positive-energy scalar
particles with Grassmann-valued electric charges plus the
electro-magnetic field we obtain both Maxwell equations and their
Hamiltonian description in non-inertial frames. Then by means of a
non-covariant decomposition we define the non-inertial radiation
gauge and we find the form of the non-covariant Coulomb potential.
We identify the coordinate-dependent relativistic inertial
potentials and we show that they have the correct Newtonian limit.

Then we study properties of Maxwell equations in non-inertial frames
like the wrap-up effect and the Faraday rotation in astrophysics.
Also the 3+1 description without coordinate-singularities of the
rotating disk and the Sagnac effect are given, with added  comments
on pulsar magnetosphere and on a relativistic extension of the
Earth-fixed coordinate system.

\end{abstract}

\maketitle

\vfill\eject

\section{Introduction}

As a consequence of many years of research devoted to try to
establish a consistent formulation of relativistic mechanics, we
have now a description of every isolated system (particles, strings,
fields, fluids), admitting a Lagrangian formulation, in arbitrary
global inertial or non-inertial frames in Minkowski space-time by
means of {\it parametrized Minkowski theories} \cite{1,2,3,4} (see
Ref.\cite{5} for a review). They allow one to get a Hamiltonian
description of the relativistic isolated systems, in which the
transition from a non-inertial (or inertial) frame to another one is
a gauge transformation generated by suitable first-class Dirac
constraints. Therefore, all the admissible conventions for clock
synchronization, identifying the instantaneous 3-spaces containing
the system and allowing a formulation of the Cauchy problem for the
equations of the fields present in the system, turn out to be gauge
equivalent.\medskip

The only known way to have a global description of non-inertial
frames is to choose an arbitrary time-like observer and a 3+1
splitting of Minkowski space-time, namely a foliation with
space-like hyper-surfaces (namely an arbitrary clock synchronization
convention) with a set of 4-coordinates (observer-dependent
Lorentz-scalar radar 4-coordinates $\sigma^A = (\tau ; \sigma^r)$,
$A = \{\tau , r\}$) adapted to the foliation and having the observer
as origin of the 3-coordinates $\sigma^r$ on each instantaneous
3-space $\Sigma_{\tau}$. The time parameter $\tau$, labeling the
leaves of the foliation, is an arbitrary monotonically increasing
function of the proper time of the observer. Each such foliation
defines a {\it global non-inertial frame centered on the given
observer} if it satisfies the M$\o$ller admissibility conditions
\cite{6}, \cite{3,5}, and if the instantaneous (in general
non-Euclidean) 3-spaces, described by the functions $z^{\mu}(\tau
,\sigma^r)$ giving their embedding in a reference inertial frame in
Minkowski space-time, tend to space-like hyper-planes at spatial
infinity \cite{3}. The 4-metric $g_{AB}(\tau ,\sigma^r) =
z^{\mu}_A(\tau ,\sigma^r)\, \eta_{\mu\nu}\, z^{\nu}_B(\tau
,\sigma^r)$, $z^{\mu}_A(\tau ,\sigma^r) = {{\partial\, z^{\mu}(\tau
,\sigma^r)}\over {\partial\, \sigma^A}}$, in the non-inertial frame
is a function of the embedding obtained from the flat metric
$\eta_{\mu\nu}$ in  inertial Cartesian 4-coordinates $x^{\mu}$ by
means of a general coordinate transformation $x^{\mu}\, \mapsto\,
\sigma^A = (\tau ; \sigma^r)$ with inverse transformation
$\sigma^A\, \mapsto\, x^{\mu} = z^{\mu}(\tau ,\sigma^r)$.
\medskip

If we couple the Lagrangian of the isolated system to an external
gravitational field, we replace the external gravitational 4-metric
with the embedding-dependent 4-metric of a non-inertial frame and we
re-express the components of the isolated system in adapted radar
4-coordinates knowing the instantaneous 3-spaces \footnote{For a
scalar field $\tilde \phi (x)$ we get $\phi (\tau ,\sigma^r) =
\tilde \phi (z(\tau ,\sigma^r))$. For the electro-magnetic potential
${\tilde A}_{\mu}(x)$ and field strength ${\tilde F}_{\mu\nu}(x)$ we
get the Lorentz-scalar fields $A_A(\tau ,\sigma^r) = {\tilde
A}_{\mu}(z(\tau ,\sigma^r))\, z^{\mu}_A(\tau ,\sigma^r)$,
$F_{AB}(\tau ,\sigma^r) = (\partial_A\, A_B -
\partial_B\, A_A)(\tau ,\sigma^r) = {\tilde F}_{\mu\nu}(z(\tau
,\sigma^r)\, z^{\mu}_A(\tau ,\sigma^r)\, z^{\nu}_B(\tau ,\sigma^r)$.
Differently from $\tilde \phi (x)$ and ${\tilde A}_{\mu}(x)$, the
fields $\phi (\tau ,\sigma^r)$ and $A_A(\tau ,\sigma^r)$ {\it know}
the whole instantaneous 3-space $\Sigma_{\tau}$. Scalar particles
are described with Lorentz-scalar 3-coordinates ${\vec \eta}_i(\tau
)$ in $\Sigma_{\tau}$ defined by $x^{\mu}_i(\tau ) = z^{\mu}(\tau
,{\vec \eta}_i(\tau ))$, $i=1,..,N$, i.e. by the intersection of
their world-lines $x^{\mu}_i(\tau )$ (parametrized not with their
proper time, but with the observer's one) with $\Sigma_{\tau}$. As a
consequence, each particle must have a well defined sign of the
energy. Both the world-lines $x^{\mu}_i(\tau )$ and the associated
4-momenta $p_i^{\mu}(\tau )$, satisfying $p_i^2(\tau ) = \sgn\,
m_i^2$ even in presence of interactions, are derived quantities. },
we get the Lagrangian of the parametrized Minkowski theory for the
given isolated system. It is a function of the matter and fields of
the isolated system (now described as Lorentz-scalar quantities in a
non-inertial frame) and of the embedding $z^{\mu}(\tau ,\sigma^r)$
of the instantaneous 3-spaces of the non-inertial frame in Minkowski
space-time. The main property of the action functional associated
with these Lagrangians is the invariance \cite{1,3,5} under
frame-preserving diffeomorphisms \footnote{Schmutzer and Plebanski
\cite{7} were the only ones emphasizing the relevance of this
subgroup of diffeomorphisms in their attempt to obtain the theory of
non-inertial frames in Minkowski space-time as a limit from
Einstein's general relativity.} : this implies that the embeddings
are {\it gauge variables}, so that all M$\o$ller-admissible clock
synchronization conventions (i.e. any definition of instantaneous
3-spaces in space-times with Lorentz signature) are gauge
equivalent.\bigskip

Inertial frames are the special class of frames connected by the
transformations of the Poincare' group (the relativity principle)
selected by the law of inertia. For every configuration of an
isolated system there is a special inertial frame intrinsically
selected by the system itself, the {\it rest frame}, whose
instantaneous 3-spaces (the Wigner 3-spaces with Wigner covariance)
are orthogonal to the conserved 4-momentum of the configuration.
This gives rise to the {\it rest-frame instant form of the
dynamics}. In Ref. \cite{8} there is a full account of the
rest-frame instant form for arbitrary isolated systems, with special
emphasis on the system of "N charged positive-energy scalar
particles with mutual Coulomb interaction plus the transverse
electro-magnetic field of the radiation gauge" \cite{9}. The
particles have Grassmann-valued electric charges (each replaced by a
two-level system, charge $+e$ - charge $-e$, described by a Clifford
algebra, after quantization)\medskip

a) to make a ultraviolet regularization of the Coulomb
self-energies;\hfill

b) to make a infrared regularization killing the emission of soft
photons and loops;\hfill

c) to allow us to have the Lienard-Wiechert transverse potential and
electric field expressible as functions only of the 3-positions and
3-momenta of the particles, independently from the chosen Green
function (retarded, advanced, symmetric, ..).

\medskip

This allows us to have a description of the one-photon exchange
diagram by means of a potential in the framework of a well defined
Cauchy problem for Maxwell equations.\bigskip

In the rest-frame instant form there are two realizations of the
Poincare' algebra:\hfill\medskip

1) An {\it external} one, in which the isolated system is simulated
by means of a {\it decoupled point particle carrying a pole-dipole
structure}: the invariant mass $M$ and the rest spin ${\vec {\bar
S}}$ of the isolated system. This decoupled point particle is
described by the canonical frozen Jacobi data of the non-covariant
external relativistic 3-center of mass: a non-covariant variable
$\vec z = Mc\, {\vec x}_{NW}(0)$ (${\vec x}_{NW}(0)$ is the Cauchy
datum of the Newton-Wigner 3-position ${\vec x}_{NW}(\tau )$) and an
adimensional 3-velocity $\vec h = \vec P/ Mc$, $\{ z^i, h^j \} =
\delta^{ij}$. This universal (i.e. independent from the isolated
system) breaking of manifest Lorentz covariance is irrelevant since
the 3-center of mass is decoupled from the internal dynamics. Since
the Poincare' generators are global quantities, the relativistic
center of mass (a known function of such generators) is a {\it
global quantity} not locally determinable (see Ref.\cite{8} for the
non-local aspects of the Newton-Wigner position). The non-covariant
canonical external 4-center of mass (or center of spin) ${\tilde
x}^{\mu}(\tau ) = ({\tilde x}^o(\tau ); {\vec {\tilde x}}(\tau ))$,
the covariant non-canonical external Fokker-Pryce 4-center of
inertia $Y^{\mu}(\tau ) = ({\tilde x}^o(\tau ); \vec Y(\tau ))$ and
the non-covariant non-canonical external M$\o$ller 4-center of
energy $R^{\mu}(\tau ) = ({\tilde x}^o(\tau ); \vec R(\tau ))$ are
known functions of $\tau$, $\vec z$, $\vec h$, $M$, ${\vec {\bar
S}}$ given in Ref.\cite{8}. All these collective variables have the
same constant 4-velocity: ${\dot Y}^{\mu}(\tau ) = {\dot {\tilde
x}}^{\mu}(\tau ) = {\dot R}^{\mu}(\tau ) = P^{\mu}/Mc = h^{\mu}$.

\bigskip

The embedding identifying the Wigner 3-spaces is ($\tau = c T$ is
the Lorentz-scalar rest time)

\beq
 z^{\mu}_W(\tau ,\sigma^u) = Y^{\mu}(\tau ) +
\epsilon^{\mu}_r(\vec h)\, \sigma^r,
 \label{1.1}
 \eeq

\noindent where $Y^{\mu}(\tau )$ is the covariant non-canonical
Fokker-Pryce external 4-center of inertia (a known function of
$\tau$, $\vec z$, $\vec h$, $M$ and ${\vec {\bar S}}$) and the 3
space-like 4-vectors $\epsilon^{\mu}_r(\vec h)$ are determined by
the standard Wigner boost $L^{\mu}{}_{\nu}(P, {\buildrel \circ \over
P})$ for time-like orbits sending the rest form ${\buildrel \circ
\over P}^{\mu} = Mc\, (1; \vec 0)$ of the total momentum into
$P^{\mu} = Mc\, u^{\mu}(P) = Mc\, \epsilon_{\tau}^{\mu}(\vec h) =
Mc\, (\sqrt{1 + {\vec h}^2}; \vec h) = Mc\, h^{\mu}$ (we collect
here the various notations used in previous papers), i.e.
$\epsilon^{\mu}_A(\vec h) = L^{\mu}{}_{\nu = A}(P, {\buildrel \circ
\over P})$. We have $\epsilon^o_{\tau}(\vec h) = \sqrt{1 + {\vec
h}^2}$, $\epsilon^i_{\tau}(\vec h) = h^i$, $\epsilon^o_r(\vec h) = -
\sgn\, h_r$, $\epsilon^i_r(\vec h) = \delta^i_r - \sgn\, {{h^i\,
h_r}\over {1 + \sqrt{1 + {\vec h}^2} }}$ (see the next Section for
the conventions on the 4-metric).
\bigskip

2) A {\it unfaithful internal} one inside the Wigner 3-spaces, whose
generators are determined by the energy-momentum tensor, obtained
from the Lagrangian of the parametrized Minkowski theory associated
with the given isolated system. The only non-vanishing generators
are $M$ and ${\vec {\bar S}}$. The vanishing of the internal
3-momentum is the {\it rest-frame condition}, while the vanishing of
the internal (interaction-dependent) Lorentz boosts {\it eliminates
the internal 3-center of mass} (this avoids a double counting of the
center of mass). As a consequence, the dynamics inside the
instantaneous Wigner 3-spaces is described {\it only by
Wigner-covariant relative variable and momenta} (${\vec \rho}_a(\tau
)$, ${\vec \pi}_a(\tau )$, $a=1,..,N-1$, for particles). The
invariant mass $M$ is the Hamiltonian for the internal Hamilton
equations. It is possible to make an orbit reconstruction \cite{4}
for the particles in the form ${\vec \eta}_i(\tau ) = {\vec
f}_i({\vec \rho}_a(\tau ), {\vec \pi}_a(\tau ))$ and to determine
the world-lines  \footnote{They turn out to be {\it covariant
non-canonical predictive coordinates}: $\{x^{\mu}_i(\tau ),
x^{\nu}_j(\tau )  \} \not= 0$ for all $i$ and $j$, $\mu$ and $\nu$.
Let us remark that this does not imply a breaking of microcausality,
which is preserved at the level of the 3-coordinates ${\vec
\eta}_i(\tau )$.},

\beq
 x^{\mu}_i(\tau ) = z^{\mu}_W(\tau, {\vec
\eta}_i(\tau )) = Y^{\mu}(\tau ) + \epsilon^{\mu}_r(\vec h)\,
f^r_i({\vec \rho}_a(\tau ), {\vec \pi}_a(\tau )).
 \label{1.2}
 \eeq

\bigskip

In this paper we study in detail the properties of global admissible
non-inertial frames in Minkowski space-time, generalizing the
notions defined in the inertial rest-frame instant form of dynamics.
We show that also in non-inertial frames every isolated system can
be described as an external decoupled non-covariant canonical center
of mass (described by frozen Jacobi data) carrying a pole-dipole
structure: the invariant mass and an effective spin. Moreover,
following the same methods developed for the inertial rest frame, we
identify the constraints eliminating the internal 3-center of mass
inside the instantaneous 3-spaces.\medskip

In the admissible non-inertial frames  the instantaneous 3-spaces
are orthogonal to a given fixed 4-vector $l^{\mu}_{(\infty )}$ at
spatial infinity \footnote{A preliminary description of particles
and of their quantization in a class of such frames was given in
Ref.\cite{10}. There we introduced an auxiliary decoupled scalar
particle whose 4-momentum coincides with $l^{\mu}_{(\infty )}$. Here
we will avoid to use this method.}.\medskip

Then we will restrict the description to the special family of
non-inertial frames, in which the instantaneous 3-spaces tend to
Wigner 3-spaces, orthogonal to the conserved 4-momentum of the
isolated system, at spatial infinity (i.e. $l^{\mu}_{(\infty )} =
h^{\mu} = P^{\mu}/Mc$): they are the {\it non-inertial rest frames},
a non-inertial extension of the inertial ones. This will allow us to
define the non-inertial rest-frame instant form of dynamics. The
non-inertial rest frame are the only ones allowed by the equivalence
principle in the treatment of canonical metric and tetrad gravity in
asymptotically flat and globally hyperbolic space-times without
super-translations as shown in Refs. \cite{5,11}.\bigskip

Even if in a non-covariant way, which is however consistent with the
coordinate-dependence of the inertial effects, we will give a
unified special relativistic description of many properties of
isolated systems in accelerated frames, which are scattered in the
literature and treated without a global interpretative framework.
\bigskip

Then, as in Ref.\cite{8}, we consider the description of the
isolated system of positive-energy scalar particles with
Grassmann-valued electric charges plus the electro-magnetic field as
a parametrized Minkowski theory. As a consequence we obtain both
Maxwell equations and their Hamiltonian description in non-inertial
frames.\medskip

By means of a non-covariant decomposition we define the {\it
non-inertial non-covariant radiation gauge}: this allows to
visualize the non-inertial dynamics of transverse electro-magnetic
fields, the electro-magnetic Dirac observables. We find the
modification of the Coulomb potential in a non-inertial frame: its
non-covariance is due to same type of coordinate-dependence present
in the relativistic inertial potentials, which are explicitly
identified for the first time and shown to have the correct
Newtonian limit. The final Dirac Hamiltonian will contain not only
the invariant mass $Mc$ but also the modifications induced by the
potentials associated with the inertial effects present in the given
non-inertial frame.

\bigskip

Then we study properties of Maxwell equations in non-inertial frames
like the wrap-up effect, the Faraday rotation in astrophysics,
pulsar magnetosphere ........... Also the 3+1 description of the
rotating disk and the Sagnac effect are given.

\bigskip

In Section II we review the admissible 3+1 splittings of Minkowski
space-time and the properties of the associated global non-inertial
frames (Subsection A), we compare them with the accelerated
coordinate systems associated with the 1+3 point of view (Subsection
B) and we define the non-covariant notations for the
electro-magnetic field in non-inertial frames (Subsection
C).\medskip

In Section III we study the description of the isolated system
"charged scalar positive-energy particles plus the electro-magnetic
field" in the framework of parametrized Minkowski theories. In
particular we show that in non-inertial frames and also in inertial
frames with non-Cartesian coordinates there is no true conservation
law for the energy-momentum tensor: like in general relativity one
could introduce a coordinate-dependent energy-momentum pseudo-tensor
describing the contribution of the foliation associated with the
admissible 3+1 splitting of Minkowski space-time. However, reverting
to inertial frames, it is possible to find the conserved (Poincare'
4-vector) 4-momentum of the isolated system.
\medskip

In Section IV we give the Hamiltonian description and the Hamilton
equations of the isolated system "charged scalar positive-energy
particles plus the electro-magnetic field" in admissible
non-inertial frames (Subsection A). Then we introduce the
non-covariant radiation gauge for the electro-magnetic field and we
find both the inertial forces and the non-inertial expression of the
coulomb potential (Subsection B). Finally we evaluate the
non-relativistic limit recovering the Newtonian apparent inertial
forces (Subsection C).\medskip

In Section V we review the determination of the internal Poincare'
generators and of the constraints eliminating the internal 3-center
of mass in the inertial rest frames (Subsection A). Then we show how
these results are modified  in the special family of the
non-inertial rest frames (Subsections B and C) and in arbitrary
admissible non-inertial frames (Subsection D) .\medskip

In Section VI we give the 3+1 point of view in admissible nearly
rigidly rotating frames of the Wrap Up effect, of the Sagnac effect
and of the inertial Faraday rotation by studying electro-magnetic
wave solutions of the non-inertial Maxwell equations.
\medskip

In the Conclusions we give an overview of the results obtained in
this paper and we identify the still open problems about
electro-magnetism in non-inertial frames.
\medskip

In Appendix A there is a review of the rotating disk and of the
Sagnac effect in the 1+3 point of view followed by their description
in the framework of the 3+1 point of view (Subsection A1) and by a
discussion on the ITRS rotating 3-coordinates fixed on the Earth
surface (Subsection A2).\medskip

In Appendix B there is the expression of the Landau-Lifschitz
non-inertial electro-magnetic fields in the 3+1 point of
view.\medskip

In Appendix C there is a comparison of the covariant and
non-covariant decompositions of the electro-magnetic field in
non-inertial frames and the definition of the non-covariant
radiation gauge.

\vfill\eject

\section{Admissible 3+1 Splittings of Minkowski Space-Time and Notations}

We use the signature convention $\eta_{\mu\nu} = \sgn\, (+---)$,
$\sgn = \pm 1$, for the flat Minkowski metric ($\sgn = +1$ is the
particle physics convention, while $\sgn = - 1$ is the one of
general relativity), since it has been used in Refs.\cite{11} for
canonical gravity. Since in Ref. \cite{8} the convention $\sgn = +
1$ was used, in this Section we also introduce the notations needed
for the treatment of the electro-magnetic field in non-inertial
frames.\bigskip

\subsection{Admissible 3+1 Splittings of Minkowski Space-Time}

Let us consider an admissible 3+1 splitting of Minkowski space-time,
whose instantaneous 3-spaces $\Sigma_{\tau}$ are identified by the
embedding $z^{\mu}(\tau ,\sigma^r)$. The radar 4-coordinates
$\sigma^A = (\tau ;\sigma^r)$ are adapted to an arbitrary time-like
observer with world-line $x^{\mu}(\tau )$ in the reference inertial
frame, chosen as the origin of the curvilinear 3-coordinates
$\sigma^r$ on each $\Sigma_{\tau}$. The Lorentz-scalar time $\tau$,
with dimensions $[\tau ] = [c\, t] = [l]$, is a monotonically
increasing function of the proper time of the observer. Therefore,
we can put the embeddings in the following form

\bea
 z^{\mu}(\tau ,\sigma^u ) &=& x^{\mu}(\tau ) + F^{\mu}(\tau ,
 \sigma^u ) = x^{\mu}_o + \epsilon^{\mu}_A\, \Big[f^A(\tau ) + F^A(\tau
 ,\sigma^u )\Big],\qquad F^{\mu}(\tau ,\vec o) = 0,\nonumber \\
 &&{}\nonumber \\
 x^{\mu}(\tau ) &=& x^{\mu}_o + \epsilon^{\mu}_A\, f^A(\tau ).
 \label{2.1}
 \eea

\noindent At spatial infinity $z^{\mu}(\tau ,\sigma^r)$ must tend in
a direction-independent way to a space-like hyper-plane with unit
time-like normal $l^{\mu}_{(\infty )} = \epsilon^{\mu}_{\tau}$: this
implies $F^{\mu}(\tau ,\sigma^s)\, \rightarrow\,
\epsilon^{\mu}_{(\infty )\, r}\, \sigma^r$ with the 3 space-like
4-vectors $\epsilon^{\mu}_{(\infty )\, r} = \epsilon^{\mu}_r$
orthogonal to $l^{\mu}_{(\infty )}$. The asymptotic orthonormal
tetrads $\epsilon^{\mu}_A$ are associated to  asymptotic inertial
observers and   satisfy $\epsilon^{\mu}_A\, \eta_{\mu\nu}\,
\epsilon^{\nu}_B = \eta_{AB}$. Let us remark that the natural
notation for the asymptotic tetrads would be $\epsilon^{\mu}_{(A)}$.
However, for the sake of simplicity we shall use the notation
$\epsilon^{\mu}_A$ for $\delta_A^{(B)}\,
\epsilon^{\mu}_{(B)}$.\medskip

The time-like observer $x^{\mu}(\tau )$, origin of the 3-coordinates
on the instantaneous 3-spaces $\Sigma_{\tau}$, has the following
unit 4-velocity and 4-acceleration (we use the notation ${\dot
x}^{\mu}(\tau ) = {{d\, x^{\mu}(\tau )}\over {d\tau}}$; it must be
$\sgn\, {\dot x}^2(\tau ) > 0$)

\begin{eqnarray*}
 u^{\mu}(\tau ) &=& {{{\dot x}^{\mu}(\tau )}\over {\sqrt{\sgn\, {\dot x}^2(\tau )}}}
 = \epsilon^{\mu}_A\, u^A(\tau ),\qquad u^2(\tau ) = \sgn,\nonumber \\
 &&{}\nonumber \\
 &&u^A(\tau ) = {{{\dot f}^A(\tau)}\over {\sqrt{\Big({\dot f}^{\tau}(\tau )\Big)^2 -
 \sum_u\, \Big({\dot f}^u(\tau )\Big)^2}}},\qquad
 \Big({\dot f}^{\tau}(\tau )\Big)^2\, >\, \sum_u\, \Big({\dot f}^u(\tau
 )\Big)^2,\end{eqnarray*}

 \bea
 a^{\mu}(\tau ) &=& {{d u^{\mu}(\tau )}\over {d\tau}} = \epsilon^{\mu}_A\, a^A(\tau
 ),\qquad a_{\mu}(\tau )\, u^{\mu}(\tau ) = 0,\nonumber \\
 &&{}\nonumber \\
 &&a^A(\tau) = {{{\ddot f}^A(\tau )\, \Big(\Big({\dot f}^{\tau}(\tau )\Big)^2 -
 \sum_u\, \Big({\dot f}^u(\tau )\Big)^2\Big) - {\dot f}^A(\tau )\,
 \Big({\dot f}^{\tau}(\tau )\, {\ddot f}^{\tau}(\tau ) - \sum_u\,
 {\dot f}^u(\tau )\, {\ddot f}^u(\tau )\Big)}\over {\Big(\Big({\dot f}^{\tau}(\tau )\Big)^2 -
 \sum_u\, \Big({\dot f}^u(\tau )\Big)^2\Big)^{3/2}}}.\nonumber \\
 &&{}
 \label{2.2}
 \eea

As a consequence we can write $u^{\mu}(\tau ) =
L^{\mu}{}_{\nu}(u(\tau ), {\buildrel \circ \over u})\, {\buildrel
\circ \over u}^{\nu}$, ${\buildrel \circ \over u}^{\mu} = \sgn\, (1;
\vec 0)$, by using the standard Wigner boost for time-like
4-vectors.
\bigskip

Eqs.(\ref{2.1}) imply

\bea
 z^{\mu}_{\tau}(\tau ,\sigma^u ) &=& \partial_{\tau}\, z^{\mu}(\tau
 ,\sigma^u) = {\dot x}^{\mu}(\tau ) + \partial_{\tau}\,
 F^{\mu}(\tau ,\sigma^u ) = \epsilon^{\mu}_A\, \Big({\dot f}^A(\tau ) +
 \partial_{\tau}\, F^A(\tau ,\sigma^u )\Big) =\nonumber \\
 &=& (1 + n(\tau ,\sigma^u ))\, l^{\mu}(\tau ,\sigma^u ) + h^{rs}(\tau
 ,\sigma^u )\, n_r(\tau ,\sigma^u )\, z^{\mu}_s(\tau ,\sigma^u ),
 \nonumber \\
  &&{}\nonumber \\
 z^{\mu}_r(\tau ,\sigma^u ) &=& \partial_r\, z^{\mu}(\tau ,\sigma^u) =
 \partial_r\, F^{\mu}(\tau , \sigma^u ) = \epsilon^{\mu}_A\,
 \partial_r\, F^A(\tau ,\sigma^u).
 \label{2.3}
 \eea

\bigskip

While the 3 independent space-like 4-vectors $z^{\mu}_r(\tau
,\sigma^u)$ are tangent to $\Sigma_{\tau}$,  the time-like 4-vector
$z^{\mu}_{\tau}(\tau ,\sigma^u)$ has been decomposed on them and on
the unit normal $l^{\mu}(\tau ,\sigma^u)$, $l^2(\tau ,\sigma^u) =
\sgn$, to $\Sigma_{\tau}$ ($l_{\mu}(\tau ,\sigma^u)\, z^{\mu}_r(\tau
,\sigma^u) = 0$). This decomposition defines the {\it lapse and
shift functions} $N(\tau ,\sigma^u) = 1 + n(\tau ,\sigma^u)
> 0$ and $N^r(\tau ,\sigma^u) = n^r(\tau ,\sigma^u)$ (we use the
notation of Ref.\cite{11}). At spatial infinity we have:
$l^{\mu}(\tau ,\sigma^u)\, \rightarrow \, l^{\mu}_{(\infty )} =
\epsilon^{\mu}_{\tau}$, $N(\tau ,\sigma^u)\, \rightarrow\, 1$
($n(\tau ,\sigma^u)\, \rightarrow\, 0$), $n^r(\tau ,\sigma^u)\,
\rightarrow 0$.\bigskip

The 4-metric induced by the 3+1 splitting is $g_{AB}(\tau ,\sigma^u)
= z^{\mu}_A(\tau ,\sigma^u)\, \eta_{\mu\nu}\, z^{\nu}_B(\tau
,\sigma^u)$ and we have

\begin{eqnarray*}
 g_{\tau\tau}(\tau ,\sigma^u) &=& \Big[z^{\mu}_{\tau}\, \eta_{\mu\nu}\,
 z^{\nu}_{\tau}\Big](\tau ,\sigma^u) = \nonumber \\
 &=& \sgn\, \Big[ \Big({\dot f}^{\tau}(\tau )+ \partial_{\tau}\,
 F^{\tau}(\tau ,\sigma^v )\Big)^2 - \sum_u\, \Big({\dot f}^u(\tau )+ \partial_{\tau}\,
 F^u(\tau ,\sigma^v )\Big)^2\Big] =\nonumber \\
 &=&\sgn\, \Big[\Big(1 + n(\tau ,\sigma^v )\Big)^2 - h^{rs}(\tau
 ,\sigma^v)\, n_r(\tau ,\sigma^v )\, n_s(\tau ,\sigma^v )\Big],
 \end{eqnarray*}

\bea
 g_{\tau r}(\tau ,\sigma^v) &=& \Big[z^{\mu}_{\tau}\, \eta_{\mu\nu}\,
 z^{\nu}_r\Big](\tau ,\sigma^v) =\nonumber \\
 &=& - \sgn\, \Big[\sum_u\, \Big({\dot f}^u(\tau ) + \partial_{\tau}\, F^u(\tau
 ,\sigma^v )\Big)\, \partial_r\, F^u(\tau ,\sigma^v ) -\nonumber \\
 &-& \Big({\dot f}^{\tau}(\tau ) + \partial_{\tau}\, F^{\tau}(\tau ,\sigma^v )\Big)\,
 \partial_r\, F^{\tau}(\tau ,\sigma^v )\Big] =\nonumber \\
 &=&  - \sgn\, n_r(\tau ,\sigma^v) = g_{rs}(\tau ,\sigma^v)\, n^s(\tau ,\sigma^v) =
 - \sgn\, h_{rs}(\tau ,\sigma^v)\, n^s(\tau ,\sigma^v),\nonumber \\
 &&{}\nonumber \\
 g_{rs}(\tau ,\sigma^v) &=& \Big[z^{\mu}_r\, \eta_{\mu\nu}\, z^{\nu}_s\Big](\tau
 ,\sigma^v) =\nonumber \\
 &=& - \sgn\, \Big[\sum_u\, \partial_r\, F^u(\tau ,\sigma^v )\, \partial_s\,
 F^u(\tau ,\sigma^v ) - \partial_r\, F^{\tau}(\tau ,\sigma^v )\, \partial_s\,
 F^{\tau}(\tau ,\sigma^v )\Big] =\nonumber \\
 &=&  - \sgn\, h_{rs}(\tau ,\sigma^v).
 \label{2.4}
 \eea

\noindent While the 3-metric $g_{rs}$ in $\Sigma_{\tau}$ and its
inverse $\gamma^{rs}$ ($\gamma^{ru}\, g_{us} = \delta^r_s$) have
signature $\sgn\, (---)$, the 3-metric $h_{rs}$ and its inverse
$h^{rs} = - \sgn\, \gamma^{rs}$ ($h^{ru}\, h_{us} = \delta^r_s$)
have signature $(+++)$.\medskip

For the inverse 4-metric $g^{AB}$ ($g^{AC}\, g_{CB} = \delta^A_B$)
we have

\bea
 g^{\tau\tau} &=& {{\sgn}\over {(1 + n)^2}},\qquad g^{\tau\tau}\, g^{rs}
 - g^{\tau r}\, g^{\tau s} = - {{h^{rs}}\over {(1 + n)^2}}, \nonumber \\
 g^{\tau r} &=& - \sgn\, {{n^r}\over {(1 + n)^2}},\qquad
 g^{rs} = - \sgn\, \Big(h^{rs} - {{n^r\, n^s}\over
 {(1 + n)^2}}\Big).
 \label{2.5}
 \eea

For the determinants we have

\bea
 &&\gamma = - \sgn\, det\, g_{rs} = det\, h_{rs} > 0,\qquad
 g = det\, g_{AB}  < 0,\quad \Rightarrow\,\, \sqrt{- g} = (1 +
 n)\, \sqrt{\gamma}.\nonumber \\
 &&{}
 \label{2.6}
 \eea

Finally the unit normal to the simultaneity surfaces $\Sigma_{\tau}$
has the expression

\begin{eqnarray*}
 l^{\mu}(\tau ,\sigma^u) &=& \Big[\eta^{\mu}{}_{\alpha\beta\gamma}\, z_1^{\alpha}\,
 z_2^{\beta}\, z_3^{\gamma}\Big](\tau ,\sigma^u) = \Big[{{1}\over {\sqrt{\gamma}}}\,
 \epsilon^{\mu}{}_{\alpha\beta\gamma}\, z_1^{\alpha}\, z_2^{\beta}\,
 z_3^{\gamma}\Big](\tau ,\sigma^u) =\nonumber \\
 &=& \epsilon^{\mu}_A\, l^A(\tau ,\sigma^v ) = \epsilon^{\mu}_A\, \eta^{AE}\,
 \Big({{\epsilon_{EBCD}}\over {\sqrt{\gamma}}}\, \partial_1\, F^B\,
 \partial_2\, F^C\, \partial_3\, F^D\Big)(\tau , \sigma^v ) =\nonumber \\
 &=& L^{\mu}{}_{\nu}(l(\tau ,\sigma^v ), {\buildrel \circ \over l})\,
 {\buildrel \circ \over l}^{\nu},\qquad {\buildrel \circ \over l}^{\mu}
 = \sgn\, (1; \vec 0),
  \end{eqnarray*}

 \bea
 l^2(\tau ,\sigma^u ) = \sgn, &&\Rightarrow  \Big(l^{\tau}(\tau
 ,\sigma^u )\Big)^2\, >\, \sum_u\, \Big(l^u(\tau ,\sigma^v
 )\Big)^2,\nonumber \\
 &&{}\nonumber \\
 \Rightarrow&& \eta_{\mu\nu} = \sgn\, \Big(l_{\mu}\, l_{\nu} -
 z_{r\mu}\, h^{rs}\, z_{s\nu}\Big)(\tau ,\sigma^v).
  \label{2.7}
  \eea
\medskip

The 3+1 splitting for which $l^{\mu}$ is constant, i.e. $\tau$- and
$\sigma^r$-independent, have the instantaneous 3-spaces
corresponding to parallel space-like hyper-planes: when the frame is
non-inertial these hyper-planes are not equally spaced due to linear
acceleration and/or have rotating 3-coordinates, so that they are
not Euclidean 3-spaces.

\medskip

The Wigner boost sending ${\buildrel \circ \over l}^{\mu}$ into
$l^{\mu}(\tau ,\sigma^u )$ ($\beta_l^i = - \sgn\, \beta_{l\, i}$)
has the following expression

\bea
 &&L^{\mu}{}_{\nu}(l(\tau ,\sigma^u ), {\buildrel \circ \over l}) =
  \begin{array}{|ll|} \gamma_l & \gamma_l\, \beta_l^i \\
 \gamma_l\, \beta_l^j & \delta^{ij} + (\gamma_l - 1)\, {{\beta_l^i\, \beta_l^j}\over
 {\sum_k\, (\beta_l^k)^2}} \end{array}(\tau ,\sigma^u ),\nonumber \\
 &&{}\nonumber \\
 &&l^{\mu}(\tau ,\sigma^u ) = L^{\mu}{}_o(l(\tau ,\sigma^u ),
 {\buildrel \circ \over l}) = \gamma_l(\tau ,\sigma^u )\, \Big(1;
 \beta_l^i(\tau ,\sigma^u )\Big) =
   \epsilon^{\mu}_A\, l^A(\tau ,\sigma^u ) {\buildrel {def}\over
 =} \epsilon^{\mu}_o(l(\tau ,\sigma^u )),\nonumber \\
 &&\epsilon^{\mu}_j(l(\tau ,\sigma^u )) {\buildrel {def}\over =}
 L^{\mu}{}_j(l(\tau ,\sigma^u ), {\buildrel \circ \over
 l}),\nonumber \\
 &&{}\nonumber \\
 &&\gamma_l = {1\over {\sqrt{1 - \sum_u\, (\beta_l^u)^2}}} = l^o
 = {1\over {\sqrt{\gamma}}}\, \epsilon^o_A\, \eta^{AE}\,
 \epsilon_{EBCD}\, \partial_1\, F^B\, \partial_2\, F^C\,
 \partial_3\, F^D,\nonumber \\
 &&\beta^i_l = \gamma_l^{-1}\, l^i = {{\epsilon^i_A\, \eta^{AE}\,
  \epsilon_{EBCD}\, \partial_1\, F^B\, \partial_2\, F^C\,
   \partial_3\, F^D}\over {\epsilon^o_A\, \eta^{AE}\,  \epsilon_{EBCD}\,
   \partial_1\, F^B\, \partial_2\, F^C\,  \partial_3\, F^D}}.
 \label{2.8}
 \eea

\noindent The orthonormal tetrads $\epsilon^{\mu}_A(l(\tau ,\sigma^u
)) = L^{\mu}{}_A(l(\tau ,\sigma^u ), {\buildrel \circ \over l})$, $\eta_{\mu\nu}\,
\epsilon^{\mu}_A(l(\tau ,\sigma^u ))\, \epsilon^{\mu}_B(l(\tau ,\sigma^u ))
= \eta_{AB}$, are the columns of the Wigner boost.
\medskip

The Wigner boosts $L^{\mu}{}_{\nu}(u(\tau ), {\buildrel \circ\over
u})$ has a similar parametrization in terms of parameters
$\beta^i_u(\tau )$.

\medskip

The M$\o$ller admissibility conditions \cite{6}, \cite{3}, implying
that the 3+1 splitting gives rise to a nice foliation of Minkowski
space-time with space-like leaves identifying the instantaneous
3-spaces $\Sigma_{\tau}$, are

\bea
 && \sgn\, g_{\tau\tau}(\tau ,\sigma^u) = \Big[(1 + n)^2 - h^{rs}\, n_r\,
n_s\Big](\tau ,\sigma^u)\,\, > 0,\qquad  \sgn\, g_{rr}(\tau
,\sigma^u ) = - h_{rr}(\tau
,\sigma^u) < 0,\nonumber \\
 &&{}\nonumber \\
 && \begin{array}{|ll|}
 g_{rr}(\tau ,\sigma^u ) & g_{rs}(\tau ,\sigma^u ) \\ g_{sr}(\tau
 ,\sigma^u ) & g_{ss}(\tau ,\sigma^u ) \end{array} = \begin{array}{|ll|}
 h_{rr}(\tau ,\sigma^u ) & h_{rs}(\tau ,\sigma^u ) \\ h_{sr}(\tau
 ,\sigma^u ) & h_{ss}(\tau ,\sigma^u ) \end{array}\, > 0, \nonumber \\
 &&{}\nonumber \\
 &&\sgn\, det\, [g_{rs}(\tau ,\sigma^u )] = - \gamma (\tau ,\sigma^u)\, <
 0,\qquad \Rightarrow det\, [g_{AB}(\tau ,\sigma^u )]\, <
 0.\nonumber \\
 &&{}
 \label{2.9}
 \eea
 \bigskip

\noindent They are restrictions on the functions $F^{\mu}(\tau
,\sigma^r)$ of Eqs.(\ref{2.1}). When they are satisfied,
Eqs.(\ref{2.1}) define a {\it global (in general non-rigid)
non-inertial frame}. While linear accelerations are not restricted
by Eqs.(\ref{2.9}), {\it rigid rotations are forbidden} \cite{3}.
The condition $\sgn\, g_{\tau\tau}(\tau ,\sigma^u) > 0$ implies that
in each point $\sigma^u$ the tangential velocity $\omega (\tau
,\sigma^u)\, r(\tau ,\sigma^u)$ is less than $c$: instead with
$\omega = \omega (\tau )$, like it happens in standard rotating
coordinate systems, we get $\sgn\, g_{\tau\tau}(\tau ,R^u) = 0$ at
the distance $R^u$ from the rotation axis where $\omega\, R = c$, so
that the time-like vector $z^{\mu}_{\tau}(\tau ,\sigma^u)$ would
become a null vector (the so-called {\it horizon problem} of the
rotating disk).\medskip

Since $1 + n(\tau ,\sigma^u ) > 0$ gives the proper time distance
from $\Sigma_{\tau}$ to $\Sigma_{\tau + d\tau}$ along the world-line
of the Eulerian observer through $(\tau ,\sigma^u )$ with tangent
vector $l^{\mu}(\tau ,\sigma^u )$, the condition $1 + n(\tau ,
\sigma^u ) > 0$ implies that $\Sigma_{\tau}$ and $\Sigma_{\tau +
d\tau}$ {\it intersect nowhere}. By continuity this implies that the
M$\o$ller-admissible 3+1 splittings are nice foliations with
space-like leaves tending to space-like hyper-planes at spatial
infinity in a direction-independent way.
\bigskip

Since the 3-metric $h_{rs}(\tau ,\sigma^u)$ is a real symmetric
matrix, it can be diagonalized with a rotation matrix
$V(\theta^i(\tau ,\sigma^u))$, $V^T = V^{-1}$ ($\theta^i(\tau
,\sigma^u)$ are Euler angles). Therefore, by using the notations of
Ref.\cite{12} for canonical gravity in the York canonical basis, we
can parametrize the 3-metric in the following form \footnote{ As
shown in Ref.\cite{12} the basic variables of tetrad gravity are not
the embedding $z^{\mu}(\tau ,\sigma^u)$ but tetrads
$E^{\mu}_{(\alpha )}(\tau ,\sigma^u)$, defined after an admissible
3+1 splitting of the space-time identifying the instantaneous
3-spaces $\Sigma_{\tau}$. The quantities $z^{\mu}_A(\tau ,\sigma^u)$
are now the {\it transition coefficients} from world components of
tensors to $\Sigma_{\tau}$-adapted components in radar coordinates
$\sigma^A = (\tau ,\sigma^u)$: $E^{\mu}_{(\alpha )} = z^{\mu}_A\,
E^A_{(\alpha )}$. The 4-metric tensor is defined by the associated
cotetrads: $g_{AB} = E^{(\alpha )}_A\, \eta_{(\alpha )(\beta )}\,
E_B^{(\beta )}$. The gauge variables of tetrad gravity in the York
canonical basis are six parameters of the Lorentz group acting on
the flat $(\alpha )$ indices of the tetrads $E^{\mu}_{(\alpha )}$,
the lapse ($1 + n$) and shift ($n_r$) functions, the Euler angles
$\theta^i$ and the momentum variable conjugate to $\phi^6 =
\gamma^{1/2}$, i.e. the trace ${}^3K$ of the extrinsic curvature of
the instantaneous 3-space $\Sigma_{\tau}$. The volume variable $\phi
= \gamma^{1/12}$ is determined by the super-hamiltonian constraint.
The momenta $\pi_i^{(\theta )}$, conjugate to $\theta^i$, are
determined by the super-momentum constraints. The symmetric 3-metric
$h_{rs} = - \sgn\, g_{rs}$ can be put in the form $h_{rs} = \sum_a\,
\lambda_a\, V_{ra}(\theta^i)\, V_{sa}(\theta^i)$, where  the
eigenvalues (assumed non degenerate) have the expression $\lambda_a
= \phi^4\, e^{2\, \sum_{\bar a}\, \gamma_{\bar aa}\, R_{\bar a}}$.
The two functions $R_{\bar a}$ describe the two physical degrees of
freedom of the gravitational field. A gauge fixing for $\theta^i$
and ${}^3K$ implies the determination of the lapse and shift
functions. \hfill\break
 Instead in non-inertial frames in Minkowski space-time, where gravity
is absent, all the functions ($n$, $n_r$, $\gamma = \phi^{12}$,
$\theta^i$, $R_{\bar a}$ ) parametrizing the components of the
4-metric $g_{AB}$ of Eq.(\ref{2.4}) are {\it gauge variables}
globally described by the embedding $z^{\mu}(\tau ,\sigma^u)$ of
Eq.(\ref{2.1}). \hfill\break
 In parametrized Minkowski theories (see the next Section), where the
embedding is the basic variable, in absence of matter the
super-hamiltonian and super-momentum constraints are replaced by the
vanishing of the momentum $\rho_{\mu}(\tau ,\sigma^u) \approx 0$,
see Eq.(\ref{3.10}), conjugated to $z^{\mu}(\tau ,\sigma^u)$. If we
fix $z^{\mu}(\tau ,\sigma^u)$ like in Eq.(\ref{4.1}), so that the
3-metric is completely fixed ($\theta^i$, $\gamma$ and $R_{\bar a}$
are given), then Eqs.(\ref{4.2}) determine the lapse and shift
functions. The extrinsic curvature is determined either from the
variation of the unit normal $l^{\mu}$ to $\Sigma_{\tau}$ or from
${}^3K_{rs} = {1\over {2\, (1 + n)}}\, (n_{r|s} + n_{s|r} -
\partial_{\tau}\, h_{rs})$.}.

\bea
  h_{rs}(\tau ,\sigma^u ) &=& - \sgn\, g_{rs}(\tau ,\sigma^u )
 = \Big(\gamma^{1/3}\, \sum_a\, Q_a^2\, V_{ra}(\theta^i)\, V_{sa}(\theta^i)
 \Big)(\tau ,\sigma^u ) = \nonumber \\
 &=& \sum_a\, e_{(a)r}(\tau ,\sigma^u )\, e_{(a)s}(\tau ,
 \sigma^u ), \nonumber \\
 &&{}\nonumber \\
 &&e_{(a)r} = \gamma^{1/6}\, Q_a\, V_{ra}(\theta^i),\qquad e^r_{(a)}
 = \gamma^{-1/6}\, Q^{-1}_a\, V_{ra}(\theta^i),\nonumber \\
 &&\gamma = det\, h_{rs},\qquad Q_a = e^{\sum_{\bar a}\, \gamma_{\bar aa}\, R_{\bar
 a}},
 \label{2.10}
 \eea

\noindent where $e_{(a)r}(\tau ,\sigma^u)$ and $e^r_{(a)}(\tau
,\sigma^u)$, ($\sum_a\, e^r_{(a)}\, e_{(a)s} = \delta^r_s$,
$\sum_r\, e^r_{(a)}\, e_{(b)r} = \delta_{ab}$) are cotriads and
triads on $\Sigma_{\tau}$, respectively. At spatial infinity we have
$e^r_{(a)}(\tau ,\sigma^u)\, \rightarrow\, \delta^r_a$,
$e_{(a)r}(\tau ,\sigma^u)\, \rightarrow\, \delta_{ra}$. To express
$e_{(a)r}$ in terms of $\partial_r\, F^A$, we must find the
eigenvalues and the eigenvectors of the matrix $h_{rs}$ in the form
given in Eqs.(\ref{2.4}).\medskip

The three eigenvalues of the 3-metric are $\lambda_a =
\gamma^{1/3}\, Q_a^2> 0$. The positivity of the eigenvalues is
implied by the M$\o$ller conditions (\ref{2.9}): $\lambda_1\,
\lambda_2\, \lambda_3 = \gamma
> 0$, $\lambda_1 + \lambda_2 + \lambda_3 = h_{11} + h_{22} + h_{33}
> 0$, $\lambda_1\, \lambda_2 + \lambda_2\, \lambda_3 + \lambda_3\,
\lambda_1 =  \begin{array}{|ll|} h_{11} & h_{12} \\ h_{21} & h_{22}
\end{array} +  \begin{array}{|ll|} h_{11} & h_{13} \\ h_{31} & h_{33}
\end{array} + \begin{array}{|ll|} h_{22} & h_{23} \\ h_{32} & h_{33}
\end{array} > 0$.\medskip

This implies that the three 4-vectors $z^{\mu}_r(\tau ,\sigma^u )$
are space-like for every $\vec \sigma$, so that the unit normal
$l^{\mu}(\tau ,\sigma^u )$ is time-like everywhere on the
instantaneous 3-spaces.
\medskip

The M$\o$ller condition $\sgn\, g_{\tau\tau}(\tau ,\sigma^u) > 0$ of
Eqs.(\ref{2.9}) implies that $z^{\mu}_{\tau}(\tau ,\sigma^u )$ is
everywhere time-like on the instantaneous 3-spaces $\Sigma_{\tau}$.

\medskip

Let us remark that while the generic 3-spaces $\Sigma_{\tau}$ have a
3-metric with 3 distinct eigenvalues, there is a family of 3+1
splittings with two coinciding eigenvalues of $h_{rs}(\tau ,
\sigma^u)$ and another family with all the 3 eigenvalues coinciding:
they correspond to the existence of symmetries corresponding to the
Killing symmetries of Einstein general relativity.

\bigskip

The lapse and shift functions have the following expressions

\bea
 1 + n(\tau ,\sigma^u ) &=& \sgn\, z^{\mu}_{\tau}(\tau ,
 \sigma^u )\, l_{\mu}(\tau ,\sigma^u ) = \Big({{\sgn}\over {\sqrt{\gamma}}}\,
 \epsilon_{\mu\alpha\beta\gamma}\, z^{\mu}_{\tau}\, z^{\alpha}_1\,
 z^{\beta}_2\, z^{\gamma}_3\Big)(\tau ,\sigma^u) =\nonumber \\
 &=& \Big({\dot f}^{\tau}(\tau ) +
 \partial_{\tau}\, F^{\tau}(\tau ,\sigma^u )\Big)\,
 l^{\tau}(\tau ,\sigma^u ) -\nonumber \\
 &-& \sum_u\, \Big({\dot f}^u(\tau ) + \partial_{\tau}\,
 F^u(\tau ,\sigma^u )\Big)\, l^u(\tau ,\sigma^u )\, > 0,\nonumber \\
 &&{}\nonumber \\
  n_r(\tau ,\sigma^u) &=& h_{rs}(\tau ,\sigma^u)\, n^s(\tau ,\sigma^u) =
  \sum_u\, \Big({\dot f}^u(\tau ) + \partial_{\tau}\, F^u(\tau ,\sigma^v )\Big)\,
 \partial_r\, F^u(\tau ,\sigma^v ) -\nonumber \\
 &-& \Big({\dot f}^{\tau}(\tau ) + \partial_{\tau}\, F^{\tau}(\tau ,\sigma^v )\Big)\,
 \partial_r\, F^{\tau}(\tau ,\sigma^v ).
 \label{2.11}
 \eea
\medskip

Let us also remark that all the information carried by
$\epsilon^{\mu}_A\, f^A(\tau )$, i.e. the velocity and acceleration
of the time-like observer $x^{\mu}(\tau )$, is hidden in the lapse
and shift functions.

\medskip
 The extrinsic curvature of the instantaneous 3-space $\Sigma{\tau}$
 can be evaluated by means of the formula ${}^3K_{rs} = {1\over {2\, (1 + n)}}\,
(n_{r|s} + n_{s|r} - \partial_{\tau}\, h_{rs})$, by using the
Christoffel symbols associated to $h_{rs}$ for the 3-covariant
derivatives $n_{r|s}$.

\bigskip

In conclusion the relevant conditions on the functions $f^A(\tau )$,
$F^A(\tau ,\sigma^u)$ of an admissible 3+1 splitting of Minkowski
space-time are $\sgn\, {\dot x}^2(\tau ) > 0$, $1 + n(\tau
,\sigma^u) > 0$, $\sgn\, g_{\tau\tau}(\tau ,\sigma^u) > 0$ and
$\lambda_a(\tau ,\sigma^u) > 0$.

\bigskip\bigskip

Finally Eq.(\ref{2.10}) suggests that it must be $z^{\mu}_r(\tau
,\sigma^u) = \Lambda^{\mu}{}_a(\tau ,\sigma^u)\, e_{(a)r}(\tau
,\sigma^u)$, where $\Lambda(\tau ,\sigma^u)$ is some Lorentz matrix,
so that $- \sgn\, g_{rs} = \sgn\, \eta_{\mu\nu}\,
\Lambda^{\mu}{}_{a}\, \Lambda^{\nu}{}_{b}\, e_{(a)r}\, e_{(b)s} = -
\sgn\, \eta_{ab}\, e_{(a)r}\, e_{(b)s} = h_{rs}$.

To find $\Lambda(\tau ,\sigma^u)$ let us remember that in tetrad
gravity in the York canonical basis (see Ref.\cite{12}) the
expression of the tetrads adapted to $\Sigma_{\tau}$ (Schwinger time
gauge) in terms of the unit normal $l^A$ and of the triads
$e^r_{(a)}$ are ${\buildrel \circ \over E}^A_{(o)} = l^A$,
${\buildrel \circ \over E}^A_{(a)} = (0; e^r_{(a)})$. In terms of
them we have ${\buildrel \circ \over V}^A = (1 + n)\, {\buildrel
\circ \over E}^A_{(o)} + e^s_{(a)}\, n_s\, {\buildrel \circ \over
E}^A_{(a)} = (1; 0)^A$. The world components of this vector are
${\buildrel \circ \over V}^{\mu} = z^{\mu}_A\, {\buildrel \circ
\over V}^A = z^{\mu}_{\tau}$, while those of ${\buildrel \circ \over
E}^A_{(a)}$ are ${\buildrel \circ \over E}^{\mu}_{(a)} = z^{\mu}_A\,
{\buildrel \circ \over E}^A_{(a)} = z^{\mu}_r\, e^r_{(a)}$, so that
we get $z^{\mu}_r = e_{(a)r}\, {\buildrel \circ \over
E}^{\mu}_{(a)}$. For the unit normal we have $l^{\mu} = z^{\mu}_A\,
l^A$.

In Minkowski space-time our parametrization of the embedding uses
the asymptotic tetrads $\epsilon^{\mu}_A$ and we have $z^{\mu}_A =
\epsilon^{\mu}_B\, \partial_A\, F^B$ and $l^{\mu} =
\epsilon^{\mu}_A\, l^A = \epsilon^{\mu}_o(l)$. Therefore a set of
tetrads adapted to $\Sigma_{\tau}$ in the point $(\tau ,\sigma^u)$
is given by the orthonormal tetrads $\epsilon^{\mu}_A(l(\tau
,\sigma^u))$ defined in Eqs.(\ref{2.8}): they replace the adapted
tetrads $l^{\mu}$, ${\buildrel \circ \over E}^{\mu}_{(a)}$ of tetrad
gravity. Therefore, consistently with Eq.(\ref{2.10}), we must have

\beq
 z^{\mu}_r(\tau ,\sigma^u) = \epsilon^{\mu}_A\, \partial_r\,
 F^A(\tau ,\sigma^u) = \epsilon^{\mu}_a(l(\tau ,\sigma^u))\,
 e_{(a)r}(\tau ,\sigma^u).
 \label{2.12}
 \eeq
 \medskip

\noindent This implies $z^{\mu}_{\tau} = \Big[(1 + n)\, l^A +
\epsilon^s_{(a)}\, n_s\, \epsilon^{\mu}_a(l)\Big](\tau ,\sigma^u) =
L^{\mu}{}_{\nu}(l(\tau ,\sigma^u), {\buildrel \circ \over l})\,
G^{\nu}(\tau ,\sigma^u)$ with $G^{\mu} = (1 + n; e^s_{(r)}\, n_s)$.
Eqs.(\ref{2.12}) are a set of non-linear partial differential
equations for $\partial_r\, F^A(\tau ,\vec \sigma )$.\bigskip

It is difficult to construct explicit examples of admissible 3+1
splittings. Let us consider the following two examples in which the
instantaneous 3-spaces are space-like hyper-planes.\medskip

A)  {\it Rigid non-inertial reference frames with translational
acceleration exist}. An example are the following embeddings

\medskip

\bea
 z^{\mu}(\tau ,\sigma^u ) &=& x^{\mu}_o +
\epsilon^{\mu}_{\tau}\, f(\tau ) + \epsilon^{\mu}_r\,
\sigma^r,\nonumber \\
 &&{}\nonumber \\
 &&g_{\tau\tau}(\tau ,\sigma^u ) = \sgn\,
 \Big({{d f(\tau )}\over {d\tau}}\Big)^2,\quad g_{\tau r}(\tau ,\sigma^u )
 =0,\quad g_{rs}(\tau ,\sigma^u ) = -\sgn\, \delta_{rs}.
 \label{2.13}
 \eea

\medskip

This is a foliation with parallel hyper-planes with normal $l^{\mu}
= \epsilon^{\mu}_{\tau} = const.$ and with the time-like observer
$x^{\mu}(\tau ) = x^{\mu}_o + \epsilon^{\mu}_{\tau}\, f(\tau )$ as
origin of the 3-coordinates. The hyper-planes have translational
acceleration ${\ddot x}^{\mu}(\tau ) = \epsilon^{\mu}_{\tau}\, \ddot
f(\tau )$, so that they are not uniformly distributed like in the
inertial case $f(\tau ) = \tau$.

\bigskip

B) As shown in Refs.\cite{3}, the simplest example of 3+1 splitting,
whose instantaneous 3-spaces are space-like hyper-planes carrying
admissible differentially rotating 3-coordinates \footnote{As shown
in Refs.\cite{3}, if we use the embedding $z^{\mu}(\tau ,\sigma^u)=
x^{\mu}(\tau ) + \epsilon^{\mu}_r\, R^r{}_s(\tau )\, \sigma^s$ such
that $\Omega^r = \Omega^r(\tau )$, then the resulting
$g_{\tau\tau}(\tau ,\sigma^u)$ violates M$\o$ller conditions,
because it vanishes at $\sigma = \sigma_R = {1\over {\Omega (\tau
)}}\, \Big[- {\dot x}_{\mu}(\tau )\, \epsilon^{\mu}_r\, R^r{}_s(\tau
)\, (\hat \sigma \times \hat \Omega (\tau ))^r + \sqrt{{\dot
x}^2(\tau ) + [{\dot x}_{\mu}(\tau )\, \epsilon^{\mu}_r\,
R^r{}_s(\tau )\, (\hat \sigma \times \hat \Omega (\tau
))^r]^2}\Big]$. We use the notations $\sigma^u = \sigma\, {\hat
\sigma}^u$, $\Omega^r = \Omega\, {\hat \Omega}^r$, ${\hat \sigma}^2
= {\hat \Omega}^2 = 1$. At this distance from the rotation axis the
tangential rotational velocity becomes equal to the velocity of
light. This is the {\it horizon problem} of the rotating disk. This
pathology is common to most of the rotating coordinate systems
quoted after Eq.(\ref{2.16}) and in Appendices A and B. Let us
remark that an analogous pathology happens on the event horizon of
the Schwarzschild black hole, where the time-like Killing vector of
the static space-time becomes light-like: in this case we do not
have a coordinate singularity but an intrinsic geometric property of
the solution of Einstein's equations. For the rotating Kerr black
hole the same phenomenon happens already at the boundary of the
ergosphere \cite{13}, as a consequence of the Killing vectors own by
this solution. Let us remark that in the existing theory of rotating
relativistic stars \cite{14}, where differential rotations are
replacing the rigid ones in model building, it is assumed that in
certain rotation regimes an ergosphere may form \cite{15}: however
in this case it is not known whether Killing vectors and a dynamical
ergosphere exist, so that the horizon problem,arising if one uses
4-coordinates adapted to the Killing vectors, could be associated to
a coordinate singularity like for the rotating disk. In the study of
the magnetosphere of pulsars the horizon of the rotating disk is
named the {\it light cylinder} (see Appendix B). }, is given by the
embedding ($\sigma = |\vec \sigma |$; $\epsilon^{\mu}_r$ are the
asymptotic space-like axes and the unit normal is $l^{\mu} =
\epsilon^{\mu}_{\tau} = const.$; $\alpha_i(\tau ,\vec \sigma ) =
F(\sigma )\, {\tilde \alpha}_i(\tau )$, $i=1,2,3$, are Euler angles;
$R^r{}_s(\alpha_i(\tau ,\sigma ))$ is a rotation matrix satisfying
the asymptotic conditions $R^r{}_s(\tau , \sigma)\,
{\rightarrow}_{\sigma \rightarrow \infty} \delta^r_s$, $\partial_A\,
R^r{}_s(\tau ,\sigma )\, {\rightarrow}_{\sigma \rightarrow
 \infty}\, 0$)

\begin{eqnarray*}
 z^{\mu}(\tau ,\sigma^u ) &=& x^{\mu}(\tau ) + \epsilon^{\mu}_r\,
R^r{}_s(\tau , \sigma )\, \sigma^s,\qquad x^{\mu}(\tau) = x^{\mu}_o +
f^A(\tau)\, \epsilon^{\mu}_A,\nonumber \\
 &&{}\nonumber \\
 R^r{}_s(\tau ,\sigma ) &=& R^r{}_s(\alpha_i(\tau,\sigma )) =
 R^r{}_s(F(\sigma )\, {\tilde \alpha}_i(\tau)),\nonumber \\
 &&{}\nonumber \\
 &&0 < F(\sigma ) < {1\over {A\, \sigma}},\qquad {{d\, F(\sigma
 )}\over {d\sigma}} \not= 0\,\, (Moller\,\, conditions),
 \end{eqnarray*}

\begin{eqnarray*}
 z^{\mu}_{\tau}(\tau ,\sigma^u) &=& {\dot x}^{\mu}(\tau ) -
 \epsilon^{\mu}_r\,  R^r{}_s(\tau
 ,\sigma )\, \delta^{sw}\, \epsilon_{wuv}\, \sigma^u\, {{\Omega^v(\tau
 ,\sigma )}\over c},\nonumber \\
  z^{\mu}_r(\tau ,\sigma^u) &=& \epsilon^{\mu}_k\, R^k{}_v(\tau
 ,\sigma )\, \Big(\delta^v_r + \Omega^v_{(r) u}(\tau ,\sigma )\,
 \sigma^u\Big),
 \end{eqnarray*}

 \bea
 \sgn\, g_{\tau\tau}(\tau ,\sigma^u) &=& \sgn\, {\dot x}^2(\tau ) -
 2\, \sgn\, {\dot x}_{\mu}(\tau )\ \epsilon^{\mu}_r\,  R^r{}_s(\tau
 ,\sigma )\, \delta^{sw}\, \epsilon_{wuv}\, \sigma^u\, {{\Omega^v(\tau
 ,\sigma )}\over c} -\nonumber \\
 &-&{1\over {c^2}}\, \sum_k\, \epsilon_{krs}\, \sigma^r\, \Omega^s(\tau ,\sigma )\,
\epsilon_{kuv}\, \sigma^u\, \Omega^v(\tau ,\sigma ),\nonumber \\
 n_r(\tau ,\vec \sigma ) &=& - \sgn\, g_{\tau r}(\tau ,\sigma ) = -
 \sgn\, {\dot x}_{\mu}(\tau )\ \epsilon^{\mu}_k\,  R^k{}_v(\tau
 ,\sigma )\, \Big(\delta^v_r + \Omega^v_{(r) u}(\tau ,\sigma )\,
 \sigma^u\Big) -\nonumber \\
 &-& \epsilon_{smn}\, \sigma^m\, {{\Omega^n(\tau ,\sigma )}\over c}\,
 \Big(\delta^s_r + \Omega^s_{(r) u}(\tau ,\sigma )\, \sigma^u\Big),
 \nonumber \\
 h_{rs}(\tau ,\sigma^u) &=& - \sgn\, g_{rs}(\tau ,\sigma^u) =
 \delta_{rs} + \Big(\Omega^r_{(s) u}(\tau ,\sigma ) + \Omega^s_{(r)
 u}(\tau ,\sigma )\Big)\, \sigma^u +\nonumber \\
 &+& \sum_w\, \Omega^w_{(r) u}(\tau ,\sigma )\, \Omega^w_{(s) v}(\tau
 ,\sigma )\, \sigma^u\, \sigma^v,
 \label{2.14}
 \eea

\noindent where $\Big(R^{-1}(\tau ,\sigma )\,
\partial_{\tau}\, R(\tau ,\sigma )\Big)^u{}_v = \delta^{um}\,
\epsilon_{mvr}\, {{\Omega^r(\tau ,\sigma)}\over c}$,
${\partial_{\tau}\, R(\tau ,\sigma)}^u{}_v = R^u{}_n(\tau ,\sigma)\,
\delta^{nm}\, \epsilon_{mvr}\, {{\Omega^r(\tau ,\sigma)}\over c}$
with $\Omega^r(\tau ,\sigma ) = F(\sigma )\, \tilde \Omega (\tau
,\sigma )\, {\hat n}^r(\tau ,\sigma )$ \footnote{${\hat n}^r(\tau
,\sigma )$ defines the instantaneous rotation axis and $0 < \tilde
\Omega (\tau ,\sigma ) < 2\, max\, \Big({\dot {\tilde \alpha}}(\tau
), {\dot {\tilde \beta}}(\tau ), {\dot {\tilde \gamma}}(\tau
)\Big)$.} being the angular velocity and with $\Omega_{(r)}(\tau
,\sigma ) = R^{-1}(\tau ,\vec \sigma )\, \partial_r\, R(\tau ,\sigma
)$. The angular velocity vanishes at spatial infinity and has an
upper bound proportional to the minimum of the linear velocity
$v_l(\tau ) = {\dot x}_{\mu}\, l^{\mu}$ orthogonal to the space-like
hyper-planes. When the rotation axis is fixed and $\tilde \Omega
(\tau ,\sigma ) = \omega = const.$, a simple choice for the function
$F(\sigma )$ is $F(\sigma ) = {1\over {1 + {{\omega^2\,
\sigma^2}\over {c^2}}}}$ \footnote{Nearly rigid rotating systems,
like a rotating disk of radius $\sigma_o$, can be described by using
a function $F(\sigma )$ approximating the step function $\theta
(\sigma - \sigma_o)$.}.\medskip

Let us remark that the unit normal is $l^{\mu}(\tau ,\sigma^u) =
\epsilon^{\mu}_{\tau} = const.$ and the lapse function is $1 +
n(\tau ,\sigma^u) = \sgn\, \Big(z^{\mu}_{\tau}\, l_{\mu}\Big)(\tau
,\sigma^u) = \sgn\, \epsilon^{\mu}_{\tau}\, {\dot x}_{\mu}(\tau )$.

\medskip

The embedding (\ref{2.14}) has been used in the first paper of
Ref.\cite{10}, on quantum mechanics in non-inertial frames, in the
form $z^{\mu}(\tau ,\sigma^u) = x^{\mu}(\tau ) + F^{\mu}(\tau
,\sigma^u) = \theta (\tau )\, \epsilon^{\mu}_{\tau} + {\cal
A}^r(\tau ,\sigma^u)\, \epsilon^{\mu}_r$ with $x^{\mu}_o = 0$,
$\theta (\tau ) = f^{\tau}(\tau )$, ${\cal A}^r(\tau ,\sigma^u) =
f^r(\tau) + R^r{}_s(\tau ,\sigma)\, \sigma^s$, describing the
freedom in the choice of the mathematical time $\tau$ and with the
world-line of the time-like observer having the expression
$x^{\mu}(\tau ) = \epsilon_{\tau}^{\mu}\, \theta (\tau ) +
\epsilon^{\mu}_r\, {\cal A}^r(\tau ,0)$, namely with $f^r(\tau ) =
{\cal A}^r(\tau ,o)$ and ${\dot f}^r(\tau ) = {{w^r(\tau )}\over
{c}}$ ($\vec w(\tau )$ is the ordinary 3-velocity). If we choose
$\theta (\tau ) = \tau$, we get from Eq.(\ref{2.2}) $u^{\mu}(\tau )
= \epsilon^{\mu}_A\, u^A(\tau ) = {{\epsilon^{\mu}_{\tau} +
\epsilon^{\mu}_r\, {{w^r(\tau)}\over c}}\over {\sqrt{1 - {{{\vec
w}^2(\tau )}\over {c^2}} }}}$, $a^{\mu}(\tau ) = \epsilon^{\mu}_A\,
u^A(\tau ) = {1\over {c^2}}\, \sum_u\, {\dot w}^u(\tau )\, {\ddot
w}^u(\tau )\, \Big(1 - {{{\vec w}^2(\tau )}\over
{c^2}}\Big)^{-3/2}\, \Big(\epsilon^{\mu}_{\tau} + \epsilon^{\mu}_r\,
{{w^r(\tau)}\over c}\Big)$. The lapse function is $1 + n(\tau ) =
{\dot f}^{\tau}(\tau )$.
\bigskip

To evaluate the non-relativistic limit for $c \rightarrow \infty$,
where $\tau = c\, t$ with $t$ the absolute Newtonian time and
$\partial_{\tau} = {1\over c}\, \partial_t$, we choose the gauge
function $F(\sigma ) = {1\over {1 + {{\omega^2\, \sigma^2}\over
{c^2}}}}\, \rightarrow_{c \rightarrow \infty}\, 1 - {{ \omega^2\,
\sigma^2}\over {c^2}} + O(c^{-4})$. This implies

\bea
 R^a{}_r(\tau ,\sigma ) &\rightarrow_{c \rightarrow \infty}&
 R^a{}_r(\tau ) - {{\omega^2\, \sigma^2}\over {c^2}}\,
 \sum_i\, {\tilde \alpha}_i(\tau )\, {{\partial\, R^a{}_r(\tau
 ,\sigma )}\over {\partial\, \alpha_i}}{|}_{F(\sigma ) = 1} +
 O(c^{-4}) =\nonumber \\
 &{\buildrel {def}\over =}& R^a{}_r(\tau ) -
 {{\omega^2\, \sigma^2}\over {c^2}}\,
 R^{(1)a}{}_r(\tau ) + O(c^{-4}),
 \label{2.15}
 \eea

\noindent and we can introduce a new 3-velocity $\vec v(\tau )$ by
means of  $w^r(\tau ) = c\, {\dot f}^r(\tau) = R^r{}_s(\tau )\,
v^s(\tau )$.  We have $\Omega^r(\tau ,\sigma) =  \tilde \Omega
(\tau) {\hat n}^r(\tau ) + O(c^{-1})$ for the angular velocity and
$\Omega_{(r)}(\tau ,\sigma) = 0 + O(c^{-2})$.
\medskip

Therefore the corrections to rigidly-rotating non-inertial frames
coming from M$\o$ller conditions are of order $O(c^{-2})$ and become
important at the distance from the rotation axis where the horizon
problem for rigid rotations appears.\medskip

Then, from Eqs. (\ref{2.14}), (\ref{2.4}), (\ref{2.7}) and
(\ref{2.11}) we get

\bea
 z^{\mu}(\tau ,\sigma^u) &\rightarrow& x^{\mu}(\tau ) +
 \epsilon^{\mu}_r\, R^r{}_s(\tau )\, \sigma^s - {{\omega^2\,
 \sigma^2}\over {c^2}}\, \epsilon^{\mu}_r\, R^{(1)r}{}_s(\tau )\,
 \sigma^s + O(c^{-4}),\nonumber \\
 &&{}\nonumber \\
 z^{\mu}_{\tau}(\tau ,\sigma^u) &\rightarrow& {\dot x}^{\mu}(\tau )
 + \epsilon^{\mu}_r\, \partial_{\tau}\, R^r{}_s(\tau )\, \sigma^s
 + O(c^{-3}) =\nonumber \\
 &=& \epsilon^{\mu}_{\tau} + \epsilon^{\mu}_r\, {\dot f}^r(\tau ) +
 {1\over c}\, \epsilon^{\mu}_r\, R^r{}_s(\tau)\, \epsilon_{suv}\,
 \Omega^u(\tau )\, \sigma^v + O(c^{-3}),\nonumber \\
 z^{\mu}_r(\tau ,\sigma^u) &\rightarrow& \epsilon^{\mu}_s\,
 \Big[R^s{}_r(\tau ) - {{\omega^2}\over {c^2}}\, R^{(1)s}{}_u(\tau )\,
 (\delta^u_r\, \sigma^2 + 2\, \sigma^u\, \sigma^v\, \delta_{vr})\Big]
 + O(c^{-4}),\nonumber \\
 &&{}\nonumber \\
 h_{rs}(\tau ,\sigma^u) &\rightarrow& \delta_{rs} - 2\, {{\omega^2}\over
 {c^2}}\, \sum_u\, R^u{}_r(\tau )\, R^{(1)u}{}_v(\tau )\, (\delta^v_s + 2\, \sigma^v\,
 \sigma^n\, \delta_{ns}) + O(c^{-4}), \nonumber \\
 &&{}\nonumber \\
 n(\tau ) &=& 0,\qquad
 n_r(\tau ,\sigma^u) \rightarrow {1\over c}\, \Big(\delta_{rs}\, v^s(\tau ) +
 \epsilon_{ruv}\, \Omega^u(\tau )\, \sigma^v\Big) +
 O(c^{-3}).\nonumber \\
 &&{}
 \label{2.16}
 \eea

\bigskip

There is the enormous amount of bibliography, reviewed in
Ref.\cite{16}, about the problems of the {\it rotating disk} and of
the {rotating coordinate systems}. Independently from the fact
whether the disk is a material extended object or a geometrical
congruence of time-like world-lines (integral lines of some
time-like unit vector field), the idea followed by many researchers
\cite{6,17,18} (in Refs.\cite{18} are quoted the attempts to develop
electro-dynamics in rotating frames) is to start from the Cartesian
4-coordinates of a given inertial system, to pass to cylindrical
3-coordinates and then to make a either Galilean (assuming a
non-relativistic behaviour of rotations at the relativistic level)
or Lorentz transformation to comoving rotating 4-coordinates (see
the locality hypothesis in the next Subsection), with a subsequent
evaluation of the 4-metric in the new coordinates. In other cases
\cite{19} a suitable global 4-coordinate transformation is
postulated, which avoids the horizon problem. Various authors (see
for instance Refs.\cite{20}) do not define a coordinate
transformation but only a rotating 4-metric. Just starting from
M$\o$ller rotating 4-metric \cite{6}, Nelson (see the second paper
in Ref.\cite{13}) was able to deduce a 4-coordinate transformation
implying it.

\medskip

See Appendix A for the description of the rotating disk and the
Sagnac effect in the 3+1 framework.

\subsection{Congruences of Time-Like Observers Associated with an
Admissible 3+1 Splitting, the 1+3 Point of View and the Locality
Hypothesis}

Each admissible 3+1 splitting of Minkowski space-time, having the
time-like observer $x^{\mu}(\tau )$ as origin of the 3-coordinates
on the instantaneous 3-spaces $\Sigma_{\tau}$, automatically
determines two time-like vector fields and therefore {\it two
congruences of (in general) non-inertial time-like observers}:

i) The time-like vector field $l^{\mu}(\tau ,\sigma^u )\,
\partial_{\mu}$ of the normals to the simultaneity surfaces $\Sigma_{\tau}$
(by construction surface-forming, i.e. irrotational), whose flux
lines are the world-lines $x^{\mu}_{l,\tau_o,\sigma^u_o}(\tau )$,
$u^{\mu}_{}(\tau) = {{{\dot x}^{\mu}_{l,\tau_o,\sigma^u_o}}\over
{\sqrt{\sgn\, {\dot x}^2_{l.\tau_o,\sigma^u_o}(\tau)}}}$,
$u^{\mu}_{l,\tau_o,\sigma^u_o}(\tau_o) = l^{\mu}(\tau_o,
\sigma^u_o)$, of the so-called (in general non-inertial) {\it
Eulerian observers}. The simultaneity surfaces $\Sigma_{\tau}$ are
(in general non-flat) Riemannian 3-spaces in which every physical
system is visualized and in each point the {\it tangent space} to
$\Sigma_{\tau}$ is the {\it local observer rest frame} of the
Eulerian observer through that point. The 3+1 viewpoint of these
observers is called {\it hyper-surface 3+1 splitting}.

ii) The time-like evolution vector field ${{z^{\mu}_{\tau}(\tau
,\vec \sigma )}\over { \sqrt{\sgn\, g_{\tau\tau}(\tau ,\vec \sigma )
} }}\, \partial_{\mu}$, which in general is not surface-forming
(i.e. it has non-zero vorticity like in the case of the rotating
disk). The observers associated to its flux lines
$x^{\mu}_{z,\sigma^u_o}(\tau ) = z^{\mu}(\tau ,\sigma^u_o)$,
$u^{\mu}_{z,\sigma^u_o}(\tau ) = {{z^{\mu}_{\tau}(\tau ,\vec \sigma
)}\over { \sqrt{\sgn\, g_{\tau\tau}(\tau ,\vec \sigma ) } }}$, have
the {\it local observer rest frames}, the tangent 3-spaces
orthogonal to the evolution vector field, {\it not tangent} to
$\Sigma_{\tau}$: there is no notion of 3-space for these observers
(1+3 point of view or {\it threading splitting}) and no
visualization of the physical system in large. However these
observers can use the notion of simultaneity associated to the
embedding $z^{\mu}(\tau ,\vec \sigma )$, which determines their
4-velocity. Like for the observer $x^{\mu}(\tau )$, their 4-velocity
is not parallel to $l^{\mu}(\tau ,\sigma^u)$. The 3+1 viewpoint of
these observers is called {\it slicing 3+1 splitting}.

\bigskip

Every 1+3 point of view considers only a time-like observer (either
$x^{\mu}(\tau )$ or $x^{\mu}_{l,\tau_o,\sigma^u_o}(\tau )$ or
$x^{\mu}_{z,\sigma^u_o}(\tau )$) and tries to give a description of
the physics in a region around the observer's world-line assumed
known. Since there is no global notion of simultaneity, namely of
instantaneous 3-space, one identifies the space-like hyper-planes
orthogonal to the observer unit 4-velocity $u^{\mu}_{obs}(\tau )$ at
every instant $\tau$ (the observer local rest frames) as local
instantaneous 3-spaces $\Sigma_{obs\, \tau}$ (strictly speaking it
is a tangent space and not a 3-space). Then one makes a choice of a
tetrad $V^{\mu}_{obs\, A}((\tau)) = \Big(u^{\mu}_{obs}(\tau );
V^{\mu}_{obs\, (r)}(\tau )\Big)$, $\eta_{\mu\nu}\, V^{\mu}_{obs\,
(A)}(\tau )\, V^{\nu}_{obs\, (B)}(\tau ) = \eta_{(A)(B)}$. The space
axes $V^{\mu}_{obs\, (r)}(\tau )$ can be chosen arbitrarily, even if
often they are chosen as the tangents to three space-like geodesics
on $\Sigma_{obs\, \tau}$ at the observer position. After parallel
transport of the tetrad to the points of $\Sigma_{obs\, \tau}$ not
on the observer world-line one tries to build an {\it accelerated
4-coordinate system} having the observer as origin of the
3-coordinates \cite{21}. In the case of the tangents to space-like
geodesics one  builds a local system of Fermi coordinates around the
observer world-line \cite{22} (see also Ref.\cite{23} for an updated
discussion of fermi-Walker and Fermi normal coordinates).

The drawback of this construction is that the $\tau$-dependent
family of hyper-planes $\Sigma_{obs\, \tau}$ will have hyper-planes
at different $\tau$'s intersecting at some distance from the
observer world-line, usually estimated by using the so-called {\it
acceleration radii} of the observer. This implies that every system
of accelerated 4-coordinates of this type will develop {\it
coordinate singularities} when the hyper-planes intersect. As a
consequence it is not possible to formulate a well-posed Cauchy
problem for Maxwell equations in these accelerated coordinate
systems: they can only be used for evaluating local
semi-relativistic inertial effects.
\medskip

At each instant $\tau$ the tetrads $V^{\mu}_{obs\, (A)}(\tau )$
coincide with some Lorentz matrix $V^{\mu}_{obs\, (A)}(\tau ) =
\Lambda^{\mu}{}_{\nu = A}(\tau )$, which connects the reference
inertial frame to the {\it instantaneous comoving inertial frame}
associated with the accelerated observer at $\tau$. A possibility is
to use the tetrads $\epsilon^{\mu}_A(u_{obs}(\tau ))$ associated
with the Wigner boost $L^{\mu}{}_{\nu}(u_{obs}(\tau ), {\buildrel
\circ \over u}_{obs})$. This fact is at the heart of the {\it
locality hypothesis} \cite{24} according to which an accelerated
observer is physically equivalent (for measurements) to a continuous
family of hypothetical momentarily comoving inertial
observers.\medskip

If we parametrize the Lorentz transformation $\Lambda (\tau )$ as
the product of a pure boost with a pure rotation $\Lambda (\tau ) =
B(\vec \beta (\tau ))\, {\cal R}(\alpha (\tau ), \beta (\tau ),
\gamma (\tau ))$ and we call $R^r_s(\tau ) = {\cal R}^r{}_s(\tau )$,
we can write (from Eq.(\ref{2.8}) we have $B^{jk}(\vec \beta (\tau
)) = \delta^{jk} + (\gamma(\tau ) - 1)\, {{\beta^j(\tau )\,
\beta^k(\tau )}\over {\sum_n\, (\beta^n(\tau ))^2}}$)

\bea
 V^\mu_{obs\, (A)}(\tau ) = \Lambda^\mu_{\nu=A}
(\tau)= \left(
\begin{array}{cc}
\frac{1}{\sqrt{1 - \vec{\beta}^2(\tau)}}& \frac{R^i_k(\tau )\,
\beta^k(\tau)}{\sqrt{1 - \vec{\beta}^2(\tau)}}\\
\frac{\beta^j(\tau)}{\sqrt{1 - \vec{\beta}^2(\tau)}}&\,
R^i_k(\tau)\, B^{jk}({\vec \beta} (\tau))
\end{array}
\right).
 \label{2.17}
\eea

Let us define the angular velocity $\omega_r(\tau )$ by means of
${{d\, R^r_s(\tau )}\over {d\tau}}\, {\buildrel {def}\over =}\,
\epsilon_{ruv}\, \omega_u(\tau)\, R^v_s(\tau )$. Even if the
observer is connected with the embedding $z^{\mu}(\tau ,\vec
\sigma)$, this angular velocity is not related to the angular
velocity defined after Eq.(\ref{2.14}).\medskip

Finally, if we write

\bea
 && {{d V^{\mu}_{obs\, (A)}(\tau )}\over {d\tau}} =
 {{\cal A}_{obs\, (A)}}^{(B)}(\tau )\, V^{\mu}_{obs\, (B)}(\tau ),
 \nonumber \\
 &&{}\nonumber \\
 &\Rightarrow & {\cal A}_{obs\, (A)(B)}(\tau ) = - {\cal
 A}_{obs\, (B)(A)}(\tau ) = {{d V^{\mu}_{obs\, (A)}(\tau
 )}\over {d\tau}}\,\eta_{\mu\nu}\,  V^\nu_{obs\, (B)}(\tau ),
 \label{2.18}
  \eea

\noindent and we introduce the definitions $a_{obs\, r}(\tau ) =
{\cal A}_{obs\, (\tau)(r)}(\tau )$, $\Omega_{obs\, r}(\tau ) =
{1\over 2}\, \epsilon_{ruv}\, {\cal A}_{obs\, (u)(v)}(\tau )$, then
the acceleration radii have the following definition \cite{24}:
$I_1(\tau) = \sum_r\, \Big(\Omega^2_{obs\, r}(\tau ) - a^2_{obs\,
r}(\tau )\Big)$, $I_2(\tau ) = \sum_r\, a_{obs\, r}(\tau )\,
\Omega_{obs\, r}(\tau )$. By means of Eq.(\ref{2.17}) they can be
expressed in terms of the parameters of the Lorentz transformation
and their $\tau$-derivatives.

\bigskip

Finally let us remark that given an admissible 3+1 splitting of
Minkowski space-time, the infinitesimal spatial length $dl$ in the
instantaneous 3-spaces $\Sigma_{\tau}$ is defined by putting $d\tau
= 0$ in the line element $ds^2 = g_{AB}(\tau ,\sigma^u)\,
d\sigma^A\, d\sigma^B$, namely we have $dl^2 = g_{rs}(\tau
,\sigma^u)\, d\sigma^r\, d\sigma^s$. This global, but
coordinate-dependent, definition has to be contrasted with the
local, but coordinate-independent, definition used in the 1+3 point
of view as it is done for instance in Landau-Lifschitz \cite{17}.
This definition is only locally valid in the local rest frame of an
observer: since there is no notion of instantaneous 3-space it
cannot be used in a global way. For a detailed comparison of these
two notions of spatial length see Section II of the first paper of
Ref.\cite{3}.

\bigskip

\subsection{Notations for the Electro-Magnetic Field in Non-Inertial Frames}

Let us add some notations for the electro-magnetic field in the
non-inertial frames, where the instantaneous 3-space is either
curved or flat but with rotating coordinates [in both cases it is
not Euclidean and has the 3-metric $h_{rs}$ of signature $(+++)$].
\medskip

The basic field is the electro-magnetic potential $A_A = (A_{\tau};
A_r)$. We have $A^A = (A^{\tau}; A^A) = g^{AB}\, A_B = g^{A\tau}\,
A_{\tau} + g^{As}\, A_s$. Instead in  inertial frames we have
$A^{\tau} = \sgn\, A_{\tau}$, $A^r = - \sgn\, A_r$.

\medskip

In non-inertial frames it is convenient to introduce the following
"Euclidean" notation: ${\tilde A}^r = h^{rs}\, A_s \not= A^r$ (in
inertial frames: ${\tilde A}^r = A_r = - \sgn\, A^r$)

\bigskip

We shall adopt the following conventions for the electric and
magnetic fields in terms of $F_{AB} = \partial_A\, A_B -
\partial_B\, A_A$ \footnote{In the inertial case, where $h_{rs} = \delta_{rs}$ implies
$V^r {\buildrel {def}\over =}\, {\tilde V}^r = V_r$ for the
components of 3-vector $\vec V$ not being the vector part of a
4-vector (like $\vec E$ and $\vec B$), we can use the vector
notation $\vec E = \{ E_r\} = \{{\tilde E}^r \}$, $\vec B = \{ B_r\}
= \{ {\tilde B}^r \}$, ${\vec E}^2 = \sum_r\, E_r^2 = \sum_r\,
({\tilde E}^r)^2$, ${\vec B}^2 = \sum_r\, B_r^2 = \sum_r\, ({\tilde
B}^r)^2$, $({\dot {\vec \eta}}_i \times \vec B)_r = \sum_{uv}\,
\epsilon_{ruv}\, {\dot \eta}^u_i\, B_v = \sum_{uv}\,
\epsilon_{ruv}\, {\dot \eta}_i^u\, {\tilde B}^v$, $(\vec E \times
\vec B)_r = \sum_{uv}\, \epsilon_{ruv}\, E_u\, B_v = \sum_{uv}\,
\epsilon_{ruv}\, {\tilde E}^u\, {\tilde B}^v$. Since ${\tilde V}^r =
h^{rs}\, V_s \not= V^r$, {\it we are not going to use the vector
notation in non-inertial frames}.} :\medskip

a) In  inertial frames we have \footnote{$\epsilon_{uvr}$ is the
Euclidean Levi-Civita tensor with $\epsilon_{123} = 1$;
$\epsilon^{uvr}$ is never introduced.}\medskip

\bea
 E_r &=& - F_{\tau r} = F^{\tau r}\, = {\tilde E}^r,\nonumber \\
 &&{}\nonumber \\
 B_r &=& {1\over 2}\, \epsilon_{ruv}\, F_{uv} = {1\over 2}\,
\epsilon_{ruv}\, F^{uv}\, = {\tilde B}^r, \qquad F_{uv} = F^{uv} =
\epsilon_{uvr}\, B_r = \epsilon_{uvr}\, {\tilde B}^r.
 \label{2.19}
 \eea
\medskip

b) In  non-inertial frames  we put the definitions

\beq
 E_r\, {\buildrel {def}\over =}\, - F_{\tau r},\qquad
 B_r\, {\buildrel {def}\over =}\, {1\over 2}\, \epsilon_{ruv}\, F_{uv},
 \qquad F_{rs} = \epsilon_{rsu}\, B_u.
 \label{2.20}
 \eeq

\noindent Since we have

\bea
 F^{AB} &=& g^{AC}\, g^{BD}\, F_{CD} = (g^{A\tau}\, g^{Br} - g^{Ar}\,
g^{B\tau})\, F_{\tau r} + g^{Ar}\, g^{Bs}\, F_{rs} =\nonumber \\
 &=&(g^{Ar}\, g^{B\tau} - g^{A\tau}\, g^{Br})\, E_r + \epsilon_{rsu}\,
g^{Ar}\, g^{Bs}\, B_u,\nonumber \\
 &&{}\nonumber \\
 F^{\tau u} &=& (g^{\tau r}\, g^{\tau u} - g^{\tau\tau}\, g^{ur})\, E_r
+ \epsilon_{rsn}\, g^{\tau r}\, g^{us}\, B_n =\nonumber \\
 &=& h^{ur}\, E_r + {1\over {(1 + n)^2}}\, \epsilon_{rsn}\, n^r\,
 h^{us}\, B_n,\nonumber \\
 F^{uv} &=& (g^{ur}\, g^{\tau v} - g^{\tau u}\, g^{vr})\, E_r +
\epsilon_{rsn}\, g^{ur}\, g^{vs}\, B_n =\nonumber \\
 &=& {{(h^{ur}\, n^v - h^{vr}\, n^u)\, E_r}\over {(1 + n)^2}} +
 \epsilon_{rsn}\, \Big(h^{ur}\, h^{vs} - {{n^r\, (n^v\, h^{us} -
 n^u\,  h^{us})}\over {(1 + n)^2}}\Big)\, B_n,
 \label{2.21}
 \eea

\noindent by analogy with inertial frames we can put

\bea
 F^{\tau r} &{\buildrel {def}\over =}& {\check E}^r,\qquad
 {\check E}^r = {\tilde E}^r + {{\epsilon_{uvn}\, n^u\,
 h^{rv}\, h_{nm}\, {\tilde B}^m}\over {(1 + n)^2}} \not= {\tilde E}^r = h^{rs}\, E_s,
 \nonumber \\
 &&{}\nonumber \\
 F^{uv} &{\buildrel {def}\over =}& \epsilon_{uvr}\, {\check B}^r,\qquad
 {\check B}^r = {2\over {(1 + n)^2}}\,
 \epsilon_{ruv}\, {\tilde E}^u\, n^v +\nonumber \\
 &+& \epsilon_{ruv}\, \epsilon_{ksn}\, \Big(h^{uk}\, h^{vs} - {{n^k\,
 (n^v\, h^{us} - n^u\, h^{vs})}\over {(1 + n)^2}}\Big)\, h_{nm}\, {\tilde
 B}^m \not= {\tilde B}^r = h^{rs}\, B_s.
 \label{2.22}
 \eea

\vfill\eject

\section{Parametrized Minkowski Theories and the
Inertial Rest-Frame Instant Form for Charged Particles plus the
Electro-Magnetic Field.}

In this Section we will give a review of the description of the
isolated system "N charged positive-energy scalar particles with
Grassmann-valued electric charges plus the electro-magnetic field"
\cite{9}  in the framework of parametrized Minkowski theories
\cite{1,5} (see also the Appendix of the first paper in
Refs.\cite{11}).
\bigskip

Let be given an admissible 3+1 splitting of Minkowski space-time
centered on a time-like observer $x^{\mu}(\tau )$. Let $\sigma^A =
(\tau ; \sigma^u)$ be the adapted observer-dependent radar
4-coordinates and $z^{\mu}(\tau ,\sigma^u)$ the embedding of the
instantaneous 3-spaces $\Sigma_{\tau}$ into Minkowski space-time as
seen from an arbitrary reference inertial observer. Let $
g_{AB}(\tau ,\sigma^u) = z^{\mu}_A(\tau ,\sigma^u)\, \eta_{\mu\nu}\,
z^{\nu}_B(\tau ,\sigma^u)$ be the associated 4-metric.\medskip

The electro-magnetic field is described by the Lorentz-scalar
potential $A_A(\tau ,\sigma^u)$ knowing the equal-time surface. The
field strength is $F_{AB}(\tau ,\sigma^u) = \Big(\partial_A\, A_B -
\partial_B\, A_A\Big)(\tau ,\sigma^u)$.\medskip

The scalar positive-energy particles are described by the
Lorentz-scalar 3-coordinates $\eta^r_i(\tau )$ defined by
$x^{\mu}_i(\tau ) = z^{\mu}(\tau ,\eta^u_i(\tau ))$, where
$x^{\mu}_i(\tau )$ are their world-lines. $Q_i$ are the
Grassmann-valued electric charges satisfying $Q^2_i = 0$, $Q_i\, Q_j
= Q_j\, Q_i \not= 0$ for $i \not= j$. Each $Q_i$ is an even bilinear
function of a complex Grassmann variable $\theta_i(\tau )$: $Q_i =
e\, \theta^*_i(\tau )\, \theta_i(\tau )$.

\bigskip

As shown in Ref.\cite{9} the description of N scalar positive-energy
particles with Grassmann-valued electric charges plus the
electro-magnetic field is done in parametrized Minkowski theories
with the action

\begin{eqnarray*}
  S &=&\int d\tau\, d^{3}\sigma \,{\cal L}(\tau ,\sigma^u) = \int d\tau\, L(\tau
),  \nonumber \\
&&\nonumber\\
 {\cal L}(\tau ,\sigma^u) &=&{\frac{i}{2}}\sum_{i=1}^{N}\,
\delta ^{3}(\sigma^u - \eta^u_i(\tau ))\, \Big[\theta _{i}^{\ast
}(\tau ){\dot{\theta}}_{i}(\tau ) -
{\dot{\theta}}_{i}^{\ast }(\tau )\theta _{i}(\tau )\Big]-  \nonumber \\
&&\nonumber\\
 &-&\sum_{i=1}^{N}\, \delta ^{3}(\sigma^u - \eta^u_i(\tau
))\, \Big[m_i\, c\, \sqrt{\sgn\, [g_{\tau \tau }(\tau ,\sigma^u) +
2\, g_{\tau r}(\tau ,\sigma^u)\, {\dot{\eta}}_i^r(\tau ) + g_{rs
}(\tau ,\sigma^u)\, {\dot{\eta}}_i^r(\tau )\, {\dot{\eta}}_i^s(\tau
)]} -  \end{eqnarray*}

\bea
 &-&{{Q_i(\tau )}\over c}\, \Big(A_{\tau }(\tau ,\sigma^u) +
A_{r}(\tau ,\sigma^u)\, {\dot{\eta}}_i^r(\tau )\Big)\Big]-  \nonumber \\
&&\nonumber\\
&-&{\frac{1}{4c}}\,\sqrt{- g(\tau ,\sigma^u)}\, g^{AC }(\tau
,\sigma^u)\, g^{BD}(\tau ,\sigma^u)\, F_{AB}(\tau ,\sigma^u)\,
F_{CD}(\tau , \sigma^u).
 \label{3.1}
 \eea

\medskip

The canonical momenta are (for dimensional convenience we introduce
a $c$ factor in the definition of the electro-magnetic momenta)

\begin{eqnarray*}
\rho _{\mu }(\tau ,\sigma^u) &=& - \sgn\,{\frac{{\partial {\cal
L}(\tau ,\sigma^u)}}{{\partial z_{\tau }^{\mu }(\tau
,\sigma^u)}}}=\nonumber \\
&&\nonumber\\
&=&\sum_{i=1}^{N}\delta ^{3}(\sigma^u - \eta^u_i(\tau ))\, m_i\,
c{\frac{{z_{\tau \mu }(\tau ,\sigma^u) + z_{{r} \mu }(\tau ,
\sigma^u)\, {\dot{\eta}}_{i}^{r}(\tau )}}{\sqrt{\sgn\, [g_{\tau \tau
}(\tau , \sigma^u) + 2\, g_{\tau {r}}(\tau,\sigma^u)\,
{\dot{\eta}}_i^r(\tau ) + g_{{r}{s}}(\tau ,\sigma^u)\, {\dot{\eta}}
_i^r(\tau )\, {\dot{\eta}}_i^s(\tau )]}}} +  \nonumber \\
&&\nonumber\\
 &+&\sgn\,{\frac{\sqrt{- g(\tau ,\sigma^u)}}{4c}}\, \Big[(g^{\tau \tau }\, z_{\tau
\mu } + g^{\tau {r}}\, z_{r\mu })\, g^{AC}\, g^{BD}\, F_{AB}\,
F_{CD}-  \nonumber \\
&&\nonumber\\
&-&2\, \Big(z_{\tau \mu }\, (g^{A\tau}\, g^{\tau C}\, g^{ BD} +
g^{AC}\, g^{B\tau }\, g^{\tau D})+  \nonumber \\
&&\nonumber\\
&+&z_{{r}\mu }\, (g^{Ar}\, g^{\tau C} + g^{A\tau }\, g^{rC})\,
g^{BD}\Big)\, F_{AB }\, F_{CD}
)\Big](\tau ,\sigma^u) =  \nonumber \\
&&\nonumber\\
&=&[(\rho _{\nu }\, l^{\nu })\, l_{\mu } + (\rho _{\nu }\,
z_{r}^{\nu })\, \gamma ^{rs}\, z_{s\mu }](\tau ,\sigma^u),
\end{eqnarray*}

\begin{eqnarray*}
 \pi ^{\tau }(\tau ,\sigma^u) &=&c\, {\frac{{\partial L}}{{\partial
\partial_{\tau }A_{\tau }(\tau ,\sigma^u)}}} = 0,  \nonumber \\
&&\nonumber\\
 \pi ^r(\tau ,\sigma^u) &=&c\, {\frac{{\partial L}}{{\partial
\partial _{\tau }A_{r}(\tau ,\sigma^u)}}} =  {{\gamma (\tau
, \sigma^u)}\over {\sqrt{- g(\tau ,\sigma^u)} }}\, \h^{rs}(\tau
,\sigma^u)\, (F_{\tau s} - n^u\, F_{us})(\tau ,\sigma^u) =\nonumber \\
 &=& - {{\sqrt{\gamma}(\tau ,\sigma^u)}\over {1 + n(\tau ,\sigma^u)}}\,
 h^{rs}(\tau ,\sigma^u)\, \Big(E_s -
\epsilon_{suv}\, n^u\,B_v\Big)(\tau ,\sigma^u),\end{eqnarray*}

\begin{eqnarray*}
 \kappa _{i{r}}(\tau ) &=&+{\frac{{\partial L(\tau )}}{{\partial {\
\dot{\eta}}_{i}^{r}(\tau )}}}= {{Q_i}\over c}\, A_{r}(\tau
,\eta^u_i(\tau )) -\nonumber \\
&&\nonumber\\
&-&\sgn\, m_i\, c\, {{g_{\tau r}(\tau ,\eta^u_i(\tau )) + g_{
rs}(\tau ,\eta^u_i(\tau ))\, {\dot{\eta}}_i^s(\tau )}\over
{\sqrt{\sgn\, [g_{\tau \tau }(\tau ,\eta^u_i(\tau )) + 2\, g_{\tau
r}(\tau ,\eta^u_i(\tau ))\, {\dot{\eta}}_i^r(\tau ) + g_{rs}(\tau
,\eta^u_i(\tau ))\, { \dot{\eta}}_i^r(\tau )\, {\dot{\eta}}_i^s(\tau
)]} }},
 \end{eqnarray*}

\bea
 \pi _{\theta \,i}(\tau ) &=&{\frac{{\partial L(\tau )}}{{\partial
{\dot{ \theta}}_{i}(\tau )}}} = - {\frac{i}{2}}\, \theta _{i}^{\ast
}(\tau ),  \qquad \pi _{\theta ^{\ast }\,i}(\tau ) =
{\frac{{\partial L(\tau )}}{{\partial {\ \dot{\theta}}_{i}^{\ast
}(\tau )}}} = - {\frac{i}{2}}\, \theta _{i}(\tau ).
 \label{3.2}
\end{eqnarray}

\noindent The following Poisson brackets are assumed

\begin{eqnarray}
&&\{z^{\mu }(\tau ,\sigma^u),\rho _{\nu }(\tau ,{\sigma}^{^{\prime
}\, u}\} = - \sgn\, \eta _{\nu }^{\mu }\, \delta ^{3}(\sigma^u -
{\sigma}^{^{\prime} u}),  \nonumber \\
&&\nonumber\\
&&\{A_{A}(\tau ,\sigma^u),\pi ^{B}(\tau,{\sigma}^{^{\prime } u})\} =
c\, \eta _A^B\, \delta ^{3}(\sigma^u - {\sigma}^{^{\prime } u}),
\qquad \{\eta _{i}^r(\tau ),\kappa _{j s}(\tau )\} = + \delta
_{ij}\, \delta _s^r,  \nonumber \\
&&\nonumber\\
&&\{\theta _{i}(\tau ),\pi _{\theta \,j}(\tau )\}=-\delta _{ij},
\qquad \{\theta _{i}^{\ast }(\tau ),\pi _{\theta ^{\ast }\,j}(\tau
)\}=-\delta_{ij}.
 \label{3.3}
\end{eqnarray}

The Grassmann momenta give rise to the second class constraints

\bea
 &&\pi_{\theta \, i}+{\frac{i}{2}}\theta^{*}_i\approx 0,\qquad \pi_{\theta^{*}\, i}
 + {\frac{i}{2} } \theta_i\approx 0,\qquad
\lbrace \pi_{\theta \, i}+{\frac{i}{2}}\theta^{*}_i,
\pi_{\theta^{*}\, j}+{\frac{i}{2}}\theta_j\rbrace =-i\delta_{ij},
 \label{3.4}
 \eea

\noindent so that $\pi _{\theta \, i}$ and $\pi_{\theta^{*}\, i}$
can be eliminated with the help of Dirac brackets

\begin{equation}
\lbrace A,B\rbrace {}^{*}=\lbrace A,B\rbrace - i\,[\lbrace
A,\pi_{\theta \, i} + {\frac{i}{2}}\theta^{*}_i\rbrace \lbrace
\pi_{\theta^{*}\, i} + {\frac{i}{2}} \theta_i,B\rbrace + \lbrace
A,\pi_{\theta^{*}\, i} + {\frac{i}{2}} \theta_i \rbrace \lbrace
\pi_{\theta \, i} + {\frac{i}{2}}\theta^{*}_i,B\rbrace ].
 \label{3.5}
\end{equation}

\noindent As a consequence, the  Grassmann variables $\theta_i(\tau
)$, $\theta^*_i(\tau )$, have the fundamental Dirac brackets ( we
will still denote it as $\lbrace .,.\rbrace$ for the sake of
simplicity)

\beq
 \{\theta _{i}(\tau ), \theta _{j}(\tau )\} = \{\theta _{i}^{\ast
}(\tau ), \theta _{j}^{\ast }(\tau )\} = 0,  \qquad \{\theta
_{i}(\tau ), \theta_{j}^{\ast }(\tau )\} = - i\, \delta _{ij}.
 \label{3.6}
\eeq

\bigskip

If we introduce the energy-momentum tensor of the isolated system
(in inertial frames we have $T_{\perp\perp} = T^{\tau\tau}$ and
$T_{\perp r} = \delta_{rs}\, T^{\tau s}$)

\begin{eqnarray*}
 T^{AB}(\tau ,\sigma^u ) &=& - {2\over {\sqrt{g(\tau ,\sigma^u
)}}}\, {{\delta\, S}\over {\delta\, g_{AB}(\tau ,\sigma^u )}},
\nonumber \\
 &&{}\nonumber \\
 T^{\mu\nu} &=& z_A^{\mu}\, z_B^{\nu}\, T^{AB} =
  l^{\mu}\, l^{\nu}\, T_{\perp\perp} + (l^{\mu}\, z^{\nu}_r +
 l^{\nu}\, z^{\mu}_r)\, \gamma^{rs}\, T_{\perp s} + z^{\mu}_r\,
 z^{\mu}_s\, T^{rs},\nonumber \\
 &&{}\nonumber \\
 &&T_{\perp\perp} = l_{\mu}\, l_{\nu}\, T^{\mu\nu} = (1 + n)^2\,
 T^{\tau\tau},\nonumber \\
 &&T_{\perp r} = l_{\mu}\, z_{r\, \nu}\, T^{\mu\nu} = - (1 + n)\,
 h_{rs}\, (T^{\tau\tau}\, n^s + T^{\tau s}),\nonumber \\
 &&T_{rs} = z_{r\, \mu}\, z_{s\, \nu}\, T^{\mu\nu} = n_r\, n_s\,
 T^{\tau\tau} + (n_r\, h_{su} + n_s\, h_{ru})\, T^{\tau u} + h_{ru}\,
 h_{sv}\, T^{uv},
 \end{eqnarray*}

\bea
  T_{\perp \perp}(\tau ,\sigma^u ) &=& \Big({1\over {2\, c\, \sqrt{\gamma}}}\,
 \Big[{1\over {\sqrt{\gamma}}}\, h_{rs}\, \pi^r\, \pi^s + {{\sqrt{\gamma}}\over
 2}\, h^{rs}\, h^{uv}\, F_{ru}\, F_{sv}\Big]\Big)(\tau ,\sigma^u)+\nonumber\\
&&\nonumber\\
&+& \sum_{i=1}^{N}\, {{\delta^3(\sigma^u - \eta^u_i(\tau ))}\over
{\sqrt{\gamma (\tau ,\sigma^u)}}}\, \Big(\sqrt{m_i^2\, c^2 +
\h^{rs}\, \Big[\kappa _{i r}(\tau ) - {{Q_i}\over c}\, A_r\Big]\,
\Big[\kappa _{i s}(\tau) - {{Q_i}\over c}\,
A_s\Big)(\tau ,\sigma^u)\Big]},\nonumber \\
&&\nonumber\\
 T_{\perp s}(\tau ,\sigma^u ) &=& \Big({{F_{rs}\, \pi^s}\over {c\,
 \sqrt{\gamma}}}\Big)(\tau ,\vec \sigma) -
 \sum_{i=1}^{N}\, {{\delta^3(\sigma^u - \eta^u_i(\tau
))}\over {\sqrt{\gamma (\tau ,\sigma^u)}}}\, \Big[\kappa _{i\,s} -
{{Q_i}\over c}\, A_s(\tau ,\sigma^u)\Big],\nonumber \\
&&\nonumber\\
 T_{rs}(\tau ,\sigma^u ) &=& \Big(h_{ru}\, h_{sv}\, \Big[- {{\pi^u\, \pi^v}\over
 {\gamma}} + {{n^u\, n^v}\over {(1 + n)^2}}\, \Big({{n_m\, \pi^m}\over {(1 + n)\,
 \sqrt{\gamma}}}\Big)^2\Big] +\nonumber \\
 &+& {1\over 2}\, h_{rs}\, \Big[{{h_{lm}\, \pi^l\, \pi^m}\over {\gamma}} +
 {1\over 2}\, h^{lm}\, h^{uv}\, F_{lu}\, F_{mv}\Big] +
  \Big[h^{lm} - {{n^l\, n^m}\over {(1 + n)^2}}\Big]\, F_{rl}\,
 F_{sm}\Big)(\tau ,\sigma^u) +\nonumber \\
 &+& \sum_{i=1}^{N}\, {{\delta^3(\sigma^u - \eta^u_i(\tau
))}\over {\sqrt{\gamma (\tau ,\sigma^u)}}}\, \Big({{\Big[\kappa
_{i\,r} - {{Q_i}\over c}\, A_r\Big]\, \Big[\kappa _{i\,s} -
{{Q_i}\over c}\, A_s\Big] }\over {\sqrt{m_i^2\, c^2 + \h^{uv}\,
\Big[\kappa_{i u}(\tau ) - {{Q_i}\over c}\, A_u\Big]\,
\Big[\kappa_{i v}(\tau) - {{Q_i}\over c}\, A_v\Big]} }}\Big)(\tau
,\sigma^u),\nonumber \\
 &&{}
 \label{3.7}
 \eea

\noindent then from Eq.(\ref{3.2}) we get

\bea
 \rho_{\mu}(\tau ,\sigma^u) &=& \Big(\sqrt{- g}\, z_{A\, \mu}\,
 T^{\tau A}\Big)(\tau ,\sigma^u) =\nonumber \\
 &=& \Big((1 + n)^2\, \sqrt{\gamma}\, T^{\tau\tau}\, l_{\mu} +
 (1 + n)\, \sqrt{\gamma}\, \Big[T^{\tau r} + T^{\tau\tau}\, n^r\Big]\,
 z_{r\, \mu}\Big)(\tau ,\sigma^u) =\nonumber \\
 &=& \Big(\sqrt{\gamma}\, \Big[l_{\mu}\, T_{\perp\perp} - z_{r\, \mu}\,
 h^{rs}\, T_{\perp s}\Big]\Big)(\tau ,\sigma^u).
 \label{3.8}
 \eea

\bigskip

Let us remark that, since all the dependence on the embeddings is in
the 4-metric, the Euler-Lagrange equations for the embeddings
$z^{\mu}(\tau ,\sigma^u )$ associated with the Lagrangian
(\ref{3.1}) are  (the symbol '${\buildrel \circ \over =}$' means
evaluated on the solutions of the equations of motion)

\bea
 {{\delta\, S }\over {\delta\, z^{\mu}(\tau ,\sigma^u)}}
 &=&\Big( {{\partial {\cal L}}\over {\partial
z^{\mu}}}-\partial_A {{\partial {\cal L}}\over {\partial
z^{\mu}_A}}\Big) (\tau ,\sigma^u ) =
 2\, \eta_{\mu\nu}\, \partial_A\, \Big[\sqrt{-g}\, T^{AB}\,
 z_B^{\nu}\Big](\tau ,\sigma^u ) =\nonumber \\
 &=& \Big(\sqrt{-g}\, z^C_{\mu}\, g_{CD}\, T^{DA}{}_{;A}\Big)(\tau
 ,\sigma^u)\, {\buildrel \circ \over =}\, 0,
 \label{3.9}
\eea

\noindent where $T^{AB}{}_{;B}(\tau ,\sigma^u)$ is the covariant
derivative associated to the 4-metric $g_{AB}(\tau ,\sigma^u)$
induced by the admissible 3+1 splitting of Minkowski
space-time.\medskip

They may be rewritten in a form valid for every isolated system
$\Big(\partial_A\, T^{AB}\, z^{\mu}_B\Big)(\tau ,\sigma^u)\,
{\buildrel \circ \over =}\, - \Big({1\over {\sqrt{-g}}}\,
\partial_A\, [\sqrt{-g}\, z^{\mu}_B]\, T^{AB}\Big)(\tau ,\sigma^u)$.
When $\partial_A\, [\sqrt{-g}\, z^{\mu}_B](\tau ,\sigma^u) = 0$, as
it happens in inertial frames in inertial Cartesian coordinates, we
get the conservation of the energy-momentum tensor
$T^{AB}{|}_{inertial}$, i.e. $\partial_A\, T^{AB}{|}_{inertial}\,
{\buildrel \circ \over =}\, 0$. Then, after integrating over a
4-volume bounded by a 3-volume $V_1$ at $\tau_1$, a 3-volume $V_2$
at $\tau_2 > \tau_1$ and a time-like 3-surface $S_{12}$ joining them
and with section $S_{\tau}$, boundary of a 3-volume $V_{\tau}$, at
$\tau$, we get ${d\over {d\tau}}\, \int_{V_{\tau}} d^3\sigma\,
T^{A\tau}{|}_{inertial}(\tau ,\sigma^u) = - \int_{S_{\tau}}
d^2\Sigma_B\, T^{AB}{|}_{inertial}(\tau ,\sigma^u)$, namely the
time-variation of the 4-momentum contained in $V_{\tau}$ is balanced
by the flux of energy-momentum through the boundary $S_{\tau}$. For
infinite volume and suitable boundary conditions we get the
conservation of the 4-momentum $P^A = \int_{\Sigma_{\tau}}
d^3\sigma\, T^{A\tau}{|}_{inertial}(\tau ,\sigma^u)$. \medskip

Otherwise, in non-inertial frames and also in inertial frames with
non-Cartesian coordinates we do not have a real conservation law,
but the equation $T^{AB}{}_{;B}(\tau ,\sigma^u) \cir 0$, which, like
in general relativity, could be rewritten as a conservation law
$\partial_B\, \Big(T^{AB} + t^{AB}\Big)(\tau ,\sigma^u) \cir 0$
involving a coordinate-dependent energy-momentum pseudo-tensor
describing the "energy-momentum" of the foliation associated to the
3+1 splitting. Moreover a quantity as $\int_{\Sigma_{\tau}}
d^3\Sigma_B\, T^{AB}{|}_{non-inertial}(\tau ,\sigma^u)$ is not a
tensor under frame-preserving diffeomorphisms (even when
$T^{AB}_{non-inertial}$ transforms correctly as a tensor density),
so that it cannot give rise to a well defined coordinate-independent
quantity. However, differently from general relativity where the
equivalence principle says that global inertial frames do not exist,
in Minkowski space-time it is always possible to revert to inertial
frames and to find the standard 4-momentum constant of motion, which
is a 4-vector under the Poincare' transformations connecting
inertial frames.

\medskip

\bigskip

At the Hamiltonian level from Eqs.(\ref{3.2}) we obtain the
following five primary constraints

\bea
 \pi^{\tau}(\tau ,\sigma^u) &\approx& 0,\nonumber \\
 &&{}\nonumber \\
 {\cal H}_{\mu }(\tau ,\sigma^u) &=& \rho _{\mu }(\tau
,\sigma^u ) - l_{\mu }(\tau ,\sigma^u)\, \sqrt{\gamma (\tau
,\sigma^u)}\, T_{\perp \perp }(\tau ,\sigma^u) +\nonumber \\
 &+& z_{r \mu }(\tau ,\sigma^u)\, \h ^{rs}(\tau
,\sigma^u)\, \sqrt{\gamma (\tau ,\sigma^u)}\, T_{\perp s}(\tau
,\sigma^u )\approx 0,
 \label{3.10}
 \eea

\medskip
The Lorentz-scalar primary constraint $\pi^{\tau}(\tau ,\sigma^u)
\approx 0$ is a consequence of the invariance of the action under
electro-magnetic gauge transformations.
\medskip

\bigskip

The canonical Hamiltonian $H_c$ is

\bea H_{c} &=&+ \sum_{i=1}^{N}\, \kappa _{i r}(\tau )\,
{\dot{\eta}}_i^r(\tau ) + \int d^3\sigma\, \Big[ {1\over c}\, \pi
^A\, \partial _{\tau }\, A_A - \rho _{\mu }\, z_{\tau }^{\mu
} - {\cal L}\Big](\tau ,\sigma^u) =\nonumber \\
&=&{1\over c}\, \int d^3\sigma\, \Big[ \partial _r\, \Big(\pi
^r(\tau , \sigma^u)\, A_{\tau }(\tau ,\sigma^u)\Big) - A_{\tau
}(\tau ,\sigma^u)\, \Gamma (\tau ,\sigma^u)\Big] = - {1\over c}\,
\int d^3\sigma\, A_{\tau }(\tau ,\sigma^u)\, \Gamma (\tau
,\sigma^u),\nonumber \\
 &&{}
 \label{3.11}
\eea

\noindent after the elimination of a surface term and the
introduction of the quantity

\beq
 \Gamma (\tau ,\sigma^u) \equiv \partial _r\, \pi ^r(\tau
,\sigma^u) + \sum_{i=1}^{N}\, Q_i\, \delta ^3(\sigma^u -
\eta^u_i(\tau )).
 \label{3.12}
\eeq

As a consequence,  the Dirac Hamiltonian is

\beq H_{D} = \int d^3\sigma\, \Big[ \lambda ^{\mu }\, {\cal H} _{\mu
} + \mu \, \pi ^{\tau } - {1\over c}\, A_{\tau }\, \Gamma \Big](\tau
,\sigma^u).
 \label{3.13}
\eeq

Here $\lambda ^{\mu }(\tau ,\sigma^u)$ and $\mu(\tau ,\sigma^u)$ are
the Dirac multipliers associated with the primary
constraints.\medskip

The requirement that the five primary constraints be $\tau
$-independent, i.e. $\{\pi ^{\tau }(\tau ,\sigma^u), H_{D}\}\approx
0$, $\{ {\cal H}^{\mu}(\tau ,\sigma^u ), H_D \} \approx 0$, implies
only  the Gauss' law secondary constraint

\beq \Gamma (\tau ,\sigma^u)\approx 0.
 \label{3.14}
\eeq

\medskip

The 6 constraints are all first class, since they satisfy the
following Poisson brackets

\bea
 &&\{\Gamma (\tau ,\sigma^u), \pi^\tau(\tau,\sigma^{'\, u})\}
 = \{\Gamma (\tau ,\sigma^u), {\cal H}_{\mu}(\tau ,\sigma^{'\, u} )\}
= \{\pi^\tau(\tau,\sigma^u), {\cal H}_{\mu}(\tau ,\sigma^{'\, u} )\} = 0\nonumber\\
 &&\nonumber\\
 &&\lbrace {\cal H}_{\mu}(\tau ,\sigma^u ),{\cal H}_{\nu}(\tau
,\sigma^{'\, u} )\rbrace = {1\over c}\,
 \Big( [l_{\mu}\, z_{{r}\nu} - l_{\nu}\, z_{{r}\mu}]\, { {\pi^{ r}}
 \over {\sqrt{\gamma }} } -\nonumber \\
 &&\nonumber\\
 &&\qquad -z_{u\mu}\, \h^{{u}{r}}\, F_{{r}{s}}
\, \h^{{s}{v}}\, z_{ v\nu}\Big)(\tau ,\sigma^u) \, \Gamma (\tau
,\sigma^u )\, \delta^3( \sigma^u - \sigma^{'\, u})\approx 0.
 \label{3.15}
 \eea

\bigskip

The constraints $\pi^{\tau}(\tau ,\sigma^u ) \approx 0$ and $\Gamma
(\tau ,\sigma^u ) \approx 0$ are the canonical generators of the
electro-magnetic gauge transformations.\medskip

Instead the constraints ${\cal H}_{\mu}(\tau ,\sigma^u ) \approx 0$
generate the gauge transformations from an admissible 3+1 splitting
of Minkowski space-time to another one. These constraints  can be
replaced with their projections ${\cal H}_r(\tau ,\sigma^u )  =
{\cal H}_{\mu}(\tau ,\sigma^u )\, z_r^{ \mu}(\tau ,\sigma^u )
\approx 0$, ${\cal H}_{\perp}(\tau ,\sigma^u ) = {\cal H}_{\mu}(\tau
,\sigma^u )\, l^{\mu}(\tau ,\sigma^u ) \approx 0$, tangent and
normal to the instantaneous 3-space $\Sigma_{\tau}$ respectively.
Modulo the Gauss law constraint $\Gamma (\tau ,\sigma^u ) \approx
0$, the new constraints satisfy the universal Dirac algebra of the
super-hamiltonian and super-momentum constraints of canonical metric
gravity (see the first paper in Refs.\cite{11}). The gauge
transformations generated by the constraint ${\cal H}_{\perp}(\tau
,\sigma^u )$ change the instantaneous 3-spaces $\Sigma_{\tau}$ (i.e.
the clock synchronization convention), while those generated by the
constraints ${\cal H}_r(\tau ,\sigma^u )$ change the 3-coordinates
on $\Sigma_{\tau}$.

\medskip

The  Hamilton-Dirac equations are

\begin{eqnarray*}
 {{\partial\, z^{\mu}(\tau ,\sigma^u)}\over {\partial\, \tau}} &=&
 \Big((1 + n)\, l^{\mu} + n^r\, z^{\mu}_r\Big)(\tau ,\sigma^u)\,\, \cir
 \,\, - \sgn\, \lambda^{\mu}(\tau ,\sigma^u),
 \end{eqnarray*}

 \begin{eqnarray*}
 {{\partial A_{\tau}(\tau ,\sigma^u )}\over {\partial \tau}} &\cir&
 \{ A_{\tau}(\tau ,\sigma^u ), H_D\} = \mu(\tau
 ,\sigma^u ),\nonumber \\
 &&{}\nonumber \\
  {{\partial A_r(\tau ,\sigma^u )}\over {\partial \tau}} &\cir&
  \{ A_r(\tau ,\sigma^u ), H_D \} =
 - \int d^3\sigma^{'}\,\Big[\Big( \lambda_\mu\, l^\mu\,
\sqrt{\gamma}\Big)(\tau,\sigma^{'\, u})\,
\{A_r(\tau, \sigma^u), T_{\perp\perp}(\tau,\sigma^{'\, u})\} -\nonumber\\
&&\nonumber\\
&-&\Big(\lambda_\mu\, z^{\mu}_u\, h^{us}\,
\sqrt{\gamma}\Big)(\tau,\sigma^{'\, u})\,
\{A_r(\tau,\sigma^u), T_{\perp s}(\tau,\sigma^{'\, u})\} +\nonumber\\
&&\nonumber\\
  &+& {1\over c}\, A_{\tau}(\tau ,\sigma^{'\, u})\, \{ A_r(\tau
  ,\sigma^u  ), \Gamma (\tau ,\sigma^{'\, u})\} \Big],
  \end{eqnarray*}

\begin{eqnarray*}
 {{\partial \pi^r(\tau ,\sigma^u )}\over {\partial \tau}} &\cir&
  \{ \pi^r(\tau ,\sigma^u ), H_D \} =
 - \int d^3\sigma^{'}\,\Big[\Big( \lambda_\mu\, l^\mu\,
\sqrt{\gamma}\Big)(\tau,\sigma^{'\, u})\,
\{\pi^r(\tau,\sigma^u), T_{\perp\perp}(\tau,\sigma^{'\, u})\} -\nonumber\\
&&\nonumber\\
&-&\Big(\lambda_\mu\,z^{\mu}_u\, h^{us}\, \sqrt{\gamma}
\Big)(\tau,\sigma^{'\, u})\, \{\pi^r(\tau,\sigma^u), T_{\perp
s}(\tau,\sigma^{'\, u})\}\Big],
 \end{eqnarray*}

\bea
 \frac{d\eta^r_i(\tau)}{d\tau}&\cir&
  \{ \eta_i^r(\tau ), H_D \} =
 - \int d^3\sigma^{'}\,\Big[\Big( \lambda_\mu\, l^{mu}\,
\sqrt{\gamma}\Big)(\tau,\sigma^{'\, u})\,
\{\eta^r_i(\tau), T_{\perp\perp}(\tau,\sigma^{'\, u})\} -\nonumber\\
&&\nonumber\\
&-&\Big(\lambda_\mu\,z^{\mu}_u\, h^{us}\, \sqrt{\gamma}
\Big)(\tau,\sigma^{'\, u})\,
\{\eta_i^r(\tau), T_{\perp s}(\tau,\sigma^{'\, u})\},\nonumber\\
&&\nonumber\\
\frac{d\kappa_{ir}(\tau)}{d\tau}&\cir&
  \{ \kappa_{ir}(\tau), H_D \} =
 - \int d^3\sigma^{'}\,\Big[\Big( \lambda_\mu\, l^{mu}\,
\sqrt{\gamma} \Big)(\tau,\sigma^{'\, u})\,
\{\kappa_{ir}(\tau), T_{\perp\perp}(\tau,\sigma^{'\, u})\} -\nonumber\\
&&\nonumber\\
&-&\Big(\lambda_\mu\, z^{\mu}_u\, h^{us}\, \sqrt{\gamma}
\Big)(\tau,\sigma^{'\, u})\,
\{\kappa_{ir}(\tau), T_{\perp s}(\tau,\sigma^{'\, u})\} +\nonumber\\
&&\nonumber\\
  &+& {1\over c}\, A_{\tau}(\tau ,\sigma^{'\, u})\, \{
  \kappa_{ir}(\tau ), \Gamma (\tau ,\sigma^{'\, u})\} \Big].
 \label{3.16}
 \eea
\medskip

The Grassmann-valued electric charges are constants of the motion,
${{d\, Q_i(\tau )}\over {d \tau}} \, \cir\, 0$.

Since the embedding variables $z^{\mu}(\tau ,\sigma^u )$ are the
only configuration variables with Lorentz indices, the ten conserved
generators of the Poincar\'{e} transformations are:

\beq
 P^{\mu } = \int d^{3}\sigma \rho ^{\mu }(\tau ,\sigma^u),
\qquad J^{\mu \nu } =  \int d^{3}\sigma (z^{\mu }\rho ^{\nu }-z^{\nu
}\rho ^{\mu })(\tau ,\sigma^u).
 \label{3.17}
\eeq
 \medskip

The determination of the radiation gauge of the electro-magnetic
field in non-inertial frames will be done in the next Section.

\vfill\eject

\section{The Hamiltonian Description of Charged Particles and the
Electro-Magnetic Field in Non-Inertial Frames }

In this Section we study the system of charged positive-energy
scalar particles plus the electro-magnetic field in a given
admissible non-inertial frame. Then we define the radiation gauge in
non-inertial frames.

\subsection{The Hamilton Equations in an Admissible Non-Inertial Frame.}

Let us choose an admissible 3+1 splitting of the type (\ref{2.1}) by
adding the gauge fixing constraints

\bea
 \chi (\tau ,\sigma^u) &=& z^{\mu}(\tau ,\sigma^u) - z^{\mu}_F(\tau
 ,\sigma^u) \approx 0,\nonumber \\
 &&{}\nonumber \\
 &&z^{\mu}_F(\tau ,\sigma^u) = x^{\mu}(\tau ) + F^{\mu}(\tau
 ,\sigma^u),\qquad F^{\mu}(\tau ,0) = 0,
 \label{4.1}
 \eea

\noindent to the first class constraints ${\cal H}_{\mu}(\tau
,\sigma^u) \approx 0$ of Eqs.(\ref{3.10}).\medskip

From the Hamilton-Dirac equations (\ref{3.16}) we have that the
Dirac multipliers $\lambda^{\mu}(\tau ,\sigma^u)$ in the Dirac
Hamiltonian (\ref{3.13}) take the form

\bea
 \lambda^{\mu}(\tau ,\sigma^u) &\cir& - \sgn\, \Big({\dot x}^{\mu}(\tau ) +
 {{\partial\, F^{\mu}(\tau ,\sigma^u)}\over {\partial\, \tau}}\Big) =
 - \sgn\, z^{\mu}_{F\, \tau}(\tau ,\sigma^u) =\nonumber \\
 &=& - \sgn\, \Big[(1 + n_F)\, l^{\mu}_F + n^r_F\, \partial_r\,F^{\mu}
 \Big](\tau ,\sigma^u),\nonumber \\
 &&{}\nonumber \\
 &&- \lambda_{\mu}\, l^{\mu}_F = 1 + n_F,\qquad \lambda_{\mu}\,
 z^{\mu}_{F\, s}\, h^{sr}_F = n^r_F.
 \label{4.2}
 \eea
 \medskip

${\cal H}_{\mu}(\tau ,\sigma^u) \approx 0$ and $\chi (\tau
 ,\sigma^u) \approx 0$ are second class constraints \footnote{We assume $\{
 {\cal H}_{\mu}(\tau ,\sigma^u_1), \chi (\tau ,\sigma^u_2) \} \not=
 0$ as a restriction of $F^{\mu}(\tau ,\sigma^u)$}, which eliminate
 the variables $z^{\mu}(\tau ,\vec \sigma )$ and $\rho_{\mu}(\tau
 ,\sigma^u)$. If we go to Dirac brackets, so that these constraints
 become strongly zero, the Dirac Hamiltonian does not depend any
 more upon the constraints ${\cal H}_{\mu}(\tau ,\sigma^u) \approx
 0$.\medskip

To find the new Dirac Hamiltonian $H_{D\, F}$ at the level of Dirac
brackets (still denoted $\{ .,.\}$) let us put the Dirac multiplier
(\ref{4.2}) in the Hamilton-Dirac equations (\ref{3.16}) for all the
variables ${\cal F} = A_{\tau}, A_r, \pi^r, \eta^r_i, \kappa_{ir}$
independent from the embeddings and their momenta

\bea
 {{\partial\, {\cal F}(..)}\over {\partial\, \tau}} &\cir& \{ {\cal
 F}(..), H_D \} =\nonumber \\
 &=& \int d^3\sigma\, \{ {\cal F}(..), \Big(\lambda^{\mu}\, {\cal
 H}_{\mu} + \mu\, \pi^{\tau} - {1\over c}\, A_{\tau}\, \Gamma\Big)(\tau
 ,\sigma^u) \} =\nonumber \\
 &&{}\nonumber \\
 &\cir& \int d^3\sigma\, \{ {\cal F}(..), \Big((1 + n_F)\, \sqrt{\gamma_F}\,
 T_{\perp\perp} + n^r_F\, \sqrt{\gamma_F}\, T_{\perp r} + \mu\, \pi^{\tau} -
 {1\over c}\, A_{\tau}\, \Gamma\Big)(\tau ,\sigma^u) \} =\nonumber \\
 &{\buildrel {def}\over =}& \{ {\cal F}(..), H_{D\, F} \}.
 \label{4.3}
 \eea

As a consequence the new Dirac Hamiltonian is

\bea
 H_{D\, F} &=& \int d^3\sigma\, \Big((1 + n_F)\, \sqrt{\gamma_F}\, T_{\perp\perp}
 + n^r_F\, \sqrt{\gamma_F}\, T_{\perp r} + \mu\, \pi^{\tau} - {1\over c}\, A_{\tau}\,
 \Gamma\Big)(\tau ,\sigma^u) =\nonumber \\
 &&{}\nonumber \\
 &=&\int d^3\sigma\, \Big((1 + n_F(\tau ,\sigma^u))\,
 \Big[\sqrt{\gamma_F(\tau ,\sigma^u)}\, T^{\prime}_{\perp\perp}(\tau ,\sigma^u) +\nonumber \\
 &+& \sum_i\, \delta^3(\sigma^u - \eta^u_i(\tau ))\, \Big(\sqrt{m_i^2\, c^2 +
\h^{rs}_F\, \Big(\kappa_{ir}(\tau ) - {{Q_i}\over c}\, A_r\Big)\,
\Big(\kappa_{is}(\tau ) - {{Q_i}\over c}\, A_s\Big)}\Big)(\tau ,\sigma^u )\,\Big]\nonumber\\
&&\nonumber\\
&+&n_F^r(\tau ,\sigma^u)\, \left[\, {1\over c}\, F_{rs}(\tau
,\sigma^u)\, \pi^s(\tau ,\sigma^u) - \sum_i\, \delta^3(\sigma^u -
\eta^u_i(\tau ))\, \Big(\kappa_{ir}(\tau ) - {{Q_i}\over c}\,
A_r(\tau ,\sigma^u)\Big)\,
\right]\nonumber\\
&&\nonumber\\
&+&\mu(\tau ,\sigma^u)\, \pi^\tau(\tau ,\sigma^u) - {1\over c}\,
A_\tau(\tau ,\sigma^u)\, \Gamma(\tau ,\sigma^u) \Big),
 \label{4.4}
 \eea

\noindent where the energy-momentum tensor is evaluated at
$z^{\mu}(\tau ,\sigma^u) = z^{\mu}_F(\tau ,\sigma^u)$

 \bea
  \Big(\sqrt{\gamma_F}\, T^{\prime}_{\perp\perp}\Big)(\tau ,\sigma^u) &=&
\frac{1}{2c}\, \Big(\frac{1}{\sqrt{\gamma_F(\tau ,\sigma^u)}}\,
\h_{F\,rs}(\tau ,\sigma^u) \pi^r(\tau ,\sigma^u)\,
\pi^s(\tau ,\sigma^u) +\nonumber \\
&+& \frac{\sqrt{\gamma_F(\tau ,\sigma^u)}}{2}\, \h^{rs}_F(\tau
,\sigma^u)\, h^{uv}_F(\tau ,\sigma^u)\, F_{ru}(\tau ,\sigma^u)\,
F_{sv}(\tau ,\sigma^u)\Big).
 \label{4.5}
 \eea

\bigskip

The Hamilton-Dirac equations for the particle positions take the
form

 \bea
  \dot{\eta}^r_i(\tau ) &\cir&
   \left(\Big(1 + n_F\Big)\,\, \frac{\h^{rs}_F\, \Big(\kappa_{is}(\tau )
- {{Q_i}\over c}\, A_s\Big)}{\sqrt{m_i^2\, c^2 + \h^{uv}_F\,
\Big(\kappa_{iu}(\tau ) - {{Q_i}\over c}\, A_u\Big)\,
\Big(\kappa_{iv}(\tau ) - {{Q_i}\over c}\, A_v\Big)}} \right) (\tau
,\eta^u_i(\tau ))
- \nonumber \\
&-&n^r_F(\tau ,\eta^u_i(\tau )),
 \label{4.6}
  \eea

\noindent which can be inverted to get

\bea
 \kappa_{ir}(\tau ) &=& \Big(\frac{\h_{F\,rs}(\tau ,\eta^u_i(\tau ))\,
 m_ic\, \Big(\dot{\eta}^s_i(\tau ) + n_F^s\Big)}
 {\sqrt{\Big(1 + n_F\Big)^2 -
\h_{F\,uv}\, \Big(\dot{\eta}^u_i(\tau ) + n_F^u\Big)\,
\Big(\dot{\eta}^v_i(\tau ) + n_F^v\Big)}}\Big)(\tau ,\eta^u_i(\tau
)) +\nonumber \\
 &+& {{Q_i}\over c}\, A_r(\tau ,\eta^u_i(\tau )).
 \label{4.7}
 \eea

For the particle momenta we get the Hamilton-Dirac equations

 \bea
 &&\frac{d}{d\tau}\kappa_{ir}(\tau )\cir\,\,  {{Q_i}\over c}\, \dot{\eta}^u_i(\tau
)\, \frac{\partial\, A_u(\tau ,\eta^u_i(\tau ))}{\partial\eta_i^r} +
{{Q_i}\over c}\, \frac{\partial\, A_\tau(\tau ,\eta^u_i(\tau
))}{\partial \eta^r_i} + {\cal
F}_{ir}(\tau ),\nonumber\\
 &&{}\nonumber \\
 &&{}\nonumber\\
 &&{\cal F}_{ir}(\tau ) = \Big({ { m_i\, c\, \Big[1 + n_F\Big]^{-1}}\over
 {\sqrt{\Big(1 + n_F\Big)^2 - \h_{F\,uv}\,
\Big(\dot{\eta}^u_i(\tau ) + n_F^u\Big)\, \Big(\dot{\eta}^v_i(\tau )
+ n_F^v\Big)} }}\Big)(\tau ,\eta^u_i(\tau ))\nonumber \\
 &&{}\nonumber \\
 &&\Big(  \frac{\partial \h_{F\,st}(\tau ,\eta^u_i(\tau
))}{\partial \eta^r_i}\,\,  \Big(\dot{\eta}^s_i(\tau ) + n_F^s(\tau
,\eta^u_i(\tau ))\Big)\, \Big(\dot{\eta}^t_i(\tau ) + n_F^t(\tau
,\eta^u_i(\tau ))\Big) - \frac{\partial n_F(\tau
,\eta^u_i(\tau ))}{\partial \eta^r_i} +\nonumber\\
 &&\nonumber\\
 &+& \frac{\partial n_F^s(\tau ,\eta^u_i(\tau ))}{\partial
\eta^r_i}\, \h_{F\,st}(\tau ,\eta^u_i(\tau ))\,\,
\Big(\dot{\eta}^t_i(\tau ) + n_F^t(\tau ,\eta^u_i(\tau ))\Big)
\,\,\Big),
 \label{4.8}
 \eea

\noindent where ${\cal F}_{ir}(\tau )$ denotes a set of {\it
relativistic inertial forces}.

As a consequence, the second order form of the particle equations of
motion implied by Eqs. (\ref{4.7}) and (\ref{4.8}) is

\begin{eqnarray*}
 &&\frac{d}{d\tau}\left(
\frac{ \h_{F\,rs}\,\, m_i\, c\, \Big(\dot{\eta}^s_i(\tau ) +
n_F^s\Big)}{\sqrt{\Big(1 + n_F\Big)^2 - \h_{F\,uv}\,
\Big(\dot{\eta}^u_i(\tau ) + n_F^u\Big)\, \Big(\dot{\eta}^v_i(\tau )
+ n_F^v\Big)}} \right)(\tau ,\eta^u_i(\tau ))\,\on\nonumber \\
 &&\cir\,\,  {{Q_i}\over c}\, \left[\dot{\eta}^u_i(\tau )\,\left( \frac{\partial
A_u(\tau ,\eta^u_i(\tau ))}{\partial\eta_i^r} - \frac{\partial
A_r(\tau ,\eta^u_i(\tau ))}{\partial\eta_i^u}\right) + \left(
\frac{\partial A_\tau(\tau ,\eta^U_i(\tau ))}{\partial \eta^r_i} -
\frac{\partial A_r(\tau ,\eta^u_i(\tau
))}{\partial\tau}\right) \right] +\nonumber\\
&&\nonumber\\
&+&{\cal F}_{ir}(\tau ),
 \end{eqnarray*}

\bea
 &&or\nonumber \\
 &&{}\nonumber \\
 &&m_i\, c\, \frac{d}{d\tau}\left(
\frac{ \dot{\eta}^s_i(\tau ) + n_F^s}{\sqrt{\Big(1 + n_F\Big)^2 -
\h_{F\,uv}\, \Big(\dot{\eta}^u_i(\tau ) + n_F^u\Big)\,
\Big(\dot{\eta}^v_i(\tau ) + n_F^v\Big)}} \right)(\tau
,\eta^u_i(\tau
))\,\cir\nonumber \\
 &&\cir\,\,  {{Q_i}\over c}\, \h_F^{sr}(\tau ,\eta^u_i(\tau ))\, \Big[
\dot{\eta}^u_i(\tau )\,\left( \frac{\partial A_u(\tau ,
\eta^u_i(\tau ))}{\partial\eta_i^r} - \frac{\partial A_r(\tau ,
\eta^u_i(\tau ))}{\partial\eta_i^u}\right) +\nonumber \\
 &+& \left( \frac{\partial A_\tau(\tau ,\eta^u_i(\tau
))}{\partial \eta^r_i} - \frac{\partial A_r(\tau ,\eta^u_i(\tau
))}{\partial\tau}\right)\Big] + {\tilde {\cal F}}^s_i(\tau ),\nonumber \\
 &&{}\nonumber \\
 &&{}\nonumber \\
 &&{\tilde {\cal F}}^s_i(\tau ) =\nonumber \\
 &=& \Big({{m_i\, c\,\, \Big(1 + n_F\Big)^{-1}\, \h_F^{sr}}\over
 {\sqrt{\Big(1 + n_F\Big)^2 -
\h_{F\,uv}\, \Big(\dot{\eta}^u_i(\tau ) + n_F^u\Big)\,
\Big(\dot{\eta}^v_i(\tau ) + n_F^v\Big)}}}\Big)(\tau ,\eta^u_i(\tau
))\nonumber \\
 &&\Big[
 \Big(\frac{\partial \h_{F\,st}(\tau ,\eta^u_i(\tau ))}{\partial
\eta^r_i}\,\,  \Big(\dot{\eta}^s_i(\tau ) + n_F^s(\tau ,
\eta^u_i(\tau ))\Big)\, \Big(\dot{\eta}^t_i(\tau ) + n_F^t(\tau
,\eta^u_i(\tau ))\Big) -\nonumber \\
 &&{}\nonumber \\
 &-& \frac{\partial n_F(\tau
,\eta^u_i(\tau ))}{\partial \eta^r_i}+
  \frac{\partial n_F^s(\tau ,\eta^u_i(\tau ))}{\partial
\eta^r_i}\, \h_{F\,st}(\tau ,\eta^u_i(\tau ))\,\,
\Big(\dot{\eta}^t_i(\tau ) + n_F^t(\tau ,\eta^u_i(\tau ))\Big)
\,\, \Big) -\nonumber \\
 &-& \Big({{\partial\, h_{F\, ru}}\over {\partial\, \tau}} + {\dot
 \eta}^v_i(\tau )\, {{\partial\, h_{F\, ru}}\over {\partial\,
 \eta^v_i}}\Big)(\tau ,\eta^u_i(\tau ))\, \Big({\dot \eta}^u_i(\tau )
 + n^u_F(\tau ,\eta^u_i(\tau ))\Big) \Big].
 \label{4.9}
 \eea
\medskip

Here ${\tilde {\cal F}}_{ir}(\tau )$ is the form of inertial forces
whose non-relativistic limit to rigid non-inertial frames is
evaluated in Subsection C.

\bigskip

If, as in Eqs.(\ref{2.20}),  we {\it define} the non-inertial
electric and magnetic fields in the form \footnote{In the {\em
inertial case} Eqs.(\ref{2.19}) and (\ref{3.2}) imply  $\pi^s\on -
\delta^{sr}\, E^r = - {\tilde E}^s$, so that the components of the
energy-momentum tensor are $T_{\tau\tau}\, \on\, \frac{1}{2c}\,
\Big( {\vec E}^2 + {\vec B}^2 \Big)$, $T_{\tau r}\, \on \, {1\over
c}\, \Big( \vec{E}\times\vec{B} \Big)_r$.}

\bea
 E_r &\byd& \left(
\frac{\partial A_\tau}{\partial \eta^r_i}-\frac{\partial
A_r}{\partial\tau}\right)
= - F_{\tau r},\nonumber\\
 &&\nonumber\\
 B_r &\byd& \frac{1}{2}\, \varepsilon_{ruv}\,F_{uv} = \epsilon_{ruv}\,
 \partial_u\, A_{\perp\, v}\,\Rightarrow\, F_{uv} = \varepsilon_{uvr}\,B_r,
 \label{4.10}
 \eea

\noindent  the homogeneous Maxwell equations, allowing the
introduction of the electro-magnetic potentials, have the standard
inertial form $\epsilon_{ruv}\, \partial_u\,B_v = 0$,
$\epsilon_{ruv}\, \partial_u\,E_v + \frac{1}{c}\, \frac{\partial\,
B_r}{\partial \tau} = 0$.

Then also Eqs.(\ref{4.9}) take the standard inertial form plus
inertial forces

 \bea
 &&\frac{d}{d\tau}\left( \frac{ \h_{F\,rs}\,\,
 m_ic\, \Big(\dot{\eta}^s_i(\tau ) + n_F^s\Big)}{\sqrt{\Big(1 + n_F\Big)^2 -
\h_{F\,uv}\, \Big(\dot{\eta}^u_i(\tau ) + n_F^u\Big)\,
\Big(\dot{\eta}^v_i(\tau ) + n_F^v\Big)}}\right)(\tau ,\eta^u_i(\tau
))  \cir \nonumber \\
 &&{}\nonumber \\
 &\cir& {{Q_i}\over c}\, \left[E_r + \epsilon_{ruv}\, {\dot \eta}^u_i(\tau ) B_v\,
 \right](\tau , \eta^u_i(\tau )) + {\cal F}_{ir}(\tau ).
 \label{4.11}
  \eea

\bigskip

The Hamilton-Dirac equations for the electro-magnetic field are

\bea
 \frac{\partial}{\partial \tau}A_\tau(\tau,\sigma^u) &\cir&
c\, \mu(\tau,\sigma^u),\nonumber \\
 \frac{\partial}{\partial
\tau}\, A_r(\tau ,\sigma^u )&\cir& \Big(\frac{\partial}{\partial
\sigma^r}\, A_\tau + \frac{1 + n_F}{\sqrt{\gamma_F}}\, \h_{F\,rs}\,
\pi^s + n_F^s\, F_{sr}\Big)(\tau ,\sigma^u ),\nonumber\\
&&\nonumber\\
\frac{\partial}{\partial \tau}\, \pi^r(\tau ,\sigma^u ) &\cir&
\sum_i\, Q_i\, \dot{\eta}_i^r(\tau )\, \delta^3(\sigma^u
- \eta^u_i(\tau )) +\nonumber\\
&+&\Big(\frac{\partial}{\partial \sigma^s}\, \Big[ (1 +
n_F)\,\sqrt{\gamma_F}\, \h_F^{rs}\, \h_F^{uv} \, F_{uv} - (n_F^s\,
\pi^r - n_F^r\, \pi^s)\Big]\Big)(\tau ,\sigma^u ).
 \label{4.12}
  \eea

Eqs.(\ref{4.12}) imply

 \bea
\pi^s(\tau, \sigma^u)&=& -\left[- \frac{\sqrt{\gamma_{F}}}{1 +
n_F}\, \h_{F}^{sr}\left( F_{\tau r} - n^v_F\,
F_{vr}\right)\right](\tau, \sigma^u)=\nonumber\\
 &&\nonumber\\
 &=& -\sqrt{-g_{F}(\tau, \sigma^u)}\, g_{F}^{\tau
A}(\tau, \sigma^u)\, g_{F}^{s B}(\tau, \sigma^u)\,F_{AB}(\tau,
\sigma^u).
 \label{4.13}
 \eea

If we introduce the charge density $\bar \rho$, the charge current
density ${\bar j}^r$ and the total charge $Q_{tot} = \sum_i\, Q_i$
on $\Sigma_{\tau}$

 \bea
  \overline{\rho}(\tau, \sigma^u)&=&
\frac{1}{\sqrt{\gamma_{F}(\tau, \sigma^u)}} \sum_{i=1}^N
Q_i\,\delta^3(\sigma^u - \eta^u_i(\tau)),\nonumber\\
 &&\nonumber\\
  \overline{J}^r(\tau, \sigma^u)&=&
\frac{1}{\sqrt{\gamma_{F}(\tau, \sigma^u)}} \sum_{i=1}^N
Q_i\,\dot{\eta}^r_i(\tau)\, \delta^3(\sigma^u - \eta^u_i(\tau)),
\nonumber \\
  &&{}\nonumber \\
  \Rightarrow&& Q_{tot} = \int d^3\sigma\, \sqrt{\gamma_{F}(\tau, \sigma^u)}\,
\overline{\rho}(\tau, \sigma^u),
 \label{4.14}
  \eea

\noindent then the last of Eqs.(\ref{4.12}) can be rewritten in form

 \bea
\frac{\partial}{\partial\sigma^r}\,\pi^r(\tau, \sigma^u)&\approx&
 - \sqrt{\gamma_{F}(\tau, \sigma^u)}\,
\overline{\rho}(\tau, \sigma^u),\nonumber\\
 &&\nonumber\\
  \frac{\partial\,\pi^r(\tau, \sigma^u)} {\partial\tau} &\,\cir\,&
\frac{\partial}{\partial\sigma^s}\,
\left[\sqrt{-g_{F}}\,\h_{F}^{sv}\,\h_{F}^{ru}\, F_{vu} - (n_F^s\,
\pi^r - n_F^r\,\pi^s)\right](\tau, \sigma^u) +\nonumber\\
 &&\nonumber\\
&+&\sqrt{\gamma_{F}(\tau, \sigma^u)}\, \overline{J}^r(\tau,
\sigma^u).
 \label{4.15}
  \eea
\medskip

If we introduce the 4-current density  $s^A(\tau ,\sigma^u)$

 \bea
s^\tau(\tau, \sigma^u)&=& \frac{1}{\sqrt{-g_{F}(\tau, \sigma^u)}}
\sum_{i=1}^N\, Q_i\,\delta^3(\sigma^u - \eta^u_i(\tau)),\nonumber\\
 &&\nonumber\\
  s^r(\tau, \sigma^u)&=&
\frac{1}{\sqrt{-g_{F}(\tau, \sigma^u)}}\, \sum_{i=1}^N\,
Q_i\,\dot{\eta}^r_i(\tau)\, \delta^3(\sigma^u - \eta^u_i(\tau)),
 \label{4.16}
  \eea

\noindent and  we use (\ref{4.13}), then Eqs.(\ref{4.15}) can be
rewritten as manifestly covariant equations for the field strengths
as in Ref.\cite{25}

 \beq
\frac{1}{\sqrt{-g_{F}(\tau, \sigma^u)}}\, \frac{\partial}{\partial
\sigma^A}\left[\sqrt{-g_{F}(\tau, \sigma^u)}\, g_{F}^{AB}(\tau,
\sigma^u)\,  g_{F}^{CD}(\tau, \sigma^u)\, F_{BD} (\tau, \sigma^u)
\right]\, \cir\, - s^C(\tau, \sigma^u).
 \label{4.17}
  \eeq

Eqs.(\ref{4.17}) imply the following continuity equation due to the
skew-symmetry of $F_{AB}$

 \bea
 &&\frac{1}{\sqrt{-g_{F}(\tau, \sigma^u)}}\,
\frac{\partial}{\partial\sigma^C}\, \left[\sqrt{-g_{F}(\tau,
\sigma^u)}\, s^C(\tau, \sigma^u)\right]\, \cir\, 0,\nonumber \\
 &&{}\nonumber \\
 &&or\nonumber \\
 &&{}\nonumber \\
 &&\frac{1}{\sqrt{\gamma_{F}(\tau, \sigma^u)}}\,
\frac{\partial}{\partial\tau}\, \left[\sqrt{\gamma_{F}(\tau,
\sigma^u)}\, \overline{\rho}(\tau, \sigma^u)\right] +
\frac{1}{\sqrt{\gamma_{F}(\tau, \sigma^u)}}\,
\frac{\partial}{\partial\sigma^r}\, \left[\sqrt{\gamma_{F}(\tau,
\sigma^u)}\,\overline{J}^r(\tau, \sigma^u)\right]\, \cir\, 0, \nonumber \\
 &&{}
 \label{4.18}
  \eea
\medskip

\noindent so that consistently we recover $\frac{d}{d\tau}\,
Q_{tot}\, \cir\, 0$.

\bigskip

See Appendix B for the expression of the Landau-Lifschitz
non-inertial electro-magnetic fields \cite{17}.

\subsection{The Radiation Gauge for the Electro-Magnetic Field in
Non-Inertial Frames.}

In Appendix C there is a general discussion about the non-covariant
decomposition of the vector potential $\vec A(\tau ,\sigma^u)$ and
its conjugate momentum $\vec \pi (\tau ,\sigma^u)$ (the electric
field) into longitudinal and transverse parts in absence of matter.
Only with this decomposition we can define a Shanmugadhasan
canonical transformation adapted to the two first class constraints
generating electro-magnetic gauge transformations and identify the
physical degrees of freedom (Dirac observables) of the
electro-magnetic field without sources. This method  identifies the
{\it radiation gauge} as the natural one from the point of view of
constraint theory. Here we extend the construction to the case in
which there are charged particles: this will allow us to find the
expression of the mutual Coulomb interaction among the charges in
non-inertial frames .

\bigskip

As in Eq.(\ref{c3}) let us introduce the non-covariant flat
Laplacian $\Delta = \sum_r\, \partial^2_r$ in the instantaneous
non-Euclidean 3-space $\Sigma_{\tau}$. We use the non-covariant
notation ${\hat \partial}^r = \delta^{rs}\, \partial_s$  relying on
the positive signature of the 3-metric $h_{F\, rs}(\tau ,\sigma^u) =
- \sgn\, g_{F\, rs}(\tau ,\sigma^u)$. Since we have:

\beq
 \Delta\, \left(-\frac{1}{4\pi}\,\frac{1}{\sqrt{\sum_u\,
(\sigma^u - \sigma^{'\, u})^2} }\right) = \delta^3(\sigma^u,
\sigma^{'\, u}),\qquad
 or\,\, {1\over {\Delta}}\, \delta^3(\sigma^u, \sigma^{'\, u})
 = -\frac{1}{4\pi}\,\frac{1}{\sqrt{\sum_u\,
(\sigma^u - \sigma^{'\, u})^2} },
 \label{4.19}
 \eeq

\noindent with $\delta^3(\sigma^u, \sigma^{'\, u})$ the delta
function for $\Sigma_{\tau}$ \footnote{The delta functions are
defined in Appendix C after Eq.(\ref{c3}).}, we can introduce the
projectors

\bea
 {\bf P}^{rs}(\sigma^u, \sigma^{'\, u}) &=& \delta^{rs}\,
\delta^3(\sigma^u ,\sigma^{'\, u}) -
 {\hat \partial}^r\, {\hat \partial}^s\, \left(-\frac{1}{4\pi}\,
 \frac{1}{\sqrt{\sum_u\, (\sigma^u - \sigma^{'\, u})^2} }\right) =
  P^{rs}_{\perp}(\sigma^u )\, \delta^3(\sigma^u
 ,\sigma^{'\, u}),\nonumber \\
 &&{}\nonumber \\
 P^{rs}_{\perp}(\sigma^u ) &=& \delta^{rs} - {{{\hat \partial}^r\,
 {\hat \partial}^s}\over {\Delta}}.
 \label{4.20}
 \eea

As a consequence the transverse part of the electro-magnetic
quantities (${\hat \partial}^r\, A_{\perp r} = \partial_r\, A_{\perp
r} = 0$, $\partial_r\, \pi^r_{\perp} = 0$) are

 \bea
 A_{\perp r}(\tau ,\sigma^u ) &=& \delta_{ru}\, \int
 d^3\sigma^{'}\, {\bf P}^{rs}(\sigma^u ,\sigma^{'\, u})\,
 A_s(\tau ,\sigma^{'\, u}) = \delta_{ru}\, P^{us}_{\perp}(
 \sigma^u )\, A_s(\tau ,\sigma^u ),\nonumber \\
 \pi^r_\perp(\tau ,\sigma^u) &=& \sum_s\,\int d^3\sigma'\,{\bf
P}^{rs}(\sigma^u, \sigma^{'\, u})\, \pi^s(\tau ,\sigma^{'\, u})\, =
\sum_s\, P^{rs}_{\perp}(\sigma^u )\, \pi^s(\tau ,\sigma^u ).
 \label{4.21}
 \eea

\bigskip

Therefore the Gauss law constraint (\ref{3.12}) implies the
following decomposition of $\pi^r(\tau ,\sigma^u )$

\beq
 \pi^r(\tau ,\sigma^u) = \pi^r_\perp(\tau ,\sigma^u) +
 {\hat \partial}^r\, \int d^3\sigma'\,\left(-\frac{1}{4\pi}\,\frac{1}{
 \sqrt{\sum_u\, (\sigma^u - \sigma^{'\, u})^2}}\right)\,
 \left(\Gamma(\tau ,\sigma^{'\, u}) - \sum_i\, Q_i\,
\delta^3(\sigma^{'\, u} , \eta^u_i(\tau ))\right).
 \label{4.22}
  \eeq

If, following Dirac \cite{26}, we introduce the variable canonically
conjugate to $\Gamma (\tau ,\sigma^u )$ (it describes a Coulomb
cloud of longitudinal photons)

\bea
 \eta_{em}(\tau ,\sigma^u) &=& - \int
d^3\sigma'\, \left(-\frac{1}{4\pi}\,\frac{1}{ \sqrt{\sum_u\,
(\sigma^u - \sigma^{'\, u})^2} }\right)\,\left(\sum_r\, {\hat
\partial}^{'\, r}\, A_r(\tau ,\sigma^{'\, u})\right),\nonumber \\
&&\nonumber\\
&&\{\eta_{em}(\tau ,\sigma^u), \Gamma(\tau ,\sigma^{'\, u})\} =
\delta^3(\sigma^u, \sigma^{'\, u}),
 \label{4.23}
  \eea

\noindent we have the following non-covariant decomposition of the
vector potential

 \beq
 A_r(\tau ,\sigma^u) = A_{\perp\,r}(\tau ,\sigma^u) - \partial_r\,
 \eta_{em}(\tau ,\sigma^u).
  \label{4.24}
 \eeq

\bigskip

If we introduce the following new Coulomb-dressed momenta for the
particles

 \bea
\check{\kappa}_{ir}(\tau ) &=& \kappa_{ir}(\tau ) + {{Q_i}\over c}\,
\frac{\partial}{\partial\eta^r_i}\, \eta_{em}(\tau
,\eta^u_i(\tau )),\nonumber \\
 &&{}\nonumber \\
 \Rightarrow && \kappa_{ir}(\tau ) - {{Q_i}\over c}\, A_r(\tau ,\eta^u_i(\tau )) =
\check{\kappa}_{ir}(\tau ) - {{Q_i}\over c}\, A_{\perp\,r}(\tau
,\eta^u_i(\tau ))
 \label{4.25}
 \eea

\noindent we arrive at the following non-covariant Shanmugadhasan
canonical transformation in non-inertial frames

\bea
 &&\begin{array}{|c|c|}
\hline
&\\
A_r(\tau ,\sigma^u)&\eta^r_i(\tau )\\
&\\
\pi^r(\tau ,\sigma^u)&\kappa_{ir}(\tau )\\
&\\
\hline
\end{array}
\mapsto
\begin{array}{|cc|c|}
\hline
&&\\
A_{\perp\,r}(\tau ,\sigma^u)&\eta_{em}(\tau ,\sigma^u)&\eta^r_i(\tau )\\
&&\\
\pi^r_\perp(\tau ,\sigma^u)&\Gamma(\tau ,\sigma^u)
\approx 0&\check{\kappa}_{ir}(\tau )\\
&&\\
\hline
\end{array},\nonumber \\
 &&{}\nonumber \\
 &&\{A_{\perp r}(\tau ,\sigma^u), \pi_\perp^s(\tau ,\sigma^{'\, u})\}
= c\, {\bf P}^{rs}(\sigma^u, \sigma^{'\, u}) = c\,
P^{rs}_{\perp}(\sigma^u )\, \delta^3(\sigma^u, \sigma^{'\, u}),
\nonumber\\
 &&\nonumber\\
 &&\{\eta^r_i(\tau ), \check{\kappa}_{is}(\tau )\} =  \delta^r_s\,\delta_{ij}.
 \label{4.26}
  \eea

The electromagnetic part of the hamiltonian (4.4) can be expressed
in terms of the new canonical variables, since we have:

\begin{eqnarray}
&&\int d^3\sigma\,\sqrt{\gamma(\tau,\sigma^u)}\left[(1 + n_F)\,
T^{\prime}_{\perp\perp}+\frac{n^r_F}{c}\,
F_{rs}\,\pi^s\right](\tau ,\sigma^u ) = \nonumber\\
&&\nonumber\\
&=& {1\over c}\, {\cal W}(\eta^u_1(\tau ),...,\eta^u_N(\tau )) +
\int d^3\sigma\, \sqrt{\gamma(\tau,\sigma^u)} \Big[(1 + n_F)\,
\check{T}_{\perp\perp} + n^r_F\, \check{T}_{\perp r}\Big](\tau
,\sigma^u)+\nonumber\\
&&\nonumber\\
&+&\frac{1}{c}\int d^3\sigma\,
a_\tau(\tau,\sigma^u)\,\Gamma(\tau,\sigma^u)+{\cal O}(\Gamma^2),
 \label{4.27}
\end{eqnarray}

\noindent where the energy-momentum tensor has the form

 \bea
 \sqrt{\gamma(\tau,\sigma^u)}\,\check{T}_{\perp\perp}(\tau ,\sigma^u)&=&
+\frac{\h_{F\,rs}(\tau ,\sigma^u)}{2\, c\, \sqrt{\gamma_F(\tau
,\sigma^u)}}\,\pi_\perp^r(\tau ,\sigma^u)\,\pi_\perp^s(\tau
,\sigma^u) +\nonumber\\
&&\nonumber\\
&+&\frac{\sqrt{\gamma_F(\tau ,\sigma^u)}}{4\, c}\, \h^{rs}_F(\tau
,\sigma^u)\, h^{uv}_F(\tau ,\sigma^u)\,
F_{ru}(\tau ,\sigma^u)\, F_{sv}(\tau ,\sigma^u),\nonumber \\
 &&{}\nonumber \\
 &&{}\nonumber \\
 \sqrt{\gamma(\tau,\sigma^u)}\,\check{T}_{\perp r}(\tau ,\sigma^u) &=& {1\over c}\,
F_{rs}(\tau ,\sigma^u)\,\pi_\perp^s(\tau ,\sigma^u).
 \label{4.28}
 \eea

In Eq.(\ref{4.27})  we have introduced the potentials ($F_{rs} =
\partial_r\, A_{\perp s} - \partial_s\, A_{\perp r}$
)

\bea
 &&{\cal W}(\eta^u_1(\tau ),...,\eta^u_N(\tau ))=\nonumber\\
&&\nonumber\\
&=&+ \int d^3\sigma\, \frac{\h_{F\,rs}(\tau ,\sigma^u)\, \Big(1 +
n_F(\tau ,\sigma^u)\Big)}{2\sqrt{\gamma_F(\tau ,\sigma^u)}}\,
\left(2\, \pi_\perp^r(\tau ,\sigma^u) + \frac{1}{4\pi}\, \sum_i\,
\frac{\partial}{\partial\sigma^r}\, \frac{Q_i}{ \sqrt{\sum_u\,
(\sigma^u - \eta_i^u(\tau ))^2} }\right)
\nonumber \\
 && \left( \frac{1}{4\pi}\, \sum_j\,
\frac{\partial}{\partial\sigma^s}\, \frac{Q_j}{ \sqrt{\sum_u\,
(\sigma^u - \eta_j^u(\tau ))^2}}\right) +\nonumber\\
&&\nonumber\\
&+&n_F^r(\tau ,\sigma^u)\, F_{rs}(\tau ,\sigma^u)\, \left(
\frac{1}{4\pi}\, \sum_j\, \frac{\partial}{\partial\sigma^s}\,
\frac{Q_j}{ \sqrt{\sum_u\, (\sigma^u - \eta_j^u(\tau ))^2} }\right),
 \label{4.29}
 \eea

\noindent and the function

\bea
 a_\tau(\tau ,\sigma^u)&=&  \int d^3\sigma'\,
 \frac{1}{4\pi\, \sqrt{\sum_u\, (\sigma^u - \sigma^{'\, u})^2}
 } \,\, \frac{\partial}{\partial\sigma^{\prime\,r}}\, \Big[
 n^s_F(\tau ,\sigma^{'\, u})\, F_{sr}(\tau ,\sigma^{'\, u})
 +\nonumber \\
 &+& \frac{(1 + n_F(\tau ,\sigma^{'\, u}))\, \h_{F\,rs}(\tau
,\sigma^{'\, })}{\sqrt{\gamma_F(\tau ,\sigma^{'\, u})}}\,
\left(\pi_\perp^s(\tau ,\sigma^{'\, u}) + \frac{1}{4\pi}\, \sum_j\,
\frac{\partial}{\partial\sigma^{\prime\,s}}\,
\frac{Q_j}{\sqrt{\sum_u\, (\sigma^u - \eta_j^u(\tau ))^2}}\right)
\Big].\nonumber\\
&&{}
 \label{4.30}
  \eea

\medskip

Then, the Dirac Hamiltonian (\ref{4.4}) has the following form in
the new variables

\bea H_{D\, F}&=&\sum_i\,\Big(1 + n_F(\tau
 ,\eta^u_i(\tau ))\Big)\times\nonumber\\
&&\nonumber\\
&&\times\sqrt{m_i^2\, c^2 + \h^{rs}_F(\tau , \eta^u_i(\tau ))\,
(\check{\kappa}_{ir}(\tau ) - {{Q_i}\over c}\, A_{\perp\,r}(\tau
,\eta^u_i(\tau )))\, (\check{\kappa}_{is}(\tau )
- {{Q_i}\over c}\, A_{\perp\,s}(\tau ,\eta^u_i(\tau )))} -\nonumber\\
&&\nonumber\\
&-&\sum_i\,n_F^r(\tau ,\eta^u_i(\tau ))\,(\check{\kappa}_{ir}(\tau )
- {{Q_i}\over c}\, A_{\perp\,r}(\tau ,\eta^u_i(\tau ))) +\nonumber\\
&&\nonumber\\
&+&{1\over c}\, {\cal W}(\eta^u_1(\tau ),...,\eta^u_N(\tau )) + \int
d^3\sigma\, \sqrt{\gamma(\tau,\sigma^u)} \Big[(1 + n_F)\,
\check{T}_{\perp\perp} + n^r_F\, \check{T}_{\perp r}\Big](\tau
,\sigma^u)\nonumber\\
&&\nonumber\\
&+&\int d^3\sigma\,\mu(\tau ,\sigma^u)\, \pi^\tau(\tau ,\sigma^u) -
{1\over c}\, \Big( A_\tau(\tau,\sigma^u) - a_\tau(\tau
,\sigma^u)\Big)\, \Gamma(\tau ,\sigma^u)\Big) + {\cal O}(\Gamma^2),
 \label{4.31}
\eea

\noindent In Eq.(\ref{4.31}) we can discard the term quadratic in
the constraint $\Gamma (\tau ,\sigma^u ) \approx 0$, because it is
strongly zero according to constraint theory: it does never
contribute to the dynamics on the constraint sub-manifold (the only
relevant region of phase space for constrained systems).\medskip

\bigskip

To get the non-covariant radiation gauge we add the gauge fixing

\beq
 \eta_{em}(\tau,\sigma^u)\approx 0,
 \label{4.32}
 \eeq

\noindent implying $A_r \approx A_{\perp r}$ due to Eq.(\ref{4.26}).
The $\tau$-constancy, $\frac{\partial\eta_{em}(\tau,
\sigma^u)}{\partial\tau} \approx 0$, of this gauge fixing, together
with the Gauss law constraint $\Gamma(\tau,\sigma^u) \approx 0$,
implies the secondary gauge fixing

\beq
 A_\tau(\tau,\sigma^u) - a_\tau(\tau,\sigma^u) \approx 0,
 \label{4.33}
 \eeq

\noindent so that we get

 \bea
&&A_\tau(\tau ,\sigma^u) \approx  \int d^3\sigma'\, \frac{1}{4\pi\,
\sqrt{\sum_u\, (\sigma^u - \sigma^{'\, u})^2} }\,\,
\frac{\partial}{\partial\sigma^{\prime\,r}}\, \Big[
n^s_F(\tau ,\sigma^{'\, u})\, F_{sr}(\tau ,\sigma^{'\, u}) +\nonumber\\
 &+&\frac{\Big(1 + n_F(\tau ,\sigma^{'\, u})\Big)\, \h_{F\,rs}(\tau
,\sigma^{'\, u})}{\sqrt{\gamma_F(\tau ,\sigma^{'\, u})}}\,
\left(\pi_\perp^s(\tau ,\sigma^{'\, u}) + \frac{1}{4\pi}\, \sum_j\,
\frac{\partial}{\partial\sigma^{\prime\,s}}\, \frac{Q_j}{
\sqrt{\sum_u\, (\sigma^{'\, u} - \eta_j^u(\tau ))^2} }\right)
\Big).\nonumber \\
 &&{}
 \label{4.34}
\eea

\bigskip

Therefore, in the radiation gauge the magnetic field of
Eqs.(\ref{2.19}) is transverse: $B_r = \epsilon_{ruv}\, \partial_u\,
A_{\perp\, v}$. But the electric field $E_r = - F_{\tau r} = -
\partial_{\tau}\, A_{\perp\, r} + \partial_r\, A_{\tau}$ is not
transverse: it has $E_{\perp\, r} = - \partial_{\tau}\, A_{\perp\,
r}$ as a transverse component. Instead the transverse quantity is
$\pi^r_{\perp}$, which coincides with $\delta^{rs}\, E_{\perp\, s}$
only in inertial frames, and whose expression in terms of the
electric and magnetic fields, determined by Eqs.(\ref{4.22}) and
(\ref{3.2}), is $ \pi^r_{\perp}(\tau ,\sigma^u) =
\Big[{{\sqrt{\gamma}}\over {1 + n}}\, h^{rs}\, (E_s -
\epsilon_{suv}\, n^u\, B_v)\Big](\tau ,\sigma^u) + {\hat
\partial}^r\, \Big(\sum_i\, {{Q_i}\over {4\pi\, \sqrt{\sum_u\,
(\sigma^u - \eta_i^u(\tau))^2} }}\Big)$.

\bigskip

The final form of the Dirac Hamiltonian in the radiation gauge
(after the elimination of the variables $\eta_{em}$, $\Gamma$,
$A_{\tau}$, $\pi^{\tau}$ by going to Dirac brackets) is

\bea
 H_{D\, F}&=&\sum_i\,\Big(1 + n_F(\tau
 ,\eta^u_i(\tau ))\Big)\times\nonumber\\
&&\nonumber\\
&&\times\sqrt{m_i^2\, c^2 + \h^{rs}_F(\tau , \eta^u_i(\tau ))\,
(\check{\kappa}_{ir}(\tau ) - {{Q_i}\over c}\, A_{\perp\,r}(\tau
,\eta^u_i(\tau )))\, (\check{\kappa}_{is}(\tau )
- {{Q_i}\over c}\, A_{\perp\,s}(\tau ,\eta^u_i(\tau )))} -\nonumber\\
&&\nonumber\\
&-&\sum_i\,n_F^r(\tau ,\eta^u_i(\tau ))\,(\check{\kappa}_{ir}(\tau )
- {{Q_i}\over c}\, A_{\perp\,r}(\tau ,\eta^u_i(\tau ))) +\nonumber\\
&&\nonumber\\
&+&{1\over c}\, {\cal W}(\eta^u_1(\tau ),...,\eta^u_N(\tau )) + \int
d^3\sigma\, \sqrt{\gamma(\tau,\sigma^u)} \Big[(1 + n_F)\,
\check{T}_{\perp \perp} + n^r_F\, \check{T}_{\perp r}\Big](\tau
,\sigma^u)
 \label{4.35}
\eea

\noindent where $\check{T}_{AB}$ is given in Eq.(\ref{4.28}). In
$H_{DF}$ the components of $g_{AB}(\tau ,\sigma^u)$ are {\it the
inertial potentials giving rise to the relativistic inertial
forces}.

\hfill

The Hamilton-Dirac equations for the particles are (${\cal
F}_{ir}(\tau )$ is defined in Eq.(\ref{4.8}))

\bea
 \dot{\eta}^r_i(\tau ) &\cir&\frac{\Big(1 + n_F(\tau ,\eta^u_i(\tau ))\Big)\,
\h^{rs}_F(\tau ,\eta^u_i(\tau ))\, \Big(\check{\kappa}_{is}(\tau ) -
{{Q_i}\over c}\, A_{\perp\,s}(\tau ,\eta^u_i(\tau
))\Big)}{\sqrt{m_i^2\, c^2 + \, h^{uv}_F(\tau ,\eta^u_i(\tau ))\,
\Big(\check{\kappa}_{iu}(\tau ) - {{Q_i}\over c}\, A_{\perp\,u}(\tau
,\eta^u_i(\tau ))\Big)\, \Big(\check{\kappa}_{iv}(\tau ) -
{{Q_i}\over c}\, A_{\perp\,v}(\tau ,\eta^u_i(\tau ))\Big)}} -\nonumber \\
&-& n^r_F(\tau ,\eta^u_i(\tau )),\nonumber \\
 &&{}\nonumber \\
 \frac{d}{d\tau}\check{\kappa}_{ir}(\tau ) &\cir&  {{Q_i}\over c}\, \dot{\eta}^u_i(\tau )\,
 \frac{\partial\, A_{\perp\,u}(\tau ,\eta^u_i(\tau ))}{\partial\, \eta_i^r} -
 {1\over c}\, \frac{\partial}{\partial \eta^r_i}{\cal
W}(\eta^u_1(\tau ),...,\eta^u_N(\tau )) + {\cal F}_{ir}(\tau ).
 \label{4.36}
  \eea

\medskip

In the second half of Eqs.(\ref{4.36}) the sum of the inertial
2-body Coulomb potentials is replaced by the non-inertial N-body
potential ${\cal W}(\eta^u_1(\tau ),...,\eta^u_N(\tau ))$ of
Eq.(\ref{4.29}), which can be shown to have the following property
due to Eq.(\ref{4.30})

 \beq
  \frac{\partial{\cal W}}{\partial\eta^r_i} = - Q_i\,
\left(\frac{\partial a_\tau}{\partial\sigma^r}\right)_{\sigma^u =
\eta^u_i} \approx - Q_i\,\left(\frac{\partial
A_\tau}{\partial\sigma^r}\right)_{\sigma^u = \eta^u_i}.
 \label{4.37}
 \eeq

In the radiation gauge the electric field of Eq.(\ref{2.19}) is $E_r
\approx - \partial_{\tau}\, A_{\perp r} + \partial_r\, A_{\tau}$.
Consistently with Eq.(\ref{4.11}) we have

\bea
 Q_i\,E_r(\tau ,\eta^u_i(\tau )) &=& - Q_i \frac{\partial
A_{\perp r}(\tau ,\eta^u_i(\tau ))}{\partial\tau} +
Q_i\left(\frac{\partial A_\tau(\tau
,\sigma^u)}{\partial\sigma^r}\right)_{\sigma^u = \eta^u_i}
\approx\nonumber \\
 &\approx& - Q_i \frac{\partial
A_{\perp r}(\tau ,\eta^u_i(\tau ))}{\partial\tau} -
\frac{\partial{\cal W}(\eta^u_1(\tau ),...,\eta^u_N(\tau
))}{\partial\eta^r_i} =\nonumber \\
 &=& Q_i\, E_{\perp r}(\tau ,\eta^u_i(\tau )) - \frac{\partial{\cal
 W}(\eta^u_1(\tau ),...,\eta^u_N(\tau
))}{\partial\eta^r_i}.
 \label{4.38}
 \eea

\bigskip

The first of Eqs.(\ref{4.36}) can be inverted to get

\bea
 \check{\kappa}_{ir}(\tau )& =&
 \Big(\frac{\h_{F\,rs}\, m_ic\,\Big(\dot{\eta}^s_i(\tau )
 + n_F^s\Big)}{\sqrt{\Big(1 + n_F\Big)^2 - \h_{F\,uv}\,
 \Big(\dot{\eta}^u_i(\tau ) + n_F^u\Big)\, \Big(\dot{\eta}^v_i(\tau )
 + n_F^v\Big)}}\Big)(\tau ,\eta^u_i(\tau ))  + \nonumber \\
 &+& {{Q_i}\over c}\, A_{\perp\,r}(\tau ,\eta^u_i(\tau )).
 \label{4.39}
  \eea

\bigskip
See the next Subsection  for its expression in a nearly
non-relativistic frame.

\bigskip

In the general case to evaluate the integral in Eq.(\ref{4.39}) we
must regularize the function $t^{rs}(\sigma^u) = \frac{1}{
\Big(\sum_u\, (\sigma^u)^2\Big)^3/2 }\,\left(\delta^{rs} - 3\,
\frac{\sigma^r\,\sigma^s}{ \Big(\sum_u\, (\sigma^u)^2\Big)
}\right)$, which is singular at $\sigma^u = 0$. By considering it as
a distribution, we must give a prescription to define the integral
$\int d^3\sigma\, t^{rs}(\sigma^u)\, f(\sigma^u)$, where
$f(\sigma^u)$ is a test function. Following Ref. \cite{27}, we
consider the sphere ${\cal S}_R$ centered in the origin and defined
by the relation $\sqrt{\sum_u\, (\sigma^u)^2} < R$ and the space
$\Omega_R$ external to it of the points such that $\sqrt{\sum_u\,
(\sigma^u)^2} \ge R$. The integral is written in the form

\beq
 \int d^3\sigma\,t^{rs}(\sigma^u)\,f(\sigma^u) =
\int_{{\cal S}_R} d^3\sigma\,t^{rs}(\sigma^u)\,f(\sigma^u)+
\int_{\Omega_R} d^3\sigma\,t^{rs}(\sigma^u)\,f(\sigma^u).
 \label{4.40}
 \eeq

The first term, containing the singularity, can be shown to have the
expression

\beq
 \lim_{R\rightarrow 0}\,\int_{{\cal S}_R}
d^3\sigma\,t^{rs}(\sigma^u)\,f(\sigma^u) =
\frac{4\pi}{3}\,\delta^{rs}\,f(0).
 \label{4.41}
 \eeq

Regarding the second term in Eq.(\ref{4.40}) we can define a
distribution $\overline{t}^{rs}(\sigma^u)$ such that the following
integral

\beq
 \lim_{R\rightarrow 0}\int_{\Omega_R}
d^3\sigma\,t^{rs}(\sigma^u)\,f(\sigma^u) = \int
d^3\sigma\,\overline{t}^{rs}(\sigma^u)\,f(\sigma^u)
 \label{4.42}
 \eeq

\noindent has no singularity in the origin. As a consequence we get

\beq
 t^{rs}(\sigma^u) = \frac{4\pi}{3}\, \delta^{rs}\, \delta^3(\sigma^u)
 + \overline{t}^{rs}(\sigma^u).
 \label{4.43}
 \eeq

\bigskip

Therefore we get

 \bea
&&{\cal W}(\eta^u_1(\tau ),...,\eta^u_N(\tau ))=\nonumber\\
&&\nonumber\\
&=&  \sum_{i\neq j}\, \int d^3\sigma\, \frac{\h_{F\,rs}(\tau
,\sigma^u)\, \Big(1 + n_F(\tau
,\sigma^u)\Big)}{2\,\sqrt{\gamma_F(\tau ,\sigma^u)}}\nonumber \\
 &&\left( \frac{1}{4\pi}\, \frac{\partial}{\partial\sigma^r}\,
\frac{Q_i}{ \sqrt{\sum_u\, (\sigma^u - \eta_i^u(\tau))^2} }\right)\,
\left( \frac{1}{4\pi}\, \frac{\partial}{\partial\sigma^s}\,
\frac{Q_j}{ \sqrt{\sum_u\, (\sigma^u - \eta_j^u(\tau))^2}
}\right) +\nonumber\\
&&\nonumber\\
&+&\int d^3\sigma\, \left[ \frac{\h_{F\,rs}\, \Big(1 + n_F\Big)}{
\sqrt{\gamma_F}}\,\pi^r_\perp + n_F^r\, F_{rs}\right](\tau
,\sigma^u)\, \left( \frac{1}{4\pi}\, \sum_j\,
\frac{\partial}{\partial\sigma^s}\, \frac{Q_j}{ \sqrt{\sum_u\,
(\sigma^u - \eta_j^u(\tau ))^2} }\right).
\nonumber \\
 &&{}
 \label{4.44}
  \eea

After some integrations by parts we get

\bea
&&{\cal W}(\eta^u_1(\tau ),...,\eta^u_N(\tau ))=\nonumber\\
&&\nonumber\\
&=&\sum_{i\neq j}\, \int d^3\sigma\, \frac{\h_{F\,rs}(\tau
,\sigma^u)\, \Big(1 + n_F(\tau ,\sigma^u)\Big)}{2\,
\sqrt{\gamma_F(\tau ,\sigma^u)}}\, \left(\frac{1}{16\pi^2}\,
\frac{Q_i\,Q_j}{ \sqrt{\sum_u\, (\sigma^u - \eta_i^u(\tau))^2}
}\right)\,t^{rs}(\sigma^u - \eta^u_j(\tau )) -\nonumber\\
&&\nonumber\\
&+&\sum_{i\neq j}\,\int d^3\sigma\,
\frac{\partial}{\partial\sigma^s}\, \left(\frac{\h_{F\,rs}(\tau
,\sigma^u)\, \Big(1 + n_F(\tau ,\sigma^u)\Big)}{2\,
\sqrt{\gamma_F(\tau ,\sigma^u)}}\right)\nonumber \\
 &&\left(\frac{1}{4\pi}\, \frac{Q_i\,Q_j}{ \sqrt{\sum_u\, (\sigma^u -
\eta_i^u(\tau))^2} }\right)\, \left(\frac{1}{4\pi}\, \frac{\sigma^r
- \eta^r_j(\tau )}
{ \Big(\sum_u\, (\sigma^u - \eta_j^u(\tau))^2\Big)^3/2}\right) -\nonumber\\
&&\nonumber\\
&-& \int d^3\sigma\, \left(\frac{1}{4\pi}\, \sum_j\, \frac{Q_j}{
\sqrt{\sum_u\, (\sigma^u - \eta_j^u(\tau))^2} }\right) \,
 \frac{\partial}{\partial\sigma^s}\, \left[ \frac{\h_{F\,rs}\,
  \Big(1 + n_F\Big)}{ \sqrt{\gamma_F}}\, \pi^r_\perp + n_F^r\,
F_{rs}\right](\tau ,\sigma^u),\nonumber \\
 &&{}
 \label{4.45}
 \eea

\noindent and then we can get the following form

 \bea
&&{\cal W}(\eta^u_1(\tau ),...,\eta^u_N(\tau )) =\nonumber\\
&&\nonumber\\
&=&\sum_{i\neq j}\, \frac{1}{12\pi}\, \sum_r\, \left(
\frac{\h_{F\,rr}(\tau ,\eta^u_j(\tau ))\, \Big(1 + n_F(\tau
,\eta_j^u(\tau ))\Big)}{2\, \sqrt{\gamma_F(\tau ,\eta^u_j(\tau ))}}
\right)\, \frac{Q_i\,Q_j}{ \sqrt{\sum_u\,
(\eta^u_j(\tau ) - \eta_i^u(\tau))^2} }+\nonumber\\
&&\nonumber\\
&+&\sum_{i\neq j}\, \int d^3\sigma\, \left( \frac{1}{4\pi}\,
\frac{Q_i\,Q_j}{ \sqrt{\sum_u\, (\sigma^u - \eta_i^u(\tau))^2}}
\right) \, \Big[ \frac{\h_{F\,rs}(\tau ,\sigma^u)\, \Big(1 +
n_F(\tau ,\sigma^u)\Big)}{2\, \sqrt{\gamma_F(\tau ,\sigma^u)}}\,
\overline{t}^{rs}(\sigma^u - \eta^u_j(\tau )) -\nonumber \\
 &+& \frac{1}{4\pi}\, \frac{\sigma^r - \eta^r_j(\tau
)}{ \Big(\sum_u\, (\sigma^u - \eta^u_j(\tau))^2\Big)^3/2 }\,
\frac{\partial}{\partial\sigma^s}\, \left(\frac{\h_{F\,rs}(\tau
,\sigma^u)\, \Big(1 + n_F(\tau ,\sigma^u)\Big)}{2\,
\sqrt{\gamma_F(\tau ,\sigma^u)}}\right)\, \Big] -\nonumber\\
&&\nonumber\\
&-&\, \int d^3\sigma\, \left(\frac{1}{4\pi}\, \sum_j\, \frac{Q_j}{
\sqrt{\sum_u\, (\sigma^u - \eta_j^u(\tau))^2} }\right) \,
 \frac{\partial}{\partial\sigma^s}\, \left[ \frac{\h_{F\,rs}\,
  \Big(1 + n_F\Big)}{ \sqrt{\gamma_F}}\, \pi^r_\perp + n_F^r\,
F_{rs}\right](\tau ,\sigma^u),\nonumber \\
 &&{}
 \label{4.46}
 \eea

\noindent which cab be checked to be explicitly symmetric in the
exchange of ${\vec \eta}_i$ with ${\vec \eta}_j$.

\bigskip

Finally the Hamilton equations for the transverse electro-magnetic
fields $A_{\perp r}$ and $\pi^r_{\perp}$ in the radiation gauge
implied by the Dirac Hamiltonian (\ref{4.35}) are

\begin{eqnarray*}
 \partial_{\tau}\, A_{\perp\, r}(\tau ,\vec \sigma ) &\cir& \{
 A_{\perp\, r}(\tau ,\vec \sigma), H_{DF} \} =\nonumber \\
 &=& \delta_{rn}\, P^{nu}_{\perp}(\vec \sigma)\, \Big[
 {{(1 + n)\, {}^3e_{(a)u}\, {}^3e_{(a)v}}\over {{}^3e}}\, \Big(\pi^v_{\perp} -
 \delta^{vm}\, \sum_i\, Q_i\, \eta_i\, {{\partial\,
 c(\vec \sigma, {\vec \eta}_i(\tau))}\over {\partial\, \sigma^m}}
 \Big) +\nonumber \\
 &+&  {\bar n}_{(a)}\,  {}^3e^v_{(a)}\, F_{vu}
 \Big](\tau ,\vec \sigma),\nonumber \\
 &&{}\nonumber \\
 \partial_{\tau}\, \pi^r_{\perp}(\tau ,\vec \sigma) &\cir& \{
 \pi^r_{\perp}(\tau ,\vec \sigma), H_{DF} \} =\nonumber \\
 &=& P^{rn}_{\perp}(\vec \sigma)\, \delta_{nm}\, \Big(
 \sum_i\, \eta_i\, Q_i\, \delta^3(\vec \sigma, {\vec
 \eta}_i(\tau))\, {}^3e^m_{(a)}(\tau ,{\vec \eta}_i(\tau))
 \nonumber \\
 &&\Big[{{(1 + n)\, {}^3e^s_{(a)}\, {\check \kappa}_{is}(\tau)}\over
 {\sqrt{m_i^2\, c^2 +
 {}^3e^r_{(a)}\, \Big({\check \kappa}_{ir}(\tau ) - {{Q_i}\over c}\,
 A_{\perp\, r}\Big)\, {}^3e^s_{(a)}\, \Big({\check \kappa}_{is}(\tau ) -
 {{Q_i}\over c}\, A_{\perp \, s}\Big)}}} -\nonumber \\
  &-& {\bar n}_{(a)}\Big](\tau ,{\vec \eta}_i(\tau)) +
 \end{eqnarray*}

  \bea
 &+&\Big[(1 + n)\, \Big({}^3e\, {}^3e^s_{(a)}\, {}^3e^v_{(b)}\,
 ({}^3e^r_{(a)}\, {}^3e^m_{(b)} - {}^3e^m_{(a)}\, {}^3e^r_{(b)})\,
 \partial_r\, F_{sv} +\nonumber \\
 &+& \partial_r\, \Big[{}^3e\,  {}^3e^s_{(a)}\, {}^3e^v_{(b)}\,
 ({}^3e^r_{(a)}\, {}^3e^m_{(b)} - {}^3e^m_{(a)}\, {}^3e^r_{(b)})\,\Big]\,
 F_{sv} \Big) +\nonumber \\
 &+& \partial_r\, n\, {}^3e\, {}^3e^s_{(a)}\, {}^3e^v_{(b)}\, ({}^3e^r_{(a)}\,
 {}^3e^m_{(b)} - {}^3e^m_{(a)}\, {}^3e^r_{(b)})\,
 F_{sv} +\nonumber \\
 &+& {\bar n}_{(a)}\, \Big({}^3e^r_{(a)}\, \partial_r\,
 \pi_{\perp}^m + \partial_r\, {}^3e^r_{(a)}\, \pi^m_{\perp} -
 \partial_r\, {}^3e^m_{(a)}\, \pi^r_{\perp} +\nonumber \\
 &+& (\partial_r\, {}^3e^r_{(a)}\, \delta^{mt} - \partial_r\,
 {}^3e^m_{(a)}\, \delta^{rt})\, \sum_i\, \eta_i\, Q_i\,
  {{\partial\, c(\vec \sigma, {\vec \eta}_i(\tau)))}\over {\partial\, \sigma^t}}
 +\nonumber \\
 &+& ({}^3e^r_{(a)}\, \delta^{mt} - {}^3e^m_{(a)}\, \delta^{rt})\,
 \sum_i\, \eta_i\, Q_i\,  {{\partial^2\, c(\vec \sigma, {\vec \eta}_i(\tau)))}
 \over {\partial\, \sigma^t\, \partial\, \sigma^r}} \Big) +\nonumber \\
 &+& \partial_r\, {\bar n}_{(a)}\, ({}^3e^r_{(a)}\, \delta^{mt} -
 {}^3e^m_{(a)}\, \delta^{rt})\, \sum_i\, \eta_i\, Q_i\,
  {{\partial\, c(\vec \sigma, {\vec \eta}_i(\tau)))}\over {\partial\, \sigma^t}}
 \Big](\tau ,\vec \sigma)\, \Big).
 \label{x}
 \eea

Here $c(\sigma^u, \sigma^{{'}\, u}) = {1\over {4\pi\, \sqrt{\sum_u\,
(\sigma^u - \sigma^{{'}\, u})^2}}}$ and, following the general
relativity notation of Ref.\cite{12}, the metric has been expressed
in terms of triads ${}^3e^r_{(a)}$ and cotriads ${}^3e_{(a)r}$ on
$\Sigma_{\tau}$ as in Eq.(\ref{2.10}): $h_{F\, rs} = \sum_a\,
{}^3e_{(a)r}\, {}^3e_{(a)s}$, $h_F^{rs} = \sum_a\, {}^3e^r_{(a)}\,
{}^3e^s_{(a)}$, $\gamma_F = {}^3e$. The shift functions of
Eq.(\ref{2.4}) are replaced by ${\bar n}_{(a)} = n^r\,
{}^3e_{(a)r}$.

\subsection{On the Non-Relativistic Limit}

Let us consider the nearly non-relativistic limit of the embedding
(\ref{2.10}) given in Eqs.(\ref{2.16}). It can be done either before
or after the choice of the radiation gauge.\medskip

\medskip

Since we have $h_{rs} = \delta_{rs} + O(c^{-2})$, we can use the
vector notation of the inertial frames for the 3-vectors: $\vec V =
\{ V_r = {\tilde V}^r \}$ (since $g_{\tau\tau} = \sgn\, \Big(1 -
\sum_r\, (n^r_F)^2\Big) + O(c^{-2}) = \sgn + O(c^{-2})$, we still
have $V^r = g^{rA}\, V_A \not= {\tilde V}^r$ for 4-vectors $V_A$).
Therefore we have $\check{\vec{\kappa}}_i = \{\check{\kappa}_i^r\}
\byd \{\check{\kappa}_{ir}\}$, $\vec E = \{ E_r = {\tilde E}^r\} +
O(c^{-2})$, $\vec B = \{ B_r = {\tilde B}^r\} + O(c^{-2})$, but
$\vec{A}_\perp = \{ A_{\perp r} = {\tilde A}^r_{\perp} \not=
A^r_{\perp}\} + O(c^{-2})$.

\hfill

In these rigidly-rotating non-inertial frames the equations of
motion (\ref{4.9}) takes the form (the Newtonian functions are
$\tilde f(t) = f(\tau = c\, t)$; $\vec \Omega (c\, t)$ has the
components $\tilde \Omega (c\, t)$ defined after Eq. (\ref{2.15}))

 \bea
m_i\, \frac{d}{dt}\, \Big[ {{d\, {\vec \eta}_i(c\, t)}\over {dt}}
 &+& \vec{v}(c\, t ) + \vec{\Omega}(c\, t )\times \vec{\eta}_i(c\, t )
\Big]\,\, \on\,\,  Q_i\, \left[\vec{E} + {1\over c}\, {{d\, {\vec
\eta}_i(c\, t)}\over {dt}} \times \vec{B}\,\right](c\, t ,{\vec
\eta}_i(c\, t )) +\nonumber \\
 &+& \vec{\cal F}_i(c\, t ),\nonumber \\
 &&{}\nonumber \\
 \vec{\cal F}_i(c\, t ) &=& - m_i\, \vec{\Omega}(c\, t ) \times
\left[ {{d\, {\vec \eta}_i(c\, t)}\over {dt}} + \vec{v}(c\, t ) +
\vec{\Omega}(c\, t ) \times \vec{\eta}_i(c\, t )\right].
 \label{4.47}
 \eea
 \medskip

As a consequence the final form of the equations of motion of the
particles is

 \bea
 m_i\, {{d^2\, \vec{\eta}_i(c\, t )}\over {dt^2}} &\on& + Q_i\, \left[\vec{E} +
 {{d\, \vec{\eta}_i(c\, t )}\over {dt}} \times \vec{B}\,\right](c\, t ,{\vec \eta}_i(c\, t ))
 + \vec{\cal F}_i^{(in)}(c\, t ),\nonumber \\
 &&{}\nonumber \\
 \vec{\cal F}_i^{(in)}(c\, t ) &=& \vec{\cal F}_i(c\, t ) +
m_i\, \frac{d}{dt}\,
\Big(\vec{v}(c\, t) + \vec{\Omega}(c\, t ) \times \vec{\eta}_i(c\, t )\Big) =\nonumber \\
 &=&- m_i\, \Big[\vec{\Omega}(c\, t ) \times \Big(\vec{\Omega}(c\, t )
\times \vec{\eta}_i(c\, t )\Big) + 2\,  \vec{\Omega}(c\, t ) \times
{{d\, \vec{\eta}_i(c\, t )}\over {dt}} + {{d\, \vec{\Omega}(c\, t
)}\over {dt}} \times \vec{\eta}_i(c\, t ) +\nonumber \\
 &+& {{d\, \vec{v}(c\, t )}\over {dt}} +
\vec{\Omega}(c\, t ) \times \vec{v}(c\, t )\Big],
 \label{4.48}
 \eea

\noindent ${\vec {\cal F}}_i^{(in)}(\tau )$ is the sum of all the
inertial forces (centrifugal, Coriolis, Jacobi, the two pieces of
the linear acceleration) present in Newtonian rigid non-inertial
frames.

\hfill

The equations of motion (\ref{4.36}), (\ref{4.29}) of the particles
in the radiation gauge become

 \bea
  m_i\, {{d^2\, \vec{\eta}_i(c\, t )}\over {dt^2}} &\on& - {{\partial}\over {\partial\,
  {\vec \eta}_i}}\, {\cal W}(\vec{\eta_1}(\tau ),...,\vec{\eta}_N(\tau ))
  + Q_i\, \left[- {1\over c}\, \frac{\partial \vec{A}_\perp}{\partial\, t}
  + {1\over c}\, {{d\, \vec{\eta}_i(c\, t )}\over {dt}} \times \vec{B}\,\right](c\, t ,{\vec
\eta}_i(c\, t ))  +\nonumber \\
  &+& \vec{\cal F}_i^{(in)}(c\, t ),
 \label{4.49}
 \eea

\noindent where the non-inertial Coulomb potential takes the form
($\tau = ct$)\footnote{ In this case from Eq.(\ref{4.30}) we get
\[
a_\tau(\tau,\vec{\sigma}) = -\left[ \sum_{k}\,
\frac{Q_k}{4\pi\mid\vec{\sigma} - \vec{\eta}_k\mid} -
{{\vec{v}}\over c} \cdot \vec{A}_\perp(\tau,\vec{\sigma}) -
{{\vec{\Omega}}\over c} \times\vec{\sigma}\cdot
\vec{A}_\perp(\tau,\vec{\sigma}) \right],
\]
}

 \bea
&&{\cal W}(\vec{\eta_1}(\tau ),...,\vec{\eta}_N(\tau )) = \nonumber\\
&&\nonumber\\
 &=& + \sum_{i>j}\, \frac{Q_i\, Q_j}{4\pi\, \mid\vec{\eta}_i(\tau )
 - \vec{\eta}_j(\tau )\mid} - \sum_i\, {{Q_i}\over c}\, \left[
\vec{v}(\tau ) \cdot \vec{A}_\perp(\tau ,\vec{\eta_i}(\tau )) +
\vec{\Omega}(\tau ) \times\vec{\eta_i}(\tau )\cdot
\vec{A}_\perp(\tau ,\vec{\eta_i}(\tau
))\right].\nonumber \\
 &&{}
 \label{4.50}
 \eea

\hfill

finally the Hamiltonian (\ref{4.35}) becomes

\bea
 \check{H}_R&=&\sum_i\, \sqrt{m_i^2\, c^2 + \Big(\check{\vec{\kappa}}_i(\tau )
 - {{Q_i}\over c}\, \vec{A}_{\perp}(\tau ,\vec{\eta}_i(\tau ))\Big)^2}
+ \sum_{i>j}\, \frac{Q_i\, Q_j}{4\pi\, c\, \mid\vec{\eta}_i(\tau ) -
\vec{\eta}_j(\tau )\mid} +\nonumber\\
 &&\nonumber\\
&+& {1\over {2 c}}\, \int d^3\sigma\, \Big( \vec{\pi}_\perp^2(\tau
,\vec{\sigma}) - \vec{A}_\perp(\tau ,\vec{\sigma}) \cdot
\left[\Delta\, \vec{A}_\perp(\tau ,\vec{\sigma})\right]\Big) +\nonumber\\
&&\nonumber\\
&-&{{\vec{v}(\tau )}\over c} \cdot
\left[\sum_i\,\check{\vec{\kappa}}_i(\tau ) - {1\over c}\, \int
d^3\sigma\, \vec{\pi}_\perp(\tau ,\vec{\sigma}) \times (\vec
\partial \times \vec{A}_\perp(\tau ,\vec{\sigma}))\right] +\nonumber\\
&&\nonumber\\
&-&{{\vec{\Omega}(\tau )}\over c} \cdot \left[\sum_i\,
\vec{\eta}_i(\tau ) \times \check{\vec{\kappa}}_i(\tau ) + \vec{\cal
J}(\tau )\right],
\nonumber \\
 &&{}\nonumber \\
  \vec{\cal J}(\tau ) &=& - {1\over c}\,  \int
d^3\sigma\, \sum_r\, \pi^r_\perp(\tau ,\vec{\sigma})\, \Big(
\vec{\sigma} \times \vec \partial\Big)\, {\tilde A}^r_\perp(\tau
,\vec{\sigma}) - \vec{A}_\perp(\tau ,\vec{\sigma}) \times
\vec{\pi}_\perp(\tau ,\vec{\sigma}),
 \label{4.51}
 \eea

\noindent where ${\vec {\cal J}}(\tau )$ is the total angular
momentum of the electro-magnetic field.
\medskip

It can be checked that this Hamiltonian generates the previous limit
of the equations of motion of the particles. In particular the first
set of Hamilton equations is

 \beq
 {1\over c}\, {{d\, \vec{\eta}_i(\tau )}\over {dt}} = \frac{\check{\vec{\kappa}}_i(\tau ) -
{{Q_i}\over c}\, \vec{A}_{\perp}(\tau ,\vec{\eta}_i(\tau
))}{\sqrt{m_i^2\, c^2 + \Big(\check{\vec{\kappa}}_i(\tau ) -
{{Q_i}\over c}\, \vec{A}_{\perp}(\tau ,\vec{\eta}_i(\tau ))\Big)^2}}
- {{\vec{v}(\tau )}\over c} -{{\vec{\Omega}(\tau )}\over c} \times
\vec{\eta}_i(\tau ).
 \label{4.52}
 \eeq

\vfill\eject

\section{The Instant Form of
Dynamics in Non-Inertial Frames and in the Inertial and Non-Inertial
Rest Frames.}

In this Section we study the problem of the separation of the
relativistic non-covariant canonical 4-center of mass of an isolated
system from the relative variables describing its dynamics. We first
recall how this problem is solved in the inertial rest-frame instant
form of dynamics \cite{1,3,4,5,8}. As said in the Introduction the
isolated system is described as a decoupled pseudo-particle
(described by the non-covariant canonical variables $\vec z$ and
$\vec h$) carrying a pole-dipole structure given by its invariant
mass and its rest spin. On each instantaneous Wigner 3-space,
centered on the inertial observer corresponding to the Fokker-Pryce
4-center of inertia, these quantities are functions of the relative
variables of the isolated system after the elimination of the
internal 3-center of mass. The double counting of the center of mass
is avoided by the presence of three pairs of second class
constraints: the rest-frame conditions, i.e. the vanishing of the
internal 3-momentum, and the vanishing of the internal
boosts.\medskip

In Subsection A we will show how to get these conditions in the
inertial rest frames starting from the embeddings (\ref{1.1}), from
the determination (\ref{3.8}) of their conjugate momenta and from
the Poincare' generators (\ref{3.17}).\medskip

In Subsection B we will extend this construction to determine the
three pairs of second class constraints in an arbitrary admissible
non-inertial frame described by the embeddings (\ref{2.1}) and
centered on an arbitrary time-like observer. Again the isolated
system can be visualized as a pole-dipole carried by the external
decoupled center of mass.

\medskip

In Subsection C we will define the special family of the {\it
non-inertial rest-frames}, centered on the inertial Fokker-Pryce
4-center of inertia, and the associated {\it non-inertial rest-frame
instant form}. They are relevant because are the only global
non-inertial frames allowed by the equivalence principle (forbidding
the existence of global inertial frames) in canonical metric and
tetrad gravity in globally hyperbolic, asymptotically flat
(asymptotically Minkowskian) space-times without super-translations,
so to have the asymptotic ADM Poincare' group \cite{11}. Also in
this case we identify the three pairs of second class constraints
eliminating the internal 3-center of mass, visualizing the isolated
system as a pole-dipole and allowing to describe the dynamics on the
instantaneous (non-Euclidean) 3-spaces only in terms of relative
variables. Then in Subsection D we show how the Hamiltonian
description of Section IV has to be modified if we take this point
of view in the description of the isolated system. We also delineate
the analogue of this procedure for the general case of Subsection B.

\subsection{The Inertial Rest-Frame Instant Form}

As said in the Introduction every configuration of an isolated
system, with associated finite Poincare' generators $P^{\mu}$,
$J^{\mu\nu}$, identifies a unique inertial frame in an intrinsic
way: the {\it inertial rest frame} whose Euclidean instantaneous
3-spaces (the Wigner 3-spaces) are orthogonal to the conserved
4-momentum $P^{\mu}$ of the configuration. The embedding
corresponding to the inertial rest frame, centered on the
Fokker-Pryce center of inertia, is given in Eq.(\ref{1.1})
\bigskip

The generators of the external realization of the Poincare' algebra
are (following footnote 10 we use only $\epsilon_{ijk}$; $M$ and
${\vec {\bar S}}$ have vanishing Poisson brackets with $\vec z$ and
$\vec h$ and we have $\{ {\bar S}^i, {\bar S}^j \} = \delta^{im}\,
\delta^{jn}\, \epsilon_{mnk}\, {\bar S}^k$)

\bea
 P^o &=& Mc\, h^{\mu},\qquad h^{\mu} = \Big(\sqrt{1 + {\vec h}^2}; \vec
 h\Big),\nonumber \\
 J^i &=& \delta^{im}\, \epsilon_{mnk}\, \Big(z^n\, h^k + {\bar S}^k\Big),\qquad K^i = -
 \sqrt{1 + {\vec h}^2}\, z^i + {{ \delta^{in}\, \epsilon_{njk}\, {\bar
 S}^j\, h^k }\over {1 + \sqrt{1 + {\vec h}^2}}},\nonumber \\
 &&{}
 \label{5.1}
 \eea
 \bigskip

\noindent while those of the unfaithful internal realization of the
Poincare' algebra determined by the energy-momentum tensor (in
inertial frames Eqs.(\ref{3.8}) imply $T_{\perp\perp} =
T^{\tau\tau}$ and $T_{\perp r} = \delta_{rs}\, T^{\tau s}$) are

\bea
 Mc &=&  \int d^3\sigma\, T^{\tau\tau}(\tau ,\sigma^u),\qquad
  {\bar S}^r = {1\over 2}\, \delta^{rs}\, \epsilon_{suv}\, \int d^3\sigma\,
  \sigma^u\, T^{\tau v}(\tau ,\sigma^u), \nonumber \\
 &&{}\nonumber \\
 {\cal P}^r &=& \int d^3\sigma\, T^{\tau r}(\tau ,\sigma^u)
 \approx 0,\qquad {\cal K}^r = - \int d^3\sigma\, \sigma^r\,
 T^{\tau\tau}(\tau ,\sigma^u) \approx 0.
 \label{5.2}
 \eea

 \medskip

The constraints ${\vec {\cal P}} \approx 0$ are the rest-frame
conditions identifying the inertial rest frame. Having chosen the
Fokker-Pryce center of inertia as origin of the 3-coordinates, the
({\it interaction-dependent}) constraints ${\vec {\cal K}} \approx
0$ are their gauge fixing: they eliminate the internal 3-center of
mass so not to have a double counting (external, internal).
Therefore the isolated system is described by the external
non-covariant 3-center of mass $\vec z$, $\vec h$, and by an {\it
internal space} of Wigner-covariant relative variables ($M$ and
${\vec {\bar S}}$ depend only upon them).\medskip

Eqs. (\ref{5.1}) and (\ref{5.2}) are obtained in the following way.
If we put the embedding (\ref{1.1}), namely $z^{\mu}(\tau ,\sigma^u)
= Y^{\mu}(0) + h^{\mu}\, \tau + \epsilon^{\mu}_r(\vec h)\, \sigma^r
= Y^{\mu}(0) + \epsilon^{\mu}_A(\vec h)\, \sigma^a$,  in
Eq.(\ref{3.8}), we get $\rho^{\mu}(\tau ,\sigma^u) \approx h^{\mu}\,
T^{\tau\tau}(\tau ,\sigma^u) + \epsilon^{\mu}_r(\vec h)\, T^{\tau
r}(\tau ,\sigma^u) = \epsilon^{\mu}_A(\vec h)\, T^{A\tau}(\tau
,\sigma^u)$. Then the first of Eqs.(\ref{3.17}) implies $P^{\mu} =
Mc\, h^{\mu}$ if $Mc = \int d^3\sigma\, T^{\tau\tau}(\tau
,\sigma^u)$ and ${\cal P}^r = \int d^3\sigma\, T^{\tau r}(\tau
,\sigma^u) \approx 0$.\medskip

The second of Eqs.(\ref{3.17}) gives $J^{\mu\nu} = \Big(Y^{\mu}(o)\,
\epsilon^{\nu}_A(\vec h) - Y^{\nu}(0)\, \epsilon^{\mu}_A(\vec
h)\Big)\, \int d^3\sigma\, T^{A\tau}(\tau ,\sigma^u) +
\epsilon^{\mu}_A(\vec h)\, \epsilon^{\nu}_B(\vec h)\, S^{AB}$ with
$S^{AB} = \int d^3\sigma\, \Big(\sigma^A\, T^{B\tau} - \sigma^B\,
T^{A\tau}\Big)(\tau ,\sigma^u)$. By using ${\cal P}^r \approx 0$ we
get $J^{\mu\nu} \approx Mc\, \Big(Y^{\mu}(0)\, h^{\nu} -
Y^{\nu}(0)\, h^{\mu}\Big) + \epsilon^{\mu}_A(\vec h)\,
\epsilon^{\nu}_B(\vec h)\, S^{AB}$ with $S^{\tau r} \approx \int
d^3\sigma\, \sigma^r\, T^{\tau\tau}(\tau ,\sigma^u)$ and $S^{rs} =
\int d^3\sigma\, \Big(\sigma^r\, T^{s\tau} - \sigma^s\,
T^{r\tau}\Big)(\tau ,\sigma^u)$. Then, by using the expression of
the Fokker-Pryce 4-center of inertia given in Eq.(2.20) of
Ref.\cite{8}, i.e. $Y^{\mu}(\tau ) = Y^{\mu}(0) + h^{\mu}\, \tau$
with $Y^{\mu}(0) = \Big(\sqrt{1 + {\vec h}^2}\, {{\vec h \cdot \vec
z}\over {Mc}};\,\, {{\vec z}\over {Mc}} + {{\vec h \cdot \vec
z}\over {Mc}}\, \vec h + {{{\vec V}_S}\over {Mc\, (1 + \sqrt{1 +
{\vec h}^2})}}\Big)$, as a function of $\tau$, $\vec z$, $\vec h$,
$Mc$ and of  ${\vec {\bar S}}$, and the expression of
$\epsilon^{\mu}_A(\vec h)$ given after Eq.(\ref{1.1}), we get:

a) $J^{ij} = z^i\, h^j - z^j\, h^i + \delta^{ir}\, \delta^{js}\,
\epsilon_{rsk}\, \int d^3\sigma\, \sigma^r\, T^{s\tau}(\tau
,\sigma^u)$, which coincides with Eq.(\ref{5.1}) if ${\vec {\bar
S}}$ has the expression given in Eq.(\ref{5.2});

b) $J^{oi} = - \sqrt{1 + {\vec h}^2}\, z^i + \delta^{in}\,
\epsilon_{njk}\, {\bar S}^j\, h^k + \epsilon^o_{\tau}(\vec h)\,
\epsilon^i_r(\vec h)\, S^{\tau r}$, which coincides with
Eq.(\ref{5.1}) if ${\cal K}^r = - S^{\tau r} \approx 0$ as in
Eqs.(\ref{5.2}).

Therefore we have $S^{AB} \approx (\delta^A_r\, \delta^B_{\tau} -
\delta^A_{\tau}\, \delta^B_r)\, {\cal K}^r + \delta^{Ar}\,
\delta^{Bs}\, \epsilon_{rsk}\, {\bar S}^k \approx \delta^{Ar}\,
\delta^{Bs}\, \epsilon_{rsk}\, {\bar S}^k $.

\bigskip

As shown in Ref.\cite{8}, the restriction of the embedding
$z^{\mu}(\tau ,\sigma^u )$ to the Wigner 3-spaces (\ref{1.1})
implies the replacement of the Dirac Hamiltonian (\ref{3.13}) with
the new one

\beq
 H_{D\, W} = Mc  + \int
 d^3\sigma\, \Big(\mu\, \pi^{\tau} - A_{\tau}\, \Gamma\Big)(\tau
 ,\sigma^u).
 \label{5.3}
 \eeq

Therefore, consistently with Eqs.(\ref{5.2}),  the effective
Hamiltonian is the invariant mass of the isolated system, whose
conserved rest spin is ${\vec {\bar S}}$. As already said, the three
pairs of second class constraints ${\vec {\cal P}} \approx 0$,
${\vec {\cal K}} \approx 0$, eliminate the internal 3-center of
mass.\medskip

\bigskip

As shown in Refs.\cite{8,9}, in the rest-frame instant form it is
possible to restrict the description of N charged positive-energy
particles plus the electro-magnetic field to the radiation gauge
(see next Section for the non-inertial case), where all the
electro-magnetic quantities are transverse. The mutual Coulomb
interaction among the particles appears in this gauge, the
Hamiltonian (\ref{5.3}) reduces to $Mc$ and we get the following
form of the internal Poincare' generators (\ref{5.2}) \footnote{In
this equation we use the notation ${\vec \kappa}_i(\tau )$ for the
Coulomb-dressed momenta ${\check {\vec \kappa}}_i(\tau ) = {\vec
\kappa}_i(\tau ) - {{\partial\, \eta_{em}(\tau ,{\vec \eta}_i(\tau
))}\over {\partial\, {\vec \eta}_i}}$ belonging to the
Shanmugadhasan canonical basis defined in Eqs.(\ref{4.26}).}

\begin{eqnarray*}
\mathcal{E}_{(int)} &=& \mathcal{P}^{\tau }_{(int)}\, c = M\, c^2 =
c\, \int d^3\sigma\, T^{\tau\tau}(\tau ,\vec \sigma ) =  \nonumber \\
&=& c\, \sum_{i=1}^{N}\, \sqrt{ m_{i}^{2}\, c^2 + \Big({\vec{
\kappa}} _i(\tau ) - {\frac{{Q_i}}{c}}\, {\vec{A }}_{\perp }(\tau
,\vec{\eta} _i(\tau ))\Big)^2} +  \nonumber \\
&+&\sum_{i\neq j}\, \frac{Q_{i}\, Q_{j}}{4\pi\, \mid
\vec{\eta}_{i}(\tau ) - \vec{\eta} _{j}(\tau )\mid } +
{\frac{1}{2}}\, \int d^{3}\sigma \, [{\vec{ \pi }}_{\perp }^{2} +
{\vec{B}}^{2}](\tau ,\vec{\sigma}) =
 \end{eqnarray*}

 \begin{eqnarray*}
 &=&c\, \sum_{i=1}^N\, \Big(\sqrt{m^2_i\, c^2 + {\vec
\kappa}^2_i(\tau )} - { \ \frac{{Q_i}}{c}}\, {\frac{{{\vec
\kappa}_i(\tau ) \cdot {\vec A} _{\perp}(\tau , {\vec \eta}_i(\tau
))}}{\sqrt{m^2_i\, c^2 + {\vec \kappa}
^2_i(\tau )}}} \Big) +  \nonumber \\
&+&\sum_{i\neq j}\, \frac{Q_{i}\, Q_{j}}{4\pi\, \mid
\vec{\eta}_{i}(\tau ) - \vec{\eta} _{j}(\tau )\mid } +
{\frac{1}{2}}\, \int d^{3}\sigma \, [{\vec{ \pi }}_{\perp }^{2} +
{\vec{B}}^{2}](\tau ,\vec{\sigma}),
 \end{eqnarray*}

\begin{eqnarray*}
 \mathcal{\vec{P}}_{(int)} &=& \int d^3\sigma\, T^{r\tau}(\tau ,\vec
\sigma ) = \sum_{i=1}^N\, {\vec{\kappa}}_i(\tau ) + {\frac{1}{c}}\,
\int d^{3}\sigma\, \lbrack {\vec{\pi}}_{\perp } \times
{\vec{B}}](\tau ,\vec{\sigma}) \approx 0,  \nonumber \\
&&{}  \nonumber \\
\mathcal{J}_{(int)}^r &=& {\bar S}^r = {\frac{1}{2}}\, \delta^{rs}\,
\epsilon_{suv}\, \int d^3\sigma\, \sigma^u\, T^{v\tau}(\tau
,\vec \sigma ) =  \nonumber \\
&=&\sum_{i=1}^{N}\,\Big(\vec{\eta}_{i}(\tau )\times
{\vec{\kappa}}_{i}(\tau ) \Big)^{r} + {\frac{1}{c}}\, \int
d^{3}\sigma (\vec{\sigma}\times \,\Big([{ \vec{\pi}}_{\perp }{\
\times }{\vec{B}}]\Big)^{r}(\tau ,\vec{\sigma}),
 \end{eqnarray*}

\bea
 \mathcal{K}_{(int)}^{r} &=&{\bar{S}}^{\tau r} = -
{\bar{S}}^{r\tau } = - \int d^3\sigma\, \sigma^r\, T^{\tau\tau}(\tau
,\vec \sigma ) =\nonumber \\
 &=& - \sum_{i=1}^{N}\, \eta^r_{i}(\tau )\, \Big( \sqrt{m^2_i\, c^2 +
{\vec \kappa}^2_i(\tau )} - {\frac{{Q_i}}{c}}\, {\frac{{{\vec
\kappa}_i(\tau ) \cdot {\vec A}_{\perp}(\tau , {\vec \eta}_i(\tau
))}}{\sqrt{m^2_i\, c^2 + {\
\vec \kappa}^2_i(\tau )}}} \Big) +  \nonumber \\
 &+& {\frac{1}{c}}\, \sum_{i=1}^{N}\, \sum_{j\not=i}^{1..N}\,
Q_{i}\, Q_{j}\, \Big[\int d^3\sigma\, {1\over {4\pi\, |\vec \sigma -
{\vec \eta}_j(\tau )|}}\, {{\partial}\over {\partial\, \sigma^r}}\,
{1\over {4\pi\, |\vec \sigma - {\vec \eta}_i(\tau )|}} +\nonumber \\
 &+&{{\eta^r_j(\tau )}\over {4\pi\, |{\vec \eta}_i(\tau ) - {\vec
\eta}_j(\tau )|}} \Big] -  \nonumber \\
 &-& {1\over c}\, \sum_{i=1}^N\, Q_{i}\, \int d^{3}\sigma\, {{{\pi}_{\perp }^{r}(\tau
,\vec{\sigma})} \over {4\pi\, |\vec{\sigma} - {\
\vec{\eta}}_{i}(\tau )|}} - {\frac{1}{2c}}\, \int d^{3}\sigma\,
\sigma ^{r}\, ({{\vec{\pi}}}_{\perp }^{2} + {{\vec{B} }} ^{2})(\tau
,\vec{\sigma}) \approx 0.
  \label{5.4}
\end{eqnarray}

\bigskip

Note that, as required by the Poincare' algebra in an instant form
of dynamics, there are interaction terms both in the internal energy
and in the internal Lorentz boosts, but not in the 3-momentum and in
the angular momentum.

\bigskip

As shown in Ref.\cite{8}, we can reconstruct the original gauge
potential ${\tilde A}_{\mu}(x)$ in the radiation gauge. It has the
following form

\beq
 {\tilde A}^{\mu}(Y^{\alpha}(\tau ) + \epsilon^{\alpha}_r(\vec h)\, \sigma^r)
= {{P^{\mu}}\over {Mc}}\, \sum_i\, {{Q_i}\over {|\vec \sigma - {\vec
\eta}_i(\tau )|}} - \epsilon^{\mu}_r(\vec h)\, A^r_{\perp}(\tau
,\sigma^u).
 \label{5.5}
 \eeq

\subsection{Amissible Non-Inertial Frames}

Let us now see whether in an arbitrary admissible non-inertial
frame, centered on an arbitrary non-inertial observer and described
by the embeddings (\ref{2.1}), we can arrive at the same picture of
an isolated system as a decoupled external canonical non-covariant
center of mass $\vec z$, $\vec h$, carrying a pole-dipole structure,
with the external Poincare' generators given by expressions like
Eqs.(\ref{5.1}) and with the dynamics described by suitable relative
variables after an appropriate elimination of the internal 3-center
of mass inside the instantaneous 3-spaces. If this is possible,
there will be a new expression for the internal invariant mass $M$,
a new effective spin ${\vec {\tilde S}}$ (supposed to satisfy the
Poisson brackets of an angular momentum and such that $J^i =
\delta^{im}\, \epsilon_{mnk}\, \Big(z^n\, h^k + {\tilde S}^k\Big)$)
and a new form of the three pairs of second class constraints
replacing the expressions given in Eqs.(\ref{5.2}) for the case of
the inertial rest frame centered on the Fokker-Pryce center of
inertia.
\bigskip

Now the embeddings (\ref{2.1}) imply the form (\ref{3.8}) for the
conjugate momenta $\rho^{\mu}(\tau ,\vec \sigma^u)$. Therefore  we
must evaluate the Poincare' generators (\ref{3.17}) by using
Eqs.(\ref{2.1}) and (\ref{3.8}). By equating the resulting
expressions with Eqs.(\ref{5.1}) we will find the new expression of
the invariant mass, of the effective spin and of the second class
constraints.
\medskip

Since the embedding (\ref{2.1}) depend on the asymptotic tetrads
$\epsilon^{\mu}_A$, we must express them in terms of the tetrads
$\epsilon^{\mu}_A(\vec h)$ determined by $P^{\mu}$ (whose expression
is given after Eq.(\ref{1.1})): $\epsilon^{\mu}_A =
\Lambda_A{}^B(\vec h)\, \epsilon^{\mu}_B(\vec h)$ with $\Lambda
(\vec h)$ a Lorentz matrix.\medskip

Then, by using Eqs.(\ref{2.1}), (\ref{3.8}) and (\ref{5.1})  the
first of Eqs.(\ref{3.17}) becomes

\bea
 P^{\mu} &=& Mc\, h^{\mu} = Mc\, \epsilon^{\mu}_{\tau}(\vec h)
 \approx \epsilon^{\mu}_A\, {\hat {\cal
 P}}^A = {\hat {\cal P}}^A\, \Lambda_A{}^B(\vec h)\,
 \epsilon^{\mu}_B(\vec h) =\nonumber \\
 &=& {\hat {\cal P}}^A\, \Big[\Lambda_A{}^{\tau}(\vec h)\, h^{\mu} +
 \Lambda_A{}^r(\vec h)\, \epsilon^{\mu}_r(\vec h)\Big],\nonumber \\
  &&{}\nonumber \\
  {\hat {\cal P}}^A &=& \int d^3\sigma\, \sqrt{\gamma (\tau ,\sigma^u)}\,
  \Big[T_{\perp\perp}\, l^A - T_{\perp s}\, h^{sr}\, \partial_r\,
  F^A\Big](\tau ,\sigma^u),
 \label{5.6}
 \eea

\noindent with $l^A(\tau ,\sigma^u)$ given in Eq.(\ref{2.7}).
\medskip

Therefore the invariant mass $M$ and the three constraints ${\tilde
{\cal P}}^r \approx 0$ replacing the rest-frame conditions are

\beq
 Mc  \approx {\hat {\cal P}}^A\, \Lambda_A{}^{\tau}(\vec h),\qquad
 {\tilde {\cal P}}^r = {\hat {\cal P}}^A\, \Lambda_A{}^r(\vec h) \approx
 0,\qquad \Rightarrow\,\, {\hat {\cal P}}^A \approx Mc\,
 \Lambda_{\tau}{}^A(\vec h).
 \label{5.7}
 \eeq

\bigskip

If we define

\bea
  {\hat S}^{AB} &=& \int d^3\sigma\, \sqrt{\gamma (\tau ,\sigma^u)}\,
 \Big[\Big(f^A(\tau ) + F^A(\tau ,\sigma^u)\Big)\, \Big(T_{\perp\perp}\, l^B -
 T_{\perp s}\, h^{sr}\, \partial_r\, F^B\Big)(\tau ,\sigma^u) -\nonumber \\
 &-& \Big(f^B(\tau ) + F^B(\tau ,\sigma^u)\Big)\, \Big(T_{\perp\perp}\,
l^A - T_{\perp s}\, h^{sr}\, \partial_r\, F^A\Big)
 (\tau ,\sigma^u)\Big] =\nonumber \\
  &=& f^A(\tau )\, {\hat {\cal P}}^B  - f^B(\tau )\, {\hat {\cal P}}^A
  + {\hat {\cal S}}^{CD}\, \Lambda_C{}^A(\vec h)\, \Lambda_D{}^B(\vec h),\nonumber \\
 &&{}\nonumber \\
 {\hat {\cal S}}^{AB} &=& \int d^3\sigma\, \sqrt{\gamma (\tau ,\sigma^u)}\,
 \Big[F^C(\tau ,\sigma^u)\, \Big(T_{\perp\perp}\, l^D -
 T_{\perp s}\, h^{sr}\, \partial_r\, F^D\Big)(\tau ,\sigma^u) -\nonumber \\
 &-& F^D(\tau ,\sigma^u)\, \Big(T_{\perp\perp}\,
l^C - T_{\perp s}\, h^{sr}\, \partial_r\, F^C\Big) (\tau
,\sigma^u)\Big]\, \Lambda_C{}^A(\vec h)\, \Lambda_D{}^B(\vec h) =\nonumber \\
  &{\buildrel {def}\over =}&
 (\delta^A_r\, \delta^B_{\tau} - \delta^A_{\tau}\,
 \delta^B_r)\, {\hat {\cal K}}^r + \delta^{Ar}\, \delta^{Bs}\,
 \epsilon_{rsk}\, {\hat S}^k,
 \label{5.8}
 \eea

\noindent then, by using Eq.(\ref{5.7}), the second of
Eqs.(\ref{3.17}) becomes

\bea
 J^{\mu\nu} &\approx& (x^{\mu}_o\, \epsilon^{\nu}_A - x^{\nu}_o\,
 \epsilon^{\mu}_A)\, {\hat {\cal P}}^A + \epsilon^{\mu}_A\, \epsilon^{\nu}_B
 \, {\hat S}^{AB} =\nonumber \\
 &=& {\hat {\cal P}}^B\, \Lambda_B{}^A(\vec h)\, \Big(x^{\mu}_o\,
 \epsilon^{\nu}_A(\vec h) - x^{\nu}_o\, \epsilon^{\mu}_A(\vec h)\Big) +
  {\hat S}^{CD}\, \Lambda_C{}^A(\vec h)\,
 \Lambda_D{}^B(\vec h)\, \epsilon^{\mu}_A(\vec h)\,
 \epsilon^{\nu}_B(\vec h) = \nonumber \\
 &&{}\nonumber \\
 &=& {\hat {\cal P}}^A\, \Lambda_A{}^D(\vec h)\, \Big[\Big(x^{\mu}_o +
 f^B(\tau)\, \Lambda_B{}^C(\vec h)\, \epsilon^{\mu}_C(\vec h)\Big)\,
 \epsilon^{\nu}_D(\vec h) -\nonumber \\
 &-& \Big(x^{\nu}_o + f^B(\tau)\, \Lambda_B{}^C(\vec h)\, \epsilon^{\nu}_C(\vec h)
 \Big)\, \epsilon^{\mu}_D(\vec h)\Big] + \epsilon^{\mu}_A(\vec h)\,
 \epsilon^{\nu}_B(\vec h)\, {\hat {\cal S}}^{AB} \approx\nonumber \\
 &\approx& Mc\, \Big[\Big(x^{\mu}_o + f^B(\tau)\, \Lambda_B{}^C(\vec h)\,
 \epsilon^{\mu}_C(\vec h)\Big)\, h^{\nu} -\nonumber \\
 &-& \Big(x^{\nu}_o + f^B(\tau)\, \Lambda_B{}^C(\vec h)\, \epsilon^{\nu}_C(\vec h)
 \Big)\, h^{\mu}\Big] + \epsilon^{\mu}_A(\vec h)\,
 \epsilon^{\nu}_B(\vec h)\, {\hat {\cal S}}^{AB}.
 \label{5.9}
 \eea
\medskip

After some algebra Eqs.(\ref{5.1}) and (\ref{5.9}) imply

\bea
 J^{ij} &=& z^i\, h^j - z^j\, h^i + \delta^{iu}\, \delta^{jv}\,
 \epsilon_{uvk}\, {\tilde S}^k \approx\nonumber \\
 &&{}\nonumber \\
 &\approx& Mc\, \Big[\Big(x^i_o + f^B(\tau )\, \Lambda_B{}^C(\vec h)\, \epsilon^i_C(\vec h)
 + {1\over {Mc}}\, \Big[\epsilon^i_r(\vec h)\, {\tilde {\cal K}}^r +
 {{\delta^{im}\, \epsilon_{mnk}\, h^n\, {\hat S}^k}\over {1 + \sqrt{1 +
 {\vec h}^2}}}\Big]\Big)\, h^j -\nonumber \\
 &-& \Big(x^j_o + f^B(\tau )\, \Lambda_B{}^C(\vec h)\, \epsilon^j_C(\vec h)
 + {1\over {Mc}}\, \Big[\epsilon^j_r(\vec h)\, {\hat {\cal K}}^r +
 {{\delta^{jm}\, \epsilon_{mnk}\, h^n\, {\hat S}^k}\over {1 + \sqrt{1 +
 {\vec h}^2}}}\Big]\Big)\, h^i \Big] +\nonumber \\
 &+& \delta^{im}\, \delta^{jn}\, \epsilon_{mnk}\, {\hat S}^k
 =\nonumber \\
 &&{}\nonumber \\
 &{\buildrel {def}\over =}& X^i\, h^j - X^j\, h^i + \delta^{iu}\, \delta^{jv}\,
 \epsilon_{uvk}\, {\hat S}^k,
 \label{5.10}
 \eea

 \bea
   J^{oi} &=& - \sqrt{1 + {\vec h}^2}\, z^i - {{\delta^{im}\, \epsilon_{mnk}\,
 h^n\, {\tilde S}^k}\over {1 + \sqrt{1 + {\vec h}^2}}} \approx \nonumber \\
 &\approx& Mc\, \Big[x^o_o + f^B(\tau )\, \Lambda_B{}^C(\vec h)\, \epsilon^o_C(\vec h)
 + {{\sum_r\, h^r\, {\hat {\cal K}}^r}\over {Mc}} \Big]\, h^i -\nonumber \\
 &-& \sqrt{1 + {\vec h}^2}\, \Big[x^i_o + f^B(\tau )\, \Lambda_B{}^C(\vec h)\,
 \epsilon^i_C(\vec h) + {1\over {Mc}}\, \Big(\epsilon^i_r(\vec h)\, {\hat {\cal K}}^r +
 {{\delta^{im}\, \epsilon_{mnk}\, h^n\, {\hat S}^k}\over {1 + \sqrt{1 +
 {\vec h}^2}}}\Big)\Big] -\nonumber \\
 &-& {{\delta^{im}\, \epsilon_{mnk}\, h^n\, {\hat S}^k}\over {1 + \sqrt{1 + {\vec h}^2}}}
 =\nonumber \\
 &&{}\nonumber \\
  &{\buildrel {def}\over =}& X^o\, h^i - \sqrt{1 + {\vec h}^2}\, X^i
  - {{\delta^{im}\, \epsilon_{mnk}\, h^n\, {\hat S}^k}\over {1 + \sqrt{1 + {\vec
  h}^2}}},
 \label{5.11}
 \eea

\noindent where in the last lines we introduced the definition of
the quantities $X^o$ and $X^i$.\medskip

This implies the reformulation of the isolated system as an external
center of mass $\vec z$, $\vec h$, plus a pole-dipole structure $M$
and ${\vec {\tilde S}}$.
\medskip

If we solve Eq.(\ref{5.11}) in $\vec z$,  we get $\vec z = \vec X -
X^o\, {{\vec h}\over {\sqrt{1 + {\vec h}^2}}} - {{({\vec {\hat S}} -
{\vec {\tilde S}}) \times \vec h}\over {\sqrt{1 + {\vec h}^2}\, (1 +
\sqrt{1 + {\vec h}^2})}}$ (we use a vector notation). If we put this
expression in Eq.(\ref{5.10}), we get the following equation:
$[({\vec {\hat S}} - {\vec {\tilde S}}) \times \vec h] \times \vec h
= \sqrt{1 + {\vec h}^2}\, (1 + \sqrt{1 + {\vec h}^2})\, ({\vec {\hat
S}} - {\vec {\tilde S}})$. It implies $({\vec {\hat S}} - {\vec
{\tilde S}}) \cdot \vec h = 0$ and then we get

\beq
 {\tilde S}^r \approx {\hat S}^r,
 \label{5.12}
 \eeq

\noindent namely the effective spin ${\vec {\tilde S}}$ is given by
${\hat S}^{rs}$ of Eqs.(\ref{5.8}).\medskip

By using Eq.(\ref{5.12}) inside Eq.(\ref{5.11}) we get three
constraints, eliminating the internal 3-center of mass and allowing
to re-express the dynamics inside the instantaneous 3-spaces only in
terms of relative variables, which are

\bea
 Mc&& \Big[x^o_o + f^B(\tau )\, \Lambda_B{}^C(\vec h)\, \epsilon^o_C(\vec h)
 + {{\sum_r\, h^r\, {\hat {\cal K}}^r}\over {Mc}} \Big]\, h^i -
  \sqrt{1 + {\vec h}^2}\, \Big[x^i_o - z^i +\nonumber \\
  &+& f^B(\tau )\, \Lambda_B{}^C(\vec h)\,
 \epsilon^i_C(\vec h) + {1\over {Mc}}\, \Big(\epsilon^i_r(\vec h)\, {\hat {\cal K}}^r +
 {{\delta^{im}\, \epsilon_{mnk}\, h^n\, {\hat S}^k}\over {1 + \sqrt{1 +
 {\vec h}^2}}}\Big)\Big] \approx 0,\nonumber \\
 &&{}\nonumber \\
 &&\Downarrow\nonumber \\
 &&{}\nonumber \\
 {\hat {\cal K}}^r &\approx& Mc\, h^r\, \Big(x^o_o + f^B(\tau )\,
 \Lambda_B{}^C(\vec h)\, \epsilon^o_C(\vec h) -
  {{\sum_u\, h^u\, \Big(x^u_o - z^u + f^B(\tau )\, \Lambda_B{}^C(\vec h)\,
 \epsilon^u_C(\vec h)\Big)}\over {1 + \sqrt{1 + {\vec h}^2}}} \Big)
 -\nonumber \\
 &-& \Big(x^r_o - z^r + f^B(\tau )\, \Lambda_B{}^C(\vec h)\,
 \epsilon^r_C(\vec h) + {{\delta^{rm}\, \epsilon_{mnk}\, h^n\, {\hat S}^k}\over
 {Mc\, (1 + \sqrt{1 + {\vec h}^2})}}\Big).
 \label{5.13}
 \eea

They replace the constraints ${\cal K}^r \approx 0$ of Subsection A.

Now we have ${\hat {\cal S}}^{AB} \approx \delta^{Ar}\,
\delta^{Bs}\, \epsilon_{rsk}\, {\hat S}^k + (\delta^A_r\,
\delta^B_{\tau} - \delta^A_{\tau}\, \delta^B_r)\, {\hat {\cal
K}}^r$.

\bigskip

Let us remark that that if we put $\Lambda_A{}^B(\vec h) =
\delta^B_A$ and $x^{\mu}_o + f^B(\tau)\, \Lambda_B{}^C(\vec h)\,
\epsilon^{\mu}_C(\vec h) = Y^{\mu}(0) + h^{\mu}\, \tau$, then we
recover the results of Subsection A for the inertial rest frame
centered on the Fokker-Pryce inertial observer.\medskip

Instead the conditions $\Lambda_A{}^B(\vec h) = \delta^B_A$ and
$f^B(\tau)\, \Lambda_B{}^C(\vec h)\, \epsilon^{\mu}_C(\vec h) =
h^{\mu}\, \tau$, identifying the inertial rest frame centered on the
inertial observer $x^{\mu}_o + h^{\mu}\, \tau$, have the constraints
${\cal K}^r \approx 0$ replaced by Eqs.(\ref{5.13}).

\bigskip

Equations of the type (\ref{5.7}), (\ref{5.12}) and (\ref{5.13})
holds not only for admissible embeddings with pure differential
rotations like the ones of Eq.(\ref{2.14}), but also for the
admissible embeddings with pure linear acceleration. If in
Eq.(\ref{2.1}) we put $F^{\tau}(\tau ,\sigma^u) = 0$, $F^r(\tau
,\sigma^u) = \sigma^r$, so that the embedding becomes $z^{\mu}(\tau
,\sigma^u) = x_o^{\mu} + \epsilon^{\mu}_{\tau}\,f^{\tau}(\tau ) +
\epsilon^{\mu}_r\, \Big(f^r(\tau ) + \sigma^r\Big)$, the
instantaneous 3-spaces are space-like hyper-planes orthogonal to
$l^{\mu} = \epsilon^{\mu}_{\tau}$ and we get $h_{rs} = \delta_{rs}$,
$1 + n(\tau ) = {\dot f}^{\tau}(\tau )$, $n_r(\tau ) = \delta_{rs}\,
{\dot f}^s(\tau )$. In the case of Eq.(\ref{2.13}), i.e. $f^r(\tau )
= 0$ and $f^{\tau}(\tau ) = f(\tau )$, we get $1 + n(\tau ) = \dot
f(\tau )$, $n_r = 0$. If $f^{\tau}(\tau ) = \tau$ and $f^r(\tau ) =
a^r = const.$, we have inertial frames centered on inertial
observers: changing $a^r$ we change the inertial observer origin of
the 3-coordinates $\sigma^r$.

\bigskip

Let us remark that the final Dirac Hamiltonian (\ref{4.35}) does not
coincide with $Mc$ due to the presence of the inertial potentials
$g_{AB}(\tau ,\sigma^u)$.

\subsection{The Non-Inertial Rest Frames}

The family of non-inertial rest frames for an isolated system
consists of all the admissible 3+1 splittings of Minkowski
space-time whose instantaneous 3-spaces $\Sigma_{\tau}$ tend to
space-like hyper-planes orthogonal to the conserved 4-momentum of
the isolated system at spatial infinity. Therefore they tend to the
Wigner 3-spaces (\ref{1.1}) of the inertial rest frame
asymptotically.\bigskip

These non-inertial frames can be centered on the external
Fokker-Pryce center of inertia like the inertial ones and  are
described by the following embeddings

\bea
 &&z^{\mu}(\tau ,\sigma^u )\approx z^{\mu}_F(\tau ,\sigma^u ) =
 Y^{\mu}(\tau ) + u^{\mu}(\vec h)\, g(\tau ,\sigma^u ) + \epsilon^{\mu}_r(\vec
 h)\, [\sigma^r + g^r(\tau ,\sigma^u )],\nonumber \\
&&\nonumber\\
&&\qquad{\rightarrow_{|\vec \sigma |\, \rightarrow \infty}}
 z^{\mu}_W(\tau ,\sigma^u ) = Y^{\mu}(\tau ) +
 \epsilon^{\mu}_r(\vec h)\, \sigma^r,\qquad
x^{\mu}(\tau ) = z^{\mu}_F(\tau , 0^u), \nonumber \\
 &&{}\nonumber \\
&&g(\tau ,0^u) = g^r(\tau ,0^u) = 0,\qquad  g(\tau ,\sigma^u )\,
\rightarrow_{|\vec \sigma | \rightarrow
 \infty}\, 0,\qquad g^r(\tau ,\sigma^u )\, \rightarrow_{|\vec \sigma | \rightarrow
 \infty}\, 0.
 \label{5.14}
 \eea

These embeddings  are a special case of Eqs.(\ref{4.1}) with
$x^{\mu}(\tau ) = Y^{\mu}(\tau )$ and $F^{\mu}(\tau ,\sigma^u ) =
\epsilon_{\tau}^{\mu}(\vec h)\, g(\tau ,\sigma^u ) +
\epsilon^{\mu}_r(\vec h)\, [\sigma^r + g^r(\tau ,\sigma^u )]\,\,$,
$\epsilon^{\mu}_{\tau}(\vec h) = h^{\mu} = {\dot Y}^{\mu}(\tau )$.

\bigskip

For the induced metric we have

\begin{eqnarray*}
 z^{\mu}_{\tau}(\tau ,\sigma^u ) &\approx& z^{\mu}_{F\,
 \tau}(\tau ,\sigma^u ) = h^{\mu}\, [1 +
 \partial_{\tau}\, g(\tau ,\sigma^u )] + \epsilon^{\mu}_r(\vec
 h)\, \partial_{\tau}\, g^r(\tau ,\sigma^u ),\nonumber \\
&&\nonumber\\
z^{\mu}_r(\tau ,\sigma^u ) &\approx& z^{\mu}_{F\, r}(\tau ,
 \sigma^u ) = h^{\mu}\, \partial_r\, g(\tau ,\sigma^u ) +
 \epsilon^{\mu}_s(\vec h)\, [\delta^s_r + \partial_r\, g^s(\tau
 ,\sigma^u )],\nonumber \\
 &&{}\nonumber \\
 \sgn\,g_{F\, \tau\tau}(\tau ,\sigma^u ) &=& [1 + \partial_{\tau}\, g(\tau
 ,\sigma^u )]^2 - \sum_r\, [\partial_{\tau}\, g^r(\tau ,\sigma^u
 )]^2 =\nonumber \\
 &=& \Big[(1 + n_F)^2 - \h_F^{rs}\, n_{F\, r}\, n_{F\, s}\Big](\tau ,\sigma^u),
 \nonumber \\
&&\nonumber\\
 \sgn\,g_{F\, \tau u}(\tau ,\sigma^u ) &=& [1 + \partial_{\tau}\,
g(\tau ,\sigma^u )]\, \partial_u\, g(\tau ,\sigma^u ) - \sum_r\,
\partial_{\tau}\, g^r(\tau ,\sigma^u )\, [\delta^r_u +
 \partial_u\, g^r(\tau ,\sigma^u )] =\nonumber \\
 &=& \Big([1 + \partial_{\tau}\, g]\, \partial_u\, g - \partial_{\tau}\, g^u
 - \sum_r\, \partial_{\tau}\, g^r\, \partial_u\, g^r\Big)(\tau
 ,\sigma^u ) = - n_{F\, u}(\tau ,\sigma^u),
 \end{eqnarray*}

\bea
\sgn\,g_{F\, uv}(\tau ,\sigma^u ) &=& - \h_{F\, uv}(\tau ,\sigma^u) =\nonumber \\
 &=& \partial_u\, g(\tau ,\sigma^u
 )\, \partial_v\, g(\tau ,\sigma^u ) - \sum_r\, [\delta^r_u +
 \partial_u\, g^r(\tau ,\sigma^u )]\, [\delta^r_v + \partial_v\,
 g^r(\tau ,\sigma^u )] =\nonumber \\
 &=& - \delta_{uv} + \Big(\partial_u\, g\, \partial_v\, g - (\partial_u\,
 g^v + \partial_v\, g^u) - \sum_r\, \partial_u\, g^r\, \partial_v\,
 g^r\Big)(\tau ,\sigma^u ),\nonumber \\
 &&{}
 \label{5.15}
 \eea

\bigskip

The admissibility conditions  of Eqs.(\ref{2.9}), plus the
requirement $1 + n_F(\tau ,\sigma^u) > 0$, can be written as
restrictions on the functions $g(\tau ,\sigma^u)$ and $g^r(\tau
,\sigma^u)$.

\bigskip

The unit normal $l_F^\mu(\tau ,\sigma^u )$ and the tangent 4-vectors
$z^\mu_{F\,r}(\tau ,\sigma^u )$ to the instantaneous 3-spaces
$\Sigma_{\tau }$ can be projected on the asymptotic tetrad
 $h^\mu = \epsilon^{\mu}_{\tau}(\vec h)$, $\epsilon^\mu_r(\vec h)$

\bea
 z^\mu_{F\,r}(\tau ,\sigma^u)&=& \Big[\partial_r\, g\, h^\mu +
 \partial_r\, g^s\,\epsilon^\mu_s(\vec h)\Big](\tau ,\sigma^u)\nonumber\\
&&\nonumber\\
 l^{\mu}_F(\tau ,\sigma^u) &=& \Big[{1\over {\sqrt{\gamma}}}\,
\epsilon^{\mu}{}_{\alpha\beta\gamma}\, z^{\alpha}_{F1}\,
z^{\beta}_{F2}\, z^{\gamma}_{F3}\Big](\tau ,\sigma^u) = \nonumber \\
 &=&{1\over {\sqrt{\gamma (\tau ,\sigma^u)}}}\, \Big[det\, (\delta^s_r
 + \partial_r\, g^s)\, h^{\mu} -\nonumber \\
 &-& \delta^{ra}\, \epsilon_{asu}\, \epsilon_{vwt}\,
\partial_v\, g\,\, \partial_w\, g^s\,\, \partial_t\, g^u\,
\epsilon^{\mu}_r(\vec h) \Big](\tau ,\sigma^u),\nonumber \\
 &&{}\nonumber \\
 1 + n_F(\tau ,\sigma^u) &=& \sgn\, z^{\mu}_{\tau}(\tau ,\sigma^u)\,
 l_{F\mu}(\tau ,\sigma^u) =\nonumber \\
 &=& {1\over {\sqrt{\gamma (\tau ,\sigma^u)}}}\, \Big[(1 +
 \partial_{\tau}\, g\, det\, (\delta^s_r + \partial_r\, g^s)
 -\nonumber \\
 &-& \partial_{\tau}\, g^r\, \epsilon_{rsu}\, \epsilon_{vwt}\,
\partial_v\, g\,\, \partial_w\, g^s\,\, \partial_t\, g^u
\Big](\tau ,\sigma^u),\nonumber \\
 &&{}\nonumber \\
  l^2_F(\tau ,\sigma^u) = \sgn,\,\, \Rightarrow\,\,
 \gamma_F(\tau ,\sigma^u) &=& \Big[\Big(det\, (\delta^s_r +
 \partial_r\, g^s)\Big)^2 -\nonumber \\
 &-& 2\, \epsilon_{vwt}\,
\partial_v\, g\,\, \partial_w\, g^s\,\, \partial_t\, g^u\,
\epsilon_{hmn}\, \partial_h\, g\,\, \partial_m\, g^s\,\,
\partial_n\, g^u \Big](\tau ,\sigma^u).\nonumber \\
 &&{}
 \label{5.16}
 \eea

 \bigskip

To define the non-inertial rest-frame instant form we must find the
form of the internal Poincare' generators replacing the ones of the
inertial rest-frame one, given in Eqs.(\ref{5.2}).\medskip

Eq.(\ref{3.8}) and  the first of Eqs.(\ref{3.17}) imply

\bea
 P^{\mu} &=& Mc\, h^{\mu} =  \int d^3\sigma\,
 \rho^{\mu}(\tau ,\sigma^u ) \approx\nonumber \\
 &\approx& h^{\mu}\, \int d^3\sigma\, \sqrt{\gamma (\tau ,\sigma^u)}\,
 \Big({{det\, (\delta^s_r + \partial_r\, g^s)}\over {\sqrt{\gamma_F}}}\,
 T_{F\, \perp\perp} -\nonumber \\
 &-& \partial_r\, g\, h_F^{rs}\,
 T_{F\, \perp s}\Big)(\tau , \sigma^u ) +\nonumber \\
 &+&  \epsilon^{\mu}_u(\vec h)\, \int d^3\sigma\, \Big(
 - {{\delta^{ua}\, \epsilon_{asr}\, \epsilon_{vwt}\,
\partial_v\, g\,\, \partial_w\, g^s\,\, \partial_t\, g^r}
\over {\sqrt{\gamma_F}}}\, T_{F\, \perp\perp} -\nonumber \\
 &-& (\delta^u_r + \partial_r\, g^u)\,
 h_F^{rs}\, T_{F\, \perp s}\Big)(\tau ,\sigma^u ) =\nonumber \\
 &{\buildrel {def}\over =}& \int d^3\sigma\, {\cal T}^{\mu}_F(\tau ,\sigma^u),
 \label{5.17}
 \eea

\noindent so that the internal mass and the rest-frame conditions
become (Eqs.(\ref{5.2}) are recovered for the inertial rest frame)

\bea
 Mc &=& \int d^3\sigma\, \Big({{det\, (\delta^s_r + \partial_r\,
 g^s)}\over {\sqrt{\gamma}}}\, T_{F\, \perp\perp} -
 \partial_r\, g\, h_F^{rs}\, T_{F\, \perp s}\Big)(\tau ,
 \sigma^u ),\nonumber \\
  &&{}\nonumber \\
  {\hat {\cal P}}^u &=& \int d^3\sigma\, \Big(
 - {{\delta^{ua}\, \epsilon_{asr}\, \epsilon_{vwt}\,
\partial_v\, g\,\, \partial_w\, g^s\,\, \partial_t\, g^r}
\over {\sqrt{\gamma_F}}}\, T_{F\, \perp\perp} -\nonumber \\
 &-& (\delta^u_r + \partial_r\, g^u)\,
 h_F^{rs}\, T_{F\, \perp s}\Big)(\tau ,\sigma^u )
 \approx 0.\nonumber \\
 &&{}
 \label{5.18}
 \eea

\bigskip

By using Eqs.(\ref{3.17}) for the angular momentum  we get
$J^{\mu\nu} \approx \int d^3\sigma\, \Big(z^{\mu}_F\, \rho_F^{\nu} -
 z_F^{\nu}\, \rho_F^{\mu}\Big)(\tau ,\sigma^u )$ with
 $\rho^{\mu}_F(\tau ,\sigma^u) = \Big[\sqrt{\gamma_F}\, \Big(T_{\perp\perp}\,
 l^{\mu}_F - T_{\perp s}\, h_F^{sr}\, z^{\mu}_{Fr}\Big)\Big](\tau
 ,\sigma^u)$, where $z^{\mu}_F$, $z^{\mu}_{Fr}$ and $l^{\mu}_F$ are
given in Eqs.(\ref{5.14}), (\ref{5.15}) and (\ref{5.16})
respectively. The description of the isolated system as a
pole-dipole carried by the external center of mass $\vec z$ requires
that we must identify the previous $J^{ij}$ and $J^{oi}$ with the
expressions like the ones given in Eqs.(\ref{5.1}), now functions of
$\vec z$, $\vec h$, $Mc$ of Eq.(\ref{5.18}) and of an effective spin
${\vec {\tilde S}}$. This identification will allow to find the
effective spin ${\vec {\tilde S}}$ and three constraints ${\tilde
{\cal K}}^r \approx 0$ eliminating the internal 3-center of mass: in
the limit of the inertial rest frame they must reproduce the
quantities in Eqs.(\ref{5.2}).\medskip

By using Eqs.(\ref{5.18}) this procedure implies (${\hat {\cal
K}}^r$ and ${\hat {\cal S}}^r$ are the analogue of the quantities
defined in Eqs.(\ref{5.8}) for the embedding (\ref{5.14}))

\begin{eqnarray*}
 J^{\mu\nu} &\approx& \int d^3\sigma\, \Big(z^{\mu}_F\, \rho_F^{\nu} -
 z_F^{\nu}\, \rho_F^{\mu}\Big)(\tau ,\sigma^u ) =\nonumber \\
 &=& Mc\, \Big(Y^{\mu}(0)\, h^{\nu} - Y^{\nu}(0)\, h^{\mu}\Big)
 + {\hat {\cal P}}^u\, \Big(Y^{\mu}(0)\, \epsilon^{\nu}_u(\vec h) -
 Y^{\nu}(0)\, \epsilon^{\mu}_u(\vec h)\Big) +\nonumber \\
 &+& \Big(\tau\, {\hat {\cal P}}^u + {\hat {\cal K}}^u\Big)\, \Big(h^{\mu}\,
 \epsilon^{\nu}_u(\vec h) - h^{\nu}\, \epsilon^{\mu}_u(\vec h)\Big)
  + \delta^{un}\, \epsilon_{nvr}\, {\hat S}^r\, \epsilon^{\mu}_u(\vec h)\,
 \epsilon^{\nu}_v(\vec h) \approx\nonumber \\
 &\approx& Mc\, \Big(Y^{\mu}(0)\, h^{\nu} - Y^{\nu}(0)\, h^{\mu}\Big)
 + {\hat {\cal K}}^u\, \Big(h^{\mu}\, \epsilon^{\nu}_u(\vec h) - h^{\nu}\,
 \epsilon^{\mu}_u(\vec h)\Big) +\nonumber \\
 &+& \delta^{un}\, \epsilon_{nvr}\, {\hat S}^r\, \epsilon^{\mu}_u(\vec h)\,
 \epsilon^{\nu}_v(\vec h),\nonumber \\
 &&{}\nonumber \\
 &&so\, that\, we\, get
 \end{eqnarray*}

 \bea
 J^{ij} &=& z^i\, h^j - z^j\, h^i + \delta^{iu}\, \delta^{jv}\,
 \epsilon_{uvk}\, {\tilde S}^k \approx\nonumber \\
 &\approx& Mc\, \Big(Y^i(0)\, h^j - Y^j(0)\, h^i\Big)
 + {\hat {\cal K}}^u\, \Big(h^i\, \epsilon^j_u(\vec h) - h^j\,
 \epsilon^i_u(\vec h)\Big) +\nonumber \\
 &+& \delta^{un}\, \epsilon_{nvr}\, {\hat S}^r\, \epsilon^i_u(\vec h)\,
 \epsilon^j_v(\vec h),\nonumber \\
 &&{}\nonumber \\
 J^{oi} &=& - \sqrt{1 + {\vec h}^2}\, z^i + {{\delta^{in}\, \epsilon_{njk}\,
 {\tilde S}^j\, h^k}\over {1 + \sqrt{1 + {\vec h}^2}}} \approx \nonumber \\
 &\approx& Mc\, \Big(Y^o(0)\, h^i - Y^i(0)\, h^o\Big)
 + {\hat {\cal K}}^u\, \Big(h^o\, \epsilon^i_u(\vec h) - h^i\,
 \epsilon^o_u(\vec h)\Big) +\nonumber \\
 &+& \delta^{un}\, \delta^{vm}\, \epsilon_{nmr}\, {\hat S}^r\, \epsilon^o_u(\vec h)\,
 \epsilon^i_v(\vec h).
 \label{5.19}
 \eea

\bigskip

As a consequence, by using the expression of $Y^{\mu}(0)$ given
after Eq.(\ref{5.2}), the constraints eliminating the 3-center of
mass and the effective spin are

\bea
 {\hat {\cal K}}^u &=& \int d^3\sigma\, \Big(g\,
 \Big[\delta^{ur}\, \partial_r\, g\, T_{F\, \perp\perp}
 - (\delta^u_r + \partial_r\, g^u)\, h_F^{rs}\, T_{F\, \perp s}\Big]
 -\nonumber \\
 &-& (\sigma^u + g^u)\, \Big[{{det\, (\delta^s_r + \partial_r\, g^s)}\over
 {\sqrt{\gamma}}}\, T_{F\, \perp\perp} - \partial_r\, g\,
 h_F^{rs}\, T_{F\, \perp s}\Big]\Big)(\tau ,\sigma^u
 ) \approx 0,\nonumber \\
 &&{}\nonumber \\
 {\tilde S}^r &\approx& {\hat S}^r
 = {1\over 2}\, \delta^{rn}\, \epsilon_{nuv}\,
 \int d^3\sigma\, \Big((\sigma^u + g^u)\, \Big[\delta^{vm}\,
 \partial_m\, g\, T_{F\, \perp\perp} - (\delta^v_r + \partial_r\, g^v)\,
 h_F^{rs}\, T_{F\, \perp s}\Big] -\nonumber \\
 &-&(\sigma^v + g^v)\, \Big[\delta^{um}\, \partial_m\, g\, T_{F\,
 \perp\perp} - (\delta^u_r + \partial_r\, g^u)\, h_F^{rs}\,
 T_{F\, \perp s}\Big] \Big)(\tau ,\sigma^u ).
 \label{5.20}
 \eea

\noindent and these formulas allow to recover Eqs.(\ref{5.2}) of the
inertial rest frame. \medskip

Therefore the non-inertial rest-frame instant form of dynamics is
well defined.

\subsection{The Hamiltonian of the Non-Inertial Rest-Frame Instant
Form}

We have now to find which is the effective Hamiltonian of the
non-inertial rest-frame instant form replacing $Mc$ of the inertial
rest-frame one. The gauge fixing (\ref{5.20}) is a special case of
Eqs.(\ref{4.1}), whose  final Dirac Hamiltonian is given in
Eq.(\ref{4.4}) [or in Eq.(\ref{4.35}) in the radiation gauge].
\medskip

To be able to impose this gauge fixing, let us put $F^{\mu}(\tau
,\sigma^u ) = h^{\mu}\, g(\tau ,\sigma^u ) + \epsilon^{\mu}_r(\vec
h)\, [\sigma^r + g^r(\tau ,\sigma^u )]$ in Eq.(\ref{4.1}), but let
us leave $x^{\mu}(\tau )$ as an arbitrary time-like observer to be
restricted to $Y^{\mu}(\tau )$ at the end. We will only assume that
$x^{\mu}(\tau )$ is canonically conjugate with $P^{\mu} = \int
d^3\sigma\, \rho^{\mu}(\tau , \sigma^u )$, $\{ x^{\mu}(\tau ),
P^{\nu} \} = - \sgn\, \eta^{\mu\nu}$.
\medskip

Due to the dependence of $F^{\mu}(\tau ,\sigma^u)$ and of
$Y^{\mu}(\tau)$ on $\vec h = \vec P/\sqrt{\sgn\, P^2}$ we must
develop a different procedure for the identification of the Dirac
Hamiltonian.

\medskip

In this case the constraints (\ref{3.10}) can be rewritten in the
following form (${\cal T}^{\mu}_F(\tau ,\sigma^u)$ is defined in
Eq.(\ref{5.17}))

 \bea
 {\cal H}^{\mu}(\tau ,\sigma^u ) &=& {\tilde {\cal H}}^{\mu}(\tau ,
 \sigma^u ) + \delta^3(\sigma^u )\, \int d^3\sigma_1\, {\cal
 H}^{\mu}(\tau, {\sigma}^u_1) \approx 0,\nonumber \\
   with \qquad &&\int d^3\sigma\,{\tilde {\cal H}}^{\mu}(\tau ,\sigma^u )
  \equiv 0,\nonumber \\
  &&{}\nonumber \\
  &&\Downarrow\nonumber \\
  &&{}\nonumber \\
  \rho^{\mu}(\tau ,\sigma^u ) &\approx& P^{\mu}\, \delta^3(
 \sigma^u ) + \Big[{\cal T}_F^{\mu}(\tau ,\sigma^u ) - \delta^3(\sigma^u )\,
 {\cal R}_F^{\mu}(\tau )\Big] =\nonumber\\
 &=& \delta^3(\sigma^u )\, H^\mu(\tau) + {\cal T}_F^{\mu}(\tau
,\sigma^u ),\nonumber \\
 &&{}\nonumber \\
 &&H^\mu(\tau) = P^\mu - {\cal R}_F^\mu(\tau ) \approx 0,\qquad {\cal
R}_F(\tau ) {\buildrel {def}\over =} \int d^3\sigma\, {\cal
T}^{\mu}_F(\tau ,\sigma^u).
 \nonumber \\
 &&{}
  \label{5.21}
 \eea

In this way the original canonical variables $z^{\mu}(\tau ,\vec
\sigma )$, $\rho^{\mu}(\tau ,\vec \sigma )$ are replaced by the
observer $x^{\mu}(\tau )$, $P^{\mu}$ and by relative variables with
respect to it.\medskip

From Eq.(\ref{5.14}) we get:\hfill\break

a) the gauge fixing to the constraints ${\tilde {\cal H}}^{\mu}(\tau
,\sigma^u ) \approx 0$ is

 \beq
 \psi^{\mu}_r(\tau ,\sigma^u ) = {{\partial\, \chi^{\mu}(\tau ,
 \sigma^u )}\over {\partial\, \sigma^r}} =
 \Big(z^{\mu}_r - \epsilon^{\mu}_s(\vec h)\, \Big[\delta^s_r +
 {{\partial\, g^s}\over {\partial\, \sigma^r}} - u^{\mu}(\vec h)\,
 {{\partial\, g}\over {\partial\, \sigma^r}}\Big]\Big)(\tau ,\sigma^u )
 \approx 0;
 \label{5.22}
 \eeq

b) the gauge fixing to the constraints $H^\mu(\tau) = P^\mu - {\cal
R}_F^\mu \approx 0$ is $\chi^{\mu}(\tau ,0) = z^{\mu}(\tau ,0) -
Y^{\mu}(\tau ) = x^{\mu}(\tau) - Y^{\mu}(\tau) \approx 0$.\medskip

The gauge fixing (\ref{5.22}) has the following Poisson brackets
with the collective variables $x^{\mu}(\tau )$, $P^{\mu}$

\bea
 &&\{ P^{\mu}, \psi^{\nu}_r(\tau, \sigma^u ) \} = 0,
 \nonumber \\
 && \{ x^{\mu}(\tau ), \psi^{\nu}_r(\tau ,\sigma^u ) \} = -
 {{\partial\, \epsilon^{\nu}_s(\vec h)}\over {\partial\, P_{\mu}}}\,
 \Big(\delta^s_r + {{\partial\, g^s(\tau ,\sigma^u )}\over {\partial\,
 \sigma^r}} \Big)- {{\partial\, \epsilon^{\nu}_{\tau}(\vec h)}\over {\partial\, P_{\mu}}}\,
 {{\partial\, g(\tau ,\sigma^u )}\over {\partial\, \sigma^r}}
 \not= 0.
 \label{5.23}
  \eea

Therefore $x^{\mu}(\tau )$ is no more a canonical variable after the
gauge fixing $\psi^{\mu}_r(\tau ,\sigma^u ) \approx 0$.
\bigskip

By introducing the notation ($\epsilon^A_\mu = \eta^{AB}\,
\epsilon_{B\mu}\, \Rightarrow\, \epsilon^\tau_\mu(\vec h) = \sgn\,
h_\mu$)

\beq
 {\cal T}_F^\mu(\tau ,\sigma^u ) \byd\, h^\mu\,
 {\cal T}_F^{\tau}(\tau ,\sigma^u ) + \epsilon^\mu_r(\vec h)\, {\cal
T}^r_F(\tau ,\sigma^u ),\qquad
 \Rightarrow {\cal T}^A_F(\tau ,\sigma^u ) = \epsilon^A_{\mu}(\vec h)\, {\cal
T}^{\mu}_F(\tau ,\sigma^u ),
 \label{5.24}
 \eeq

\noindent the angular momentum generator of Eq.(\ref{3.17}) takes
the form

\bea
 J^{\mu\nu} &=& x^{\mu}(\tau )\, P^{\nu} - x^{\nu}(\tau)\, P^{\mu} +
 S^{\mu\nu},\nonumber \\
 &&{}\nonumber \\
 S^{\mu\nu} &\approx& \epsilon^{\mu}_r(\vec h)\,
 \epsilon^{\nu}_s(\vec h)\, \int d^3\sigma\, \Big[(\sigma^r + g^r)\, {\cal T}^s -
 (\sigma^s + g^s)\, {\cal T}^r\Big](\tau ,\sigma^u ) +\nonumber \\
&&\nonumber\\
&+& \Big(\epsilon^{\mu}_r(\vec h)\, \epsilon^{\nu}_{\tau}(\vec h) -
 \epsilon^{\nu}_r(\vec h)\, \epsilon^{\mu}_{\tau}(\vec h)\Big)\,
 \int d^3\sigma\, \Big[(\sigma^r + g^r)\, {\cal T}^{\tau} + g\, {\cal T}^r\Big](\tau
 ,\sigma^u ) =\nonumber \\
&&\nonumber\\
&=& \epsilon^{\mu}_A(\vec h)\, \epsilon^{\nu}_B(\vec h)\,
 S^{AB},\nonumber \\
 &&{}\nonumber \\
 S^{rs} &=& \int d^3\sigma\, \Big[(\sigma^r + g^r)\, {\cal T}^s -
 (\sigma^s + g^s)\, {\cal T}^r\Big](\tau ,\sigma^u )\byd
 \delta^{rn}\, \epsilon_{nsu}\,{\cal J}^u,\nonumber \\
 S^{\tau r} &=& - S^{r\tau} = -  \int d^3\sigma\, \Big[(\sigma^r +
 g^r)\, {\cal T}^{\tau} + g\, {\cal T}^r\Big](\tau ,\sigma^u )\byd{\cal
 K}^r,
 \label{5.25}
 \eea

\noindent where only the constraints $\tilde{\cal H}^\mu(\tau
,\sigma^u) \approx 0$ have been used.

 \medskip

 Since we have

 \bea
 \{ x^{\mu}(\tau ), S^{\alpha\beta} \} &=& 0,\nonumber \\
 &&{}\nonumber \\
 \{ {{\partial\, z^{\mu}(\tau ,\sigma^u )}\over
 {\partial\, \sigma^r}}, S^{\alpha\beta} \} &=& \Big({{\partial\, z^{\beta}}\over
 {\partial\, \sigma^r}}\,\eta^{\mu\alpha} - {{\partial\, z^{\alpha}}\over
 {\partial\, \sigma^r}}\,\eta^{\mu\beta} \Big)(\tau ,\sigma^u )
 \approx\nonumber \\
 &\approx& \Big(\Big[\epsilon^{\beta}_s(\vec h)\, (\delta^s_r + {{\partial\,
 g^s}\over {\partial^r}}) + h^{\beta}\, {{\partial\, g}\over {\partial\,
 \sigma^r}}\Big]\, \eta^{\mu\alpha}\Big)(\tau ,\sigma^u ),
 \label{5.26}
 \eea

\noindent after the gauge fixing the new canonical variable for the
observer becomes

\beq
 {\tilde x}^{\mu}(\tau ) = x^{\mu}(\tau ) - {1\over 2}\,
 \epsilon_{\sigma\, A}(\vec h)\, {{\partial\, \epsilon^A_{\rho}(\vec
 h)}\over {\partial\, P_{\mu}}}\, S^{\sigma\rho},\qquad
 \{ {\tilde x}^{\mu}(\tau ), \psi^{\nu}_r(\tau ,\vec \sigma )\} =
 0.
 \label{5.27}
 \eeq
 \medskip

If we eliminate the relative variables by going to Dirac brackets
with respect to the second class constrainta ${\tilde {\cal
H}}^{\mu}(\tau ,\sigma^u ) \approx 0$, $\psi^{\mu}_r(\tau , \sigma^u
) \approx 0$, the canonical variables $z^{\mu}(\tau ,\sigma^u )$,
$\rho_{\mu}(\tau ,\sigma^u )$ are reduced to the canonical variables
${\tilde x}^{\mu}(\tau )$, $P^{\mu}$.\medskip

By defining ${\cal R}_F(\tau ) = \sgn\, h_{\mu}\, {\cal
R}^{\mu}_F(\tau ) \approx Mc = \sqrt{\sgn\, P^2}$, the remaining
constraints are

\bea
 H^\mu(\tau)&=&h^\mu\, \Big(\sqrt{\sgn\,P^2} - {\cal
R}_f(\tau)\Big) + \epsilon^\mu_r(\vec{h})\, {\hat {\cal P}}^r,
\nonumber\\
 &&{}\nonumber \\
 or&& \sgn\,h^\mu\, H_\mu(\tau) = \sqrt{\sgn\,P^2} - {\cal
R}_F(\tau)\approx 0,\qquad
 \epsilon^r_\mu(\vec{h})H^\mu(\tau) = {\hat {\cal P}}^r \approx 0.
 \nonumber \\
 &&{}
 \label{5.28}
  \eea
\bigskip

Like in Eqs.(\ref{4.1}) and (\ref{4.2}), after this reduction the
Dirac multiplier $\lambda^{\mu}(\tau ,\sigma^u )$ in the Dirac
Hamiltonian (\ref{3.16}) becomes

\bea
 \lambda_\mu(\tau,\sigma^u)&=& \sgn\, h_\mu\,
 \Big(\lambda_\tau(\tau) - \frac{\partial
g(\tau,\sigma^u)}{\partial\tau}\Big) + \sgn\,
\epsilon_{\mu\,r}(\vec{h})\, \Big(\lambda^r(\tau) - \frac{\partial
g^r(\tau,\sigma^u)}{\partial\tau}\Big)\on \nonumber \\
 &\cir& - \sgn\,\frac{\partial z^\mu_F(\tau ,\sigma^u)}{\partial\tau}
 \label{5.29}
 \eea

At this stage the Dirac Hamiltonian depends only on the residual
Dirac multipliers $\lambda_{\tau}(\tau )$ and $\vec \lambda (\tau )$

 \beq
 H_D = \lambda_{\tau}(\tau )\, (\sqrt{\sgn\,P^2} - {\cal R}_F)
 - \vec \lambda (\tau ) \cdot {\hat {\vec {\cal P}}} +
 \int d^3\sigma\, \Big({{\partial\, g^r}\over {\partial\, \tau}}\,
 {\cal T}_{F\, r} + \frac{\partial g}{\partial\tau}\,
 {\cal T}_{F\, \tau}\Big)(\tau ,\sigma^u ),
 \label{5.30}
 \eeq

\noindent where we introduced the notation ${\cal T}_{F\, A}(\tau
,\sigma^u ) \byd \sgn\,\epsilon_{\mu\,A}(\vec h)\, {\cal
T}_F^{\mu}(\tau ,\sigma^u )$ so that ${\cal T}_F^\tau = {\cal
T}_{F\, \tau},\qquad {\cal T}_F^r = - \sgn\, {\cal T}_{F\, r}$.
\bigskip

To implement the gauge fixing $x^{\mu}(\tau) - Y^{\mu}(\tau) \approx
0$ requires two other steps:\medskip

1) Firstly we impose the gauge fixing  ${\tilde x}^{\mu}(\tau )\,
h_{\mu} = \sgn\, \tau$. It implies $\lambda_{\tau}(\tau ) = -1$ and
$\sqrt{\sgn\,P^2} = Mc \equiv  {\cal R}_F$. The Dirac Hamiltonian
becomes

 \bea
 H_{F\, D} &=& {\cal M}\, c - \vec \lambda (\tau ) \cdot {\hat {\vec {\cal P}}}+
 \int d^3\sigma\, \Big[\mu\, \pi^\tau - A_\tau\,\Gamma\Big](\tau, \sigma^u )
,\nonumber \\
 &&{}\nonumber \\
 {\cal M}\, c &=& Mc + \int d^3\sigma\, \Big({{\partial\,
 g^r}\over {\partial\, \tau}}\,
 {\cal T}_{F\, r} + \frac{\partial g}{\partial\tau}\,
 {\cal T}_{F\, \tau}\Big)(\tau ,\sigma^u ).
 \label{5.31}
 \eea

\medskip

2) Then we add the gauge fixing ${\hat {\cal K}}^r \approx 0$
 to the rest-frame conditions ${\hat {\cal
P}}^r \approx 0$. In this way we get $x^{\mu}(\tau) \approx
Y^{\mu}(\tau)$ and we also eliminate the internal 3-center of mass.
Having chosen the Fokker-Pryce external 4-center of inertia
$Y^{\mu}(\tau)$ as origin of the 3-coordinates the constraints
${\hat {\cal K}}^r \approx 0$ correspond to the requirement $S^{\tau
r} \approx 0$.

\medskip

In conclusion the effective Hamiltonian ${\cal M}\, c$ (modulo
electro-magnetic gauge transformations) of the non-inertial
rest-frame instant form is not the internal mass $Mc$, since it
describes the evolution from the point of view of the asymptotic
inertial observers. There is an additional term interpretable as an
inertial potential producing relativistic inertial effects (see
Eqs.(\ref{5.16}) for $1 + n_F(\tau ,\sigma^u)$ and Eqs.(\ref{5.15})
for $n_{F\, r}(\tau ,\sigma^u)$)

\bea
 {\cal M}\, c&=& Mc + \int d^3\sigma\, \Big({{\partial\, g^r}\over
{\partial\, \tau}}\,
 {\cal T}_{F\, r} + \frac{\partial g}{\partial\tau}\,
 {\cal T}_{F\, \tau}\Big)(\tau ,\sigma^u )=\nonumber\\
&&\nonumber\\
&=&\int d^3\sigma\,\sgn\,\Big(\Big[\,h_\mu\Big(1+\frac{\partial
g}{\partial\tau}\Big)+ \epsilon_{\mu\,r}\,\frac{\partial
g^r}{\partial\tau}\,\Big]{\cal T}_F^\mu
\Big)(\tau ,\sigma^u) =\nonumber\\
&&\nonumber\\
&=&\int d^3\sigma\, \sqrt{\gamma (\tau, \sigma^u)}\, \Big((1 +
n_F)\, T_{F\, \perp\perp} + n_F^r\, T_{F\, \perp\,r}\Big)(\tau
,\sigma^u )
 \label{5.32}
  \eea

\noindent where

\bea
 &&\sqrt{\gamma (\tau, \sigma^u)}\, T_{F\, \perp\perp}(\tau ,\sigma^u)=
 \sqrt{\gamma (\tau, \sigma^u)}\, T'_{F\, \perp\perp}(\tau ,\sigma^u) +\nonumber \\
 &&\qquad + \sum_i\,\delta(\sigma^u - \eta^u_i)\,\sqrt{m_i^2\, c^2 +
\h^{rs}_F(\tau ,\sigma^u)\,(\kappa_{ir}(\tau ) - Q_i\,A_r(\tau
,\sigma^u))\, (\kappa_{is}(\tau ) - Q_i\,A_s(\tau ,\sigma^u))},\nonumber \\
 &&\sqrt{\gamma (\tau, \sigma^u)}\, T_{F\, \perp\,r}(\tau ,\sigma^u) =  F_{rs}(\tau
,\sigma^u)\,\pi^s(\tau ,\sigma^u) - \sum_i\,\delta(\sigma^u -
\eta^u_i)\, (\kappa_{ir}(\tau )
 - Q_i\,A_r(\tau ,\sigma^u)),
 \label{5.33}
 \eea

\noindent with $T'_{\perp\perp}$ given in Eq. (\ref{4.28}).

\bigskip

Let us remark that a similar procedure should be applied also to the
gauge fixing (\ref{4.1}) if we want to reproduce the results of
Subsection B for arbitrary non-inertial frames. We do not add these
calculations, because they agrees substantially with the results of
this Subsection and do not alter the conclusions of Section IV.

\vfill\eject

\section{Non-Inertial Maxwell Equations in Nearly Rigid Rotating Frames}

In the 3+1 point of view the Maxwell equations (\ref{4.17}) in an
arbitrary inertial frame are identical to the Maxwell equations in
general relativity, but now the 4-metric is describing only the
inertial effects present in the given frame. Therefore we can adapt
the techniques used in general relativity to non-inertial frames,
for instance the definition of electric and magnetic fields done in
Ref.\cite{28} (see Appendix B) or the geometrical optic
approximation to light rays of Ref.\cite{29}.

For the 1+3 point of view on this topic see for instance
Ref.\cite{30} and its bibliography. In particular, for the treatment
of electromagnetic wave in rotating frame by means of Fermi
coordinates \cite{31} and for the determination of the
helicity-rotation coupling, as a special case of spin-rotation
coupling \cite{32,33}. In all these calculations the locality
hypothesis is used.

In the case of linear acceleration an analysis of the inertial
effects has been done in Ref.\cite{34}. The same non-inertial
4-metric has been used in Ref.\cite{35} to study the optical
position meters constituents of the laser interferometers on ground
used for the detection of gravitational waves. However the 4-metric
used has a bad behavior at spatial infinity, so that the conclusions
on the electro-magnetic waves in these frames (even if supposed to
hold at distances smaller than those where there are coordinate
singularities) are questionable because the Cauchy problem for
Maxwell equations is not well posed.

\medskip

In this Section we study some properties of electro-magnetic waves
and of geometrical optic approximation to light rays in the
radiation gauge in the admissible rotating non-inertial frame
defined by the embedding (\ref{2.14}), ensuring a well-posed Cauchy
problem, at small distances from the rotation axis where the
$O(c^{-1})$ deviations from rigid rotations is governed by
Eqs.(\ref{2.15}) and (\ref{2.16}). Even if we will ignore these
deviations, doing the calculations in the radiation gauge in locally
rigidly rotating frames, they could be taken into account in a more
refined version of the subsequent calculations base on the 3+1 point
of view, which is free from coordinate singularities. This would
also allow to verify the validity of the locality hypothesis. In
particular we consider the Phase Wrap Up effect \cite{31,36}, the
Sagnac effect \cite{37,31a} and the Faraday Rotation \cite{38}.

\subsection{The 3+1 Point of View on Electro-Magnetic Waves and Light Rays  in
Nearly Rigidly Rotating Non-Inertial Frames.}

Let us consider a non-inertial frame of the type (\ref{2.14}) with
vanishing linear acceleration and $\tau$-independent angular
velocity and centered on an inertial observer. In the notation of
Eqs.(\ref{2.15}), (\ref{2.16}) and (\ref{4.47}), we have
$x^{\mu}(\tau) = \epsilon^{\mu}_{\tau}\, \tau$, i.e. $\vec v(\tau )
= \vec w(\tau) = 0$, and $\vec \Omega (\tau) = \vec \Omega = const.$
(whose components are ${\tilde \Omega}^r = const.$). We will ignore
the higher order terms, so that locally we have a rigidly rotating
frame, but with more effort small deviations from rigid rotation
could be taken into account.

\medskip

In this case the Hamiltonian (\ref{4.35}), or (\ref{4.51}), gives
the following Hamilton equations for the transverse electro-magnetic
field (${\vec A}_{\perp} = \{A_{\perp\, r} = {\tilde A}^r_{\perp}
\not= A^r_{\perp} \} + O(c^{-2})$)

 \bea
\frac{\partial {\tilde
A}^r_\perp(\tau,\vec{\sigma})}{\partial\tau}&=&
\pi^r_\perp(\tau,\vec{\sigma}) -
 {1\over c}\, \int d^3\sigma'\, \left[ - \vec{\Omega} \cdot
\vec{\sigma}' \times {\vec \partial}'\,
\vec{A}_\perp(\tau,\vec{\sigma}') + \vec{\Omega} \times
\vec{A}_\perp(\tau,\vec{\sigma}')
\right]^s\,{\bf P}^{sr}(\vec{\sigma}',\vec{\sigma}),\nonumber\\
&&\nonumber\\
\frac{\partial \pi^r_\perp(\tau,\vec{\sigma})}{\partial\tau}&=&
\Delta {\tilde A}^r_\perp(\tau,\vec{\sigma}) -
 {1\over c}\, \int d^3\sigma'\, \left[ - \vec{\Omega} \cdot \vec{\sigma}'
 \times {\vec \partial}'\, \vec{\pi}_\perp(\tau,\vec{\sigma}') + \vec{\Omega}
\times \vec{\pi}_\perp(\tau,\vec{\sigma}')
\right]^s\, {\bf P}^{sr}(\vec{\sigma}',\vec{\sigma}) +\nonumber\\
&&\nonumber\\
&+&\sum_i\, Q_i\, \left(\dot{\vec{\eta}}_i(\tau ) + \vec{\Omega}
\times \vec{\eta}_i(\tau )\right)^s\, {\bf
P}^{sr}(\vec{\eta}_i,\vec{\sigma}).
 \label{6.1}
 \eea

\hfill

For the study of homogeneous solutions of these equations, i.e. for
incoming electro-magnetic waves propagating in regions where there
are no charged particles, these equations can be replaced with the
following ones (we use the vector notation of Subsection C of
Section IV)

 \bea
\frac{\partial \vec{A}_\perp(\tau,\vec{\sigma})}{\partial\tau}&=&
\vec{\pi}_\perp(\tau,\vec{\sigma}) -
  {1\over c}\, \left[ - \vec{\Omega} \cdot \vec{\sigma} \times \vec
\partial\, \vec{A}_\perp(\tau,\vec{\sigma}) + \vec{\Omega}
\times \vec{A}_\perp(\tau,\vec{\sigma})\right],\nonumber\\
&&\nonumber\\
\frac{\partial \vec{\pi}_\perp(\tau,\vec{\sigma})}{\partial\tau}&=&
\Delta \vec{A}_\perp(\tau,\vec{\sigma}) -
 {1\over c}\, \left[ - \vec{\Omega} \cdot \vec{\sigma} \times \vec
\partial\, \vec{\pi}_\perp(\tau,\vec{\sigma}) + \vec{\Omega} \times
\vec{\pi}_\perp(\tau,\vec{\sigma})\right].
 \label{6.2}
 \eea

\hfill

As shown in Appendix B, this result allows to recover the form given
by Schiff in Appendix A of ref.\cite{28} for the Landau-Lifschitz
non-inertial electro-magnetic fields \cite{17}.
\bigskip

Let us look at solutions of Eqs.(\ref{6.2}) in the following two
ways.

\subsubsection{Going back to an Inertial Frame}

Let us look at solution by reverting to an inertial frame.\medskip

By introducing the 3-coordinates

 \beq
 X^a(\tau) = R^a{}_r(\tau)\, \sigma^r,
  \label{6.3}
 \eeq

\noindent at each value of $\tau$ by means of a $\tau$-dependent
rotation (it would become also point-dependent if we go beyond rigid
rotations) we can go from the rigidly rotating non-inertial frame
with radar 4-coordinates $(\tau ; \sigma^u)$ to an instantaneously
comoving inertial frame, centered on the same inertial observer,
with 4-coordinates $(\tau; X^a)$.
\medskip

Let us assume that the non-inertial transverse electromagnetic
potential $A_{\perp\, r}(\tau ,\sigma^u)$ can be obtained from the
instantaneously comoving inertial transverse potential
$A^{(com)}_{\perp\, a}(\tau , X^a(\tau))$ defined by

 \beq
 A_{\perp\, r}(\tau, \sigma^u) = A^{(com)}_{\perp\, a}\Big(\tau, X^a(\tau) =
 R^a{}_s(\tau)\, \sigma^s\Big)\, R^a{}_r(\tau),
 \label{6.4}
  \eeq

\noindent which by definition satisfies the inertial Maxwell
equations in the radiation gauge (obtainable by putting
Eqs.(\ref{6.4}) into Eqs.(\ref{6.2}))

 \beq
  \frac{\partial^2\, A^{(com)}_{\perp\, a}(\tau, X^b)}{\partial\tau^2} -
  \Delta_X\, A^{(com)}_{\perp\, a}(\tau, X^b) = 0,\qquad
\sum_a\,\frac{\partial}{\partial X^a}\, A^{(com)}_{\perp\, a}(\tau,
X^b) = 0.
 \label{6.5}
  \eeq

\bigskip

This result is in accord with the general covariance of non-inertial
Maxwell equations and is also consistent with the locality
hypothesis (see Subsection B of Section II) of the the 1+3 approach.

\bigskip
If we consider the following plane wave solution with constant $F_a$
and ${\hat K}_a$ and $\sum_a\, {\hat K}_a\, F_a = 0$ (transversality
condition)

 \beq
 A^{(com)}_{\perp\, a}(\tau, X^b) = \frac{1}{\omega}\, F_a\,
 e^{i\, \frac{\omega}{c}\, \left(\tau - \sum_a\, {\hat{K}}_a\, X^a\right)},
 \label{6.6}
  \eeq

\noindent we get the following expression for the non-inertial
solution

   \bea
 A_{\perp\, r}(\tau, \sigma^u) &=& F_a\, R^a{}_r(\tau)\, e^{
i\, \frac{\omega}{c}\, \Phi(\tau, \sigma^u)},\nonumber \\
 &&{}\nonumber \\
 \Phi(\tau, \sigma^u) &=& \tau - \hat{K}_a\, R^a{}_r(\tau)\, \sigma^r
 \approx{|}_{\vec \Omega = const.}\, \tau\, \Big(1 + {{\vec \Omega}\over c}\,
 \cdot \vec \sigma \times \hat K\Big) -
 {\hat K} \cdot \vec \sigma + O(\Omega^2/c^2).
 \label{6.7}
 \eea

\subsubsection{Eikonal Approximation}

Let us now look at solutions  by making the following eikonal
approximation (without any commitment with the locality hypothesis)

 \beq
  A_{\perp\, r}(\tau, \sigma^u) =
\frac{1}{\omega}\, a_r(\tau, \sigma^u)\, e^{i\, {{\omega}\over c}\,
\Phi(\tau, \sigma^u)} + O(1/\omega^2).
 \label{6.8}
  \eeq

\noindent and by putting this expression in Eqs.(\ref{6.2}).
\medskip

Let us consider the case in which we have $\omega/c
>> 1$ e $\Omega/c << 1$, so that Eqs.(\ref{6.2}) become a power
series in $\omega/c$. By neglecting terms in $\Omega^2/c^2$ and
terms in $(c/\omega)^{-k}$ for $k \geq 0$, the dominant terms are:

a) at the order $\omega/c$ the equation for the phase $\Phi$, named
{\em eikonal equation};

b) at the order $(\omega/c)^o=1$ the equation for the amplitude
$a_r$, named {\em first-order transport equation}.
\medskip

These equations have the following form ($\vec a = \{ a_r \}$)

 \bea
 &&\Big[\left(\frac{\partial\Phi}{\partial\tau}\right)^2 - 2\,{{\vec{\Omega}}\over c} \cdot
 \vec{\sigma} \times \vec \partial\, \Phi - \left(\vec
 \partial\, \Phi\right)^2\Big](\tau ,\sigma^u) + O(\Omega^2/c^2)
 = 0\nonumber\\
&&\nonumber\\
&&\nonumber\\
&&\Big[\frac{\partial\Phi}{\partial\tau}\,\left(\frac{\partial\vec{a}}{\partial\tau}
+ {{\vec{\Omega}}\over c} \times \vec{a} - {{\vec{\Omega}}\over c}
\cdot \vec{\sigma} \times\vec{\partial} \,\vec{a}\right) -
\frac{\partial\vec{a}}{\partial\tau}\, {{\vec{\Omega}}\over c} \cdot
\vec{\sigma} \times \vec
\partial\, \Phi - \left(\vec \partial\, \Phi \cdot \vec
\partial \right)\, \vec{a}\Big](\tau ,\sigma^u) =\nonumber\\
\nonumber\\
&=&- \frac{1}{2}\, \left(\frac{\partial^2\Phi}{\partial\tau^2} -
2\,(\vec{\Omega} \times\vec{\sigma} \cdot \vec{\partial})\,
\frac{\partial\Phi}{\partial\tau} - \triangle\Phi\right)(\tau
,\sigma^u) + O(\Omega^2/c^2)
\nonumber\\
&&\nonumber\\
&&\nonumber\\
&& \Big[\vec a \cdot \vec \partial\, \Phi\Big](\tau ,\sigma^u) =
0\quad (transversality\, condition).
 \label{6.9}
 \eea

Let us look for solutions of the  {\em eikonal equation} for $\Phi$
of the form

 \beq
\Phi(\tau, \sigma^u) = \tau + F(\sigma^u),
 \label{6.10}
  \eeq

\noindent where we have chosen the boundary condition

 \beq
  \frac{\partial\Phi}{\partial\tau}=1.
   \label{6.11}
 \eeq

This condition implies that the solution of Eq.(\ref{6.8}) describes
a ray emitted from a source having a characteristic frequency
$\omega$ when it is at rest in the non-inertial frame. Let us remark
that in more general cases this type of boundary conditions are
possible only if the {\em 3-metric} $h_{rs}$ and the  {\em lapse},
$n$, and {\em shift}, $n^r$, functions are stationary in the
non-inertial frame.\medskip

 An expansion in powers of $\Omega/c$ of $F(\sigma^u)$, namely
$F(\sigma^u) = F_o(\sigma^u) + \frac{\Omega}{c}\, F_1(\sigma^u) +
O\Big(\frac{\Omega^2}{c^2}\Big)$, gives the following form of the
eikonal equation

 \beq
\Big[1 - \Big(\vec{\partial}\, F_o(\sigma^u)\Big)^2\Big] -
\frac{2\Omega}{c}\, \Big[ \hat{\Omega} \cdot \vec{\sigma}
\times\vec{\partial}F_o(\sigma^u) + \vec{\partial}\, F_o(\sigma^u)
\cdot \vec{\partial}\, F_1(\sigma^u) \Big] +
O\Big(\frac{\Omega^2}{c^2}\Big) = 0,
 \label{6.12}
 \eeq

\noindent implying:

a) the equation $1 - \Big(\vec{\partial}\, F_o(\sigma^u)\Big)^2 = 0$
at the order zero in $\Omega$. If $\hat k$ is an arbitrary unit
vector (the propagation direction of the plane wave in the inertial
limit $\Omega \mapsto 0$), its solution is

 \beq
 F_o(\sigma^u) = - \hat{k} \cdot \vec{\sigma}.
 \label{6.13}
 \eeq

\medskip

b) the equation $\hat{k} \cdot \vec{\partial}\, F_1(\sigma^u) = -
\hat{\Omega} \cdot \vec{\sigma} \times \hat{k}$ for $F_1(\sigma^u)$,
after having used Eq.(\ref{6.13}), at the order one in $\Omega$.
Since we have $(\hat{k} \cdot \vec{\partial})\, (\hat{\Omega} \cdot
\vec{\sigma} \times \hat{k}) = 0$ and $(\hat{k} \cdot
\vec{\partial})\, (\hat{k} \cdot \vec{\sigma}) = 1$, the solution
for $F_1(\sigma^u)$ is

 \beq
  F_1(\sigma^u) = - \Big(\hat{\Omega}
\cdot \vec{\sigma} \times \hat{k}\Big)\, (\hat{k} \cdot
\vec{\sigma}).
 \label{6.14}
 \eeq

\medskip

Therefore the solution for $\Phi$ is

 \beq
\Phi(\tau,\sigma^u) = \tau - \hat{k} \cdot \vec{\sigma}\, \Big(1 +
\vec{\Omega} \cdot\vec{\sigma} \times \hat{k}\Big).
 \label{6.15}
  \eeq

\bigskip

The phases in the solutions (\ref{6.7}) and (\ref{6.15}) of
Eqs.(\ref{6.2}) are different since the solutions have different
boundary conditions. The solution (\ref{6.7}) satisfies also the
eikonal equation but not the boundary condition (\ref{6.11}), since
we have ${{\partial\, \Phi}\over {\partial\, \tau}} = 1 - {\hat
K}_a\, R^a{}_r(\tau )\, \epsilon_{ruv}\, {\tilde \Omega}^u\,
\sigma^v \not= 1$.

\bigskip

Let us remark that both the solutions (\ref{6.7}) and (\ref{6.15})
have the following structure

 \beq
\tilde{A}^r_\perp(\tau, \sigma^u) \sim\, {\cal A}^r(\tau,
\sigma^u)\, e^{i\, \varphi(\tau, \sigma^u)},
 \label{6.16}
  \eeq

\noindent where ${\cal A}^r(\tau, \sigma^u) \sim\,  O(1/\omega)$ is
the {\em amplitude} and $\varphi(\tau, \sigma^u) \sim  O(\omega)$ is
the {\em phase}. The only difference is that the solution
(\ref{6.7}) holds for every value of $\omega$ (also for the small
values corresponding to the radio waves of the GPS system), while
the solution (\ref{6.15}) for the phase of the eikonal approximation
(\ref{6.8}) holds only for higher values of $\omega$, corresponding
to visible light.

\subsubsection{Light Rays}

Given the phase of Eq.(\ref{6.16}), the trajectories of the light
rays are defined as the lines orthogonal (with respect to the
4-metric $g_{AB}$ of the 3+1 splitting) to the hyper-surfaces
$\varphi(\tau, \sigma^u) = \,const.$. Therefore the trajectories
$\sigma^A(s)$ ($s$ is n affine parameter) satisfy the equation

 \beq
\frac{d\sigma^A(s)}{ds} = g^{AB}(\sigma(s))\, \,
\frac{\partial\varphi}{\partial\sigma^B}(\sigma(s)).
 \label{6.17}
 \eeq

For instance in the case of our rigidly rotating foliation, for
which Eqs.(\ref{2.14})-(\ref{2.16}) imply $g^{\tau\tau} = 1$,
$g^{\tau r} = - (\vec{\Omega} \times \vec{\sigma})^r$, $g^{rs} = -
\delta^{rs} + O(\Omega^2/c^2)$, Eqs.(\ref{6.17}) take the form

 \bea
\frac{d\tau(s)}{ds}&=&\omega\, + \vec{k} \cdot
\left(\frac{\vec{\Omega}}{c}
\times \vec{\sigma}\right) + O(\Omega^2/c^2),\nonumber\\
&&\nonumber\\
\frac{d\sigma^r(s)}{ds}&=&\omega\, \left(\frac{\vec{\Omega}}{c}
\times \vec{\sigma}\right)^r + k^r\, \left(1 +
\frac{\vec{\Omega}}{c} \times \vec{\sigma} \cdot \hat{k}\right) -
\left(\frac{\vec{\Omega}}{c} \times \hat{k}\right)^r\, (\vec{k}
\cdot \vec{\sigma}) + O(\Omega^2/c^2),\nonumber \\
 &&{}
 \label{6.18}
 \eea

\noindent whose solution has the  form

 \beq
\vec{\sigma}(\tau) - \vec{\sigma}(0) = \hat{k}\, \tau +
\left(\frac{\vec{\Omega}}{c} \times \hat{k}\right)\, \tau^2 +
O(\Omega^2/c^2).
 \label{6.19}
  \eeq

This equation shows that in the rotating frame the ray of light
appears to deviate from the {\it inertial} trajectory
$\vec{\sigma}(\tau) = \hat{k}\, \tau$ due to the {\it centrifugal
correction} $\vec{c}(\tau) = \left(\frac{\vec{\Omega}}{c} \times
\hat{k}\right)\, \tau^2 + O(\Omega^2/c^2)$ implying $\hat{k} \cdot
\vec{c}(\tau) = 0 + O(\Omega^2/c^2)$.

\subsection{Sources and Detectors}

To connect the previous solutions to the interpretation of observed
data we need a schematic description of {\it sources} and {\it
detectors}.

In many applications {\em sources} and {\em detectors} are described
from point-like objects, which follow a prescribed world-line $
\zeta^A(\tau) = (\tau, \eta^u(\tau))$ with unit 4-velocity
$v^A(\tau) = \frac{d \zeta^A(\tau)}{d\tau}\,
\left(g_{CD}(\zeta(\tau))\, \frac{d \zeta^C(\tau)}{d\tau}\, \frac{d
\zeta^D(\tau)}{d\tau}\right)^{-1/2} $.

This description is enough for studying the influence of the
relative motion between source and detector on the frequency emitted
from the source and that observed by the detector (it works equally
well for the Doppler effect and for the gravitational redshift in
presence of gravity). With solutions like Eq.(\ref{6.16}) the
frequency emitted by a source located in $\zeta_s{}^A$ and moving
with 4-velocity  $v_s{}^A$ and that observed by a detector in
$\zeta_r{}^A$ and moving with 4-velocity  $v_r{}^A$ are $\omega_s =
v_s{}^A\, \partial_A\, \varphi(\zeta_s)$ and $\omega_r = v_r{}^A\,
\partial_A\, \varphi(\zeta_r)$, respectively.

This justifies the boundary condition (\ref{6.11}), because sources
at rest in the rotating frame with coordinates $(\tau,\sigma^r)$
have 4-velocity $v^A = (1,0)$.

\medskip

However, to measure the electro-magnetic field in assigned (spatial)
polarization direction we must assume that the detector is endowed
with a tetrad orthonormal with respect to the 4-metric of the 3+1
splitting, such that the time-like 4-vector is the unit 4-velocity
of the detector: in 4-coordinates adapted to the 3+1 splitting they
are ${\cal E}^A_{(\alpha)}(\tau) = \Big({\cal E}^A_{(o)}(\tau ) =
v^A(\tau); {\cal E}^A_{(i)}(\tau)\Big)$, $g_{AB}(\zeta_r(\tau))\,
{\cal E}^A_{(\alpha)}(\tau)\, {\cal E}^B_{(\beta)}(\tau) =
\eta_{(\alpha)(\beta)}$ (see Subsection B of Section II for the 1+3
point of view). A detector measures the following field strengths
along the spatial polarization directions ${\cal E}^A_{(i)}(\tau)$:
$\check{E}_{(i)} = F_{AB}\, v^A\, {\cal E}^B_{(i)}$ and
$\check{B}_{(i)} = (1/2)\, \epsilon_{(i)(j)(k)}\, F_{AB}\, {\cal
E}^A_{(j)}\, {\cal E}^B_{(k)}$.

\medskip

Let us consider the following two cases.

\subsubsection{Detectors at Rest in an Inertial Frame}

A detector at rest in the instantaneous inertial frame with
coordinates $(\tau; X^a(\tau))$ follows the straight world-line
$\zeta^{\mu}_{r,in}(\tau) = \tau\, \epsilon^{\mu}_{\tau} +
\epsilon^{\mu}_a\, \eta_{in}^a$ with $\eta_{in}^a = const.$ and has
the 4-velocity $u^{\mu} = \epsilon^{\mu}_{\tau}$. If the reference
asymptotic tetrad $\epsilon^{\mu}_A$ of the foliation is related by
$\epsilon^{\mu}_A = \Lambda^{\mu}_{(o)}{}_{\nu}\, e^{\nu}_{(A)}$ to
a tetrad $e^{\mu}_{(A)} = \delta^{\mu}_A$  aligned to the axes of
the inertial frame in Cartesian coordinates, then a generic
time-independent non-rotating tetrad associated with the detector
will be ${\cal G}^{\mu}_{(A)} = \Lambda_{(A)}{}^{(B)}\,
e^{\mu}_{(B)} = \Lambda_{(A)}{}^{(\mu)}$ if ${\cal G}^{\mu}_{(\tau)}
= u^{\mu}$. Here the $\Lambda$'s denote Lorentz transformations. The
detector will measure the standard electric and magnetic fields
$\check{E}_{(i)} = F_{\mu\nu}\, u^{\mu}\, {\cal G}^{\nu}_{(i)}$ and
$\check{B}_{(i)} = (1/2)\, \epsilon_{(i)(j)(k)}\, F_{\mu\nu}\, {\cal
G}^{\mu}_{(j)}\, {\cal G}^{\nu}_{(k)}$.

\subsubsection{Sources and Detectors at Rest in Inertial and Rotating Frames}

Lt us now consider sources and detectors at rest in the nearly rigid
rotating frame described by the embedding $z^{\mu}(\tau ,\sigma^u) =
\epsilon^{\mu}_{\tau}\, \tau + \epsilon^{\mu}_r\, R^r{}_s(\tau)\,
\sigma^s + O(\Omega^2/c^2)$, so that $z^{\mu}_{\tau}(\tau ,\sigma^u)
= \epsilon^{\mu}_{\tau} + \epsilon^{\mu}_r\, {\dot R}^r{}_s(\tau)\,
\sigma^s + O(\Omega^2/c^2)$ and $z^{\mu}_r(\tau ,\sigma^u) =
\epsilon^{\mu}_s\, R^s{}_r(\tau) + O(\Omega^2/c^2)$.
\medskip

The world-line of these objects will have the form $\zeta^{\mu}(\tau
) = \tau\, \epsilon^{\mu}_{\tau} + \epsilon^{\mu}_r\, R^r{}_s(\tau
)\, \eta_o^s + O(\Omega^2/c^2) = \epsilon^{\mu}_A\, \zeta^A(\tau)$
with $\eta^r_o = const.$. We have $\zeta^{\tau}(\tau) = \tau$ and
$\zeta^r(\tau) = R^r{}_s(\tau )\, \eta_o^s + O(\Omega^2/c^2)$.
Therefore these objects coincide with some of the observers
belonging at the non-surface forming congruence generated by the
evolution vector field as said in Subsection B of Section II. Since
the world-lines of the Eulerian observers of the other congruence
are not explicitly known, it is not possible to study the behavior
of objects coinciding with some of these observers.
\medskip

Therefore the unit 4-velocity $u^{\mu}(\tau) = \epsilon^{\mu}_A\,
v^A(\tau)$ will have the components $v^A(\tau)$  proportional to
${\dot \zeta}^A(\tau) = \Big(1; {\dot R}^r{}_s(\tau )\, \eta_o^s +
O(\Omega^2/c^2)\Big) \approx{|}_{\vec \Omega = const.}\,\, \Big(1;
R^r{}_s(\tau)\, ({\vec \eta}_o \times \vec {{\Omega}\over c})^s
+O(\Omega^2/c^2)\Big)$, where the definitions after Eq.(\ref{2.14})
have been used.\medskip

We can also write $u^{\mu}(\tau) = {\tilde u}^A(\tau)\,
z^{\mu}_A(\tau, \eta_o^u)$ by using the non-orthonormal tetrads
$z^{\mu}_A(\tau ,\sigma^u)$. Then we get  $v^{\tau}(\tau) = {\tilde
u}^{\tau}(\tau) + O(\Omega^2/c^2)$ and $v^r(\tau) = {\tilde
u}^{\tau}(\tau)\, {\dot R}^r{}_s(\tau)\, \eta_o^s + R^r{}_s(\tau)\,
{\tilde u}^s(\tau) + O(\Omega^2/c^2)$. While the quantities
$v^A(\tau)$ give the description of the 4-velocity with respect to
the asymptotic non-rotating inertial observers, the quantities
${\tilde u}^A(\tau)$ explicitly show the effect of the rotation at
the position $\eta_o^r$ of the object. Therefore it should be
${\tilde u}^A(\tau) = (1; 0)$ at the lowest order: indeed we get
${\tilde u}^{\tau}(\tau) = 1 + O(\Omega^2/c^2)$ and ${\tilde
u}^r(\tau) = v^s(\tau)\, R_s{}^r(\tau) - {\tilde u}^{\tau}\,
\Big(R^{-1}(\tau)\, {\dot R}(\tau)\Big)^r{}_s\, \eta^s_o   = 0 +
O(\Omega^2/c^2)$.\medskip

For the constant unit normal to the instantaneous 3-spaces we get
$l^{\mu} = \epsilon^{\mu}_{\tau} = {\tilde l}^A(\tau, \eta_o^r)\,
z^{\mu}(\tau, \eta_o^r)$ with ${\tilde l}^{\tau}(\tau ,\eta_o^r) = 1
+ O(\Omega^2/c^2)$ and ${\tilde l}^r(\tau, \eta_o^u) = - {\tilde
l}^{\tau}(\tau ,\eta_o^u)\, \Big(R^{-1}(\tau)\, {\dot
R}(\tau)\Big)^r{}_s\, \eta^s_o = - ({\vec \eta}_o \times {{\vec
\Omega}\over c})^r + O(\Omega^2/c^2)$.\medskip

Let us introduce an orthonormal tetrad ${\cal W}^{\mu}_{(\alpha)}$,
$\eta_{\mu\nu}\, {\cal W}^{\mu}_{(\alpha)}\, {\cal
W}^{\nu}_{(\beta)} = \eta_{(\alpha)(\beta)}$, whose time-like
4-vector is $l^{\mu}$, i.e. We have ${\cal W}^{\mu}_{(o)} = l^{\mu}
= \epsilon^{\mu}_{\tau} = {\cal W}^A_{(o)}\, \epsilon^{\mu}_A =
{\tilde {\cal W}}^A_{(o)}(\tau ,\eta_o^u)\, z^{\mu}_A(\tau
,\eta_o^u)$ with .${\cal W}^A_{(o)} = (1; 0)$ and ${\tilde {\cal
W}}^A_{(o)}(\tau ,\eta_o^u) = {\tilde l}^A(\tau ,\eta_o^u) = \Big(1;
- ({\vec \eta}_o \times {{\vec \Omega}\over c})^r\Big) +
O(\Omega^2/c^2)$. The spatial axes ${\cal W}^{\mu}_{(i)} = {\cal
W}^A_{(i)}\, \epsilon^{\mu}_A = {\tilde {\cal W}}^A_{(i)}(\tau
,\eta_o^u)\, z^{\mu}_A(\tau ,\eta_o^u)$ with $l_{\mu}\, {\cal
W}^{\mu}_{(i)} = [{\tilde l}^A\, g_{AB}\, {\tilde {\cal W}}^B](\tau
,\eta_o^u) = 0$ must be non-rotating with respect to the observer
with 4-velocity proportional to $z^{\mu}_{\tau}(\tau ,\eta_o^u)$.
Therefore we must have ${\tilde {\cal W}}^A_{(i)} = \Big(0; {\tilde
{\cal W}}^r_{(i)}\Big)$ with ${\tilde {\cal W}}^r_{(i)} = const.$.
As a consequence we have ${\cal W}^A_{(i)}(\tau) = \Big(0;
R^r{}_s(\tau)\, {\tilde {\cal W}}^s_{(i)} \Big) + O(\Omega^2/c^2)$.

\medskip

The polarization axes of sources and detectors will be defined by a
tetrad ${\cal E}^{\mu}_{(\alpha)}(\tau ,\eta_o^r) = {\cal
E}^A_{(\alpha)}(\tau ,\eta_o^r)\, \epsilon^{\mu}_A = {\tilde {\cal
E}}^A_{(\alpha)}(\tau ,\eta_o^r)\, z^{\mu}_A(\tau ,\eta_o^r)$,
$\eta_{\mu\nu}\, {\cal E}^{\mu}_{(\alpha)}\, {\cal
E}^{\nu}_{(\beta)} = \eta_{(\alpha)(\beta)}$ with the following
properties:\medskip

a) the time-like 4-vector ${\cal E}^{\mu}_{(o)}(\tau ,\eta_o^r)$ is
such that its components ${\tilde {\cal E}}^A_{(o)}(\tau ,\eta_o^r)$
coincide with the components ${\tilde u}^A(\tau) = (1; 0) +
O(\Omega^2/c^2)$ of the 4-velocity $u^{\mu}(\tau)$ of the object
located at $\zeta^{\mu}(\tau) = z^{\mu}(\tau ,\eta_o^r)$: as a
consequence we have ${\cal E}^{\mu}_{(o)}(\tau ,\eta_o^r) =
z^{\mu}_{\tau}(\tau ,\eta_o^r) + O(\Omega^2/c^2) = u^{\mu}(\tau)$;

b) the spatial axes ${\cal E}^{\mu}_{(i)}(\tau ,\eta_o^r)$,
orthogonal to the 4-velocity $u^{\mu}(\tau)$, must be at rest in the
rotating frame: we have to identify their components ${\tilde {\cal
E}}^A_{(i)}(\tau ,\eta_o^r)$.
\medskip

If at the observer position we consider the Lorentz transformation
sending $l^{\mu}$ to $u^{\mu}(\tau )$, i.e. $L^{\mu}{}_{\nu}(l
\mapsto u(\tau))$, its projection  $L^A{}_B(\vec \beta) \,
{\buildrel {def}\over =}\, \epsilon^A_{\mu}\, L^{\mu}{}_{\nu}(l
\mapsto u(\tau))\, \epsilon^{\nu}_B$ is a Wigner boost, see
Eq.(\ref{2.8}), with parameter $\vec \beta = \{ \beta^r =
R^r{}_s(\tau )\, \Big({\vec \eta}_o \times {{\vec \Omega}\over
c}\Big)^s$ (so that $\gamma = \sqrt{1 - {\vec \beta}^2} = 1 +
O(\Omega^2/c^2)$). Therefore the transformation sending the
components ${\tilde l}^A(\tau ,\eta_o^u)$ of the unit normal into
the components ${\tilde u}^A(\tau)$ of the 4-velocity modulo terms
of order $O(\Omega^2/c^2)$ is

\bea
 {\tilde {\cal E}}^A_{(o)}(\tau, \eta_o^u) &=& {\tilde u}^A(\tau) =
(1; 0) + O(\Omega^2/c^2) =\nonumber \\
 &=&\Big(z^A_{\mu}\, \epsilon^{\mu}_C\, L^C{}_D(\vec \beta)\,
\epsilon_{\nu}^D\, z^{\nu}_B\, {\tilde l}^B_{(o)}\Big)(\tau
,\eta_o^u) + O(\Omega^2/c^2) =\nonumber \\
 &=&\Big(z^A_{\mu}\, \epsilon^{\mu}_C\, L^C{}_D(\vec \beta)\,
 \epsilon_{\nu}^D\, z^{\nu}_B\, {\tilde {\cal W}}^B_{(o)}\Big)(\tau ,\eta_o^u)
 + O(\Omega^2/c^2),\nonumber \\
 &&{}\nonumber \\
 {\tilde {\cal E}}^A_{(i)}(\tau, \eta_o^u) &=&\Big(z^A_{\mu}\, \epsilon^{\mu}_C\,
 L^C{}_D(\vec \beta)\, \epsilon_{\nu}^D\, z^{\nu}_B\,
 \Big)(\tau ,\eta_o^u)\, {\tilde {\cal
W}}^B_{(i)}.
 \label{6.20}
 \eea

This complete the construction of the non-rotating tetrads ${\cal
E}^{\mu}_{(\alpha)}(\tau ,\eta_o^u)$ for the objects at rest at
$\eta_o^r$.\medskip

A detector endowed of such a non-rotating tetrad will measure the
following projections of the electro-magnetic field strength on its
polarization directions

 \beq
\hat{E}_{(i)} = F_{AB}\, {\tilde u}^A\, {\tilde {\cal
E}}^B_{(i)},\qquad \hat{B}_{(i)} = {1\over 2}\,
\epsilon_{(i)(j)(k)}\, F_{AB}\, {\tilde {\cal E}}^A_{(j)}\, {\tilde
{\cal E}}^B_{(k)}.
 \label{6.21}
 \eeq

These quantities have to be confronted with the non-inertial
electric and magnetic fields $E_r$ and $B_r$, whose projections on
the non-rotating spatial axes ${\tilde {\cal W}}^A_{(i)} = (0;
{\tilde {\cal W}}^r_{(i)})$ inside the instantaneous 3-space are

 \beq
{E}_{(i)} = E_r\, {\tilde {\cal W}}^r_{(i)},\qquad {B}_{(i)} = B_r\,
{\tilde {\cal W}}^r_{(i)}.
 \label{6.22}
  \eeq

Eqs.(\ref{6.20}) imply the following connection among these
quantities

 \bea
  \hat{E}_{(i)}&=& {E}_{(i)} + {\cal
O}(\Omega^2/c^2),\nonumber\\
&&\nonumber\\
\hat{B}_{(i)}&=& {B}_{(i)} - \epsilon_{ijk}\, {\tilde {\cal
W}}^r_{(j)}\, \delta_{rs}\, \Big({\vec \eta}_o \times {{\vec
\Omega}\over c}\Big)^s\, {E}_{(k)} + {\cal O}(\Omega^2/c^2).
 \label{6.23}
 \eea

\medskip

For radio wave (like in the case of GPS) the directions ${\cal
G}_{(i)}^a$ or ${\tilde {\cal E}}_{(i)}^r$ are realized by means of
antennas attached to both emitters and receivers. In the optical
range the antennas are replaced by components of the macroscopic
devices used for the emission and the detection.

\subsection{The Phase Wrap Up Effect}

The  {\em phase wrap up} is a  modification of the phase when a
receiver in rotational motion analyzes the circularly polarized
radiation emitted by a source at rest in an inertial frame. Till now
the effect has been explained by using the 1+3 point of view and the
locality hypothesis in Refs.\cite{31}, where it shown that it is a
particular case of helicity-rotation coupling (the spin-rotation
coupling for photons). It has been verified experimentally, in
particular in GPS \cite{36}, where the receiving antenna on the
Earth surface is rotating with Earth.

\medskip

We will explain the effect by using the non-inertial solution
(\ref{6.7}) and an observer at rest in an inertial frame endowed of
the tetrad ${\cal G}^{\mu}_{(A)}$ defined in Subsubsection 1 of
Subsection B. We rewrite the spatial axes in the form $ {\cal
G}^a_{(i)} = \Big(I^a_{(1)}, I^a_{(2)}, \hat{K}^a\Big)$ with the
vectors satisfying $\vec{I}_{(1)} \cdot \vec{I}_{(2)} = 0$,
$\vec{I}_{(\lambda)} \cdot \hat{K} = 0$ ($\lambda = 1,2$),
$\vec{I}_{(\lambda)}^{\,2} = 1$. Then we pass to a circular basis by
introducing the vectors $\vec{I}_{(\pm)} = \frac{\vec{I}_{(1)} + i\,
\vec{I}_2}{\sqrt{2}}$, which satisfy $\hat{K} \cdot \vec{I}_{(\pm)}
= 0$, $\vec{I}_{(\pm)}^{\,2} = 0$ and $\vec{I}_{(+)} \cdot
\vec{I}_{(-)} = 1$.\medskip

In the rotating non-inertial frame a right-circularly polarized
wave, emitted in the inertial frame, will have the form (\ref{6.7})
($\hat{K} \cdot \vec{I}_{(+)} = 0$ is the transversality condition)

\beq A_{\perp r}(\tau,\vec{\sigma}) = \frac{F}{\omega}\, I_{(+)a}\,
R^a{}_r(\tau)\, e^{i\, {{\omega}\over c}\, \Phi(\tau,\vec{\sigma})}.
 \label{6.24}
 \eeq
\medskip

Let us remark that in the circular basis we have ${\vec A}_{\perp} =
A_n\, \hat n + A_+\, {\vec I}_{(+)} + A_{-}\, {\vec I}_{(-)}$, but
the components $A_n$, $A_{\pm}$, coincide with either linearly or
circularly polarized states of the electro-magnetic field only for
$\hat{n} = \hat{k}$, since $\hat K = {{\omega}\over c}\, \hat k$
(${\hat K}^2 = {{\omega^2}\over {c^2}}$) is the wave vector.

From Eqs(\ref{6.24}) we obtain the following non-inertial magnetic
and electric fields (\ref{2.19})

 \bea
B_r&=&-\, \frac{F}{c}\, I_{(+)a}\, R^a{}_r(\tau)\, e^{i\,
{{\omega}\over c}\, \Phi(\tau,\vec{\sigma})}\byd B_{o}\,
I_{(+)a}(K)\, R^a{}_r(\tau)\, e^{i\, {{\omega}\over
c}\, \Phi(\tau,\vec{\sigma})},\nonumber\\
&&\nonumber\\
E_r&=&- i\, \frac{F}{c}\,\, I_{(+)a}\, R^a{}_r(\tau)\, e^{i\,
{{\omega}\over c}\, \Phi(\tau,\vec{\sigma})} + \frac{1}{c}\,
(\vec{\Omega}\times\vec{\sigma}) \times \vec{B} =\nonumber \\
 &{\buildrel {def}\over =}& E_{o}\, I_{(+)a}\, R^a{}_r(\tau)\, e^{i\,
 {{\omega}\over c}\, \Phi(\tau,\vec{\sigma})}
 +\, E_\ell\, \hat{K}_a\, R^a{}_r(\tau)\,\, e^{i\,
{{\omega}\over c}\, \Phi(\tau,\vec{\sigma})},\nonumber \\
 &&{}\nonumber \\
  B_o = - \frac{F}{c},&& E_o = - i\, \frac{F}{c} + \frac{1}{c}\,
(\vec{\Omega}\times\vec{\sigma}) \times \vec{B} \cdot \vec{I}_{(-)},
\quad E_\ell = \frac{1}{c}\, (\vec{\Omega} \times \vec{\sigma})
\times \vec{B} \cdot \hat{K}.
 \label{6.25}
 \eea

\bigskip

Let us now consider a receiver at rest in the rotating frame. Since
its 4-velocity is ${\tilde u}^A = (1; 0)$, it can be endowed with
the non-rotating tetrad ${\tilde {\cal W}}^A_{(\alpha)}$ of
Subsubsection 2 of Subsection B. If $\hat n$ is the unit vector in
the direction of the rotation axis (), i.e. if $\vec \Omega =
\Omega\, \hat n$, we can choose the spatial axes ${\tilde {\cal
W}}^r_{(i)} = (\epsilon^r_{(1)}, \epsilon^r_{(2)}, \hat{n}^r)$ with
$\vec{\epsilon}_{(1)} \cdot \vec{\epsilon}_{(2)} = 0$,
$\vec{\epsilon}_{(\lambda)} \cdot \hat{K} = 0$,
$\vec{\epsilon}_{(\lambda)}^{\,2} = 1$. If we introduce the circular
basis $\vec{\epsilon}_{(\pm)} = \frac{\vec{\epsilon}_{(1)} + i\,
\vec{\epsilon}_2}{\sqrt{2}}$, we have $\hat{n} \cdot
\vec{\epsilon}_{(\pm)} = 0$, $\vec{\epsilon}_{(\pm)}^{\,2} = 0$,
$\vec{\epsilon}_{(+)} \cdot \vec{\epsilon}_{(-)} = 1$ and
$R^a{}_r(\tau)\, \epsilon_{(\pm)}{}^r = \epsilon_{(\pm)}^a\,\,
e^{\left[\pm\, i\, {{\Omega}\over c}\, \tau\right]}$.
\bigskip

The receiver will measure the following magnetic and electric fields

 \begin{eqnarray*}
B_n&=&B_r\,\hat{n^r}=B_o
\,\left(\vec{I}_{(+)a} \hat{n}^a\right)\, \exp\,\left[i {{\omega}\over c}\,\Phi\right],\nonumber\\
&&\nonumber\\
B_{(\pm)}&=&B_r\,\epsilon^r_{(\mp)}=B_o \, \left(\vec{I}_{(+)a}
{\epsilon}^a_{(\mp)}\right)\,\exp\,\left[{i\over c}\, \left(\mp
\Omega\, \tau + \omega\, \Phi(\tau,\vec{\sigma})\right)\right],
 \end{eqnarray*}

 \bea
E_n&=&E_r\,\hat{n^r}=\left[E_o
\,\left(\vec{I}_{(+)a} \hat{n}^a\right)+E_\ell\,\hat{K}_a\hat{n}^a\right]\,
\exp\,\left[i {{\omega}\over c}\,\Phi\right],\nonumber\\
&&\nonumber\\
E_{(\pm)}&=&E_r\,\epsilon^r_{(\mp)}=\left[E_o \,
\left(\vec{I}_{(+)a}
{\epsilon}^a_{(\mp)}\right)+E_\ell\,\hat{K}_a\,\epsilon^a_{(\pm)}\right]\,
\exp\,\left[{i\over c}\, \left(\mp \Omega\, \tau + \omega\,
\Phi(\tau,\vec{\sigma})\right)\right].
 \label{6.27}
 \eea

\bigskip

In the case $\hat{n}^a = \hat{K}^a$  we find

 \begin{eqnarray*}
B_n&=&B_{(-)}=0\nonumber\\
&&\nonumber\\
B_{(+)}&=&B_o \,e^{\,\left[{i\over c}\, \left( (\omega-\Omega) \tau
+ \vec{K}\cdot\vec{\sigma}\right)\right]},
 \end{eqnarray*}

 \bea
E_n&=&E_\ell\, e^{\,\left[i {{\omega}\over c}\, \left(
\omega \tau + \vec{K}\cdot\vec{\sigma}\right)\right]},\qquad E_{(+)}=0\nonumber\\
&&\nonumber\\
E_{(+)}&=&E_o\, e^{\,\left[{i\over c}\, \left( (\omega-\Omega) \tau
+ \vec{K}\cdot\vec{\sigma}\right)\right]}.
 \label{6.28}
 \eea

Therefore the components $B_{(+)},E_{(+)}$ have the frequency
modified to $\omega \mapsto \omega - \Omega$: this is the phase wrap
up effect. These are same results as in Ref.\cite{31} at the lowest
order in $\Omega/c$. The only new fact is the presence of the
component $E_n\neq 0$.

It would be interesting to make the calculation of the deviations of
order $O(\Omega^2/c^2)$ from rigid rotation, to see whether the
result $\omega \mapsto \gamma\, (\omega \pm \Omega)$ ($\gamma$ is a
Lorentz factor), found in Ref.\cite{31} by using the locality
hypothesis and supporting the interpretation with the
helicity-rotation coupling, is confirmed.

\subsection{The Sagnac Effect}

Following a suggestion of Ref.\cite{37} let us consider the solution
(\ref{6.8}) in the eikonal approximation, which describes the
propagation of the radiation along a ray of light whose trajectory
is given in Eq.(\ref{6.19}). This solution allows to get a
derivation of the Sagnac effect (described in Appendix A) along the
lines of Ref.\cite{31a}.
\medskip

Let us consider two receivers $A$ and $B$ at rest in the rotating
frame and characterized by the 3-coordinates $\eta^r_A$ and
$\eta^r_B$ respectively. Let us assume that $A$ and $B$ lie in the
same 2-plane containing the origin $\sigma^r = 0$ and orthogonal  to
$\vec \Omega$. Therefore we have $\vec{\Omega} \cdot \vec{\eta}_A =
\vec{\Omega} \cdot \vec{\eta}_B=0$. Let us assume that $A$ and $B$
are both on the trajectory of a ray of light, so that
Eq.(\ref{6.19}) implies the existence of a time $\tau_{AB}$ such
that we have

 \beq
\vec{\eta}_B-\vec{\eta}_A=\hat{k}\,\tau_{AB}+
\left(\frac{\vec{\Omega}}{c}\times\hat{k}\right)\,\tau^2_{AB}+{\cal
O}(\Omega^2/c^2).
 \label{6.29}
  \eeq
\medskip

The phase difference between $A$ and $B$ at the same instant $\tau$
is

 \bea
 \Delta\varphi_{AB}&=&\frac{\omega}{c}\, \left[\Phi(\tau,\vec{\eta}_B)
- \Phi(\tau,\vec{\eta}_A)\right] =\nonumber\\
&&\nonumber\\
&=&- \frac{\omega}{c}\, \left[\hat{k} \cdot (\vec{\eta}_B -
\vec{\eta}_A) + (\hat{k} \cdot \vec{\eta}_B)\,
\left(\frac{\vec{\Omega}}{c} \cdot \vec{\eta}_B \times
\hat{k}\right) - (\hat{k} \cdot \vec{\eta}_A)\,
\left(\frac{\vec{\Omega}}{c} \cdot \vec{\eta}_A \times
\hat{k}\right)\right] +\nonumber\\
&&\nonumber\\
&+&O(\Omega^2/c^2).
 \label{6.30}
 \eea

Eq.(\ref{6.29}) implies

 \beq
\vec{\eta}_B = \vec{\eta}_A + \hat{k}\, \tau_{AB} + O(\Omega/c)\,
\Rightarrow\, \frac{\vec{\Omega}}{c} \cdot \vec{\eta}_B \times
\hat{k} = \frac{\vec{\Omega}}{c} \cdot \vec{\eta}_A \times \hat{k} +
O(\Omega^2/c^2),
 \label{6.31}
 \eeq

\noindent so that we get

 \beq
\Delta\varphi_{AB} = - \frac{\omega}{c}\, \left[\hat{k} \cdot
(\vec{\eta}_B - \vec{\eta}_A) + \hat{k} \cdot (\vec{\eta}_B -
\vec{\eta}_A)\, \left(\frac{\vec{\Omega}}{c} \cdot \vec{\eta}_A
\times \hat{k}\right) + O(\Omega^2/c^2) \right].
 \label{6.32}
 \eeq
\medskip

Since  Eq.(\ref{6.29}) also implies $\vec{\eta}_B - \vec{\eta}_A =
\mid\vec{\eta}_B - \vec{\eta}_A\mid\, \hat{k} + O(\Omega^2/c^2)$, we
arrive at the result

 \beq
\Delta\varphi_{AB} = - \frac{\omega}{c}\, \left[\mid\vec{\eta}_B -
\vec{\eta}_A\mid + \frac{\vec{\Omega}}{c} \cdot \vec{\eta}_A \times
(\vec{\eta}_B - \vec{\eta}_A) \right] + O(\Omega^2/c^2).
 \label{6.33}
 \eeq

\medskip

If $A_{BAO}$ is the area of the triangle BAO in the 2-plane
orthogonal to $\vec \Omega$, we have $\frac{\vec{\Omega}}{c} \cdot
\vec{\eta}_A \times (\vec{\eta}_B - \vec{\eta}_A) = \pm 2\,
\frac{\Omega}{c}\, A_{BAO}$ (the choice of $\pm$ depends on the
direction of motion of the ray). As a consequence, the phase
difference is the sum of the following two terms

 \beq
\Delta\varphi_{AB} = - \frac{\omega}{c} \mid\vec{\eta}_B -
\vec{\eta}_A\mid + \delta\varphi_{AB} + O(\Omega^2/c^2).
 \label{6.34}
 \eeq

While the first term, $- \frac{\omega}{c} \mid \vec{\eta}_B -
\vec{\eta}_A\mid$, is present also in the inertial frames, the
second term

 \beq
\delta\varphi_{AB} = \mp \frac{2\, \omega\, \Omega}{c^2}\, A_{BAO},
 \label{6.35}
 \eeq

\noindent is the extra phase variation due to the rotation of the
frame. This is the Sagnac effect.

\subsection{The Inertial Faraday Rotation}

Let us give the derivation of the rotation of the polarization of an
electro-magnetic wave in a rotating frame, named inertial Faraday
rotation, which is important in astrophysics \cite{38}, were it is
induced by the gravitational field (due to the equivalence principle
only non-inertial frames are allowed in general relativity). Our
approach is analogous to the one of Ref.\cite{29} in the case of
Post-Newtonian gravity.
\bigskip

Let us consider the amplitude $\vec{a}$ of the solution (\ref{6.8})
in the eikonal approximation: it carries the information about the
polarization of a ray of light. To study the first-order transport
equation for it, the second of Eqs.(\ref{6.9}), let us make the
series expansion

 \beq
\vec{a}(\tau,\vec{\sigma}) = \vec{a}_o(\tau,\vec{\sigma}) +
\frac{\Omega}{c}\, \vec{a}_1(\tau,\vec{\sigma}) +
O\Big(\frac{\Omega^2}{c^2}\Big),
 \label{6.36}
 \eeq

\noindent and let us make the ansatz (in an inertial frame it
corresponds to a plane wave)

 \beq
\vec{a}_o(\tau,\vec{\sigma}) = \vec{a}_o = \mbox{ const.}, \quad
\Rightarrow \quad \frac{\partial\vec{a_o}}{\partial\tau} = 0,\qquad
\partial_r\,\vec{a}_o = 0.
 \label{6.37}
 \eeq
\medskip

This ansatz implies the following form of the second and third
equation in Eqs.(\ref{6.9})

 \bea
  &&\frac{\Omega}{c}\, \left[
\left(\frac{\partial\vec{a}_1}{\partial\tau} + \hat{\Omega} \times
\vec{a}_o\right) - (\hat{k} \cdot \vec{\partial})\, \vec{a}_1
\right] + O\Big(\frac{\Omega^2}{c^2}\Big) = 0,\nonumber\\
&&\nonumber\\
&&\vec{a}_o \cdot \hat{k} + \frac{\Omega}{c}\, \left[\vec{a}_o \cdot
\Big(\,\hat{k}\, (\hat{\Omega} \cdot \vec{\sigma} \times \hat{k}) -
(\hat{k} \cdot \vec{\sigma})\, (\hat{\Omega} \times \hat{k}) \,\Big)
 + \vec{a}_1 \cdot \hat{k}\right] + O\Big(\frac{\Omega^2}{c^2}\Big) = 0.
 \label{6.38}
 \eea

To study these equations, let us assume that each rotating receiver
is endowed with a tetrad of the type given in Eq.(\ref{6.20}): the
spatial axes ${\tilde {\cal W}}^r_{(i)} = (R^r_{1}(k), R^r_{2}(k),
\hat{k}^r)$ with $\vec{R}_{\lambda}(k) \cdot \vec{R}_{\lambda'}(k) =
\delta_{\lambda\lambda'}$, $\vec{R}_\lambda(k) \cdot \hat{k} = 0$.
\medskip

The second of Eqs.(\ref{6.38}) for the unknown $\vec{a}_o$,
$\vec{a}_1$ is the transversality condition and it implies

\bea
 order\, 0\, in\, \Omega &\rightarrow& \vec{a}_o \cdot \hat{k}
 = 0\, \Rightarrow\, \vec{a}_o = a_o^\lambda\, \vec{R}_\lambda(k),
 \nonumber \\
 order\, 1\, in\, \Omega &&\nonumber \\
 &&{}\nonumber \\
 \vec{a}_1 \cdot \hat{k}&=&- \vec{a}_o \cdot \Big(\,\hat{k}\, (\hat{\Omega}
 \cdot \vec{\sigma} \times \hat{k}) + (\hat{k} \cdot \vec{\sigma})\,
 (\hat{\Omega} \times \hat{k})\,\Big)=\nonumber\\
&&\nonumber\\
&=&- a_o^\lambda\, \vec{R}_\lambda(k) \cdot (\hat{\Omega} \times
\hat{k})\, (\hat{k} \cdot \vec{\sigma}).
 \label{6.39}
 \eea

\medskip

Due to the ansatz (\ref{6.37}) the first of Eqs.(\ref{6.38}) is of
order 1 in $\Omega$ and gives the following condition on $\vec{a}_1$

 \beq
\frac{\partial\vec{a}_1}{\partial\tau} - \hat{\Omega} \times
\vec{a}_o + (\hat{k} \cdot \vec{\partial})\, \vec{a}_1 = 0.
 \label{6.40}
 \eeq

If we project this equation on the directions $\hat{k}$,
$\vec{R}_\lambda(k)$, we get

 \bea
&&\frac{\partial}{\partial\tau}\, (\vec{a_1} \cdot \hat{k}) -
\hat{\Omega} \times \vec{a}_o \cdot \hat{k} + (\hat{k} \cdot
\vec{\partial})\, (\vec{a}_1 \cdot \hat{k}) = 0,\nonumber\\
&&\nonumber\\
&&\frac{\partial a_1^\lambda}{\partial\tau} - \hat{\Omega} \times
\vec{R}_{\lambda'}(k) \cdot \vec{R}_\lambda(k)\, a_o^{\lambda'} +
(\hat{k} \cdot \vec{\partial})\, a_1^\lambda = 0.
 \label{6.41}
 \eea

While the first of Eqs.(\ref{6.41}) is automatically satisfied, the
second one is an equation for the components $a_1^\lambda$. The
simplest solutions are obtained with the following ansatz

 \beq
\frac{\partial a_1^\lambda}{\partial\tau} = 0,\quad \Rightarrow
\quad a_1^\lambda(\tau) = \left[\hat{\Omega} \times
\vec{R}_{\lambda'}(k) \cdot\vec{R}_\lambda(k)\right]\, \hat{k} \cdot
\vec{\sigma}\, a_o^{\lambda'}.
 \label{6.42}
 \eeq

\medskip

The final solution for the transverse electro-magnetic potential is

 \bea
\vec{A}_\perp&=&\frac{a_o{}^1}{\omega}\, \left[\vec{R}_{1} +
\theta(\vec{\sigma})\, \vec{R}_{2}(k) - \frac{\hat{k} \cdot
\vec{\sigma}}{c}\, (\vec{\Omega} \cdot \vec{R}_2(k))\, \hat{k}
\right]\, e^{\,\left(\,i\, \frac{\omega}{c}\, \Phi\right)} +\nonumber\\
&&\nonumber\\
&+&\frac{a_o{}^2}{\omega}\, \left[\vec{R}_{2}(k) -
\theta(\vec{\sigma})\, \vec{R}_{1}(k) + \frac{\hat{k} \cdot
\vec{\sigma}}{c}\, (\vec{\Omega} \cdot \vec{R}_1(k))\, \hat{k}
\right]\, e^{\,\left(\,i\, \frac{\omega}{c}\, \Phi\right)} +
O(1/\omega^2),\nonumber \\
 &&{}\nonumber \\
 &&with\nonumber \\
 &&{}\nonumber \\
 \theta(\vec{\sigma}) &=& \frac{1}{c}\, (\hat{k} \cdot
\vec{\sigma})\, (\vec{\Omega} \cdot \hat{k}).
 \label{6.43}
 \eea
\medskip

The resulting non-inertial magnetic and electric fields are
($\vec{B} = \{B_r\}$, $\vec{E} = \{E_r\}$)

 \begin{eqnarray*}
  \vec{B}&=&- \frac{i\, a_o{}^1}{c}\, \left[\,\vec{R}_{2}(k)
  - \theta(\vec{\sigma})\, \vec{R}_{1}(k)\,\right]\,
e^{\,\left(\,i\, \frac{\omega}{c}\, \Phi\right)} -\nonumber\\
&&\nonumber\\
&-&\frac{i\, a_o{}^1}{c}\, \left[\left(\frac{\vec{\Omega}}{c} \times
\vec{\sigma} \cdot \hat{k}\right)\, \vec{R}_{2}(k) - \frac{\hat{k}
\cdot \vec{\sigma}}{c}\, (\vec{\Omega} \cdot \vec{R}_1(k))\,
\hat{k}\, \right]\, e^{\,\left(\,i\, \frac{\omega}{c}\, \Phi\right)}
+\nonumber\\
&&\nonumber\\
&+&\frac{i\, a_o{}^2}{c}\, \left[\,\vec{R}_{1}(k) +
\theta(\vec{\sigma})\, \vec{R}_{2}(k)\,\right]\,
e^{\,\left(\,i\, \frac{\omega}{c}\, \Phi\right)} +\nonumber\\
&&\nonumber\\
&+&\frac{i\, a_o{}^2}{c}\, \left[\left(\frac{\vec{\Omega}}{c} \times
\vec{\sigma} \cdot \hat{k}\right)\, \vec{R}_{1}(k) + \frac{\hat{k}
\cdot \vec{\sigma}}{c}\, (\vec{\Omega} \cdot \vec{R}_2(k))\, \hat{k}\,
\right]\, e^{\,\left(\,i\, \frac{\omega}{c}\, \Phi\right)} +\nonumber\\
&&\nonumber\\
&+&O(1/\omega) + O(\Omega^2/c^2) =\nonumber\\
&&\nonumber\\
&\byd&\,b(\vec{\sigma})\,\, e^{\,\left(\,i\, \frac{\omega}{c}\,
\Phi\right)} + O(1/\omega) + O(\Omega^2/c^2),
\end{eqnarray*}

\bea
 \vec{E}&=&- \frac{i\, a_o{}^1}{c}\, \left[\vec{R}_{1}(k) + \theta(\vec{\sigma})\,
 \vec{R}_{2}(k) - \frac{\hat{k} \cdot \vec{\sigma}}{c}\, (\vec{\Omega}
 \cdot \vec{R}_2(k))\, \hat{k}\right]\, e^{\,\left(\,i\, \frac{\omega}{c}\,
 \Phi\right)} -\nonumber\\
&&\nonumber\\
&-&\frac{i\, a_o{}^2}{c}\, \left[\vec{R}_{2}(k) -
\theta(\vec{\sigma})\, \vec{R}_{1}(k) + \frac{\hat{k} \cdot
\vec{\sigma}}{c}\, (\vec{\Omega} \cdot \vec{R}_1(k))\, \hat{k}
\right]\, e^{\,\left(\,i\, \frac{\omega}{c}\, \Phi\right)} +\nonumber\\
&&\nonumber\\
&+&O(1/\omega) + O(\Omega^2/c^2) =\nonumber\\
&&\nonumber\\
&\byd&\,f(\vec{\sigma})\,\, e^{\,\left(\,i\, \frac{\omega}{c}\,
\Phi\right)} + O(1/\omega) + O(\Omega^2/c^2).
 \label{6.44}
 \eea

\medskip

As in the case of the Sagnac effect let us consider two receivers
$A$ and $B$ at the endpoints of the same light ray described by Eqs,
(\ref{6.19}) and (\ref{6.29}). The magnetic field observed by $A$,
$\vec{B}(\tau,\vec{\eta}_A)$, differs from the one observed by $B$,
$\vec{B}(\tau,\vec{\eta}_B)$. Since the phase changes have been
already analyzed for the Sagnac effect, let us concetrate on the
amplitudes $\vec{b}(\vec{\eta}_A)$ and $\vec{b}(\vec{\eta}_B)$.
Since Eq.(\ref{6.29}) gives $\vec{\eta}_B - \vec{\eta}_A = \hat{k}\,
\tau_{AB} + O(\Omega/c)$, we find

 \bea
  \vec{b}(\vec{\eta}_B) - \vec{b}(\vec{\eta}_A)&=&
\frac{i\, a_o{}^1}{c}\, \delta\theta_{BA}\, \vec{R}_{1}(k) +
\frac{i\, a_o{}^2}{c}\, \delta\theta_{BA}\, \vec{R}_{2}(k) +\nonumber\\
&&\nonumber\\
&+&\frac{i\, a_o{}^1}{c}\, \left[\frac{\mid\vec{\eta}_B -
\vec{\eta}_A\mid}{c}\, (\vec{\Omega} \cdot \vec{R}_1(k))\,
\hat{k}\,\right] +\nonumber\\
&&\nonumber\\
&+&\frac{i\, a_o{}^2}{c}\, \left[\frac{\mid\vec{\eta}_B -
\vec{\eta}_A\mid}{c}\, (\vec{\Omega} \cdot \vec{R}_2(k))\,
\hat{k}\,\right] + O(\Omega^2/c^2),\nonumber \\
 &&{}\nonumber \\
 &&with\nonumber \\
 &&{}\nonumber \\
 \delta\theta_{BA}&=&\theta(\vec{\eta}_B) - \theta(\vec{\eta}_A) =
\frac{1}{c}\, \mid\vec{\eta}_B - \vec{\eta}_A\mid\, (\vec{\Omega}
\cdot \hat{k}) + O(\Omega^2/c^2).
 \label{6.45}
 \eea

\noindent $\delta\theta_{BA}$ is the angle of the {\it inertial
Faraday rotation} (in this case it is small, $\delta\theta_{AB} \sim
\Omega/c$). It agrees with Eq.(4) of Ref.\cite{38}, where it has the
form $\delta\theta_{AB} = - \frac{1}{2}\, \int_A^B\,
\sqrt{g_{\tau\tau}}\, (\nabla \times \vec{n}) \cdot d\vec{\sigma}$
as a line integral along the spatial trajectory of the light ray.
This formula agrees with our result, because, due to the
approximations we have done, we have $g_{\tau\tau} = 1$, $(\nabla
\times \vec{n}) = - \frac{2\, \vec{\Omega}}{c}$ and our ray
trajectory is $\vec{\sigma}(\tau) = \hat{k}\, \tau + \vec{\sigma}_o
+ O(\Omega^2/c^2)$.

\medskip

To make the {\it rotation} explicit, let us write the components
along the two polarization directions: $b_{(\lambda)}(\vec{\eta}_A)
= \vec{b}(\vec{\eta}_A) \cdot \vec{R}_\lambda(k)$ and
$b_{(\lambda)}(\vec{\eta}_B) = \vec{b}(\vec{\eta}_B) \cdot
\vec{R}_\lambda(k)$. In this way we get

 \bea
b_{(1)}(\vec{\eta}_B)&=&b_{(1)}(\vec{\eta}_A) + \delta\theta_{AB}\,
\frac{i\, a_o{}^1}{c} + O(\Omega^2/c^2) = b_{(1)}(\vec{\eta}_A) -
\delta\theta_{AB}\, b_{(2)}(\vec{\eta}_A) + O(\Omega^2/c^2),\nonumber\\
&&\nonumber\\
b_{(2)}(\vec{\eta}_B)&=&b_{(2)}(\vec{\eta}_A) + \delta\theta_{AB}\,
\frac{i\, a_o{}^2}{c} + O(\Omega^2/c^2) = b_{(2)}(\vec{\eta}_A) +
\delta\theta_{AB}\, b_{(1)}(\vec{\eta}_A) + O(\Omega^2/c^2).
 \label{6.46}
 \eea

\noindent This is just a small angle rotation with
$b_{(\lambda)}(\vec{\eta}_B) = R_{\lambda}{}^{\lambda'}(k)\,
(\delta\theta_{AB})\, b_{(\lambda')}(\vec{\eta}_A)$.

\medskip

The electric field may be treated in the same way.

\vfill\eject

\section{Conclusions}

In this paper we have defined the general theory of {\it
non-inertial frames} in Minkowski space-time. It is based on
M$\o$ller-admissible 3+1 splittings of Minkowski space-time (they
give conventions for clock synchronization, i.e. for the
identification of instantaneous 3-spaces) and on parametrized
Minkowski theories for isolated systems admitting a Lagrangian
description. The transition from a non-inertial frame to every other
one is formalized as a gauge transformation, so that physical
results do not depend on how the clock are synchronized.\medskip

The M$\o$ller conditions, implying the absence of rotational
velocities higher than the velocity of light $c$ and requiring that
the three eigenvalues of the non-inertial 3-metric inside the
instantaneous Riemannian 3-spaces has three non-null positive
eigenvalues, have to be implemented with the following two extra
conditions:

a) the lapse function must be positive definite in each point of the
instantaneous 3-space, so to avoid the intersection of 3-spaces at
different times;

b) the space-like hyper-surfaces corresponding to the Riemannian
3-spaces must become space-like hyper-planes (Euclidean 3-spaces) at
spatial infinity with a direction-independent unit normal
$l^{\mu}_{(\infty )}$ (asymptotic inertial observers to be
identified with the fixed stars).\bigskip

Among the admissible non-inertial frames we identified the {\it
non-inertial rest frames}, generalizing the inertial rest frames and
relevant for canonical gravity \cite{5,11,12}.\bigskip

All the properties of the inertial rest-frame instant form of
dynamics, studied in details in Refs.\cite{8}, have been extended to
non-inertial frames. Again every isolated system may be described as
a decoupled non-covariant external center of mass carrying a
pole-dipole structure: the internal mass of the system and an
effective spin (becoming the rest spin in the inertial rest frame).
In particular we have found the non-inertial generalization of the
second class constraints eliminating the internal 3-center of mass
inside the instantaneous 3-spaces.\bigskip

This theory of non-inertial frames is free by construction from the
coordinate singularities of all the approaches to accelerated frames
based on the 1+3 point of view, in which the instantaneous 3-spaces
are identified with the local rest frames of the observer. The
pathologies of this approach are either the horizon problem of the
rotating disk (rotational velocities higher than $c$), which is
still present in all the calculations of pulsar magnetosphere in the
form of the light cylinder, or the intersection of the local rest
3-spaces. The main difference between the 3+1 and 1+3 points of view
is that the M$\o$ller conditions forbid {\it rigid rotations} in
relativistic theories. The simplest example of 3+1 splitting with
{\it differential rotations} is given and the 3+1 point of view for
the rotating disk and the Sagnac effect is evaluated. This splitting
is also used to give a special relativistic generalization of the
non-relativistic non-inertial International Terrestrial Reference
System (ITRS) used to describe fixed coordinates on the surface of
the rotating Earth in the conventions IERS2003 \cite{39}.

\bigskip

We have done a detailed study of the isolated system of
positive-energy scalar particles with Grassmann-valued electric
charges plus the electro-magnetic field extending to non-inertial
frames its Hamiltonian description given in the inertial rest frame
in Ref.\cite{8}.\medskip

By using a non-covariant (i.e. coordinate-dependent) decomposition
of the electro-magnetic potential we obtained the {\it non-inertial
radiation gauge}, in which the electro-magnetic field is described
by means of transverse quantities (the Dirac observables). This
allowed us to find the non-inertial expression of the Coulomb
potential, which is now dependent also on the field strengths and
the inertial potentials. The non-covariance of the description is
natural due to the presence in the Hamiltonian of the {\it
relativistic inertial potentials}, namely the components
$g_{AB}(\tau ,\sigma^r)$ of the 4-metric induced by the 3+1
splitting, which are {\it intrinsically coordinate dependent}. The
non-relativistic limit of the inertial potentials reproduces the
standard (again coordinate-dependent) Newtonian ones. The
Hamiltonian in non-inertial frames turns out to be the sum of the
invariant mass (now coordinate-dependent due to its dependence on
the 4-metric) of the system plus terms in the inertial potentials
disappearing in the inertial rest frame.
\bigskip

Then we re-examined some properties of the electro-magnetic wave
solutions of non-inertial Maxwell equations, which till now were
described only by means of the 1+3 point of view, in the 3+1
framework, where there is a well-posed Cauchy problem due to the
absence of coordinate singularities. By considering admissible
nearly rigid rotating frames we recover the results of the 1+3
approach and open the possibility to make these calculations in
presence of deviations from rigid rotations.

\bigskip

A still open problem are the constitutive equations for
electrodynamics in material media in non-inertial systems. For
linear isotropic media see the Wilson-Wilson experiment in
Refs.\cite{40} and Refs.\cite{37,41}, while for an attempt towards a
general theory in arbitrary media (including the premetric extension
of electro-magnetism) see Refs.\cite{42}

\bigskip

In conclusion we have now a good understanding of particles and
electro-magnetism in non-inertial frames in Minkowski space-time,
where the 4-metric induced by the admissible 3+1 splitting describes
all the inertial effects. Going to canonical gravity in the York
canonical basis of Ref.\cite{12} is possible to see which components
remain inertial effects and which become dynamical tidal effects
(the physical degrees of freedom of the gravitational field);
moreover the inertial 3-volume element and some inertial components
of the extrinsic curvature of the instantaneous 3-spaces become
complicated functions of both general relativistic inertial and
tidal effects, because they are determined by the solution of the
super-Hamiltonian constraint (the Lichnerowicz equation) and of the
super-momentum constraints.

 \vfill\eject

\appendix

\section{The Rotating Disk and the Sagnac Effect}

In this Appendix we give the description of a rotating disk and of
the Sagnac effect starting from an admissible 3+1 splitting of
Minkowski space-time of the type of Eqs.(\ref{2.14}). An enlarged
exposition with a rich bibliography is given in Section I Subsection
D and E and in Section VI Subsections B and C of the first paper in
Ref.\cite{3}.

\bigskip

While at the non-relativistic level one can speak of a rigid (either
geometrical or material) disk put in global rigid rotatory motion,
the problem of the relativistic rotating disk is still under debate
(see Refs.\cite{16,43}) after one century from the enunciation of
the Ehrenfest paradox about the 3-geometry of the rotating disk. The
problems arise when one tries to define measurements of length, in
particular that of the circumference of the disk. Einstein \cite{44}
claims that while the rods along the radius $R_o$ are unchanged
those along the rim of the disk are Lorentz contracted: as a
consequence more of them are needed to measure the circumference,
which turns out to be greater than $2\pi\, R_o$ (non-Euclidean
3-geometry even if Minkowski space-time is 4-flat) and not smaller.
This was his reply to Ehrenfest \cite{45}, who had pointed an
inconsistency in the accepted special relativistic description of
the disk \footnote{If $R$ and $R_o$ denote the radius of the disk in
the rotating and inertial frame respectively, then we have $R = R_o$
because the velocity is orthogonal to the radius. But the
circumference of the rim of the disk is Lorentz contracted so that
$2\pi\, R < 2\pi\, R_o$ inconsistently with Euclidean geometry. } in
which it is the circumference to be Lorentz contracted: as a
consequence this fact was named the {\it Eherenfest paradox} (see
the historical paper of Gr$\o$n in Ref.\cite{46}).

\medskip

Since relativistic rigid bodies do not exist, at best we can speak
of {\it Born rigid motions} \cite{47} and {\it Born reference
frames} \footnote{A reference frame or platform is {\it Born-rigid}
\cite{48} if the expansion $\Theta$ and the shear $\sigma_{\mu\nu}$
of the associated congruence of time-like observers vanish, i.e. if
the spatial distance between neighboring world-lines remains
constant. }. However Gr$\o$n \cite{46} has shown that the
acceleration phase of a material disk is not compatible with Born
rigid motions and, moreover, we do not have a well formulated and
accepted relativistic framework to discuss a relativistic elastic
material disk.
\medskip

 As a consequence most of the authors treating the rotating disk
(either explicitly or implicitly) consider it as a {\it geometrical
entity} described by a congruence of time-like world-lines (helices
in Ref.\cite{49}) with non-zero vorticity, i.e. non-surface forming
and therefore non-synchronizable (see for instance Ref.\cite{50}).
This means that there is no notion of instantaneous 3-space where to
visualize the disk (see Ref.\cite{43} for the attempts to define
rods and clocks associated to this type of congruences): every
observer on one of these time-like world-lines can only define the
local rest frame and try to define a local accelerated reference
frame as said in Section IIB.\medskip

In the 3+1 point of view the disk is considered to be a relativistic
isolated system (either a relativistic material body or a
relativistic fluid or a relativistic dust as a limit case
\footnote{As an example of a congruence simulating a geometrical
rotating disk we can consider the relativistic dust described by
generalized Eulerian coordinates of Ref.\cite{51} after the gauge
fixing to a family of differentially rotating parallel
hyper-planes.}) with compact support always contained in a finite
time-like world-tube $W$, which in the Cartesian 4-coordinates of an
inertial system is a time-like cylinder of radius R. Each admissible
3+1 splitting of Minkowski space-time, centered on an arbitrary
time-like observer, gives a visualization of the disk in its
instantaneous 3-spaces $\Sigma_{\tau}$: at each instant $\tau$ the
points of the disk in $W \cap \Sigma_{\tau}$ are synchronized and
through each one of them pass an Eulerian observer. Instead the
irrotational congruence of the disk is described by the second
congruence (whose unit 4-velocity is $z^{\mu}_{\tau}(\tau ,\sigma^u
)/ \sqrt{\sgn\, g_{\tau\tau}(\tau ,\sigma^u )}$  and whose observers
follow generalized helices $\sigma^u = \sigma^u_o$) associated to
the admissible 3+1 splitting: each of the observers of this
congruence, whose world-lines are inside $W$, has no intrinsic
notion of synchronization.\medskip

As a consequence, each instantaneous 3-space $\Sigma_{\tau}$ of an
admissible 3+1 splitting has a well defined (in general Riemannian)
notion of 3-geometry and of spatial length: the radius and the
circumference of the disk are defined in $W \cap \Sigma_{\tau}$, so
that the disk 3-geometry is 3+1 splitting dependent. When the
material disk can be described by means of a parametrized Minkowski
theory, all these 3-geometry are gauge equivalent like the notions
of clock synchronization.\bigskip

The other important phenomenon connected with the rotating disk is
the {\it Sagnac effect} (see the recent review in Ref.\cite{52} for
how many interpretations of it exist), namely the phase difference
generated by the difference in the time needed for a round-trip by
two light rays, emitted in the same point, one co-rotating and the
other counter-rotating with the disk \footnote{For monochromatic
light in vacuum with wavelength $\lambda$ the fringe shift is
$\delta z = 4\, \vec \Omega \cdot \vec A / \lambda\, c$, where $\vec
\Omega $ is the Galilean velocity of the rotating disk supporting
the interferometer and $\vec A$ is the vector associated to the area
$|\vec A|$ enclosed by the light path. The time difference is
$\delta t = \lambda\, \delta z /c = 4\, \vec \Omega \cdot \vec
A/c^2$, which agrees, at the lowest order, with the proper time
difference $\delta \tau = (4\, A\, \Omega /c^2)\, (1 - \Omega^2\,
R^2/c^2)^{-1/2}$, $A = \pi\, R^2$, evaluated in an inertial system
with the standard rotating disk coordinates. This proper time
difference is twice the time lag due to the {\it synchronization
gap} predicted for a clock on the rim of the rotating disk with a
non-time orthogonal metric. See Refs.\cite{37,52,53} for more
details. See also Ref.\cite{36} for the corrections included in the
GPS protocol to allow the possibility of making the synchronization
of the entire system of ground-based and orbiting atomic clocks in a
reference local inertial system. Since  usually, also in GPS, the
rotating coordinate system has $t^{'} = t$ ($t$ is the time of an
inertial observer on the axis of the disk) the gap is a consequence
of the impossibility to extend Einstein's convention of the inertial
system also to the non-inertial one rotating with the disk: after
one period two nearby synchronized clocks on the rim are out of
synchrony.}. This effect, which has been tested (see the
bibliography of Refs.\cite{52,54}) for light, X rays and matter
waves (Cooper pairs, neutrons, electrons and atoms), has important
technological applications and must be taken into account for the
relativistic corrections to space navigation, has again an enormous
number of theoretical interpretations (both in special and general
relativity) like for the solutions of the Ehrenfest paradox. Here
the lack of a good notion of simultaneity leads to problems of {\it
time discontinuities or desynchronization effects} when comparing
clocks on the rim of the rotating disk.
\bigskip

Another area which is in a not well established form is
electrodynamics in non-inertial systems either in vacuum or in
material media ({\it problem of the non-inertial constitutive
equations}). Its clarification is needed both to derive the Sagnac
effect from Maxwell equations without gauge ambiguities \cite{37}
and to determine which types of experiments can be explained by
using the locality principle to evaluate the electro-magnetic fields
in the comoving system (see the Wilson experiment and the associated
controversy \cite{40} on the validity of the locality principle)
without the need of a more elaborate treatment like for the
radiation of accelerated charges. It would also help in the tests of
the validity of special relativity (for instance on the possible
existence of a preferred frame) based on Michelson-Morley - type
experiments \cite{34,55}.

\bigskip

Instead (see also Ref.\cite{37}) we remark that the Sagnac effect
and the Foucault pendulum are {\it experiments which signal the
rotational non-inertiality of the frame}. The same is true for
neutron interferometry \cite{56}, where different settings of the
apparatus are used to {\it detect either rotational or translational
non-inertiality of the laboratory}. As a consequence a null result
of these experiments can be used to give a definition of {\it
relativistic quasi-inertial system}.

\bigskip

Let us remark that the disturbing aspects of rotations are rooted in
the fact that there is a deep difference between translations and
rotations at every level both in Newtonian mechanics and special
relativity:  the generators of translations satisfy an Abelian
algebra, while the rotational ones a non-Abelian algebra. As shown
in Refs.\cite{57}, at the Hamiltonian level we have that the
translation generators are the three components of the momentum,
while the generators of rotations are a pair of canonical variables
($L^3$ and $arctg\, {{L^2}\over {L^1}}$) and an unpaired variable
($|\vec L|$). As a consequence we can separate globally the motion
of the 3-center of mass of an isolated system from the relative
variables, but we cannot separate in a global and unique way three
Euler angles describing an overall rotation, because the residual
vibrational degrees of freedom are not uniquely defined.
\bigskip

We will now give the 3+1 point of view on these topics (Subsection
1), followed by a discussion on the rotating 3-coordinates fixed to
the Earth surface (Subsection 2).

\subsection{The 3+1 Point of View on the Rotating Disk and
the Sagnac Effect.}

Let us describe an abstract geometrical disk with an admissible 3+1
splitting of the type (\ref{2.14}), in which the instantaneous
3-spaces are parallel space-like hyper-planes with normal $l^{\mu}$
centered on an inertial observer $x^{\mu}(\tau ) = l^{\mu}\, \tau$

\beq
 z^{\mu}(\tau ,\vec \sigma ) = l^{\mu}\, \tau +
 \epsilon^{\mu}_r\, R^r_{(3)\, s}(\tau ,\sigma )\,
 \sigma^s.
 \label{a1}
 \eeq

\medskip

The rotation matrix $R_{(3)}$ describes a differential rotation
around the fixed axis "3" (we take a constant $\omega$, but nothing
changes with $\omega(\tau )$)

 \bea
 &&R^r_{(3)\, s}(\tau ,\sigma ) = \left( \begin{array}{ccc} \cos\,
 \theta (\tau ,\sigma ) & - \sin\,  \theta (\tau ,\sigma )& 0\\
 \sin\,  \theta (\tau ,\sigma )& \cos\,  \theta (\tau ,\sigma )& 0\\
 0& 0& 1 \end{array} \right),\nonumber \\
 &&{}\nonumber \\
 &&\theta (\tau ,\sigma ) = F(\sigma )\, \omega\, \tau,\quad
 F(\sigma ) < {c\over {\omega\, \sigma}},\nonumber \\
 &&{}\nonumber \\
 &&\Omega^r{}_s(\tau ,\sigma ) = \left(R^{-1}_{(3)}\, {{d R_{(3)}}\over
 {d\tau}}\right){}^r{}_s(\tau ,\sigma ) = \omega\, F(\sigma )\,
 \left( \begin{array}{ccc} 0& -1& 0\\ 1& 0& 0\\ 0& 0& 0
 \end{array} \right),\nonumber\\
 &&\nonumber\\
 &&\Omega (\tau ,\sigma ) = \Omega (\sigma ) = \omega\,
 F(\sigma ).
 \label{a2}
 \eea

\bigskip

A simple choice for the gauge function $F(\sigma )$ is $F(\sigma ) =
{1\over {1 + {{\omega^2\, \sigma^2}\over {c^2}}}}$ (in the rest of
the Section we put $c = 1$), so that at spatial infinity we get
$\Omega (\tau ,\sigma ) = {{\omega}\over {1 + {{\omega^2\,
\sigma^2}\over {c^2}}}} \rightarrow_{\sigma \rightarrow \infty}\,
0$.

\bigskip

By introducing cylindrical 3-coordinates $r$, $\varphi$, $h$ by
means of the equations $\sigma^1 = r\, \cos\, \varphi$, $\sigma^2 =
r\, \sin\, \varphi$, $\sigma^3 = h$, $\sigma = \sqrt{r^2 + h^2}$, we
get the following form of the embedding and of its gradients

\begin{eqnarray*}
 z^{\mu}(\tau ,\vec \sigma ) &=& l^{\mu}\, \tau +
 \epsilon^{\mu}_1\, [\cos\, \theta (\tau ,\sigma )\, \sigma^1 -
 \sin\, \theta (\tau ,\sigma )\, \sigma^2] +\nonumber \\
 &&\nonumber\\
 &+& \epsilon^{\mu}_2\, [\sin\, \theta (\tau ,\sigma )\, \sigma^1 +
 \cos\, \theta (\tau ,\sigma )\, \sigma^2] + \epsilon^{\mu}_3\,
 \sigma^3 =\nonumber \\
&&\nonumber\\
 &=& l^{\mu}\, \tau + \epsilon^{\mu}_1\, r\, \cos\, [\theta (\tau
 ,\sigma ) + \varphi ] + \epsilon^{\mu}_2\, r\, \sin\, [\theta
 (\tau ,\sigma ) + \varphi ] + \epsilon^{\mu}_3\, h,
 \end{eqnarray*}

 \bea
  \frac{\partial
z^\mu(\tau,\vec{\sigma})}{\partial\tau}&=&
 z^{\mu}_{\tau}(\tau ,\vec \sigma ) = l^{\mu} - \omega\, r\,
 F(\sigma )\, \Big( \epsilon^{\mu}_1\, \sin\, [\theta
 (\tau ,\sigma ) + \varphi ] - \epsilon^{\mu}_2\, \cos\, [\theta
 (\tau ,\sigma ) + \varphi ]\Big),\nonumber \\
&&\nonumber\\
\frac{\partial z^\mu(\tau,\vec{\sigma})}{\partial\varphi}&=&
 z^\mu_\varphi(\tau,\vec{\sigma})=
-\epsilon^\mu_1\,r\,\sin\,[\theta(\tau,\sigma )+\varphi]
+\epsilon^\mu_2\,r\,\cos\,[\theta(\tau,\sigma )+\varphi]
\nonumber\\
&&\nonumber\\
\frac{\partial z^\mu(\tau,\vec{\sigma})}{\partial r}&=&
 z^\mu_{(r)}(\tau,\vec{\sigma})=
-\epsilon^\mu_1\,\left((\cos\,[\theta(\tau,\sigma )+\varphi]
-\frac{r^2\omega\tau}{\sqrt{r^2+h^2}}\,\frac{dF(\sigma)}{d\sigma}
\,\sin\,[\theta(\tau,\sigma )+\varphi]\right)+\nonumber\\
&&\nonumber\\
&+&\epsilon^\mu_2\,\left(\sin\,[\theta(\tau,\sigma )+\varphi]
+\frac{r^2\omega\tau}{\sqrt{r^2+h^2}}\,\cos\,[\theta(\tau,\sigma
)+\varphi]\right)
\nonumber\\
&&\nonumber\\
\frac{\partial z^\mu(\tau,\vec{\sigma})}{\partial h}&=&
 z^\mu_h(\tau,\vec{\sigma})=\epsilon^\mu_3
-\epsilon^\mu_1\,\left(\frac{rh\omega\tau}{\sqrt{r^2+h^2}}\,\frac{dF(\sigma)}{d\sigma}
\,\sin\,[\theta(\tau,\sigma )+\varphi]\right)+\nonumber\\
&&\nonumber\\
&+&\epsilon^\mu_2\,\left(\frac{rh\omega\tau}{\sqrt{r^2+h^2}}\,\frac{dF(\sigma)}{d\sigma}
\,\cos\,[\theta(\tau,\sigma )+\varphi]\right),
 \label{a3}
 \eea

\noindent where we have used the notation $(r)$ to avoid confusion
with the index $r$  used as 3-vector index (for example in
$\sigma^r$).
\bigskip

In the cylindrical 4-coordinates $\tau$, $r$, $\varphi$ and $h$ the
4-metric is

 \begin{eqnarray*}
 \sgn\,g_{\tau\tau}(\tau ,\vec \sigma ) &=& 1 - \omega^2\, r^2\,
 F^2(\sigma ),\qquad \sgn\,g_{\tau\varphi}(\tau ,\vec \sigma ) =
 -\omega\,r^2\,F(\sigma),\qquad \sgn\,g_{\varphi\varphi}(\tau ,\vec \sigma )
 = - r^2,\nonumber\\
 &&{}\nonumber\\
 \sgn\,g_{\tau (r)}(\tau ,\vec \sigma ) &=& -\frac{\omega^2\,r^3\,\tau}{\sqrt{r^2+h^2}}
\,F(\sigma)\,\frac{dF(\sigma)}{d\sigma},\qquad
 \sgn\,g_{\tau h}(\tau ,\vec \sigma ) = -\frac{\omega^2\,r^2\,h\,\tau}{\sqrt{r^2+h^2}}
\,F(\sigma)\,\frac{dF(\sigma)}{d\sigma},
 \end{eqnarray*}

\begin{eqnarray*}
 \sgn\,g_{(r)(r)}(\tau ,\vec \sigma ) &=&-1-\frac{r^4\,\omega^2\,\tau^2}{r^2+h^2}
\left(\frac{dF(\sigma)}{d\sigma}\right)^2,\nonumber\\
 &&\nonumber\\
 \sgn\,g_{hh}(\tau ,\vec \sigma ) &=&-1-\frac{r^2\,h^2\,\omega^2\,\tau^2}{r^2+h^2}
\left(\frac{dF(\sigma)}{d\sigma}\right)^2,\nonumber\\
 &&\nonumber\\
 \sgn\,g_{(r)\varphi}(\tau ,\vec \sigma ) &=&- \frac{\omega\,r^3\,\tau}{\sqrt{r^2+h^2}}
\,\frac{dF(\sigma)}{d\sigma},\qquad
 \sgn\,g_{h\varphi}(\tau ,\vec \sigma ) = - \frac{\omega^2\,r^2\,h\,\tau}{\sqrt{r^2+h^2}}
\,\frac{dF(\sigma)}{d\sigma},\nonumber\\
 &&\nonumber\\
 \sgn\,g_{h(r)}(\tau ,\vec \sigma ) &=& - \frac{r^3\,h\,\omega^2\,\tau^2}{r^2+h^2}
\left(\frac{dF(\sigma)}{d\sigma}\right)^2,
 \end{eqnarray*}

\bea
 &&with\, inverse\nonumber \\
 &&{}\nonumber \\
 \sgn\, g^{\tau\tau}(\tau ,\vec \sigma ) &=& 1,
 \qquad \sgn\, g^{\tau\varphi}(\tau ,\vec \sigma ) = - \omega\,
 F(\sigma ),\nonumber \\
 &&{}\nonumber \\
 \sgn\, g^{\tau (r)}(\tau ,\vec \sigma ) &=& \sgn\, g^{\tau h}(\tau ,\vec \sigma )
 = 0,\qquad \sgn\, g^{(r)(r)}(\tau ,\vec \sigma ) = \sgn\, g^{hh}(\tau ,\vec \sigma )
 = - 1,\nonumber \\
 &&{}\nonumber \\
 \sgn\, g^{\varphi\varphi}(\tau ,\vec \sigma ) &=& - {{1 + \omega^2\, r^2\, [\tau^2\,
 ({{dF(\sigma )}\over {d\sigma}})^2 - F^2(\sigma )}\over
 {r^2}},\nonumber \\
 &&{}\nonumber \\
 \sgn\, g^{\varphi (r)}(\tau ,\vec \sigma ) &=& {{\omega\, r\,
 \tau}\over {\sqrt{r^2 + h^2}}}\, {{dF(\sigma )}\over
 {d\sigma}},\qquad \sgn\, g^{\varphi h}(\tau ,\vec \sigma ) = {{\omega\, h\,
 \tau}\over {\sqrt{r^2 + h^2}}}\, {{dF(\sigma )}\over
 {d\sigma}}.
 \label{a4}
 \eea
\bigskip

It is easy to observe that the congruence of (non inertial)
observers defined by the 4-velocity field

 \beq
  {{z^{\mu}_\tau(\tau
,\vec \sigma )}\over {\sqrt{\sgn\, g_{\tau\tau}(\tau ,\vec \sigma
)}}} = {{l^{\mu} - \omega\, r\, F(\sigma )\, \Big(
\epsilon^{\mu}_1\, \sin\, [\theta
 (\tau ,\sigma ) + \varphi ] - \epsilon^{\mu}_2\, \cos\, [\theta
 (\tau ,\sigma ) + \varphi ]\Big)}\over
{1 - \omega^2\, r^2\, F^2(\sigma )}},
 \label{a5}
  \eeq

\noindent has the observers moving along the world-lines

 \bea
 &&x^{\mu}_{{\vec\sigma}_o}(\tau ) = z^{\mu}(\tau ,{\vec \sigma}_o)
= \nonumber\\ &&\nonumber\\ &=&l^{\mu}\, \tau + r_o\, \Big(
\epsilon^{\mu}_1\, \cos\, [\omega\, \tau\, F(\sigma_o) + \varphi_o]
+ \epsilon^{\mu}_2\, \sin\, [\omega\, \tau\, F(\sigma_o) +
\varphi_o]\Big) + \epsilon^{\mu}_3\, h_o.
 \label{a6}
  \eea

The world-lines (\ref{a6}) are labeled by their initial value
$\vec{\sigma} = {\vec \sigma}_o = (\varphi_o,r_o,h_o)$ at $\tau =
0$.
\bigskip

In particular for $h_o=0$ and $r_o=R$ these world-lines are {\it
helices} on the {\em  cylinder} in the Minkowski space

 \beq
 \epsilon_3^\mu\,z_\mu = 0,\qquad
\left(\epsilon_1^\mu\,z_\mu\right)^2 +
\left(\epsilon_2^\mu\,z_\mu\right)^2 = R^2,
 \qquad or \qquad r = R, \qquad h = 0.
   \label{a7}
    \eeq

These helices are defined the equations $\varphi = \varphi_o,\, r =
R,\,h = 0$ if expressed in the embedding adapted coordinates
$\varphi,r,h$. Then the congruence of observers (\ref{a5}), defined
by the foliation (\ref{a1}), defines on the cylinder (\ref{a7}) the
{\em rotating observers} usually assigned to the rim of a rotating
disk, namely  observes running along the helices $
x^{\mu}_{{\vec\sigma}_o}(\tau ) = l^{\mu}\, \tau + R\, \Big(
\epsilon^{\mu}_1\, \cos\, [\Omega(R)\, \tau\, + \varphi_o] +
\epsilon^{\mu}_2\, \sin\, [\Omega(R)\, \tau\,  + \varphi_o]\Big)$
after having put $\Omega(R)\equiv\omega\,F(R)$.
\bigskip

On the cylinder (\ref{a7}) the line element is obtained from the
line element $ds^2 = g_{AB}\, d\sigma^A\, d\sigma^B$ for the metric
(\ref{a4}) by putting $dh = dr = 0$ and $r = R$, $h = 0$. Therefore
the cylinder line element is

 \beq
  \sgn\,(ds_{cyl})^2 = \Big[1 - \omega^2\, R^2\,
F^2(R)\Big]\, (d\tau )^2 - 2\, \omega\, R^2\, F(R)\, d\tau d\varphi
- R^2\, (d\varphi )^2.
 \label{a8}
  \eeq
\medskip

We can define the {\it light rays on the cylinder}, i.e. the null
curves on it, by solving the equation

\begin{equation}
 \sgn\,(ds_{cyl})^2=(1-R^2\,\Omega^2(R))\,d\tau^2-2\,R^2\,\Omega(R)\,d\tau\,d\varphi-
R^2\,d\varphi^2=0,
 \label{a9}
 \end{equation}

\noindent which implies

\begin{equation}
 R^2\,\left(\,\frac{d\varphi(\tau)}{d\tau}\,\right)^2+
2\,R^2\,\Omega(R)\,\left(\,\frac{d\varphi(\tau)}{d\tau}\,\right)-
(1-R^2\,\Omega(R))=0.
 \label{a10}
 \end{equation}

\bigskip

The two solutions

\begin{equation}
\frac{d\varphi(\tau)}{d\tau}=\pm\,\frac{1}{R}-\Omega(R),
 \label{a11}
 \end{equation}

\noindent define the world-lines on the cylinder for {\em clockwise
or anti-clockwise} rays of light.

\begin{equation}
 \begin{array}{l}
\Gamma_1:\qquad \varphi (\tau ) - \varphi_o =
\left(+\frac{1}{R}-\Omega(R)\right)\,\tau ,\\ \\
 \Gamma_2:\qquad
\varphi (\tau ) - \varphi_o =
\left(-\frac{1}{R}-\Omega(R)\right)\,\tau
\end{array}.
 \label{a12}
 \end{equation}

\bigskip

This is the {\em geometric origin} of the {\em Sagnac Effect}. Since
$\Gamma_1$ describes the world-line of the ray of light emitted at
$\tau=0$ by the rotating observer $\varphi=\varphi_o$ in the
increasing sense of $\varphi$ (anti-clockwise), while $\Gamma_2$
describes that of the ray of light emitted at $\tau=0$ by the same
observer in the decreasing sense of $\varphi$ (clockwise), then the
two rays of light will be re-absorbed by the same observer at {\it
different $\tau$-times} \footnote{Sometimes the {\em proper time of
the rotating observer} is used: $d{\cal
T}_o=d\tau\sqrt{1-\Omega^2(R)\,R^2}$.} $\tau_{(\pm\, 2\pi ) }$,
whose value, determined by the two conditions $\varphi (\tau_{(\pm\,
2\pi )}) - \varphi_o= \pm\, 2\pi$, is

\beq
 \begin{array}{l}
\Gamma_1:\qquad \tau_{(+2\pi )} =
\frac{2\pi\,R}{1-\Omega(R)\,R},\qquad \Gamma_2:\qquad \tau_{(-2\pi
)}=\frac{2\pi\,R}{1+\Omega(R)\,R}.
\end{array}
 \label{a13}
 \eeq

The {\it time difference} between the re-absorption of the two light
rays is\medskip

\begin{equation}
\Delta\tau = \tau_{(+2\pi )} - \tau_{(-2\pi)} =
\frac{4\pi\,R^2\,\Omega(R)}{1-\Omega^2(R)\,R^2} =
\frac{4\pi\,R^2\,\omega\, F(R)}{1-\omega^2\, F^2(R)\,R^2},
 \label{a14}
 \end{equation}

\noindent and it corresponds to the phase difference named the {\em
Sagnac effect}

\begin{equation}
 \Delta\Phi=\Omega\,\Delta\tau,\qquad \Omega = \Omega (R) = \omega\, F(R) .
 \label{a15}
 \end{equation}
\medskip

We see that we recover the standard result if we take a function
$F(\sigma )$ such that $F(R) = 1$. In the non-relativistic
applications, where $F(\sigma ) \rightarrow 1$, the correction
implied by admissible relativistic coordinates is totally
irrelevant.

\bigskip

With an admissible notion of simultaneity, all the clocks on the rim
of the rotating disk lying on a hyper-surface $\Sigma_{\tau}$ are
automatically synchronized. Instead for rotating observers of the
irrotational congruence there is a {\em desynchronization effect or
synchronization gap} because they cannot make a global
synchronization of their clocks: {\it usually a discontinuity in the
synchronization of clocks is accepted and taken into account} (see
Ref.\cite{36} for the GPS).
\medskip

To clarify this point and see the emergence of this gap, let us
consider a reference observer $(\varphi_o = const., \tau )$ and
another one $(\varphi = const. \not= \varphi_o, \tau )$. If $\varphi
> \varphi_o$ we use the notation $(\varphi_R, \tau )$, while for
$\varphi < \varphi_o$ the notation $(\varphi_L, \tau )$ with
$\varphi_R - \varphi_o = - (\varphi_L - \varphi_o)$.

Let us consider the two rays of light $\Gamma_{R\, -}$ and
$\Gamma_{L\, -}$, with world-lines given by Eqs.(\ref{a12}), emitted
in the right and left directions at the event $(\varphi_o,\tau_{-})$
on the rim of the disk and received at $\tau$ at the events
$(\varphi_R,\tau)$ and $(\varphi_L,\tau )$ respectively. Both of
them are  reflected towards the reference observer, so that we have
two rays of light $\Gamma_{R\, +}$ and $\Gamma_{L\, +}$ which will
be absorbed at the event $(\varphi_o,\tau_{+})$. By using
Eq.(\ref{a13}) for the light propagation, we get

\bea
 && \Gamma_{R\, -}:\,\,  (\varphi  -
\varphi_o) = \frac{1-R\Omega(R)}{R}\,(\tau - \tau_{-}),\qquad
\Gamma_{R\, +}:\,\, (\varphi  - \varphi_o) =
\frac{1+R\Omega(R)}{R}\,(\tau_{+} - \tau),\nonumber \\
 &&{}\nonumber \\
 &&\Gamma_{L\, -}:\,\, (\varphi  - \varphi_o) =
-\frac{1+R\Omega(R)}{R}\,(\tau - \tau_{-}),\qquad \Gamma_{L\,
+}:\,\, (\varphi  - \varphi_o) = -\frac{1-R\Omega(R)}{R}\,(\tau_{+}
- \tau).\nonumber \\
 &&{}
 \label{a16}
 \eea

\medskip

As shown in Section II, Eqs.(2.17) and (2.18), of the first paper in
Ref.\cite{3}, in a neighborhood of the observer $(\varphi_o ,\tau )$
[$(\varphi ,\tau )$ is an observer in the neighborhood] we can only
define the following local synchronization \footnote{See
Ref.\cite{53} for a derivation of the Sagnac effect in an inertial
system by using Einstein's synchronization in the locally comoving
inertial frames on the rim of the disk and by asking for the
equality of the one-way velocities in opposite directions.}

\begin{equation}
 c\, \Delta\,\widetilde{\cal T}= \sqrt{1-R^2\,\Omega^2(R)}\,\Delta
\tau_E= \sqrt{1-R^2\,\Omega^2(R)}\,\Delta \tau-
\frac{R^2\,\Omega^2(R)}{\sqrt{1-R^2\Omega^2(R)}} \,\Delta \varphi.
 \label{a17}
 \end{equation}
\medskip

If we try to extend this local synchronization  to a global one for
two distant observers $(\varphi_o, \tau )$ and $(\varphi ,\tau )$ in
the form of an Einstein convention  (the result is the same both for
$\varphi = \varphi_R$ and $\varphi = \varphi_L$)

\begin{equation}
\tau_E=\frac{1}{2}\,(\tau_{+} + \tau_{-}) = \tau-
\frac{R^2\Omega(R)}{1-R^2\,\Omega^2(R)}(\varphi  - \varphi_o),
 \label{a18}
 \end{equation}

\noindent we arrive at a contradiction, because the curves defined
by $\tau_E=constant$ are {\it not closed}, since they are helices
that assign the {\it same time} $\tau_E$ to different events on the
world-line of an observer $\varphi_o=constant$. For example
$(\varphi_o,\tau)$ and $\left(\varphi_o,\tau+2\pi
\,\frac{R^2\Omega(R)}{1-R^2\,\Omega^2(R)}\right)$ are on the same
helix $\tau_E=constant$. As a consequence we get the synchronization
gap.

\bigskip

As shown in both papers of Ref.\cite{3}, by using the global
synchronization on the instantaneous 3-spaces $\Sigma_{\tau}$ we can
define a generalization of Einstein's convention for clock
synchronization by using the radar time $\tau$. If an accelerated
observer A emits a light signal at $\tau_{-}$, which is reflected at
a point P of the world-line of a second observer B and then
reabsorbed at $\tau_+$, then the B clock at P has to be synchronized
with the following instant of the A clock [$n = +$ for $\varphi =
\varphi_R$, $n = -$ for $\varphi = \varphi_L$]

\bea
 \tau(\tau_{-}, n, \tau_{+}) &=&
\frac{1}{2}(\tau_{+} + \tau_{-}) - \frac{n\,R\,\Omega(R)}{2}\,
(\tau_{+} - \tau_{-})\, {\buildrel {def}\over =}\, \tau_{-} +
 {\cal E}(\tau_{-}, n, \tau_{+})\, (\tau_{+} - \tau_{-}),\nonumber \\
 &&{}\nonumber \\
 with &&\qquad {\cal E}(\tau_{-}, n, \tau_{+}) = {{1 - n\, R\,
 \Omega (R)}\over 2},\qquad \Omega (R) = \omega\, F(R).
 \label{a19}
 \eea
\bigskip

Finally in the first paper of Ref.\cite{3} [see Eqs.(6.37)-(6.47) of
Section VI] there is the evaluation of the radius and the
circumference of the rotating disk. If we choose the spatial length
of the instantaneous 3-space $\Sigma_{\tau}$ of the admissible
embedding (\ref{a1}), we get aa Euclidean 3-geometry, i.e. a
circumference $2\pi\, R$ and a radius $R$ at each instant $\tau$
independently from the choice of the gauge function $F(\sigma )$.
With other admissible 3+1 splittings we would get non-Euclidean
results: as said they are gauge equivalent when the disk can be
described with a parametrized Minkowski theory. Instead the use of a
local notion of synchronization from the observers of the
irrotational congruence located on the rim of the rotating disk
implies a local definition of spatial distance based on the 3-metric
${}^3\gamma_{uv} = - \sgn\, \Big(g_{uv} - \frac{g_{\tau u}\,g_{\tau
v}}{g_{\tau\tau}}\Big)$, i.e. a non-Euclidean 3-geometry. In this
case the radius is $R$, but the circumference is $2\pi\, R/ \sqrt{1
- R^2\, \Omega^2(R)}$. However this result holds only in the local
rest frame of the observer with the tangent plane orthogonal to the
observer 4-velocity (also called the abstract relative space)
identified with a 3-space (see Section IIB).
\bigskip

See Subsection D of Section VI for a derivation of the Sagnac effect
in nearly rigid rotating frames.

\subsection{The Rotating ITRS 3-Coordinates fixed on the Earth Surface.}

The embedding (\ref{2.14}), describing admissible differential
rotations in an Euclidean 3-space, can be used to improve the
conventions IERS2003 (International Earth Rotation and Reference
System Service) \cite{39} on the non-relativistic transformation
from the 4-coordinates of the Geocentric Celestial Reference System
(GCRS) to the International Terrestrial Reference System (ITRS), the
Earth-fixed geodetic system of the new theory of Earth rotation
replacing the old precession-nutation theory. It would be a special
relativistic improvement to be considered as an intermediate step
till to a future development leading to a post-Newtonian (PN)
general relativistic approach unifying the existing non-relativistic
theory of the geo-potential below the Earth surface with the GCRS PN
description of the geo-potential outside the Earth given by the
conventions IAU2000 (International Astronomical Union) \cite{39} for
Astrometry, Celestial Mechanics and Metrology in the relativistic
framework.
\bigskip

In the IAU 2000 Conventions the Solar System is described in the
Barycentric Celestial Reference System (BCRS) as a {\it
quasi-inertial} frame, centered on the barycenter of the Solar
System, with respect to the Galaxy. BCRS is parametrized with
harmonic PN 4-coordinates $x^{\mu}_{BCRS} = \Big(x^o_{BCRS} = c\,
t_{BCRS}; x^i_{BCRS}\Big)$, where $t_{BCRS}$ is the barycentric
coordinate time and the mutually orthogonal spatial axes are {\it
kinematically non-rotating} with respect to fixed radio sources.
This a nearly Cartesian 4-coordinate system in a PN Einstein
space-time and there is an assigned 4-metric, determined modulo
$O(c^{-4})$ terms and containing the gravitational potentials of the
Sun and of the planets, PN solution of Einstein equations in
harmonic gauges: in practice it is considered as a special
relativistic inertial frame with nearly Euclidean instantaneous
3-spaces $t_{BCRS} = const.$ (modulo $O(c^{-2})$ deviations) and
with Cartesian 3-coordinates $x^i_{BCRS}$. This frame is used for
space navigation in the Solar System. The geo-center (a fictitious
observer at the center of the earth geoid) has a world-line
$y^{\mu}_{BCRS}(x^o_{BCRS}) = \Big(x^o_{BCRS};
y^i_{BCRS}(x^o_{BCRS})\Big)$, which is approximately a straight
line.

\bigskip

For space navigation near the Earth (for the Space Station and near
Earth satellites using NASA coordinates) and for the studies from
spaces of the geo-potential one uses the GCRS, which is defined
outside the Earth surface as a local reference system centered on
the geo-center. Due to the earth rotation of the Earth around the
Sun, it deviates from a nearly inertial special relativistic frame
on time scales of the order of the revolution time. Its harmonic
4-coordinates $x^{\mu}_{GCRS} = \Big(x^o_{GCRS} = c\, t_{GCRS};
x^i_{GCRS}\Big)$, where $t_{GCRS}$ is the geocentric coordinate
time, are obtained from the BCRS ones by means of a PN coordinate
transformation which may be described as a special relativistic pure
Lorentz boost without rotations (the parameter is the 3-velocity of
the geo-center considered constant on small time scales) plus
$O(c^{-4})$ corrections taking into account the gravitational
acceleration of the geo-center induced by the Sun and the planets.
As a consequence the GCRS spatial axes are kinematically
non-rotating in BCRS and the relativistic inertial forces (for
instance the Coriolis ones) are hidden in the geodetic precession;
the same holds for the aberration effects and the dependence on
angular variables. A PN 4-metric, determined modulo $O(c^{-4})$
terms, is given in IAU2000: it also contains the GCRS form of the
geo-potential and the inertial and tidal effects of the Sun and of
the planets. Again the instantaneous 3-spaces are considered nearly
Euclidean (modulo $O(c^{-2})$ deviations) 3-spaces $t_{GCRS} =
const.$.\bigskip

In IAU200 the coordinate times $t_{BCRS}$ and $t_{GCRS}$ are then
connected with the time scales used on Earth: SI Atomic Second, TAI
(International Atomic Time), TT (Terrestrial Time), $T_{EPH}$
(Ephemerides Time), UT and UT1 and UTC (Universal Times for civil
use), GPS (Mastr Time), ST (Station Time).\bigskip

Finally we need a 4-coordinate system fixed on the Earth crust. It
is the ITRS with 4-coordinates $x^{\mu}_{ITRS} = \Big(x^o_{ItRS}
{\buildrel {def}\over =}\, c\, t_{GCRS}; x^i_{ITRS}\Big)$, which
uses the same coordinate time as GCRS. It is obtained from GCRS by
making a suitable set of non-relativistic time-dependent {\it rigid}
rotations on the nearly Euclidean 3-spaces $t_{GCRS} = const.$. The
geocentric rectangular 3-coordinates $x^i_{ITRS}$ match the
reference ellipsoid WGS-84 (basis of the terrestrial coordinates
(latitude, longitude, height) obtainable from GPS) used in geodesy.
As shown in IERS2003, we have $x^i_{ITRS} = \Big( W^T(t_{GCRS})\,
R^T_3(- \theta )\, C \Big)^i{}_j\, x^j_{GCRS}$, where $C =
R^T_3(s)\, R^T_3(E)\, R^T_2(- d)\, R^T_3(- E)$ and $W(t_{GCRS}) =
R_3(- s^{'})\, R_2(x_p)\, R_1(y_p)$ are rotation matrices. This
convention is based on the new definition of the Earth rotation axis
($\theta$ is the angle of rotation about this axis): it is the line
through the geo-center in direction of the Celestial Intermediate
Pole (CIP) at date $t_{GCRS}$, whose position in GCRS is $n^i_{GCRS}
= \Big(sin\, d\, cos\, E, sin\, d\, sin\, E, cos\, d\Big)$. The new
non-rotating origin (NLO) of the rotation angle $\theta$ on the
Earth equator (orthogonal to the rotation axis) is a point named the
Celestial Intermediate Origin (CIO), whose position in CGRS requires
the angle $s$, called the CIO locator. Finally in the rotation
matrix $W^T(t_{GCRS})$ (named the polar motion or wobble matrix) the
angles $x_p$ and $y_p$ are the angular coordinates of CIP in ITRS,
while the angle $s^{'}$ is connected with the re-orientation of the
pole from the ITRS z-axis to the CIP plus a motion of the origin of
longitude from the ITRS x-axis to the so-called Terrestrial
Intermediate Origin (TIO), used as origin of the azimuthal angle.

\bigskip

Let us now consider the embedding $z^{\mu}(\tau ,\sigma^u) =
x^{\mu}(\tau) + \epsilon^{\mu}_r\, R^r{}_s(\tau ,\sigma )\,
\sigma^s$ of Eq.(\ref{2.14}). Let us identify $x^{\mu} =
z^{\mu}(\tau ,\sigma^u)$ with the GCRS 4-coordinates
$x^{\mu}_{GCRS}$ centered on the world-line of the geo-center
assumed to move along a straight line. then, if we identify the
space-like vectors $\epsilon^{\mu}_r$ with the GCRS non-rotating
spatial axes, we have $x^{\mu}(\tau ) = \epsilon^{\mu}_{\tau}\, \tau
= l^{\mu}\, \tau$, where $l^{\mu}$ is orthogonal to the nearly
Euclidean 3-spaces $t_{GCRS} = const.$. the proper time $\tau$ of
the geo-center coincides with $c\, t_{GCRS}$ modulo $O(c^{-2})$
corrections from the GCRS PN 4-metric.

Then a special relativistic definition of ITRS can be done by
replacing the rigidly rotating 3-coordinates $x^i_{ITRS}$ with the
differentially rotating 3-coordinates $\sigma^r$. The rotation
matrix $R(\tau ,\sigma )$, with the choice $F(\sigma ) = {1\over {1
+ {{\omega^2\, \sigma^2}\over {c^2}}}}$ for the gauge function
($\omega$ can be taken equal to the mean angular velocity for the
Earth rotation), will contain three Euler angles determined by
putting $R(\tau ,\sigma ){|}_{F(\sigma) = 1} = C^T\, R_3(- \theta)\,
W(t_{GCRS})$.

In this way a special relativistic version of ITRS can be given as a
preliminary step towards a PN general relativistic formulation of
the geo-potential inside the Earth to be joined consistently with
GCRS outside the Earth. Even if this is irrelevant for geodesy
inside the geoid, it could lead to a refined treatment of effects
like geodesic precession taking into account a model of
geo-potential interpolating smoothly between inside and outside the
geoid and the future theory of heights over the reference ellipsoid
under development in a formulation of relativistic geodesy based on
the use of  the new generation of microwave and optical atomic
clocks both on the Earth surface and in space.

\vfill\eject

\section{The Landau-Lifschitz Non-Inertial Electro-Magnetic Fields}

Sometimes, see for instance Ref.\cite{17}, the following {\it
generalized non-inertial electric and magnetic fields} are
introduced

 \bea
  {\cal E}^{s}_{(F)}(\tau, \sigma^u)&=&
- \left[\frac{\sqrt{\gamma_{F}}}{\sqrt{1 + n_F}}\, \h_{F}^{sr}\left(
F_{\tau r} - n_{F}^v\, F_{vr}\right)\right](\tau, \sigma^u)
\,\cir\,\pi^s(\tau, \sigma^u)),\nonumber\\
 &&\nonumber\\
  {\cal B}_{(F)}^{w}(\tau, \sigma^u)&=& \frac{1}{2}\,
  \delta^{wt}\, \epsilon_{tsr}\,
\left[(1 + n_F)\,\sqrt{\gamma_{F}}\, \h_{F}^{sv}\,\h_{F}^{ru}\, F_
{vu}- (n_F^s\,\pi^r - n_F^r\,\pi^s) \right](\tau, \sigma^u),
 \label{b1}
  \eea

\medskip

They allow us to rewrite the Hamilton-Dirac Eqs.(\ref{4.15}) in the
following form (we use a vector notation as in the 3-dimensional
Euclidean case)

 \bea
  \partial_r\, {\cal E}^r{}_{(F)}(\tau, \sigma^u)&=&
\sqrt{\gamma_{F}(\tau, \sigma^u)}\,
\overline{\rho}(\tau, \sigma^u),\nonumber\\
 &&\nonumber\\
  \epsilon_{ruv}\, \partial_u\, {\cal B}^v{}_{(F)}(\tau, \sigma^u)
  - \frac{\partial{\cal E}_{(F)}^{r}(\tau, \sigma^u)} {\partial\tau}
 &=& \sqrt{\gamma_{F}(\tau, \sigma^u)}\,
\overline{J}^r(\tau, \sigma^u),
 \label{b2}
  \eea

\noindent namely {\it in the same form of  the usual source-
dependent Maxwell equations in an inertial frame}.
\bigskip

Since Eqs.(\ref{b1}) can be rewritten in the form

 \bea
  {\cal E}^{s}_{(F)}(\tau, \sigma^u)&=&
\left[+\frac{\sqrt{\gamma_{F}}}{\sqrt{(1 + n_F)}}\, \h_{F}^{sr}
\,E_r -\frac{\sqrt{\gamma_{F}}}{\sqrt{(1 + n_F)}}\, \h_{F}^{sr}\,
 \epsilon_{ruv}\, n_F^u\, B_v \right](\tau, \sigma^u),
\nonumber\\
 &&\nonumber\\
  {\cal B}_{(F)}^{w}(\tau, \sigma^u)&=& \delta^{wt}\,
  \epsilon_{tsr}\, \left[\frac{1}{2}\,
 (1 +n_F)\,\sqrt{\gamma_{F}}\, \h_{F}^{sv}\,\h_{F}^{ru}\,
\epsilon_{vu\ell}\,B_\ell + n^s_F\, E_r \right](\tau, \sigma^u),
 \label{b3}
  \eea
\medskip

\noindent we get the following form of the Maxwell equations for the
field strengths $E_r$ and $B_r$

  \bea
\partial_r\, E_r(\tau, \sigma^u)&=&
\sqrt{\gamma_{F}(\tau, \sigma^u)}\,\Big[\, \overline{\rho}(\tau,
\sigma^u) - \overline{\rho}_R(\tau, \sigma^u)\,\Big],\nonumber\\
 &&\nonumber\\
 \epsilon_{suv}\, \partial_u\, B_v(\tau , \sigma^u )
 - \frac{\partial\, E_s(\tau, \sigma^u)}{\partial\tau}
 &=& \delta_{sr}\, \sqrt{\gamma_{F}(\tau, \sigma^u)}\,
\Big[\,\overline{J}^r(\tau, \sigma^u)- \overline{J}^r_R(\tau,
\sigma^u)\,\Big],
 \label{b4}
  \eea

\noindent where the new charge and current densities are the
following functions only of the metric tensor and of the fields
$E_r$, $B_r$

\bea
 \overline{\rho}_R(\tau, \sigma^u)&=&
\frac{1}{\sqrt{\gamma_{F}(\tau, \sigma^u)}}\, \partial_r\,
\left({\cal E}^r_{(F)}(\tau, \sigma^u) - \delta^{rs}\, E_s(\tau,
\sigma^u)\right),\nonumber\\
 &&\nonumber\\
  \overline{J}^r_R(\tau, \sigma^u)&=&
\frac{1}{\sqrt{\gamma_{F}(\tau, \sigma^u)}}\, \Big[ -
\frac{\partial}{\partial \tau}\, \left({\cal E}^r_{(F)}(\tau,
\sigma^u) - \delta^{rs}\,
E_s(\tau, \sigma^u)\right) +\nonumber \\
 &+& \delta^{rs}\, \epsilon_{suv}\, \partial_u\, \left({\cal
B}^v{}_{(F)} - \delta^{vk}\, B_k\right)(\tau ,\sigma^u)\Big].
 \label{b5}
  \eea

\medskip

Instead, as a consequence of Eqs.(\ref{4.10}), the homogeneous
equations take the form

\beq
 \epsilon_{ruv}\, \partial_u\, E_v(\tau ,\sigma^s) = - {{\partial\,
 B_r(\tau ,\sigma^s)}\over {\partial\, \tau}},\qquad \epsilon_{ruv}\,
 \partial_u\, B_v(\tau ,\sigma^s) = 0.
 \label{b5a}
 \eeq

\bigskip

By using Eq.(\ref{6.2}) of Section VI we find the results of the
Appendix A of Ref.\cite{18}

 \bea
  \vec{\cal E}_{(F)}(\tau, \sigma^u)&=&\vec{E}(\tau, \sigma^u)
+ ({{\vec{\Omega}(\tau)}\over c} \times \vec{\sigma})
\times\vec{B}(\tau, \sigma^u),\nonumber\\
 &&\nonumber\\
  \vec{\cal B}{}_{(F)}(\tau, \sigma^u)&=& \vec{B}
+ ({{\vec{\Omega}(\tau)}\over c} \times \vec{\sigma}) \times
\vec{E}(\tau, \sigma^u) + ({{\vec{\Omega}(\tau)}\over c} \times
\vec{\sigma}) \times[ ({{\vec{\Omega}(\tau)}\over c} \times
\vec{\sigma}) \times \vec{B}(\tau, \sigma^u)].
 \label{b6}
 \eea

\bigskip

In absence of sources Eqs.(\ref{4.17}) are  the generally covariant
equations $\nabla_{\nu}\, F^{\mu\nu} \cir 0$, suggested by the
equivalence principle, in the 3+1 point of view after having taken
care of the asymptotic properties at spatial infinity.
\medskip

Let us remark that in the case of the nearly rigid limit of the
foliation (\ref{2.14}) (see Section VI) and with $\vec \Omega (\tau)
= (0,0, \tilde \Omega = const.)$  Eqs.(\ref{b4}) and (\ref{b5a})
coincide with Eqs.(9) of Schiff \cite{28} if we identify ${\bar
\rho}_R$ with $\sigma$ and ${\bar j}^r_R$ with $j^r$. This is due to
the fact that Schiff's fields $\vec E$, $\vec B$, have the
components coinciding with the covariant fields $E_r$ and $B_r$ of
Eqs.(\ref{4.10}); these fields obviously differ from the fields
(\ref{b3}) defined in Ref.\cite{17}.

\medskip

Eqs. (\ref{b4}) and (\ref{b5}), with the metric associated to the
admissible notion of simultaneity (\ref{2.14}), should be the
starting point for the calculations in the magnetosphere of pulsars,
where one always assumes a rigid rotation $\omega$ with the
consequent appearance of the so-called {\it light cylinder}  for
$\omega\, R = c$ (the horizon problem of the rotating disk). See
Refs.\cite{58} based on Schiff's equations \cite{28} (\ref{b4}) and
(\ref{b6}) or the more recent literature of Refs. \cite{59}. Instead
in Refs.\cite{60} the light cylinder is avoided using the rotating
coordinates of  Refs.\cite{19}, but at the price of a bad behavior
at spatial infinity.

\bigskip

These equations also show that the non-inertial electric and
magnetic fields ${\vec {\cal E}}_{(F)}$ and ${\vec {\cal B}}_{(F)}$
are {\it not}, in general, {\it equal} to the fields obtained from
the inertial ones ${\vec E}$ and ${\vec B}$ with a Lorentz
transformations to the comoving inertial system like it is usually
assumed following Rohrlich \cite{61} and the locality hypothesis.

\vfill\eject

\section{Covariant and Non-Covariant Decompositions of the
Electro-Magnetic Field and the Radiation Gauge in Non-Inertial Rest
frames.}

In inertial frames the identification of the physical degrees of
freedom (Dirac observables) of the free electro-magnetic field was
done in Refs. \cite{26,62,63,64} by means of the Shanmugadhasan
canonical transformation adapted to the first class constraints
$\pi^{\tau}(\tau ,\sigma^u ) \approx 0$ and $\Gamma (\tau , \sigma^u
) = \partial_r\, \pi^r(\tau ,\sigma^u ) \approx 0$. The final
canonical basis identifies the {\it radiation gauge} with its
transverse fields as the natural one from the point of view of
constraint theory.
\bigskip

In the parametrized Minkowski theories of Setion III Subsection A,
due to the last two lines of Eqs.(\ref{3.15}), we see that two
successive gauge transformations, of generators $G_i(\tau ,\sigma^u
) = \lambda^{\mu}_i(\tau ,\sigma^u )\, {\cal H}_{\mu}(\tau ,\sigma^u
)$, $i=1,2$, do not commute but imply an electro-magnetic gauge
transformation. Since the effect of the $i=1,2$ gauge
transformations is to modify the notions of simultaneity, also the
definition of the Dirac observables of the electro-magnetic field
will change with the 3+1 splitting. In general, given two different
3+1 splittings, the two sets of Dirac observables associated with
them will be connected by an electro-magnetic gauge transformation.

\medskip

Since it is not clear whether it is possible to find a
quasi-Shanmugadhasan canonical transformation adapted to ${\cal
H}_r(\tau ,\sigma^u ) = {\cal H}_{\mu}(\tau ,\sigma^u )\,
z^{\mu}_r(\tau ,\sigma^u ) \approx 0$, $\pi^{\tau}(\tau ,\sigma^u )
\approx 0$, $\Gamma (\tau ,\sigma^u ) \approx 0$ \footnote{${\cal
H}_{\perp}(\tau ,\sigma^u ) = {\cal H}^{\mu}(\tau ,\sigma^u )\,
l_{\mu}(\tau ,\sigma^u ) \approx 0$, like an ordinary Hamiltonian,
can be included in the adapted Darboux-Shanmugadhasan basis only in
case of integrability of the equations of motion.}, the search of
the electro-magnetic Dirac observables must be done with the
following strategy:

i) make the choice of an admissible 3+1 splitting by adding four
gauge-fixing constraints determining the embedding $z^{\mu}(\tau
,\sigma^u )$, so that the induced 4-metric $g_{AB}(\tau ,\sigma^u )$
becomes a numerical quantity and is no more a configuration
variable;

ii) find the Dirac observables on the resulting completely fixed
simultaneity surfaces $\Sigma_{\tau}$ with a suitable Shanmugadhasan
canonical transformation adapted to the two remaining
electro-magnetic constraints.

\medskip

Let us remark that a similar scheme has to be followed also in the
canonical Einstein-Maxwell theory: only after having fixed a 3+1
splitting (a system of 4-coordinates on the solutions of Einstein's
equations) we can find the Dirac observables of the electro-magnetic
field.

\medskip

This strategy is induced by the fact that, while the Gauss law
constraint $\Gamma (\tau ,\sigma^u ) = \partial_r\, \pi^r(\tau ,
\sigma^u ) \approx 0$ is a scalar under change of admissible 3+1
splittings \footnote{$\pi^r(\tau ,\sigma^u )$ is a vector density
like in canonical metric gravity.},  the gauge vector potential
$A_r(\tau ,\sigma^u )$ is the pull-back to the base of a connection
one-form and can be considered as a tensor only with topologically
trivial surfaces $\Sigma_{\tau}$ (like in the case we are
considering). Since a Shanmugadhasan canonical transformation
adapted to the Gauss law constraint transforms $\Gamma (\tau ,
\sigma^u )$  in one of the new momenta, it is not clear how to
define a conjugate gauge variable $\eta_{em}(\tau ,\sigma^u )$ such
that $\{ \eta_{em}(\tau ,\sigma^u ), \Gamma (\tau ,{\sigma}^u_1) \}
= \delta^3(\sigma^u , {\sigma}^u_1)$ and two conjugate pairs of
Dirac observables having vanishing Poisson brackets with both
$\eta_{em}(\tau ,\sigma^u )$ and  $\Gamma (\tau ,\sigma^u )$ when
the 3-metric on $\Sigma_{\tau}$ is not Euclidean ($g_{rs}(\tau ,
\sigma^u ) \not= -\sgn\, \delta_{rs}$).
\medskip

With every fixed type of instantaneous 3-space $\Sigma_{\tau}$ with
non-trivial 3-metric, $g_{rs}(\tau ,\sigma^u ) \not= -\sgn\,
\delta_{rs}$, we have to find suitable gauge variable
$\eta_{em}(\tau ,\sigma^u )$ and the Dirac observables replacing
$A^r_{\perp}(\tau ,\sigma^u )$ and $\pi^r_{\perp}(\tau ,\sigma^u )$.
\bigskip

Let us consider an arbitrary admissible non-inertial frame
identified by the embedding $z^{\mu}_F(\tau ,\sigma^u) =
x^{\mu}(\tau ) + F^{\mu}(\tau ,\sigma^u)$ of Eq.(\ref{4.1}). In it
the fields $A_r(\tau ,\sigma^u)$ and $\pi^r(\tau ,\sigma^u)$ admit
both a {\it covariant} and a {\it non-covariant} decomposition.
\bigskip

The {\it covariant decomposition} \cite{65} is

\bea
 &&\pi^r(\tau,\sigma^u)={\hat \pi}^r_{\perp}(\tau,\sigma^u) + {\hat \pi}^r_L(\tau,\sigma^u)\nonumber \\
 &&{}\nonumber \\
 &&{\hat \pi}^r_{\perp}(\tau ,\sigma^u) = \Big(\delta^r_s - \nabla^r_{F}\, {1\over
 {\Delta_{F}}}\, \nabla_{F\,s}\Big)\, \pi^s(\tau ,\sigma^u) = \Big(\delta^r_s - \nabla^r_{F}\, {1\over
 {\Delta_{F}}}\, \partial_s\Big)\, \pi^s(\tau ,\sigma^u),\nonumber \\
  \qquad&& \Rightarrow\,\nabla_{F\,r}{\hat \pi}^r_{\perp}(\tau ,\sigma^u) = 0,\nonumber \\
&&\nonumber\\
 &&{\hat \pi}^r_L(\tau ,\sigma^u) =  \nabla^r_{F}\, {1\over
 {\Delta_{F}}}\, \nabla_{F\,s}\, \pi^s(\tau ,\sigma^u) = \nabla^r_{F}\, {1\over
 {\Delta_{F}}}\, \partial_s\, \pi^s(\tau ,\sigma^u),\nonumber \\
 &&{}\nonumber \\
 &&{}\nonumber \\
&& A_r(\tau,\sigma^u) = {\hat A}_{\perp r}(\tau,\sigma^u) + {\hat A}_{L\, r}(\tau,\sigma^u),\nonumber\\
&&\nonumber\\
 &&{\hat A}_{\perp r}(\tau ,\sigma^u) = \Big(\delta^s_r - \nabla_{F\, r}\, {1\over {\Delta_{F}}}\,
 \nabla^r_{F})\, A_r(\tau ,\sigma^u)\,\Rightarrow\,\nabla^r_{F}{\hat A}_{\perp\,r}(\tau
 ,\sigma^u) = 0,\nonumber \\
&&\nonumber\\
 &&{\hat A}_{L\, r}(\tau ,\sigma^u) = \nabla_{F\, r}\, {1\over {\Delta_{F}}}\,
 \nabla^s_{F}\, A_s(\tau ,\sigma^u).
 \label{c1}
  \eea

Here $\nabla^r_{F}$ and $\triangle_{F} = \nabla^r_{F}\, \nabla_{F\,
r} = {1\over { \sqrt{\gamma_{F}(\tau ,\sigma^u )}}}\,
\partial_r\, \Big( \sqrt{ \gamma_{F}(\tau ,\sigma^u ) }\,
\gamma_{F}^{rs}(\tau ,\sigma^u )\, \partial_s\Big)$ are the
covariant derivative and the Laplace-Beltrami operator associated to
the positive 3-metric $h_{F\, rs}(\tau ,\vec \sigma^u)$,
respectively. The inverse of Laplace-Beltrami operator
$(1/\Delta_F)$ is defined by the {\em fundamental solution} of the
Laplace-Beltrami operator $G(\sigma^u,\sigma^{\,\prime\,u})$
\footnote{His existence is assured by existence's theorem (see for
example Ref.\cite{66},  but a closed analytic form is not known. A
general property of these fundamental solutions is a {\em
singularity} when the geodesic distance
$s(\sigma^u,\sigma^{\,\prime\,u})$ between $P=\{\sigma^u\}$ and
$Q=\{\sigma^{\,\prime\,u}\}$ goes to zero $\lim_{s\mapsto 0}\,
G(\sigma^u, \sigma^{\,\prime\,u}) \mapsto\, - \frac{1}{4\pi}\,
\frac{1}{s(\sigma^u,\sigma^{\,\prime\,u})} $.}
$f(\sigma^u)=\frac{1}{\Delta_F}\,g(\sigma^u)\byd\int
d^3\sigma^{\,\prime}\, \sqrt{\gamma(\sigma^{\,\prime\,u})}\,
G(\sigma^u,\sigma^{\,\prime\,u})\, g(\sigma^{\,\prime\,u})$, such
that $\Delta_F\,f(\sigma^u) = g(\sigma^u)$.

\bigskip

Since $\pi^r(\tau ,\sigma^u)$ is a vector density, we have
$\partial_r\, \pi^r(\tau ,\sigma^u) = \nabla_{F\, r}\, \pi^r(\tau
,\sigma^u)$: this quantity is a {\it 3-scalar density} on
$\Sigma_{\tau}$.

\bigskip

Instead the {\it non-covariant decomposition} \cite{1,5,9,59} in a
transverse and a longitudinal part (${\hat \partial}^r\, {\buildrel
{def}\over =}\,  \delta^{rs}\, \partial_r$, $\triangle =
\partial_r\, {\hat \partial}^r = {\vec \partial}^2$)  is

\bea
 &&\pi^r(\tau,\sigma^u) = \pi^r_{\perp}(\tau,\sigma^u) + \pi^r_L(\tau,\sigma^u),\nonumber \\
 &&{}\nonumber \\
 &&\pi^r_{\perp}(\tau ,\sigma^u) =
 \Big(\delta^r_s - {\hat \partial}^r\, {1\over {\Delta}}\,
 \partial_s \Big)\, \pi^s(\tau ,\sigma^u)\,\Rightarrow\,\partial_r\,
 \pi^r_{\perp}(\tau ,\sigma^u) = 0,\nonumber\\
 &&\nonumber\\
 &&\pi^r_L(\tau ,\sigma^u) = {\hat \partial}^r\, {1\over {\Delta}}\,
 \partial_s\,\pi^s(\tau ,\sigma^u),\nonumber\\
 &&{}\nonumber \\
 &&{}\nonumber \\
 &&A_r(\tau,\sigma^u) = A_{\perp\,r}(\tau,\sigma^u) + A_{L\,r}(\tau,\sigma^u),\nonumber \\
 &&{}\nonumber \\
 &&A_{\perp\,r}(\tau ,\sigma^u) =
 \Big(\delta^s_r - \partial_r\, {1\over {\Delta}}\,
  {\hat \partial}^s\Big)\, A_s(\tau ,\sigma^u)\,\Rightarrow\,
  {\hat\partial}^r\, A_{\perp\,r}(\tau ,\sigma^u) = 0,\nonumber\\
 &&\nonumber\\
 &&A_{L\,r}(\tau ,\sigma^u) = \,\partial_r {1\over {\Delta}}\,
 \,{\hat \partial}^s A_s(\tau ,\sigma^u).
 \label{c2}
 \eea

In Eq.(\ref{c2}) ${\hat \partial}^r\, A_r = \triangle\, \eta_{em}$
is a {\it non-covariant quantity}.\medskip

Here the inverse of Laplacian is defined used the standard
(Euclidean-like) fundamental solution: $c(\sigma^u -
\sigma^{\,\prime\,u}) = - \frac{1}{4\pi}\,
\frac{1}{\sqrt{\sum_{u=1}^3\,(\sigma^u - \sigma^{\,\prime\,u})^2}}$,
so that $f(\sigma^u) = \frac{1}{\Delta}\, g(\sigma^u)\byd \int
d^3\sigma^{\,\prime}\, c(\sigma^u-\sigma^{\,\prime\,u})
g(\sigma^{\,\prime\,u})$ and $\Delta\, f(\sigma^u) =
\Big(\sum_{r=1}^3\, {\hat \partial}^r\, \partial_r\Big)\,
f(\sigma^u) = g(\sigma^u)$.

\bigskip

Eq.(\ref{c2}) allow us to define the following non-covariant
Shanmugadhasan canonical transformation

\bea
  &&\begin{minipage}[t]{1cm}
\begin{tabular}{|l|} \hline
$A_A$ \\  \hline
 $\pi^A$ \\ \hline
\end{tabular}
\end{minipage} \ {\longrightarrow \hspace{.2cm}} \
\begin{minipage}[t]{2 cm}
\begin{tabular}{|l|l|l|} \hline
$A_{\tau}$ & $\eta_{em}$   & $A_{\perp\, r}$   \\ \hline
$\pi^{\tau}\approx 0$& $\Gamma \approx 0$ &$\pi^r_{\perp}$ \\
\hline
\end{tabular}
\end{minipage}\nonumber \\
 &&{}\nonumber \\
 &&{}\nonumber \\
 A_r(\tau ,\sigma^u )&=& - {
{\partial}\over {\partial \sigma^r} }\, \eta_{em} (\tau ,\sigma^u
) + A_{\perp\,r}(\tau ,\sigma^u ),\nonumber \\
 &&{}\nonumber \\
 \pi^r(\tau ,\sigma^u )&=&\pi^r_{\perp}(\tau
,\sigma^u ) + {1\over {\Delta}}\, {\hat \partial}^r\,
\Gamma (\tau ,\sigma^u ),\nonumber \\
 &&{}\nonumber \\
  \eta_{em}(\tau ,\sigma^u )&=&- {\hat \partial}^r\, A_r(\tau,\sigma^u),\nonumber \\
 &&{}\nonumber \\
A_{\perp\,r}(\tau ,\sigma^u) &=&
 \Big(\delta^s_r - \partial_r\, {1\over {\Delta}}\,
  {\hat \partial}^s\Big)\, A_s(\tau ,\sigma^u),\nonumber \\
&&\nonumber\\
  \pi^r_{\perp}(\tau ,\sigma^u) &=&
 \Big(\delta^r_s - {\hat \partial}^r\, {1\over {\Delta}}\,
 \partial_s \Big)\, \pi^s(\tau ,\sigma^u),\nonumber \\
 &&{}\nonumber \\
  &&\lbrace \eta_{em}(\tau ,\sigma^u ),\Gamma
(\tau ,{\sigma}^{\,\prime\,u} ) \rbrace = \delta^3(
\sigma^u, {\sigma}^{\,\prime\,u}),\nonumber \\
 &&{}\nonumber \\
 &&\lbrace A_{\perp\,r}(\tau ,
\sigma^u ),\pi^s_{\perp}(\tau ,{\sigma} ^{\,\prime\,u})\rbrace =
c\,(\delta^{rs} - {{\partial_r\, {\hat \partial}^s} \over
{\Delta}})\, \delta^3(\sigma^u, { \sigma}^{\,\prime\,u}).
 \label{c3}
 \eea
\medskip

If we add the gauge fixing $\eta_{em}(\tau ,\sigma^u) \approx 0$,
then its $\tau$-constancy implies $A_{\tau}(\tau ,\sigma^u) \approx
0$ and we get a non-inertial realization of the {\it non-covariant
radiation gauge}.

\bigskip

While with the non-covariant decomposition we can easily find a
Shanmugadhasan canonical transformation adapted to the Gauss law
constraint with the standard canonically conjugate (but
non-covariant) Dirac observables ${\vec A}_{\perp}$ and ${\vec
\pi}_{\perp}$ of the radiation gauge, it is not clear whether the
covariant decomposition can produce such a canonical basis. In any
case, as shown in Ref.\cite{65}, the radiation gauge formalism is
well defined in both cases if we add suitable gauge fixings.

\bigskip

In the inertial rest-frame instant form reviewed in Section III
Subsection B the 3-metric inside the Wigner 3-spaces is $g_{rs}(\tau
,\sigma^u ) = - \sgn\, h_{rs}(\tau ,\sigma^u ) = - \sgn\,
\delta_{rs}$ and the two decompositions coincide.

\bigskip

In Subsection B of Section IV there is the non-covariant
Shanmugadhasan canonical transformation in non-inertial frames in
presence of charged particles.
\bigskip

Let us remark that on the non-Euclidean 3-space we are using a delta
function $\delta^3(\sigma^u, \sigma^{\,\prime\,u})$, with the
properties $\delta^3(\sigma^u, \sigma^{\,\prime\,u}) =
\delta^3(\sigma^{\,\prime\,u}, \sigma^u)$ and
$\frac{\partial}{\partial\sigma^r}\, \delta^3(\sigma^u,
\sigma^{\,\prime\,u}) = - \frac{\partial}{\partial
\sigma^{\,\prime\,r}}\, \delta^3(\sigma^u, \sigma^{\,\prime\,u})$,
such that $d^3\sigma^\prime\, \delta^3(a^u,
\sigma^{\,\prime\,u})\,f(\sigma^{\,\prime\,u}) = f(a^u)$, and not a
covariant one $D^3(\sigma^u, \sigma^{\,\prime\,u}) =
\frac{1}{\sqrt{\gamma(\tau,\sigma^{\,\prime\,u})}}\,
\delta^3(\sigma^u, \sigma^{\,\prime\,u}) =
\frac{1}{\sqrt{\gamma(\tau,\sigma^{u})}}\, \delta^3(\sigma^u,
\sigma^{\,\prime\,u})$ such that $\int d^3\sigma^\prime\,
\sqrt{\gamma(\tau,\sigma^{\,\prime\,u})}\, D^3(a^u,
\sigma^{\,\prime\,u})\, f(\sigma^{\,\prime\,u}) = f(a^u)$.

\vfill\eject


\begin{thebibliography}{}

\bibitem{1}L. Lusanna, {\it The N- and 1-Time Classical Descriptions of N-Body
Relativistic Kinematics and the Electromagnetic Interaction}, Int.
J. Mod. Phys. {\bf A12}, 645 (1997).

\bibitem{2} D.Alba, L.Lusanna and M.Pauri, \textit{New Directions in
Non-Relativistic and Relativistic Rotational and Multipole
Kinematics for N-Body and Continuous Systems} (2005), in
\textit{Atomic and Molecular Clusters: New Research}, ed.Y.L.Ping
(Nova Science, New York, 2006) (hep-th/0505005).

\bibitem{3}D. Alba and L.Lusanna, {\it Simultaneity, Radar 4-Coordinates
and the 3+1 Point of View about Accelerated Observers in Special
Relativity} (2003) (gr-qc/0311058); {\it Generalized Radar
4-Coordinates and Equal-Time Cauchy Surfaces for Arbitrary
Accelerated Observers} (2005), Int.J.Mod.Phys. {\bf D16}, 1149
(2007) (gr-qc/0501090).

\bibitem{4}D.Alba, H.W.Crater and L.Lusanna, \textit{Hamiltonian
Relativistic Two-Body Problem: Center of Mass and Orbit
Reconstruction}, J.Phys. {\bf A40}, 9585 (2007) (gr-qc/0610200).

\bibitem{5}L.Lusanna, {\it The Chrono-Geometrical Structure of Special and
General Relativity: A Re-Visitation of Canonical Geometrodynamics},
lectures at 42nd Karpacz Winter School of Theoretical Physics:
Current Mathematical Topics in Gravitation and Cosmology, Ladek,
Poland, 6-11 Feb 2006, Int.J.Geom.Methods in Mod.Phys. {\bf 4}, 79
(2007). (gr-qc/0604120).

\bibitem{6}C.M. M$\o$ller, {\it The Theory of Relativity} (Oxford
Univ.Press, Oxford, 1957).\hfill\break
 C.M$\o$ller, {\it Sur la dinamique des syste'mes ayant
un moment angulaire interne}, Ann.Inst.H.Poincare' {\bf 11}, 251
(1949).

\bibitem{7}E.Schmutzer and J.Plebansi, {\it  Quantum Mechanics in
Noninertial Frames of Reference}, Fortschr.Phys. {\bf 25}, 37
(1978).


\bibitem{8}D.Alba, H.W.Crater and L.Lusanna, {\it Towards Relativistic
Atom Physics. I. The Rest-Frame Instant Form of Dynamics and a
Canonical Transformation for a system of Charged Particles plus the
Electro-Magnetic Field} (arXiv: 0806.2383).

\bibitem{9}H.W.Crater and L.Lusanna, \textit{The Rest-Frame Darwin
Potential from the Lienard-Wiechert Solution in the Radiation
Gauge}, Ann.Phys. (N.Y.) \textbf{289}, 87 (2001).

\bibitem{10}D.Alba and L.Lusanna, \textit{Quantum Mechanics in Noninertial
Frames with a Multitemporal Quantization Scheme: I. Relativistic
Particles}, Int.J.Mod.Phys. \textbf{A21}, 2781 (2006)
(hep-th/0502194).\hfill\break
 D.Alba, {\it Quantum Mechanics in Non-Inertial Frames with a
Multi-Temporal Quantization Scheme: II) Non-Relativistic Particles},
Int.J.Mod.Phys. {\bf A21}, 3917 (2006) (hep-th/0504060).

\bibitem{11}L.Lusanna, {\it The Rest-Frame Instant Form of Metric Gravity},
Gen.Rel.Grav. {\bf 33}, 1579 (2001)(gr-qc/0101048).\hfill\break
 L.Lusanna and S.Russo, {\it A New Parametrization for Tetrad Gravity},
Gen.Rel.Grav. {\bf 34}, 189 (2002)(gr-qc/0102074).\hfill\break
R.DePietri, L.Lusanna, L.Martucci and S.Russo, {\it Dirac's
Observables for the Rest-Frame Instant Form of Tetrad Gravity in a
Completely Fixed 3-Orthogonal Gauge}, Gen.Rel.Grav. {\bf 34}, 877
(2002) (gr-qc/0105084).\hfill\break
 J.Agresti, R.De Pietri, L.Lusanna and L.Martucci, {\it
Hamiltonian Linearization of the Rest-Frame Instant Form of Tetrad
Gravity in a Completely Fixed 3-Orthogonal Gauge: a Radiation Gauge
for Background-Independent Gravitational Waves in a Post-Minkowskian
Einstein Spacetime}, Gen.Rel.Grav. {\bf 36}, 1055 (2004)
(gr-qc/0302084).\hfill\break
 J.Agresti, R.De Pietri, L.Lusanna and
L.Martucci, {\it Hamiltonian Linearization of the Rest-Frame Instant
Form of Tetrad Gravity in a Completely Fixed 3-Orthogonal Gauge: a
Radiation Gauge for Background-Independent Gravitational Waves in a
Post-Minkowskian Einstein Spacetime}, Gen.Rel.Grav. {\bf 36}, 1055
(2004) (gr-qc/0302084).


\bibitem{12}D.Alba and L.Lusanna, {\it The York Map as a Shanmugadhasan
Canonical Transformationn in Tetrad Gravity and the Role of
Non-Inertial Frames in the Geometrical View of the Gravitational
Field}, Gen.Rel.Grav. {\bf 39}, 2149 (2007) (gr-qc/0604086, v2; see
v1 for an expanded version).

\bibitem{13}R.M.Wald, {\it General Relativity} (Chicago Univ.
Press, Chicago, 1984).\hfill\break
 M.Heusler, {\it Black Hole Uniqueness Theorems} (Cambridge
 Univ.Press, Cambridge, 1996); {\it Stationary Black Holes: Uniqueness and Beyond},
 Living Reviews in Relativity 1998
 (www.livingreviews.org/Articles/Volume1/1998-6heusler).

\bibitem{14}N.Stergioulas, {\it Rotating Stars in Relativity},
Living Reviews in Relativity 2003
(www.livingreviews.org/lrr-2003-3).

\bibitem{15}J.M.Bardeen and R.Wagoner, {\it Relativistic Disks.
I. Uniform Rotation}, Ap.J. {\bf 167}, 359 (1971).\hfill\break
 E.M.Butterworth and J.R.Ipser, {\it On the Structure and Stability
 of Rapidly Rotating Fluid Bodies in General Relativity. I. The Numerical
 Method for Computing Structure and its Application to Uniformly
 Rotating Homogeneous Bodies}, Ap.J. {\bf 204}, 200
 (1976).\hfill\break
 N.Comins and B.F.Schutz, {\it On the Ergoregion Instability},
 Proc. R. Soc. London {\bf A364}, 211 (1978); {\it On the Existence of
 Ergoregions in Rotating Stars}, Mon.Not.R.astr.Soc. {\bf 182}, 69
 (1978).\hfill\break
 J.L.Friedman, {\it Ergosphere Instability}, Commun.Math.Phys.
 {\bf 63}, 243 (1978).

\bibitem{16}G.Rizzi and M.L.Ruggiero eds., {\it Relativity in
Rotating Frames. Relativistic Physics in Rotating Reference Frames.}
(Kluwer, Dordrecht, 2003).

\bibitem{17} L.Landau and E.Lifschitz, {\it The Classical Theory of Fields}
(Addison-Wesley, Cambridge, 1951).


\bibitem{18}W.M.Irvine, {\it Electrodynamics in a Rotating System of
Reference}, Physica {\bf 30}, 1160 (1964).\hfill\break
 See the rich bibliography in K.T. McDonald, {\it Electrodynamics of
 Rotating Frames}, 2008 (http://cosmology.princeton.edu/~mcdonald/examples/).
 \hfill\break
 Z.N.Osmanov, G.Z.Machabeli and A.D.Rogava, {\it Electromagnetic
 Waves in a Rigidly Rotating Frame}, Phys.Rev. {\bf A66}, 042103
 (2002).\hfill\break
 J.C.Hauck and B.Mashhoon, {\it Electromagnetic Waves in a
 Rotating Frame of Reference}, Ann.Phys. (Leipzig) {\bf 12}, 275 (2003)
  (gr-qc/0304069).




\bibitem{19}G.Trocheris, {\it Electrodynamics in a Rotating Frame of
Reference}, Philos.Mag. {\bf 40}, 1143 (1949).\hfill\break
 H.Takeno, {\it On Relativistic Theory of Rotating Disk},
 Prog.Thor.Phys. {\bf 7}, 367 (1952).

\bibitem{20} J.F.Corum, {\it Relativistic Rotation and the Anholonomic Object},
 J.Math.Phys. {\bf 18}, 770 (1977).  {\it Relativistic Covariance
 and Rotational Electrodynamics}, J.Math.Phys. {\bf 21}, 2360
 (1980).\hfill\break
 B.Chakraborty and S.Sarkar, {\it Physics in Rotating Frames. I.
 On Uniform Rotation about a Fixed Axis in Some Holonomic and
 Anholonomic Frames}, Ann.Phys. (N.Y.) {\bf 163}, 167
 (1985).\hfill\break
  P.N.Arendt jr, {\it Electromagnetic Forces and Fields in a
 Rotating Reference Frame}, (astro-ph/9801194).\hfill\break
 V.Bashkov and M.Malakhaltsev, {\it Relativistic Mechanics
 on Rotating Disk} (gr-qc/0104078); {\it Geometry of Rotating Disk
 and the Sagnac Effect} (gr-qc/0011061).

\bibitem{21}W.T.Ni and M.Zimmermann, {\it Inertial and
Gravitational Effects in the Proper Reference Frame of an
Accelerated, Rotating Observer}, Phys.Rev. {\bf D17}, 1473 (1978).
\hfill\break
 R.A.Nelson, {\it Generalized Lorentz Transformation for
an Accelerated, Rotating Frame of Reference}, J.Math.Phys. {\bf 28},
2379 (1987) [erratum J.Math.Phys. {\bf 35}, 6224
(1994)].\hfill\break
 R.N.Henriksen and L.A.Nelson, {\it Clock Synchronization
 by Accelerated Observers: Metric Construction for Arbitrary Congruences
 of Worldlines}, Can.J.Phys. {\bf 63}, 1393 (1985).


\bibitem{22}K.P.Marzlin, {\it Fermi Coordinates for Weak Gravitational Fields},
Phys.Rev. {\bf D50}, 888 (1994).\hfill\break
 B.Mashhoon and U.Muench, {\it Length Measurement in Accelerated
Systems}, Ann.Phys. (Leipzig) {\bf 11}, 532 (2002).\hfill\break
 D.Bini, L.Lusanna and B.Mashhoon, {\it Limitations of Radar
Coordinates for Accelerated Observers}, Int.J.Mod.Phys. {\bf D14},
1413 (2005) (gr-qc/0409052).


\bibitem{23}D.Klein and P.Collas, {\it General Transformation
Formulas for Fermi-Walker Coordinates}, 2008 (arXiv: 0712.3838).


\bibitem{24}B.Mashhoon, {\it Limitations of Spacetime Measurements}, Phys.Lett.
{\bf A143}, 176 (1990). {\it The Hypothesis of Locality in
Relativistic Physics}, Phys.Lett. {\bf A145}, 147 (1990). {\it
Measurement Theory and General Relativity}, in {\it Black Holes:
Theory and Observation}, Lecture Notes in Physics 514, ed. F.W.Hehl,
C.Kiefer and R.J.K.Metzler (Springer, Heidelberg, 1998), p.269. {\it
Acceleration-Induced Nonlocality}, in {Advances in General
Relativity and Cosmology}, ed. G.Ferrarese (Pitagora, Bologna, 2003)
(gr-qc/0301065). {\it The Hypothesis of Locality and its
Limitations}, (gr-qc/0303029).


\bibitem{25}S.Weinberg, {\it Gravitation and Cosmology}, ch.5 (Wiley, New York, 1972).


\bibitem{26}P.A.M.Dirac, {\it Gauge Invariant Formulation of Quantum
Electrodynamics}, Can.J.Phys. {\bf 33}, 650 (1955).

\bibitem{27}See for instance H.Bethe and E.E.Salpeter, {\em Quantum Mechanics
of One and Two Electron Atoms} (Springer, Berlin, 1957) and
O.Keller, {\it On the Theory of Spatial Photon Localization},
 Phys.Rep. {\bf 411}, 1 (2005).


\bibitem{28}L.I.Schiff, {\it A Question in General Relativity},
Proc.Nat.Acad.Sci. {\bf 25}, 391 (1939).

\bibitem{29}J.Plebanski, {\it Electromagnetic Waves in Gravitational
Fields}, Phys.Rev. {\bf 111}, 1396 (1960).


\bibitem{30}C.G.Tsagas, {\it Electromagnetic Fields in Curved Spacetimes},
Class.Quantum Grav. {\bf 22}, 393 (2005)
(gr-qc/0407080).\hfill\break
 J.D.Barrow and C.G.Tsagas, {\it }, Class.Quantum Grav. {\bf 14},
 2539 (1997)(gr-qc/9704015).


\bibitem{31} J.C.Hauck and B.Mashhoon, {\it Electromagnetic Waves
in a Rotating Frame of Reference}, Ann.Phys. (Leipzig) {\bf 12}, 275
(2003) (gr-qc/0304069).\hfill\break
 B.Mashhoon, R.Neutze, M.Hannam and G.E.Stedman, {\it Observable
 Frequency Shifts via Spin-Rotation Couplings}, Phys.Lett. {\bf
 A249}, 161 (1998) (gr-qc/9808077).\hfill\break
 B.Mashhoon, {\it Spin-Gravity Coupling}, Acta Phys,Polon. Suppl.
 {\bf 1}, 113 (2008) (arXiv: 0801.2134).



\bibitem{32}B.Mashhoon, {\it Nonlocal Electrodynamics of Accelerated
Systems}, Phys.Lett. {\bf A366}, 545 (2007). {\it Modification of
the Doppler Effect due to the Helicity-Rotation Coupling},
Phys.Lett. {\bf A306}, 66 (2002). {\it Nonlocal Electrodynamics of
Linearly Accelerated Systems}, Phys.Rev. {\bf A70}, 062103 (2004).
{\it Nonlocal Electrodynamics of Rotating Systems}, Phys.Rev. {\bf
A72}, 052105 (2005).\hfill\break
 J.D.Anderson and B.Mashhoon, {\it Pioneer Anomaly and the
Helicity-Rotation Coupling}, Phys.Lett. {\bf A315}, 199
(2003).\hfill\break
 U.Muench, F.W.Hehl and B.Mashhoon, {\it Acceleration-Induced
Nonlocal Electrodynamics in Minkowski Spacetime}, Phys.Lett. {\bf
A271}, 8 (2000).


\bibitem{33}B.Mashhoon, {\it Influence of Gravitation on the Propagation
of Electromagnetic Radiation}, Phys.Rev. {\bf D11}, 2679 (1975).
{\it Can Einstein's Theory of Gravitation be tested beyond the
Geometric Optics Limit?}, Nature {\bf 250}, 316 (1974). {\it
Gravitational Coupling of Intrinsic Spin}, Class.Quant.Grav. {\bf
17}, 2399 (2000)(gr-qc/0003022). {\it On the Spin-Rotation Gravity
Coupling }, Gen.Rel.Grav. {\bf 31}, 681 (1999).


\bibitem{34}R.Nutze and G.E.Stedman, {\it Detecting the Effects of
Linear Acceleration on the Optical Response of Matter}, Phys.Rev.
{\bf A58}, 82 (1997).

\bibitem{35}S.P.Tarabrin and A.A.Seleznyov, {\it Optical Position
Meters analyzed in the Non-Inertial References Frames}, (arXiv:
0804.4292).

\bibitem{36} N.Ashby, {\it Relativity in the Global Positioning System},
 Living Reviews in Relativity (http://www.livingreviews.org).
\hfill\break
 N.Ashby and J.J.Spilker, {\it Introduction to
Relativistic Effects on the Global Positioning System}, in {\it
Global Positioning System: Theory and Applications}, Vol.1, eds.
B.W.Parkinson and J.J.Spilker (American Institute of Aeronautics and
Astronautics, 1995).

\bibitem{37}E.J.Post, {\it Sagnac Effect}, Rev.Mod.Phys. {\bf 39},
475 (1967).

\bibitem{31a} B.Mashhoon, R.Neutze, M.Hannam and G.E. Stedman, {\it Observable
Frequency Shifts via Spin-Rotation Coupling}, Phys.Lett. {\bf A249},
161 (1998).



\bibitem{38}M.Sereno, {\it Gravitational Faraday Rotation in a Weak
Gravitational Field}, Phys.Rev. {\bf D69}, 087501
(2004).\hfill\break
 M.Giovannini and K.E.Kunze, {\it Faraday
Rotation, Stochastic Magnetic Fields and CMB Maps}, arXiv 0804.3380
\hfill\break
 V.Faraoni, {\it The Rotation of Polarization by Gravitational
Waves}, New Astronomy {\bf 13}, 178 (2008) (arXiv
0709.0386).\hfill\break
 M.Halilsoy and O.Gurtug, {\it Search for Gravitational Waves through
the Electromagnetic Faraday Rotation}, Phys.Rev. {\bf D75}, 124021
(2007).\hfill\break
 V.Perlick and W.Hasse, {\it Gravitational Faraday Effect in
Conformally Stationary Spacetimes}, Clas.Q.Grav. {\bf 10}, 147
(1993).\hfill\break
 P.Nag, S.Bharadwaj and S.Kar, {\it Can the Rotation of the Dark
Matter Halo of our Galaxy be detected through its Effect on the
Cosmic Microwave Background Polarization?}, arXiv astro-ph/0506009






\bibitem{39}{\it IERS Conventions (2003)}, eds. D.D.McCarthy and
G.Petit, IERS TN 32 (2004), Verlag des BKG.\hfill\break
 M.Soffel, S.A.Klioner, G.Petit, P.Wolf, S.M.Kopeikin, P.Bretagnon,
V.A.Brumberg, N.Capitaine, T.Damour, T.Fukushima, B.Guinot, T.Huang,
L.Lindegren, C.Ma, K.Nordtvedt, J.Ries, P.K.Seidelmann,
D.Vokroulicky', C.Will and Ch.Xu, {\it The IAU 2000 Resolutions for
Astrometry, Celestial Mechanics and Metrology in the Relativistic
Framework: Explanatory Supplement} Astron.J., \textbf{126},
pp.2687-2706, (2003) (astro-ph/0303376).\hfill\break
  G.H.Kaplan, {\it The IAU Resolutions on Astronomical
Reference Systems, Time Scales and Earth Rotation Models}, U.S.Naval
Observatory circular No. 179 (2005) (astro-ph/0602086).


\bibitem{40}M.Wilson and H.A.Wilson, {\it On the Electric Effect of
Rotating a Magnetic Insulator in a Magnetic Field}, Proc.R.Soc.
London {\bf A89}, 99 (1913).\hfill\break
 G.N.Pellegrini and A.R.Swift, {\it Maxwell's Equations in a
 Rotating Medium" Is there a Problem?}, Am.J.Phys. {\bf 63}, 694
 (1995).\hfill\break
T.A.Weber, {\it Measurements on a Rotating Frame in Relativity
 and the Wilson and Wilson Experiment}, Am.J.Phys. {\bf 65}, 946
 (1997).\hfill\break
 C.T.Ridgely, {\it Applying Relativistic Electrodynamics to a
 Rotating Material Medium}, Am.J.Phys. {\bf 66}, 114 (1998). {\it
 Applying Covariant versus Contravariant Electromagnetic Tensors
 to Rotating Media}, Am.J.Phys. {\bf 67}, 414 (1998).\hfill\break
 K.T.McDonald, {\it The Wilson-Wilson Experiment}, 2008
(http://cosmology.princeton.edu/~mcdonald/examples/).


\bibitem{41}E.J.Post and D.D.Bahulikar, {\it Note on the
Electrodynamics of Accelerated Systems}, J.Math.Phys. {\bf 12}, 1098
(1971).\hfill\break
 E.J.Post, {\it Kottler-Cartan-van Dantzig (KCD) and Noninertial
 Systems}, Found.Phys. {\bf 9}, 619 (1979).\hfill\break
 B.M.Bolotovskii and S.N.Stolyarov, {\it Current Status of the
 Electrodynamics of Moving Media (Infinite Media)}, Sov.Phys.Usp.
 {\bf 17}, 875 (1975).


\bibitem{42}F.W.Hehl, {\it Maxwell's Equations in Minkowski's World:
their Premetric Generalization and the Electromagnetic
Energy-Momentum Tensor}, Annalen der Physik... 2008 (arXiv:
0807.4249).\hfill\break
 F.W.Hehl and Y.Obukov, {\it Electrodynamics of Moving
 Magnetoelectric Media: Variational Approach}, Phys.Lett. {\bf
 A371}, 11 (2007); {\it Forces and Momenta caused by Electromagnetic
 Waves in Magnetoelectric Media} 2007 (arXiv: 0710.2219); {\it
 Foundations of Classical Electrodynamics: Charges, Flux and Metric}
 (Birkhauser, Boston, 2003).\hfill\break
  V.A.De Lorenci and G.P.Goulart, {\it Magnetoelectric Birifrangence
  Revisited}, (arXiv: 0806.4685).



\bibitem{43}G.Rizzi and M.L.Ruggiero, {\it Space Geometry of
 Rotating Platforms: an Operational Approach}, (gr-qc/0207104).
\hfill\break
 A.Tartaglia, {\it Lengths on Rotating Platforms},
Found.Phys.Lett. {\bf 12}, 17 (1999).\hfill\break
 P.Dombrowski, J.Kuhlmann and U.Proff, {\it On the Spatial Geometry
of a Non-Inertial Observer in Special Relativity}, in {\it Global
Riemannian Geometry}, eds. T.J.Willmore and N.J.Hitchin (Horwood,
Wiley, New York, 1984).





\bibitem{44}A.Einstein, {\it Zum Ehrenfestschen Paradoxon},
Phys.Z. {\it 12}, 509 (1911); {\it Die Grunlage der Allgemeinen
Relativitatstheorie}, Annalen der Physik {\it 49}, 769 (1916); {\it
The Meaning of Relativity} (Princeton Univ.Press, Princeton, 1950).

\bibitem{45}P.Ehrenfest, {\it Gleichf\"ormige Rotation starrer
K\"orper und Relativit\"atheorie}, Phys.Z. {\bf 10}, 918 (1909).


\bibitem{46}$\O$. Gr$\o$n, {\it Rotating Frames in Special
Relativity}, Int.J.Theor.Phys. {\bf 16}, 603 (1977). {\it
Relativistic Description of a Rotating Disk}, Am.J.Phys. {\bf 43},
869 (1975). {\it Covariant Formulation of Hooke's Law}, Am.J.Phys.
{\bf 49}, 28 (1981).


\bibitem{47}M.Born, {\it Die Theorie des starren Elektrons in
der Kinematik des Relativit\"atsprinzipe}, Ann.Phys.(Leipzig) {\bf
30}, 1 (1909).

\bibitem{48}M.H.Soffel, {\it Relativity in Astrometry, Celestial Mechanics
and Geodesy} (Springer, Berlin, 1989).


\bibitem{49} J.L.Synge, {\it Time-like Helices in Flat Space-Time},
Proc. Royal Irish Acad. A {\bf 65}, 27 (1967). \hfill\break
 E.Honig, E.L.Schuking and C.V.Vishveshwara, {\it Motion of Charged Particles
in Homogeneous Electro-magnetic Fields}, J.Math.Phys. {\bf 15}, 774
(1974). \hfill\break
 B.R.Iyer and C.V.Vishveshwara, {\it The
Frenet-Serret Formalism and Black Holes in Higher Dimensions},
Class.Quantum Grav. {\bf 5}, 961 (1988); {\it The Frenet-Serret
Description of Gyroscopic Precession}, Phys.Rev. {\bf D48}, 5706
(1993).

\bibitem{50}W.A.Rodrigues jr and M.Sharif, {\it Rotating Frames in
SRT: the Sagnac Effect and Related Issues}, Found.Phys. {\bf 31},
1767 (2001); {\it Equivalence Principle and the Principle of Local
Lorentz Invariance}, Found.Phys. {\bf 31}, 1785 (2001) [erratum
Found.Phys. {\bf 32}, 811 (2002)].



\bibitem{51}D.Alba and L.Lusanna, {\it Generalized Eulerian
Coordinates for Relativistic Fluids: Hamiltonian Rest-Frame Instant
Form, Relative Variables, Rotational Kinematics}, to appear in
Int.J.Mod.Phys. (hep-th/0209032).\hfill\break
 L.Lusanna and D.Nowak-Szczepaniak, {\it The Rest-Frame Instant
Form of Relativistic Perfect Fluids with Equation of State $\rho
=\rho (n,s)$ and of Non-Dissipative Elastic Materials}, Int. J. Mod.
Phys. {\bf A15}, 4943 (2000) (hep-th/0003095).



\bibitem{52}G.E.Stedman, {\it Ring-Laser Tests of Fundamental
Physics and Geophysics}, Rep.Prog.Phys. {\bf 60}, 615 (1997).


\bibitem{53}G.Rizzi and M.L.Ruggiero, {\it The Relativistic
Sagnac Effect: two Derivations} (gr-qc/0305084).

\bibitem{54}A.Tartaglia, {\it General Relativistic Corrections to
the Sagnac Effect}, Phys.Rev. {\bf D58}, 064009 (1998).




\bibitem{55}A.Brillet and J.L.Hall, {\it Improved Laser Test of
the Isotropy os Space}, Phys.Rev.Lett. {\bf 42}, 549
(1979).\hfill\break
 C.Braxmaier, H.M\"uller, O.Pradl, J.Mlynek and A.Peters, {\it
 Tests of Relativity Using a Cryogenic Optical Resonator},
 Phys.Rev.Lett. {\bf 88}, 010401 (2002).\hfill\break
 J.A.Lipa, J.A.Nissen, S.Wang, D.A.Stricker and D.Avaloff, {\it
 New Limit on Signals of Lorentz Violation in Electrodynamics},
 Phys.Rev.Lett. {\bf 90}, 060403 (2003).\hfill\break
 P.Wolf, S.Bize, A.Clairon, A.N.Luiten, G.Santarelli and
 M.E.Tobar, {\it Tests of Relativity using a Microwave Resonator},
 (gr-qc/0210049).


\bibitem{56} H.Rauch and S.A.Werner, {\it Neutron Interferometry: Lessons in
 Experimental Quantum Mechanics} (Clarendon Press,
 Oxford, 2000).

\bibitem{57}D. Alba, L. Lusanna and M. Pauri, {\it Dynamical
Body Frames, Orientation-Shape Variables and Canonical Spin Bases
for the Nonrelativistic N-Body Problem },
 J. Math. Phys. {\bf 43}, 373 (2002) (hep-th/0011014).




\bibitem{58}W.M.Fawley, J.Arons and E.T.Scherlemann, {\it
Potential Drops above Pulsar Polar Caps: Acceleration on Nonneutral
Beams from the Stellar Surface}, Astrop.J. {\bf 217}, 227 (1977).


\bibitem{59}F.C.Michel, {\it Theory of Pulsar Magnetosphere},
Rev.Mod.Phys. {\bf 54}, 1 (1982); {\it The State of Pulsar Theory},
Adv.Space Research {\bf 31}, 542 (2004).\hfill\break
 S.P.Goodwin, J.Mestel, L.Mestel and G.A.E.Wright, {\it An Idealized
Pulsar Magnetosphere: the Relativistic Force-Free Approximation},
Mon.Not.R.Astron.Soc. {\bf 349}, 213 (2004).\hfill\break
 A.G.Muslimov and A.K.Harding, {\it Effects of Rotation and
 Relativistic Charge Flow on Pulsar Magnetospheric Structure},
 Astron.J. {\bf 630}, 454 (2005).\hfill\break
 Z.Osmanov, O.Dolakishvili and G.Machabeli, {\it On the Reconstruction
 of a Magnetosphere nearby the Light Cylinder Surface}, Mon.Not.
 R.Astron. Soc. {\bf 383}, 1007 (2008).



\bibitem{60}S.Kichenassamy and R.A.Krikorian, {\it The
Relativistic Rotation Transformation and the Corotating Source
Model}, Astrop.J. {\bf 371}, 277 (1991); {\it The Relativistic
Rotation Transformation and Pulsar Electrodynamics}, Astrop.J. {\bf
431}, 715 (1994).\hfill\break
 O.V.Chedia, T.A.Kahniashvili, G.Z.Machabeli and I.S.Nanobashvili,
 {\it On the Kinematics of a Corotating Relativistic Plasma Stream
 in the Perpendicular Rotator Model of a Pulsar Magnetosphere},
 Astrophys.Space Science {\bf 239}, 57 (1996).




\bibitem{61}F.Rohrlich, {\it The Principle of Equivalence},
Ann.Phys.(N.Y.) {\bf 22}, 169 (1963).







\bibitem{62}S.Shanmugadhasan, {\it Canonical Formalism for
Degenerate Lagrangians}, J.Math.Phys. {\bf 14}, 677 (1973).

\bibitem{63}L.Lusanna, {\it The Shanmugadhasan Canonical
Transformation, Function Groups and the Extended Second Noether
Theorem}, Int.J.Mod.Phys. {\bf A8}, 4193 (1993).


\bibitem{64}L.Lusanna, {\it Towards a Unified Description of the Four
Interactions in Terms of Dirac-Bergmann Observables}, invited
contribution to the book {\it Quantum Field Theory: a 20th Century
Profile}, of the Indian National Science Academy, ed.A.N.Mitra,
forewards by F.J.Dyson (Hindustan Book Agency, New Delhi, 2000)
(hep-th/9907081).





\bibitem{65}S.Deser, {\it Covariant Decomposition of Symmetric
Tensors and the Gravitational Cauchy Problem}, Ann.Inst.H.Poincare'
{\bf VII}, 146 (1967).\hfill\break
 R.Arnowitt, S.Deser and C.W.Misner, {\it Gravitational Electromagnetic
 Coupling and the Classical Self-Energy Problem}, Phys.Rev. {\bf
 120}, 313 (1960).




\bibitem{66}G.F.D. Duff, {\em Partial Differential
Equations} (Univ. of Toronto Press, Toronto, 1956)),





















\end{thebibliography}
\end{document}